\documentclass{MyBook}

\begin{document}

\pagestyle{empty}
\firstpage

\newpage
% Choose the good geometry depending on the version
\ifprinted
\newgeometry{top=1cm, bottom=1cm, right=1cm, left=2.5cm}
\else
\newgeometry{top=1cm, bottom=1cm, left=1.775cm, right=1.775cm}
\fi

\includegraphics[height=2.45cm]{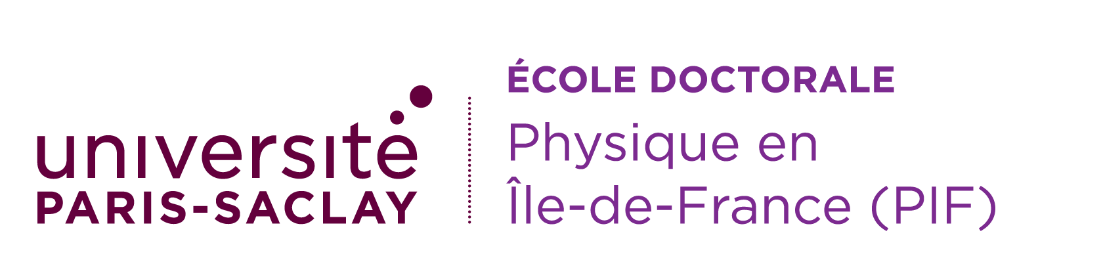}

\begin{Abst}
%\begin{small}
	\noindent%
	\textbf{Title:} 
	Field Theories on Quantum Space-Times: Towards the Phenomenology of	Quantum Gravity	
	
	\noindent%

	\begin{small}
	\noindent%
	\textbf{Abstract:}
	Noncommutative geometry is a mathematical framework that expresses the structure of space-time in terms of operator algebras. By using the tools of quantum mechanics to describe the geometry, noncommutative space-times are expected to give rise to quantum gravity effects, at least in some regime. This manuscript focuses on the physical aspects of these so-called quantum space-times, in particular through the formalism of field and gauge theories. Scalar field theories are shown to possibly trigger mixed divergences in the infra-red and ultra-violet for the $2$-point function at one loop. This phenomenon is generically called UV/IR mixing and stems from a diverging behaviour of the propagator. The analysis of such divergences differs from the commutative case because the momentum space is now also noncommutative. From another perspective, a gauge theory on $\dpkM$-\namefont{Minkowski}, a quantum deformation of the \namefont{Minkowski} space-time, is derived. A first perturbative computation is shown to break the gauge invariance, a pathological behaviour common to other quantum space-times. A causality toy model is also developed on $\dpkM$-\namefont{Minkowski}, in which an analogue of the speed-of-light limit emerges. The phenomenology of quantum gravity arising from quantum space-times is discussed, together with the actual constraints it imposes. Finally, a toy model for noncommutative gravity is tackled, using the former $\dpkM$-\namefont{Minkowski} space-time to describe the tangent space. It necessitates the notion of noncommutative partition of unity specifically defined there.
\end{small}
\end{Abst}

\begin{Abst}
	\noindent%
	\textbf{Titre :}
	Th\'{e}ories de Champs en Espace-Temps Quantiques : vers la Ph\'{e}nom\'{e}nologie de la Gravit\'{e} Quantique

	\noindent%
	\textbf{Mots-clefs :}
	g\'{e}om\'{e}trie noncommutative, espace-temps quantiques, th\'{e}orie quantique des champs, th\'{e}orie de jauge, ph\'{e}nom\'{e}nologie de la gravit\'{e} quantique

	\begin{small}
	\noindent%
	\textbf{R\'{e}sum\'{e} :}	
	La g\'{e}om\'{e}trie noncommutative est un formalisme math\'{e}matique qui exprime la structure de l'espace-temps avec des alg\`{e}bres d'op\'{e}rateurs. On s'attend \`{a} ce que les espace-temps noncommutatifs fassent \'{e}merger des effets de gravit\'{e} quantiques, au moins dans un certain r\'{e}gime, notamment parce qu'ils utilisent les outils de la m\'{e}canique quantique pour d\'{e}crire la g\'{e}om\'{e}trie. Ce manuscrit se concentre sur les aspects physiques de ces espace-temps quantiques, tout particuli\`{e}rement \`{a} travers le formalisme des th\'{e}ories de champs et de jauge. Il est montr\'{e} que les th\'{e}ories de champs scalaires engendrent possiblement des divergences dans l'infra-rouge et l'ultra-violet pour la fonction $2$-point \`{a} une boucle. Ce ph\'{e}nom\`{e}ne s'appelle g\'{e}n\'{e}riquement le m\'{e}lange UV/IR et d\'{e}coule de la divergence du propagateur. L'analyse de ces divergences diff\`{e}rent du cas commutatif car l'espace des moments y est noncommutatif. D'autre part, une th\'{e}orie de jauge sur $\dpkM$-\namefont{Minkowski}, une d\'{e}formation quantique de l'espace de \namefont{Minkowski}, est construite. Un premier calcul perturbatif produit une brisure de l'invariance de jauge, un comportement pathologique commun \`{a} d'autres espace-temps quantiques. Un mod\`{e}le-jouet de causalit\'{e} est aussi d\'{e}velopp\'{e} sur $\dpkM$-\namefont{Minkowski}, dans lequel appara\^{i}t un analogue de la vitesse de lumi\`{e}re comme vitesse limite. La ph\'{e}nom\'{e}nologie de la gravit\'{e} quantique \'{e}mergeant des espace-temps quantiques est abord\'{e}e, avec les contraintes qu'elle impose. Finalement, un mod\`{e}le-jouet de gravit\'{e} noncommutative, utilisant $\dpkM$-\namefont{Minkowski} pour d\'{e}crire l'espace tangent, est trait\'{e}. Il n\'{e}cessite le concept de partition de l'unit\'{e} noncommutative sp\'{e}cialement d\'{e}fini dans ce contexte.
\end{small}
\end{Abst}

\restoregeometry

\frontmatter
\pagestyle{frontpsty}

\chapter{Remerciements (Acknowledgements)}
% Command for headers of a * chapter
\markboth{Remerciements (Acknowledgements)}{}
\paragraph{}
J'ai eu de nombreux r\'{e}cits de doctorants, laiss\'{e}s \`{a} l'abandon, pour qui l'encadrant est une vague figure lointaine que l'on aperçoit une fois par semaine, au mieux, et de qui on doit avoir l'autorisation pour tout. Ce n'est pas le cas de Jean-Christophe, qui n'a pas manqu\'{e} \`{a} l'appel t\'{e}l\'{e}phonique quotidien pour savoir \enquote{Quelles sont les \emph{news}?}. Pour m'avoir trait\'{e} comme un collaborateur plus que comme un \'{e}l\`{e}ve, \`{a} tous les \'{e}gards, je te remercie beaucoup.

\paragraph{}
Le laboratoire m'a \'{e}galement \'{e}t\'{e} d'un charmant accueil, avec un panel de coll\`{e}gues tout aussi diff\'{e}rents qu'int\'{e}ressants et tous pr\`{e}s \`{a} discuter de science \`{a} n'importe quel moment. Pour cette chaleur quotidienne, j'aimerais remercier toutes les personnes du b\^{a}timent 210 que j'ai pu c\^{o}toyer (par ordre alphab\'{e}tique comme il se doit): Donald, Eduardo, Gatien, Gioacchino, Giulia, Ioannis, Jay, L\'{e}onard, Mathieu, M\'{e}ril, Michele, Nicolas, $\pi$anaioti, Param, Paul, Salva, Simon, Teseo, Valentine. J'esp\`{e}re n'oublier personne ! Je salue notamment ceux qui ont pris le temps de r\'{e}pondre \`{a} toutes les questions que j'avais, aussi basiques soit-elles. Je souhaite bon courage \`{a} ceux qui ont commenc\'{e} le fran\c{c}ais et \`{a} qui j'ai pu servir d'aide.

Je remercie \'{e}galement l'amabilit\'{e} de certains permanents, soucieux du sort des petits, ou simplement d\'{e}sireux de discuter: Adam, Asmaa, C\'{e}cile, Christos, Evgeni, Gatien, Karim, Olcyr, Vincent, Yann. Merci \'{e}galement \`{a} Gatien et Pierre pour avoir \'{e}t\'{e} un comit\'{e} de suivi de th\`{e}se parfaitement adapt\'{e} \`{a} mes attentes.

Je suis \'{e}galement reconnaissant envers Marie qui m'a accueilli au sein du laboratoire avec toute l'attention et la bienveillance qu'elle porte. 

Ma th\`{e}se n'aurait pas non-plus \'{e}t\'{e} la m\^{e}me sans ma tr\`{e}s ch\`{e}re co-bureau, Valentine, \`{a} qui je dois une relecture compl\`{e}te et minutieuse de cette th\`{e}se.

Je souhaite remercier le laboratoire IJCLab pour avoir financ\'{e} mes conf\'{e}rences et \`{a} Pascale pour m'avoir aid\'{e} dans ces d\'{e}marches. Je suis \'{e}galement reconnaissant envers l'\'{e}cole doctorale EDPIF pour m'avoir permis de me concentrer sur mes recherche sans avoir \`{a} me soucier de l'argent.

\paragraph{}
The PhD adventure is also full of international encounters which I would like to salute: P.~Ashieri, J.~Barrett, P.~Bieliavsky, P.~Bosso, M.~Dimitrievi\'{c}-Ciri\'{c}, R.~Fioresi, S.~Franchino-Vi\~{n}as, R.~Iseppi, F.~Lizzi, S.~Koren, J.~Kowalski-Glikman, S.~Majid, P.~Martinetti, T.~Masson, K.~Rejzner, A.~Sitarz, E.~Skvortsov, H.~Steinacker and P.~Vitale.

I would also like to thank the jury of this thesis who accepted to spend hours on my work.

L'aventure de la th\`{e}se se passe aussi parfois \`{a} domicile et j'aimerais saluer mes coll\`{e}gues de physique th\'{e}orique avec qui j'ai pu partager des bons moments: Antoine, Gaston, Gloria, R\'{e}mi, Salvatore, Samy.

\paragraph{}
Pour finir, je suis reconnaissant envers le RER B pour m'avoir convoy\'{e} tous les jours. D'autre-part je souhaite remercier les personnes ayant cr\'{e}\'{e}, maintenu et particip\'{e} au podcast \enquote{Sant\'{e} vous bien, ici on ne se th\`{e}se pas !} pour m'avoir aider \`{a} avoir une th\`{e}se saine.

Enfin et surtout, j'aimerais remercier du fond du c\oe{}ur tous mes proches sans qui ces trois ann\'{e}es auraient \'{e}t\'{e} vachement moins fun.

\tableofcontents
% Command for headers of a * chapter
\markboth{\tocname}{}
\addcontentsline{toc}{chapter}{\tocname}

\mainmatter

\pagestyle{frontpsty}
\chapter*{Introduction}
% Command for headers of a * chapter
\markboth{Introduction}{}
% Adds a line in the toc
\addcontentsline{toc}{chapter}{Introduction}

\paragraph{}
Suppose you want to drink a can. You would first open the can and then drink its content. To do the opposite, that is to try to drink the can first and then open it, makes few sense. In the first case, you have drunk something and in the second you have not. Two processes that cannot be performed interchangeably are said to be \emph{noncommutative}\footnote{
	This illustration to explain noncommutativity with drinks is actually due to \namefont{Connes} in a \href{http://www2.cnrs.fr/sites/communique/fichier/la_geometrie_non_commutative.pdf}{communiqu\'{e} from CNRS} [in French].
}.
Let us try to understand where this noncommutativity comes from. We have just compared the ordering importance of two \emph{actions} (\enquote{trying to drink} and \enquote{opening}) applied to the same \emph{system} (the can). One could think of another action that could be performed interchangeably with \enquote{trying to drink} the can, such as \enquote{measuring the length} of the can. The can has the same length whether it is full or empty. In this case, these two processes are said to be commutative.

The noncommutativity of the first scenario lies in the fact that the action of \enquote{opening} has changed the can from closed to open: it has affected our system. We say that the can has changed of \emph{state}. Drinking while in the state \enquote{open} is not the same thing as drinking while in the state \enquote{closed}. The second scenario is fully commutative because the action of measuring is assumed not to change the state of the system. However, one of the radical changes \qM brought, is precisely that performing a measure on a system changes it. At the scale of the can, the change produced by the measurement process are negligible, but not for a quantum system. Therefore, the actions of \enquote{trying to drink} and \enquote{measuring the length} could be noncommutative for a would-be quantum can. This noncommutativity has been implemented in the model of \qM precisely through the use of observables and \opalg{s}.

In this thesis, we study noncommutative (also called quantum) space-times, which are defined as \opalg{s}. In the same spirit as above, the introduction of noncommutativity on the \sT structure expresses the fact that some processes can change the \sT itself. In particular, for the \qST{s} studied in this manuscript, the measure of space and time distances cannot be performed independently: one necessarily affects the other.

\paragraph{}
The previous hypothesis on \sT measurements follows from the idea that \sT should loose its smooth structure at the \namefont{Planck} scale. The puzzle regarding the fate of \sT at small scales arises from the theoretical inconsistencies between the frameworks of \qM and \gR. In this sense, \qG corresponds to a would-be theory of gravity that accommodates the geometrical interpretation of gravity with the probabilistic viewpoint of the quantum world. Despite many efforts and the development of many theoretical frameworks for \qG, there is no observed effects due to \qG.

The lack of a consistent theoretical framework urged some authors to opt for a bottom-up approach. In order to grasp more insights, \qG phenomenology postulates properties which would be carried by a complete theory of \qG and that makes testable predictions.

\paragraph{}
The study of the symmetries of \qST{s}, often called quantum or deformed symmetries, has triggered a promising pool of \qG phenomenological frameworks. Some focus have been made on deformations of the \namefont{Poincar\'{e}} group that imposes both an observer independent speed and an observer independent length. Those are called \dSR theories. The \sT carrying such a \dSR cannot be smooth and naturally writes as a \qST. The possible noncommutativity, induced by such theories, also spread to the momentum space, which is therefore now curved. The framework of curved momentum space and \namefont{Born} geometries is heavily studied for \qG phenomenology. The implementation of a minimal length can also be performed by generalising the \namefont{Heisenberg} uncertainties that stems from the noncommutativity of the position and momentum operators. This minimal length turns field theories with a \gUP to be ultra-violet finite and therefore with strong predictive power. Note that \gUP{s} may appear on some \qST{s}. From another perspective, the deformation of a classical symmetry implies that this classical symmetry is broken at some scale. The deformations of \namefont{Poincar\'{e}} symmetry may thus induce \LIV or \CPTV. Both phenomenological frameworks have already attracted lot of attention, especially in the context of \SM extensions.

\paragraph{}
As all these considerations on deformed symmetries suggest, the physics of \qST{s} could provide even more insights into the nature of \qG effects. Yet, the physics of \qST{s} remains poorly known.

The field and gauge theories on such space-times have been explored up to the one-loop level. Already in this loop order, these theories have been shown to exhibit a behaviour called the \UVIR. This behaviour corresponds to the appearance of mixed infra-red and ultra-violet divergences and is thought to spoil the usual perturbative renormalisation. The mixing was first experienced on the \Moy space and later in other \qST{s}, but there is, so far, no general notion of when and how such a mixing occurs. The latter problem is tackled in \cite{Hersent_2024a}. The possible impact of the \UVIR on particle physics or gravity on \qST{s} has been poorly studied. The appearance of higher loops phenomenons, as well as the fate of renormalisability, has yet not been deepened. In the case of (\namefont{Yang-Mills}-like) gauge theories, it was shown that the gauge invariance is broken after quantisation, already at the level of the one-point function. This \enquote{gauge anomaly} has been noticed but not characterised. The unitarity of these fields theories with higher derivatives is also an open question.

Furthermore, there is no consensual notion of vacuum states and vacuum energy. The formalism of \qg{s} allows one to perform non-linear change of coordinates of momentum so that expressions of the form $p^2$, for $p$ is a momentum, is not covariant anymore. One even struggles to define multi-particle states in the noncommutative framework \cite{Arzano_2023}. 

Finally, the classical notion of causality is thought break at the \namefont{Planck} scale, since superimposed states of mass implies superimposed \sT geometries and so competing causalities. Accordingly, the first studies of causality on \qST{s} points to a conceptual change for a would-be quantum causality. Yet, the noncommutative causality frameworks already developed mainly evolve around \enquote{flat} causality since they were defined on deformations of \namefont{Minkowski}. Moreover, they lack of global coherence and common theoretical grounds.

\paragraph{}
The aim of this manuscript is threefold. First, it introduces and motivates the use of \nCG in the context of physics study. Second, it summarises the state of the art in field and gauge theories on \qST{s}, focusing on the more recent developments proposed by the author. Finally, it underlines the importance of \nCG for \qG phenomenology and advocates that the study of \qST{s} could shed even more light on this topic.

The \chapref{gnc} focuses on defining what \nCG is, and the different mathematical frameworks it encompasses. The physical motivations for such geometries are also discussed. The way in which field and gauge theories are implemented on \qST{s} is reviewed in \chapref{ncft} together with the recent approach \cite{Hersent_2024a} based on momentum space analysis. The promising deformations of \Minkt are discussed in \chapref{kM}. The focus is made on the \kMt deformation for which the first quantum properties of a gauge theory are analysed. Considerations of (quantum) gravity appear in \chapref{qg}, where the phenomenology is introduced together with the theoretical motivations for studying \qG. A toy model of noncommutative gravity, based on a \kMt tangent space, is tackled. Finally, more mathematical content is gathered in the Appendices. The \appref{ha} deals with \qg{s} formalism and the \appref{oa} presents the necessary notions of \Csalg{s}.

\paragraph{}
In all the manuscript, we apply \namefont{Einstein} summation convention of repeated indices. This convention is only broken for the labelling of open covers and algebra covers in \secref{qg_ncg}, as specified in the latter Section. We also work in natural units, \ie $c = \hbar = G = 1$, where $c$ is the speed of light, $\hbar$ the \namefont{Planck} constant and $G$ the \namefont{Newton} constant.

\section*{Contribution of the author}
\markboth{Contribution of the author}{}
\addcontentsline{toc}{chapter}{Contribution of the author}
\paragraph{}
The present manuscript tries to have a global viewpoint of the physics of and on noncommutative space-times. Therefore, it gathers well-established results as well as new contributions due, fully or partly, to the author. These new contributions were obtained during a 3-year Ph.D.~study from October 2021 to October 2024, under the supervision of Jean-Christophe \namefont{Wallet}, at IJCLab in Orsay, France.

\paragraph{}
The new material contained in this manuscript, in which the author has taken an active part, corresponds to 
\begin{nclist}
	\Item{Hersent_2022a}
		{K.~Hersent, P.~Mathieu \andbib J.-C.~Wallet}
		{Quantum instability of gauge theories on $\kappa$-Minkowski space}
		{\PRD}
		{105}
		{2022}
		{106013}
		{10.1103/PhysRevD.105.106013}
		{2107.14462}
	\Item{Hersent_2022b}
		{K.~Hersent, P.~Mathieu \andbib J.-C.~Wallet}
		{Algebraic structures in $\kappa$-Poincar\'{e} invariant gauge theories}
		{\IJGMMP}
		{19}
		{2022}
		{2250078}
		{10.1142/S0219887822500785}
		{2110.10763}
	\Item{Hersent_2022c}
		{K.~Hersent \andbib J.-C.~Wallet}
		{Gauge theory models on $\kappa$-Minkowski space: Results and prospects}
		{PoS CORFU2021}
		{406}
		{2022}
		{286}
		{10.22323/1.406.0286}
		{2203.12706}
	\Item{Hersent_2023a}
		{K.~Hersent, P.~Mathieu \andbib J.-C.~Wallet}
		{Gauge theories on quantum spaces}
		{\PR}
		{1014}
		{2023}
		{1-83}
		{10.1016/j.physrep.2023.03.002}
		{2210.11890}
	\Item{Hersent_2023b}
		{K.~Hersent}
		{Quantum properties of $U(1)$-like gauge theory on $\kappa$-Minkowski}
		{PoS}
		{CORFU2022}
		{2023}
		{328}
		{10.22323/1.436.0328}
		{2302.03998}
	\Item{Franco_2023}
		{N.~Franco, K.~Hersent, V.~Maris \andbib J.-C.~Wallet}
		{Quantum causality in $\kappa$-Minkowski and related constraints}
		{\CQG}
		{40}
		{2023}
		{164001}
		{10.1088/1361-6382/ace588}
		{2302.10734}
	\Item{Hersent_2023c}
		{K.~Hersent \andbib J.-C.~Wallet}
		{Field theories on $\rho$-deformed Minkowski space-time}
		{\JHEP}
		{07}
		{2023}
		{031}
		{10.1007/JHEP07(2023)031}
		{2304.05787}
	\Item{Hersent_2024a}
		{K.~Hersent}
		{On the UV/IR mixing of Lie algebra-type noncommutatitive $\phi^4$-theories}
		{\JHEP}
		{2024}
		{2024}
		{23}
		{10.1007/JHEP03(2024)023}
		{2309.08917}
	\Item{Hersent_2023d}
		{K.~Hersent \andbib J.-C.~Wallet}
		{$\kappa$-Minkowski as tangent space I: quantum partition of unity}
		{arXiv preprint}
		{None}
		{2024}
		{None}
		{None}
		{2311.12584}
\end{nclist}

\paragraph{}
The review \cite{Hersent_2023a} gathers general \gT constructions on \qST{s} as well as detailed construction for \Moy, $\Rcl$ and \kMt. The general setting for \gT[ies] as been put in \chapref{ncft} and the \gT on \kMt is presented in \secref{kM_gt}.

The paper tackling \phif on arbitrary \Lieat-type noncommutative space-times \cite{Hersent_2024a} has been summarised in \secref{ncft_p4}. It also proposes an unambiguous definition of the \UVIR and hints toward a possible criterion for \phif to trigger the mixing or not.

In \cite{Hersent_2022b}, the space-time dimension constraint obtained via gauge invariance of the \nCYM action on \kMt has been shown to be strong. The \gT, from which this action is derived, is reviewed in \cite{Hersent_2022c} and is elucidated in \subsecref{kM_gt_td}.

Given the latter \gT on \kMt, the computation of the one-loop one-point function (tadpole) has been performed in \cite{Hersent_2022a} and discussed more closely in \cite{Hersent_2023b}. The results are grouped in \subsecref{kM_gt_tad}.

The causality toy model on \kMt space-time is constructed in \secref{kM_c} and has been computed in \cite{Franco_2023}.
  
A \phif on the \rMt space-time is studied in \cite{Hersent_2023c}, as briefly mentioned in \secref{kM_rM}.

A toy model for gravity on \qST{s} having \kMt as tangent space is constructed in \cite{Hersent_2023d}. It necessitates the notion of noncommutative partition of unity which is defined there. The latter model is detailed in \secref{qg_ncg}.
\pagestyle{mainpsty}
\chapter{The principle of noncommutative geometry}
\label{ch:gnc}
\paragraph{}
\NCG is a broad field of mathematics and physics that tries to have an algebraic point of view on geometry. Its main goal is to use the language of \qM (expressed by \Csalg{s}) to encode the geometry. Therefore, there is a strong motivation to use the framework of \nCG to develop toy models of \qG. But the motivations for studying \oq\nCST{s}\cq capture other aspects of new physics, such as beyond the \SM studies.

\paragraph{}
As we will see in the following, the \nCG framework is too wide to be tackled in a single manuscript chapter. Therefore, this part of the thesis is mainly aimed at understanding the basic principles and the physical motivations for studying this field.

The most detailed aspects of \nCG that this manuscript deals with mostly evolves around \nCFT. Other aspects of \nCG are still mentioned in this \chapref{gnc}. A curious reader can find more details in the references.

\section{Foundations of noncommutative geometry}
\label{sec:gnc_found}
\paragraph{}
All of modern theoretical physics is governed by two main models: \gR and \qFT. Both \gR and classical field theory are expressed in the same mathematical language of \dg. The goal of \nCG is to generalise \dg into a global framework that allows for more complex geometries, such as discrete geometries or fractals. As noncommutativity is a key component of \qM, the \nCG[ies] are also thought to encode quantum fields and \enquote{\qST{s}} into only geometrical notions.

Here we develop the main features of non-commutative geometry and the different paths it has taken, from a very physical point of view. 

\paragraph{}
In order to generalise \dg, one has to start with \dg. The three main components of a \dg one has to generalise are the topology, the differential structure and the fiber bundles.

The topology corresponds to the part of space you can continuously  deform, \ie without closing holes, opening holes, tearing, gluing, or passing through itself. It gives information about the space in terms of its number of \enquote{pieces} (connected components) or holes and allows to define distances between objects. Still, it does not give information about how the space is curved.

The differential structure comes with the smoothness of the space and allows to define derivatives along curves, or \vf{s}. A (smooth) field can be derived only if a differential structure is defined on the space. This implies that one can construct the set of forms on the space and also to define its curvature.

Finally, the fiber bundles correspond to an additional structure one sets on the space and that allows to define fermionic (spinor) fields or gauge fields. 

\paragraph{}
First, let us generalise the topology. Let $\manft{M}$ be a \sT and $\opaft{A} = \func^\infty(\manft{M})$ the space of observables. Then, an observable $f \in \opaft{A}$ attaches to each point $x \in \manft{M}$ a number $f(x) \in \Real$ or $\Cpx$. The observables $\opaft{A}$ form an algebra (see \defref{ha_as_alg}) using the product of $\Real$ or $\Cpx$ through the point-wise product
\begin{align}
	(f \cdot g)(x)
	= f(x)\ g(x).
	\label{eq:gnc_found_cpdt}
\end{align}
This algebra is further commutative, that is $f \cdot g = g \cdot f$, because the products in $\Real$ and $\Cpx$ are commutative.

Now, consider the space of states, noted $\Spst{\opaft{A}}$, as the functions over the space of observables $\opaft{A}$. Explicitly, it corresponds to the set of elements $\psi$ which, to a given observable $f$, associates a number $\psi(f) \in \Real$ or $\Cpx$. Then, the commutative \nameref{thm:oa_rt_cGN} states that the space of states $\Spst{\opaft{A}}$ is in one-to-one correspondence with the \sT $\manft{M}$ itself. Explicitly, given a point $x \in \manft{M}$, there exists one and only one state $\psi_x$ such that $\psi_x(f) = f(x)$. One can underline the correspondence with the \oq bracket\cq notation. If one writes
\begin{align}
	\braket{f}{x}
	= f(x), &&
	\braket{f}{\psi}
	= \psi(f),
	\label{eq:gnc_GN_dual}
\end{align}
then the correspondence is
\begin{align}
	\ket{\psi_x}
	= \ket{x}.
	\label{eq:gnc_GN_corr}
\end{align}
This picture shows how the \nameref{thm:oa_rt_cGN} relates the notion of \sT point to the notion of state.

\paragraph{}
From there, the point of view of \nCG is to consider the quantum version of this picture, where $f$ is a quantum observable and $\psi$ is a quantum state. In order to do so, one turns $\opaft{A}$ into a noncommutative algebra. In this case, the \nCST (the would-be set of point $x$) is pictured as being the (pure) states $\psi$ through an extrapolated correspondence of the form \eqref{eq:gnc_GN_corr}.

\paragraph{}
Thus, the way to generalise the topological picture is to consider an algebra, which is not necessarily commutative\footnote{
	In fact, the algebraic generalisation of geometry was undertook before \nCG came into play, with algebraic geometry. However, this setting requires commutativity and cannot account for some geometric spaces which are noncommutative (like the quantum mechanics space, see \secref{gnc_phys}). We refer to \namefont{Connes}' book \cite{Connes_1994} for more details. It should be mentioned that there are many attempts into constructing noncommutative algebraic geometry, but this goes far beyond the scope of this thesis and of the author's personal knowledge.
}
and which stands as the algebra of smooth function over a \oq\nCST\cq. Understanding the properties and characteristics of this kind of \sT is a central question in \nCG.

\paragraph{}
More quantitatively, one can construct a noncommutative algebra of functions by starting with a commutative algebra of functions over some \sT, say the one of \eqref{eq:gnc_found_cpdt}, and deform the latter with a new product that is noncommutative, generically called the \spdtt. This procedure is called \dq.

\begin{Emph}{Deformation quantisation}
	The \dq setting aims at finding an unambiguous quantisation procedure and thus at classifying the different quantisation methods. Inspired by the formalism of quantum mechanics, one can deform the smooth functions of a (\namefont{Poisson}) manifold into an algebra of functions with a noncommutative product, called the \spdtt, generically denoted by $\star$ in this manuscript. This product expands over a small parameter, called the \dpt, similar to $\hbar$ for quantum mechanics. In this sense, and as it is made explicit in \secref{gnc_phys} through equation \eqref{eq:gnc_qm}, the \spdtt formalism is linked to canonical quantisation. The standard definition of \dq and some ways of constructing \spdtt{s} will be given in \secref{gnc_dq}.
\end{Emph}

\paragraph{}
The second step is to generalise the differential structure, and thus the geometry, in the sense that one defines the metric thanks to this differential structure.

One can hint that this task will be much more difficult. Indeed, the topology gathers \oq global\cq properties of a space and so functions defined on the whole space allow to recover the latter properties. However, the differential structure is local and even defined point by point. But the algebra of functions cannot render points by itself, as it does not contain the \Ddf. This smearing out of the \nCST, due to its definition through its algebra of functions, is the reason \nCST{s} are sometimes called \oq fuzzy\cq \sT{s} in the physics community. This also echoes some physical interpretation of fuzziness discussed in \subsecref{gnc_phys_fuz}.

There are several ways to generalise the differential structure which are presented below: the \st with its quantised calculus, the \qg approach and the \dbdc.

\begin{Emph}{Spectral triple}
	The most commonly acknowledged approach toward the generalisation of the differential structure is the \namefont{Connes}' \st approach to \nCG. Its starting point is to consider the \Do as containing information on the geometry. Indeed, one has, in a local coordinate chart,
	\begin{align}
		\Dirop 
		= -i \gamma^\mu (\partial_\mu + \omega_\mu), &&
		\left\{ \gamma^\mu, \gamma^\nu \right\}
		= 2 \metft{g}^{\mu\nu} \Matid{d+1},
	\end{align}
	where $\metft{g}$ is the \sT metric and $\omega_\mu$ is the spin connection. Therefore, the \Do gathers the differential and the metric (through the gamma matrices).
	
	At first, there was hope that the Laplacian and its spectrum could bring all the needed information. To this end, \namefont{Kac} asked the question \oq Can one hear the shape of a drum?\cq \cite{Kac_1966}. But it turns out we cannot \cite{Gordon_1992}. The missing information of the Laplace operator can be recovered when considering a \Hsp on which we (faithfully) represent the noncommutative algebra. We then need a triplet for our noncommutative geometry which is called the \st: the algebra, the \Hsp and the \Do. More details about this structure are gathered in \secref{gnc_st}.

	From there, using the so-called quantised calculus, one can define a differential calculus on the \st. The main idea is to represent the function of the algebra as operators on the \Hsp and then introduce the differential given by the commutator of operators. The master piece of this setting then resides in \nameref{thm:gnc_st_rt} which states that for any compact oriented Riemannian manifold, one can recover the usual \dc thanks to the triple.
\end{Emph}

\begin{Emph}{\Qg{s}}
	\Qg{s} are generalisations of the usual notion of group and their applications are broader than only \nCG. Their definition may also vary since different approaches to \qg{s} were considered. In our case, and throughout this manuscript, the notion of \qg will be associated with the structure of \Hfat and both terms will be used equivalently.

	In the context of \nCG, \qg{s} are mainly associated with quantum symmetries, in a similar way as groups are associated with symmetries for commutative geometries. Therefore, one can define from the universal \dc a calculus that satisfies this quantum symmetry. From there, a full picture analogue to Riemannian geometry with \qg{s} can be constructed \cite{Beggs_2020}.

	However, the link between \qg{s} and geometry is deeper thanks to their structure: the dual of a \Hfat is also a \Hfat. Thus, the \nCST, seen as a dual to its quantum symmetries, also have a \Hfat structure. In other words, the correspondence \eqref{eq:gnc_GN_corr} associates similar structure to the \sT and its observables via a \Hfat formalism. In the context of \nCST{s}, this situation is actually occurring. More details will be given in \secref{gnc_qg} and the mathematical definitions are postponed to \appref{ha}.
\end{Emph}

\begin{Emph}{\Dbdc}
	Another approach to noncommutative differential geometry focuses on the tangent space. Indeed, the \dc over a \sT is based on the notion of \vf. If one were to generalise the \vf{s} to a noncommutative setting, then all the usual definitions of tensors, connections or curvature could be algebraically copy-pasted from their \dg expressions. The key point is to observe that \vf{s} are in one-to-one correspondence with the derivations of the algebra of smooth functions. Then, one could define the analogues of noncommutative \vf{s} to be the derivations of the noncommutative algebra.

	This construction is called the \dbdc \cite{Dubois-Violette_2001}. It is the main topic of \secref{gnc_dbdc}.
\end{Emph}

\begin{Rmk}{}{gnc_diff_approach}
	An important remark to point out is that these different approaches to noncommutative geometry are not necessarily in competition with one another. One can find bridges linking all these settings, but there is not, at this date, a coherent language to gather all. We here make some links explicit.

	\paragraph{}
	First, the \dq setting allows one to go from any smooth \sT to a noncommutative algebra of functions with a \spdtt. Thus, \qST can be constructed through \dq and then combined with any of the other three formalisms by adding more structure, explicitly a \st, a \Hfat of symmetries, or the derivations of the algebra.

	\paragraph{}
	From any algebra, one can build a \dc called the universal \dc. It is called universal in the sense that one can recover any graded \dc over the algebra from the universal one. Therefore, the quantised \dc and the \dbdc can both be seen as specific restrictions of the same universal \dc.

	\paragraph{}
	In the case of a \nCST on which acts a \Hfat of symmetries, one can read from the \Hfat structure some derivations over the \sT algebra. Indeed, as it is made clear in the following, a trivial symmetry element with a trivial coproduct acts as a derivation over the algebra. This makes a link between \dbdc and \Hfat{s} as it will be used in \chapref{ncft} and \ref{ch:kM}.

	\paragraph{}
	The spectral structure of quantum groups was thought not to exists but it was exhibited a \st for the quantum group $\SUqg{2}$ \cite{Chakraborti_2003}.
\end{Rmk}

\paragraph{}
Now that the differential part has been set up, there remains the question on how to build a connection and its corresponding curvature on a \enquote{noncommutatve fiber bundle}. This question is central in both gravity theory and \qFT\footnote{
	In \qFT, the connection corresponds to the gauge field $A$ and its curvature to the field strength $F$.	
},
as connection and curvature encode physical quantities. The key ingredient to generalise the notion of fiber bundle to the noncommutative setting is the \namefont{Serre-Swan} theorem. This theorem relates the geometric notion of vector bundles to the algebraic concept of projective modules (see \subsecref{gnc_qg_qst} for more details on modules and how it arises in \nCG), so that one can consider the modules as a generalisation of vector bundles in a noncommutative geometry. The definitions of connections and their curvature will then follow from their usual expressions exported to the case of the module.

This observation is the starting point of \nCFT and \gT as detailed in \chapref{ncft}.

\paragraph{}
Considering the coordinate function, that is the function $x^\mu(x) = x^\mu$, on a \nCST, one has that $x^\mu \star x^\nu \neq x^\nu \star x^\mu$ by definition of noncommutativity. Here, the $\star$ denotes the noncommutative product of the space, be it generated by \dq or not. The previous statement will be most of the time written under the form
\begin{align}
	[x^\mu, x^\nu]_\star
	= x^\mu \star x^\nu - x^\nu \star x^\mu 
	\neq 0
	\label{eq:gnc_found_qst}
\end{align}
as the bracket $[\cdot,\cdot]_\star$ precisely measures the amount of \enquote{noncommutativity}.

In terms of \qM[al] interpretation, the relation \eqref{eq:gnc_found_qst} states that coordinates cannot be measured simultaneously and that the measurement of $x^\mu$ affects the measurement of $x^\nu$, and \latint{vice versa}. For this precise reason, the \nCST{s} are often called \qST{s} and both terminology will be used in an interchangeable fashion in this  manuscript.

\section{Physical motivations of noncommutative geometry}
\label{sec:gnc_phys}
\paragraph{}
Beyond the mathematical purpose of a generalised setting for geometric objects, the \nCG framework has several motivations coming from physics. The primary motivation and first example of such a \nCG is the phase space of \qM. 

\begin{Emph}{The geometry of \qM}
	One of the cornerstones of \qM is that the measurement of the system affects the post-measurement system by projecting it to a certain state. Therefore, the order of a series of measurements becomes of primordial importance since this order may change the latter measurements. In other words, measuring a quantity $\hat{A}$ and then a quantity $\hat{B}$ might not amount to measuring $\hat{B}$ and then $\hat{A}$. We say that the measurement process is noncommutative. Furthermore, one can quantify the difference between the two measurements by computing the quantity $[\hat{A}, \hat{B}]$, which is non-zero by definition. In \qM, the noncommutativity is applied to the phase space coordinates $x$ and $p$, that is why, roughly speaking, \qM is a noncommutative geometry.

	\paragraph{}
	Let us be a bit more explicit. Consider the phase space of classical mechanics in one space dimension, that is (a subset of) $\Real^2$, with coordinates $x$ and $p$. The following discussion can be extended to any dimension, that is on (a subset of) $\Real^{2n}$, straightforwardly. From there, $x$ and $p$ are numbers that stand for the position and momentum of a certain particle, respectively. If one wants to impose commutation relation such as $[x, p] \neq 0$, it cannot be done with only numbers, since any numbers commute. Therefore, we need to implement another structure.
	
	Taking the observable point of view, like in \secref{gnc_found}, we can consider the functions $x, p \in \func^\infty(\Real^2)$ such that $x(x,p) = x$ and $p(x,p) = p$. The space $\func^\infty(\Real^2)$ has now the structure of an algebra, thanks to the product \eqref{eq:gnc_found_cpdt}, but it is still commutative. Therefore, one cannot implement the commutation relation $[x, p] \neq 0$ even at the level of functions.
	
	\paragraph{}
	The procedure, to obtain the previous commutation relations, is to perform a quantisation. To do so, one needs a quantisation map \cite{Landsmann_1998} $\Qmap: \func^\infty(\Real^2) \to \opaft{A}$, where $\opaft{A}$ is called the \opalg or \Csalg. For example, the position function $x$ is turned into the position operator\footnote{
		We adopt here the usual physics hatted notation for operators, but we drop the hat outside this discussion.
	}
	$\Qmap(x) = \hat{x}$. There are several ways to construct such a quantisation map, see \secref{gnc_dq} for more details. One usually represents $\opaft{A}$ on a \Hsp $\manft{H}$ through a (representation) map $\pi: \opaft{A} \to \Hilbft{L}(\Hilbft{H})$, where $\Hilbft{L}(\Hilbft{H})$ is the set of linear operators on $\Hilbft{H}$. To exemplify the previous objects, let us take a physical example. Given a state $\ket{\psi} \in \Hilbft{H}$, the position and momentum operators act on $\ket{\psi}$ through\footnote{
		Note that we chose $\Hilbft{H} = L^2(\Real)$ here, that is the space of square integrable functions on $\Real$. It corresponds to the usual \namefont{Schr\"{o}dinger} representation of \qM , where the \enquote{sum} of probabilities corresponds to the integral.	
	}
	\begin{align}
		\pi(\hat{x}) \ket{\psi(x)}
		&= x \ket{\psi(x)}, &
		\pi(\hat{p}) \ket{\psi(x)}
		&= - \iCpx \frac{\td}{\td x} \ket{\psi(x)}
		\label{eq:gnc_qm_rep}
	\end{align}
	where the $\pi$ is generally omitted in both physics and mathematical literature. The \opalg $\opaft{A}$ can be viewed as the \enquote{\nCG} of \qM.
	
	\paragraph{}
	In the \dq formalism, one can go back from the \opalg $\opaft{A}$ to a space of functions, noted $\algft{A}$, using $\Qmap^{-1}$. An operator $\hat{f}$ is thus transformed back into a function on $\Real^2$, $f$, and the product of $\opaft{A}$ is also brought back as a new (star) product $\star_\hbar$ via
	\begin{subequations}
		\label{eq:gnc_qm}
	\begin{align}
		f \star_\hbar g
		&= \Qmap^{-1} \big( \Qmap(f) \Qmap(g) \big).
		\label{eq:gnc_qm_spdt_gen}
	\end{align}
	The space $\algft{A}$ together with $\star_\hbar$ forms an algebra. Moreover, this \spdtt is, by construction, noncommutative. One can obtain an explicit expression for the previous \spdtt \cite{Weyl_1927, von_Neumann_1931, Durhuus_2013}, called the \Moy product in \namefont{Weyl} ordering, 
	\begin{equation}
		(f \star_\hbar g)(x,p)
		= \frac{1}{(\pi \hbar)^2} \int \td{y} \td{q} \td{z} \tdl{}{s} f(x+y, p+q)\ g(x+z,p+s)\ e^{\frac{2 \iCpx}{\hbar} (qz - ys)},
		\label{eq:gnc_qm_spdt}
	\end{equation}
	where $f$ and $g$ are functions of $\Real^2$ and $x$ and $p$ the coordinates of $\Real^2$. This \spdtt transforms the functions on $\Real^2$ into a noncommutative algebra of functions, denoted generically by $\algft{A} = \func^\infty(\Real^2)\llbracket \hbar \rrbracket$. This notation means that any such functions can be written formally as a power series expansion in $\hbar$. Defining the position function $x(x,p) = x$ and the momentum function $p(x,p) = p$ as above, one obtains
	\begin{equation}
		[x, p]_{\star_\hbar}
		= x \star_\hbar p - p \star_\hbar x
		= \iCpx \hbar.
		\label{eq:gnc_qm_bracket}
	\end{equation}
	\end{subequations}
	In this sense, $\algft{A}$ can be viewed as the \enquote{\nCG} of \qM, and the \spdtt formalism as a straightforward way to implement canonical quantisation.
\end{Emph}
One can find more examples of \nCG[ies] in \cite{Connes_2006}.

\paragraph{}
The \spdtt formalism structurally generates semi-classical and phenomenological frameworks. Indeed, the \spdtt deformation is usually expressed as power series expansions of the \dpt. The latter parameter is called $\kbar$ in a general context. The product of two functions write, in generic form,
\begin{align}
	f \star_{\kbar} g 
	= f \cdot g + \sum_{n=1}^\infty \kbar^n\, C_n(f,g)
	\label{eq:gnc_spdt}
\end{align}
where $C_n$ is a bilinear differential operator. The appearance of the usual commutative product of functions $\cdot$ \eqref{eq:gnc_found_cpdt} in the \spdtt $\star_{\kbar}$ \eqref{eq:gnc_spdt} precisely account for the fact that $\star_{\kbar}$ is a deformation of $\cdot$, a deformation which is controlled by the \dpt $\kbar$. In other words, the product $\cdot$ correspond to the zeroth order in $\kbar$ of the product $\star_{\kbar}$.

One can consider to expend $\star_{\kbar}$ only to first order in $\kbar$ and thus have a linear correction to the usual commutative theory. This would give a semi-classical theory, in which $\kbar$ is a free parameter to be constrained by experiments. As we will see in the following, the \dpt is most of the time dimensionful (like $\hbar$) and would thus give a scale below which the commutative theory is valid.

\paragraph{}
Another general argument on how \nCG could hint new physics concerns deformations of symmetries. As the \sT structure is deformed, the corresponding symmetries will also be deformed. Most of the deformed symmetries encountered in this manuscript takes the structure of a \Hfat, therefore underlying that these symmetries are quantum. In the context of \spdtt deformation, the deformation of the symmetry is also controlled by the \dpt $\kbar$. This implies that the classical symmetries are broken and replaced by quantum ones. However, the classical symmetries can be recovered at some scale, corresponding to the limit $\kbar \to 0$.

The violations of some symmetries are intensely studied in the context of (high energy) new physics, like \LIV, \Pdef or \CPTV \cite{Addazi_2022}. These phenomenological frameworks can naturally arise in the context of \nCG because of symmetry deformations. This is more discussed in \secref{qg_ph}.

\paragraph{}
Beyond the broad arguments above for considering \nCG as an inspiration for new physics, more specific arguments can be made for aspects of physics that relies on quantum geometries like the \SM of particle physics or some model of \qG.

\subsection{The geometry of particle physics}
\label{subsec:gnc_phys_pp}
\paragraph{}
The main achievement of \nCG concerning particle physics is the noncommutative \SM \cite{Chamseddine_2007}. All the \SM fields, minimally coupled to gravity, are encoded in a single (noncommutative) geometry expressed as a \st. The axioms of the \st yield the gauge fields and the \namefont{Higgs} field. Both emerge and are treated on the same footing. The \namefont{Higgs} potential as well as the \namefont{Einstein-Hilbert} action and the \SM action in curved space appear when computing the so-called spectral action. One of the main asset of this model is that it predicts the \namefont{Higgs} mass.

\begin{Emph}{Noncommutative \SM}
	Let $\manft{M}$ be a Riemannian spin geometry (a space without time) and let $\algft{A} = \func^\infty(\manft{M}) \otimes (\Cpx \oplus \Mat{2}{\Cpx} \oplus \Mat{3}{\Cpx})$. This algebra encodes the functions on a \nCG having four different parts: $\func^\infty(\manft{M})$ only encodes the functions over the space $\manft{M}$ and thus the geometry of $\manft{M}$, $\Cpx$ imposes that functions are complex-valued and so takes into account the electric charge, $\Mat{2}{\Cpx}$ accounts for the isospin doublets and $\Mat{3}{\Cpx}$ stands for the three colors of quarks. $\algft{A}$ can be viewed as the set of functions which associates to a point of space $\manft{M}$ a complex number and two complex matrices of size $2$ and $3$. Those are called \enquote{almost commutative} spaces.
	
	One represents this algebra on $\Hilbft{H} = L^2(\manft{M}, \manft{S}) \otimes \Cpx^{96}$, where $L^2(\manft{M}, \manft{S})$ stands for square integrable functions on the spinor bundle $\manft{S}$, which encodes the fermionic (spinor) fields, and $\Cpx^{96}$ describes the latter fermions counting the isospin doublet (up/down or electron/neutrino, with $2$ choices), the $3$ generations or flavours, the $2$ chiralities (left or right), if it encodes a particle or an anti-particle ($2$ choices) and if the fermion is a quark or a lepton (the quarks have $3$ colors and the lepton do not, so $3 + 1$ choices): therefore, there are $ 2 \times 2 \times 3 \times 2 \times 2 \times (3 + 1) = 96$ fields. Finally, the \Do writes
	\begin{align}
		\Dirop = \slashed{\partial} \otimes \Matid{96} + \gamma^5 \otimes 
		\begin{pmatrix}
			0 & M & M_R & 0 \\
			M^\dagger & 0 & 0 & 0 \\
			M_R^\dagger & 0 & 0 & \overline{M} \\
			0 & 0 & \tp{M} & 0
		\end{pmatrix}
		\label{eq:gnc_phys_Do_SM}
	\end{align}
	where $\slashed{\partial}$ is the \Do on the space $\manft{M}$,  $M$ contains the quarks, leptons, neutrinos \namefont{Yukawa} couplings, as well as the quarks and neutrinos mixing parameters and $M_R$ contains the \namefont{Majorana} neutrinos mass.
	
	The \st $(\algft{A}, \Hilbft{H}, \Dirop)$ encodes the geometry of the \SM of particle physics minimally coupled to gravity. 
	
	\paragraph{}
	From there, one can compute the gauge group, the gauge bosons and the full action associated with that geometry. The gauge group corresponds to the unitary elements of the algebra $\algft{A}$, that are here $\Diff(\manft{M}) \ltimes \mathrm{Map} \big( \manft{M}, \Ug{1} \times \SUg{2} \times \SUg{3} \big)$, where $\Diff(\manft{M})$ stands for the diffeomorphism group of the space $\manft{M}$, that is the group of coordinate change, and $\mathrm{Map}(\manft{M}, G)$ stands for the function from $\manft{M}$ to $G$. The previous notation means that each point of the space $\manft{M}$ has its own $\Ug{1} \times \SUg{2} \times \SUg{3}$ gauge symmetry, which adapts to the choice of coordinates thanks to $\Diff(\manft{M})$.
	
	The gauge bosons can be read in the fluctuation of the \Do,  corresponding to \eqref{eq:gnc_st_gt}, and their gauge transformation through the first order axiom of the \st formalism \eqref{eq:gnc_st_1st}. More explicitly, if the \Do \eqref{eq:gnc_phys_Do_SM} is noted $\Dirop = \slashed{\partial} \otimes \Matid{} + \gamma^5 \otimes \Dirop_\mathrm{F}$, then one computes
	\begin{align}
		\begin{aligned}
			f [\Dirop, g]
			&= -i \gamma^\mu \otimes f \partial_\mu g + \gamma_5 \otimes f [\Dirop_F, g] \\
			&= \gamma^\mu \otimes A_\mu + \gamma_5 \otimes \phi
		\end{aligned}
		\label{eq:gnc_phys_SM_g}
	\end{align}
	where $A_\mu$ have all properties of the gauge fields and $\phi$ corresponds to the \namefont{Higgs} boson.
	
	Finally, the action is computed thanks to the spectral action for the \Do \eqref{eq:gnc_phys_Do_SM}. The space part of the \Do gives rise to the \namefont{Einstein-Hilbert} action, which is summed with the \SM action in curved space with a spontaneously broken \namefont{Higgs} $\phi$.
	
	\paragraph{}
	The noncommutative \SM may give new phenomenological insights on particle physics. First, the \namefont{Higgs} mechanism, that is the \namefont{Higgs} boson and its potential, is implemented naturally. Then, the model imposes mass relations between the fermions and predicts the \namefont{Higgs} mass. Finally, the see-saw mechanism can be implemented within the framework. The phenomenology of this model is still under study.
\end{Emph}

This manuscript does not expand too much on the noncommutative \SM as it was not part of the author's work. Moreover, one should note that this model is still under construction, as their actual formulation are purely classical (in the sense that there was no second/\pIQ done) and Riemannian, therefore without a causal structure\footnote{
	For Lorentzian extensions of the noncommutative \SM, see for example \cite{Barrett_2007, Devastato_2020}.
}.
Nevertheless, the \rmkref{gnc_diff_approach} is to be underlined as some part of what is contained in this thesis may, one day, be linked to this model. Relevant textbook on the noncommutative \SM are, for example, \cite{Connes_1994, van_Suijlekom_2015}.

\paragraph{}
The study of field theories on \nCST, called generically \nCFT, have shown a peculiar behaviour which might be of physical interest, called the ultra-violet/infra-red mixing, or \UVIR.

The first idea of studying field theory on \qST, goes back to \namefont{Snyder} \cite{Snyder_1947} who expected that the noncommutativity would regularize the ultra-violet divergences occurring in \qFT, before renormalisation came into play. The hope of \namefont{Snyder} was that the \dpt could play the role of a ultra-violet cut-off. However, this phenomenon happens only to very specific cases and \nCFT also experiences divergence, some similar to the commutative case and some new ones. For example, in the \Moy space a \phif gives rise to the usual UV divergence for so-called planar diagrams \eqref{eq:gnc_mix_cut_p}, but exhibits UV finiteness and IR singularity for non-planar diagrams \eqref{eq:gnc_mix_cut_np}. The presence or absence of divergences has not been characterised yet for general \nCFT.

\begin{Emph}{UV/IR mixing}
	The first appearance of the \UVIR occurred on the \Moy space \cite{Minwalla_2000} (see \secref{ncft_Moy} for more details on the \Moy space). It starts by considering a deformed \phif, with an action given by
	\begin{align}
		S
		= \int \tdl{4}{x} 
		\frac{1}{2} \big(\partial_\mu \phi \star_\dpMoy \partial^\mu \phi \big)(x)
		+ \frac{m^2}{2} \big(\phi \star_\dpMoy \phi \big)(x)
		+ \frac{\cC^2}{4!} \big(\phi \star_\dpMoy \phi \star_\dpMoy \phi \star_\dpMoy \phi \big)(x)
		\label{eq:gnc_phi4_act}
	\end{align}
	where $\star_\dpMoy$ is the \Moy \spdtt \eqref{eq:ncft_Moy_spdt}, $m$ is the mass of the field $\phi$ and $g$ the coupling constant. By performing \pIQ, one finds that the \npf{2} at one-loop has two contributions, one coming from planar diagrams and one from non-planar diagrams (see \figref{gnc_phi4}).
	\begin{Figure}%
		[label={fig:gnc_phi4}]%
		{
			The two contributions to the \npf{2} at one-loop of the \phif \eqref{eq:gnc_phi4_act} on \Moy space.
		}%
		\begin{Subfigure}
			[label={fig:gnc_phi4_plan}]
			{2}
			{
				Planar diagram
			}
			\begin{tikzpicture}[scale = 1.3]
        	\draw[black] (-.7, 0) node[left]{$\phi$} to ( .7, 0) node[right]{$\phi$};
        	\draw[black] (0, .4) circle (.4);
\end{tikzpicture}
		\end{Subfigure}%
		\begin{Subfigure}
			[label={fig:gnc_phi4_nplan}]
			{2}
			{
				Non-planar diagram
			}
			\begin{tikzpicture}[scale = 1.3]
	\draw[black] (-.7, 0) node[left]{$\phi$} to (.7, 0) node[right]{$\phi$};
	\draw[black] (-.4, 0) arc (180:8:.4);
	\draw[black] (-.4, 0) arc (-180:-8:.4);
\end{tikzpicture}
		\end{Subfigure}
	\end{Figure}
	which reads
	\begin{subequations}
		\label{eq:gnc_mix}
	\begin{align}
		\big\langle \phi(p) \phi(q) \big\rangle_{\text{1-loop}}^{\text{p}}
		&= \frac{\cC^2}{3} \, \delta(p + q) \int \tdl{4}{k} \frac{1}{k^2 + m^2}
		\label{eq:gnc_mix_p} \\
    	\big\langle \phi(p) \phi(q) \big\rangle_{\text{1-loop}}^{\text{np}}
		&= \frac{\cC^2}{6}  \, \delta(p + q) \int \tdl{4}{k} \frac{e^{i k_\mu \Theta^{\mu\nu} p_\nu}}{k^2 + m^2}
		\label{eq:gnc_mix_np}
	\end{align}
	\end{subequations}%
	where subscripts $\text{p}$ and $\text{np}$ stands for \enquote{planar} and \enquote{non-planar} respectively and $p$ and $q$ are the external (incoming and outgoing) momenta. The constant matrix $\Theta^{\mu\nu}$ corresponds to the deformation of the commuting coordinates, \ie $[x^\mu, x^\nu]_{\star_\theta} = \iCpx \Theta^{\mu\nu}$.
	
	One can analyse the behaviour of these integrals by first going into \namefont{Schwinger} parametrization $\frac{1}{k^2 + m^2} = \int_0^{+\infty} \tdl{}{\alpha} e^{-\alpha(k^2 + m^2)}$ which turns \eqref{eq:gnc_mix} into Gaussian integrals. Then, by multiplying by a regulator $e^{- 1 / (\alpha \Lambda^2)}$, where $\Lambda$ stands as a UV scale, one obtains
	\begin{subequations}
		\label{eq:gnc_mix_cut}
	\begin{align}
		\big\langle \phi(p) \phi(q) \big\rangle_{\text{1-loop}}^{\text{p}}
		&= \frac{\cC^2}{3} \, \delta(p + q) \, \left( \Lambda^2 - m^2 \log\left( \frac{\Lambda^2}{m^2}\right) + \mathcal{O}(1) \right)
		\label{eq:gnc_mix_cut_p} \\
		\big\langle \phi(p) \phi(q) \big\rangle_{\text{1-loop}}^{\text{np}}
		&= \frac{\cC^2}{6} \, \delta(p + q) \, \left( \Lambda^2_{\text{eff}} - m^2 \log\left( \frac{\Lambda^2_{\text{eff}}}{m^2}\right) + \mathcal{O}(1) \right)
		\label{eq:gnc_mix_cut_np}
	\end{align}
	with
	\begin{align}
		\Lambda^2_{\text{eff}} 
		= \frac{1}{ 1/\Lambda^2 + (p\Theta)^2/4}
		\label{eq:gnc_mix_cut_Leff}
	\end{align}
	\end{subequations}
	Here, $(p\Theta)^2 = p_\mu \Theta^{\mu\nu} p_\rho \Theta^{\rho\sigma} \metft{g}_{\nu\sigma}$. From there, one observes that the planar contribution behaves as the commutative \phif and is UV divergent. However, for non-vanishing external momentum $p$, the non-planar contribution now has a \enquote{noncommutative} regulator $\Lambda_{\text{eff}} = 4 / (p\Theta)^2$, which makes it UV finite. Whenever $p \to 0$, this regulator diverges and the non-planar contribution also diverges in a similar fashion as the planar contribution. The fact that the IR limit $p \to 0$ makes the non-planar contribution diverge in the UV is precisely what is called the \UVIR.
	
	\paragraph{}
	More details about how the mixing arises and how it is linked to a noncommutative context will be given in \secref{ncft_p4}.
\end{Emph}

\paragraph{}
Some authors have studied how the \UVIR, specific to the noncommutative case, could hint for new physics.

First, a deep insight and physical reasoning on the \UVIR was proposed in \cite{Craig_2020}. The previous authors pointed out that the \UVIR phenomenon could not be accounted for in an \eFT, as the \eFT cannot generate IR contributions from UV regularization. In some toy models, they showed how a low mass scale could arise from the regularization of a UV divergence. This could hint for a testable solution to the hierarchy problem.

Second, in the context of \nCYM, the \UVIR also arises for the gauge fields on the \Moy space. The theory of \nCYM is taken to be a $\Ug{n}$ \gT rather then $\SUg{n}$ since the \Lieat $\SUa{n}$ is not closed in the noncommutative theory, but $\Ua{n}$ is. Then, the gauge group can be split into $\Ug{n} \simeq \SUg{n} \times \Ug{1}$ and one can show that the \UVIR is actually generated by the extra $\Ug{1}$ part. The basis of emergent gravity \cite{Steinacker_2007} in this context is to interpret this extra $\Ug{1}$ part as a coupling between the $\SUg{n}$ gauge field and an emerging gravitational field.

\subsection{Quantum gravity and noncommutativity}
\label{subsec:gnc_phys_qg}
\paragraph{}
Most of the work done on \nCST arises in the context of \qG, since \nCG could be a hint for the formulation of \qG or at least for some \qG effects. The distinction that \nCST could be used to tackle beyond \gR experiments and not necessarily emerge as a full model of \qG, is the very motivation for this thesis. Indeed, most of the \nCST{s} discussed in this manuscript are not necessarily thought of as the one \sT of nature, but are rather studied for their new properties and their phenomenology.

\paragraph{}
The recent interests in \qST{s} come from its appearance in string theory \cite{Seiberg_1999}. \NCST has been shown to emerge in limiting regime of string theory and matrix (M) theory with magnetic background. More precisely, the noncommutative setting was used to compute \namefont{Veneziano} amplitude of bosonic open strings. Furthermore, a noncommutative version of \namefont{Yang–Mills} theory arises naturally when studying closed strings in some limit natural for M theory.

However, the link to other \qG models has been made. The main example is the relationship between the \nCFT on a deformation of $\Real^3$, called $\Rcl$, and the group field theory. It has first been pointed out by considering matter coupled to 3-dimensional quantum gravity \cite{Freidel_2006}, which, upon integrating out the gravity fluctuations, yields a \nCFT on $\Rcl$, interpreted as the effective theory of the dynamics of matter.

More details about the link between \nCFT and these frameworks can be found in \cite{Hersent_2023a}.

\paragraph{}
There is a well-known argument as to why the \sT structure should lose its continuous property close to the \namefont{Planck} scale. In other words, in a theory of quantum gravity, the \sT should not be a manifold.

Suppose we want to measure a \sT distance $\Delta x$ as precisely as possible. The \HUP, coming from \qM, specifies the minimal amount of energy to deliver to a region of volume $(\Delta x)^4$, explicitly $\Delta E \geqslant 1/\Delta x$, in order to have such a precision. Decreasing the distance $\Delta x$ will thus make the minimal energy input grow. In a continuous \sT, $\Delta x$ can be arbitrarily small. Therefore, one can reach a $\Delta x$ where the minimal energy corresponds to the \Pmast $\Pmas $. For this distance $1/\Pmas$, a black hole is created, which hampers one to probe smaller distances. Therefore, by mixing \qM and \gR, one ends up in a \sT with a smallest measurable distance\footnote{
	One should note that this argument involves classical and quantum properties of space-time. Therefore, a black hole is not expected to pop up \emph{in reality}. This argument is just here to put forward inconsistencies when merging both models.
	
	Besides, one could also argue that such a black hole evaporates. However, evaporation would make the energy flow away and precision is lost. A more striking argument would be to consider the evaporating time, which writes in this case $t_{\text{e}} = 5120\pi \frac{G^2}{c^4 \hbar} \Pmas^3 = 1280\pi \frac{\Plen}{c} \sim 10^{-40}$ s, where $\Plen$ is the \Plent and $c$ is the speed of light. This consists thus also in a smallest scale of time.
}.

To avoid generating a black hole when doing high resolution microscopy, \namefont{Doplicher}, \namefont{Fredenhagen} and \namefont{Roberts} \cite{Doplicher_1995} developed a \sT framework which generates \sT uncertainties of the form
\begin{align}
	\Delta x^\mu \Delta x^\nu
	\geqslant \frac{1}{2} |\Theta^{\mu\nu}|.
	\label{eq:gnc_phys_Mup}
\end{align}
This implies that $\Delta x^\mu$ and $\Delta x^\nu$ cannot be arbitrarily small independently\footnote{
	\Aprio this does not prevent one from having an arbitrarily small distance, only to have arbitrarily small area. Indeed, one can take $\Delta x^\mu$ to be arbitrarily small if $\Delta x^\nu$ is very large. The set-up for minimal length does rather appear in the framework of \gUP \cite{Kempf_1995}. We refer to \secref{qg_ph} for more details on \qG phenomenology.
}.
There is a minimal area, modelled by $\Theta$, that one can probe in such a \sT. This relation could be enforced by considering a \qST satisfying
\begin{align}
	[x^\mu, x^\nu]
	= \iCpx \Theta^{\mu\nu}
	\label{eq:gnc_phys_Moy}
\end{align}
which corresponds to the \Moy space (see \secref{ncft_Moy} for details).

\paragraph{}
The previous framework showed how a \dpt (here $\dpMoy$) can be introduced as a physical limit for a coherent approach of both \qM and \gR. The \dpt plays the role of a new dimensionful physical constant the \sT has to accommodate for. This approach features new possibilities for a \sT \enquote{beyond \gR}, but is rather arbitrary in the sense that this new physical quantity is magically imposed. 

However, one could also introduce the \dpt directly through the symmetries of the \sT by deforming the classical symmetries, via \qg{s}. In doing so, the \dpt becomes the scale at which these deformed symmetries should replace the usual symmetries. For example, the \kMt space-time, a deformation of the \Minkt space-time, has \kPt symmetries, which correspond to a deformation of the \Pogt. In this space-time the \dpt is $\dpkM$ and has the dimension of an energy. Therefore, considering a \nCFT on \kMt only makes sense only at energies similar to $\dpkM$, or above. For energies far below $\dpkM$, one can  consider a field theory on the \Minkt space-time.

The \kMt spaces and the \kPt algebra are presented in detail in \secref{kM_kP}. Still, we want to emphasise here that the \dpt actually indicates whether the \sT is to be regarded as noncommutative or not, and the symmetries as deformed or not. This is exactly the behaviour expected for a theory of quantum gravity that does stand at the \namefont{Planck} scale, but from which one can recover \gR at low energies.

\paragraph{}
Finally, the physics of \nCST has sparked a tremendous amount of phenomenological framework, most of them handling around \qG \cite{Addazi_2022}. Here, we cite the main examples as a motivation, but more details about \qG phenomenology will be given in \secref{qg_ph}.

First, if one considers a \qST with deformed symmetries, then the classical symmetries would no more stand and thus be violated above some scale. Therefore, the \qg framework allows one to consider deformations of the \Logt and leads to \LIV. The phenomenological framework of \LIV is used both in the search of \qG and is some \SM extensions. The main advantage of using deformations through \qg{s}, in this context, is that one can recover the usual \namefont{Lorentz} symmetry of the \SM and of local \gR by considering a scale far below the scale induced by the \dpt.

The noncommutativity of the \sT can also affect the dispersion relation of a particle. The propagation of a single particle is given by the d'Alembertian operator\footnote{
	In a mathematical context, the d'Alembertian corresponds to the Laplacian in a pseudo-Riemannian context (that is on a manifold having a time coordinate). The d'Alembertian is thus sometimes called Laplacian even when the space considered is pseudo-Riemannian, \ie when the considered object is a \sT.
}
on the considered \sT. From the point of view of infinitesimal quantities (that is the point of view of the \Lieat rather then the \Liegt), the d'Alembertian is given by the Casimir operator. Moreover, in a physical context, this d'Alembertian can be written in terms of the \sT symmetries. Therefore, when considering deformed symmetries via the formalism of \qg{s}, the Casimir operator is also deformed, which means that the particles propagate differently in a \qST. The deformation of the particle propagation is studied in the context of \mDR{s}.

The \Logt symmetry was first implemented in special relativity in order to have invariance under the change of inertial frame and conservation of the speed of light. If one wants further that an energy scale is conserved under the change of \enquote{inertial frame}, where inertial has now a new meaning, one should also consider deformed symmetries. The framework, in which two observer-independent upper scale, one of speed, one of energy, are implemented, is called \dSR\footnote{
	It was also named deformed special relativity.
} \cite{Amelino-Camelia_2002}.
It was first set up to ease a mismatch between the contraction and dilatations of length in a special relativity context and the hypothetical minimal \Plent of \qG. In this context, one cannot find an alternative \Liegt to the \Logt in order to have such a frame transformation. Still, one can realise a \dSR via the \Hfat deformation of the \Logt or of the \Pogt, as for example \kPt.

The minimal length scenario was also not compatible with the \HUP and therefore led to consider \gUP{s}. A \gUP takes the form
\begin{subequations}
	\label{eq:gnc_phys_gup}
\begin{align}
	\Delta x \Delta p
	\geqslant \frac{\hbar}{2}\, \big(1 + f(p^2) \big),
	\label{eq:gnc_phys_gup_p}
\end{align}
where $f$ is some smooth function vanishing at zero. This relation can follow from a deformation of the commutator
\begin{align}
	[x, p]
	= \iCpx \hbar\, \big(1 + f(p^2) \big),
	\label{eq:gnc_phys_gup_c}
\end{align}
\end{subequations}
which, in turn, can be realised in a \nCG.

Finally, other phenomenological frameworks can be inherited from a \nCG. Most of the time, the noncommutativity of coordinates \eqref{eq:gnc_found_qst} also implies the noncommutativity of the law for momentum addition. This is made explicit in the \secref{ncft_p4}. This noncommutativity of momenta implies that the momentum space is curved. It is also believed that \CPTV arises in a noncommutative momentum framework, but this claim is still debated. Indeed, in such a phase space, there are ambiguities in defining the discrete symmetries, \ie the charge conjugation C, the parity P and the time reversal T.

\subsection{When objects are fuzzy}
\label{subsec:gnc_phys_fuz}
\paragraph{}
The \nCG framework handles both continuous and discrete geometries so that a \nCST can be continuous, discrete or both. This \enquote{undetermined shape} of a \nCG pushed the literature toward calling the \qST{s} \enquote{fuzzy space-times}. This denomination can actually make sense physically and mathematically. But before addressing these points let us discuss the example of $\Rcl$.

\begin{Emph}{Description of $\Rcl$}
\noindent%
\begin{minipage}{.49\textwidth}
	\paragraph{}
	One can define the space $\Rcl$ to be a pile of matrices of different sizes, that is 
	\begin{align}
		\Rcl
		= \bigoplus_{n = 0}^{\infty} \Mat{n}{\Cpx}
		\label{eq:gnc_R3l}
	\end{align}
	where $\Mat{n}{\Cpx}$ corresponds to the space of $n \times n$ matrices with complex entries. In terms of \nCG, the space of matrices $\Mat{n}{\Cpx}$ is more often called the fuzzy sphere. Therefore, $\Rcl$ is a pile of fuzzy sphere and was sometimes dubbed the \enquote{fuzzy onion}.
\end{minipage}%
\hfill%
\begin{minipage}{.49\textwidth}
	\begin{Figure}
		[label={fig:R3l}]%
		{
			A schematic view of $\Rcl$.
		}%
		\begin{tikzpicture}[scale=.4]
	\draw[->, dashed] (-4,0) -- (4,0) node[above right] {$x^1$};
	\draw[->, dashed] (0,-4) -- (0,4) node[above right] {$x^2$};
	\foreach \x in {.05, 1, 2, 3}
		\draw[] (0,0) circle (\x);
\end{tikzpicture}
	\end{Figure}
\end{minipage}

As a deformation of $\Real^3$, the latter space has another definition in terms of \Lieat of coordinates
\begin{align}
	[x^j, x^k]_{\star_\lambda}
	= \iCpx \dpRcl \, \tensor{\varepsilon}{^{jk}_l} \, x^l
	\label{eq:gnc_R3l_coord}
\end{align}
where $\dpRcl$ is the deformation parameter and $\varepsilon$ is the \namefont{Levi-Civita} fully anti-symmetric tensor. The \Lieat \eqref{eq:gnc_R3l_coord} corresponds to $\SUa{2}$. Considering one of the position operators $x^j$, its spectrum is made of the integers $n \in \NInt$, where the different $n$ correspond to the sizes of the fuzzy spheres that $x^j$ crosses. However, each sphere has its own continuous $\SOg{3}{\Real}$ symmetry stemming from the action of $\SUg{2}$ on $\SUa{2}$. Therefore, this space contains both discrete and continuous structures.

In view of the discussion of \subsecref{gnc_phys_qg}, one can note that the \dpt $\dpRcl$ has a length dimension.
\end{Emph}

\paragraph{}
Beyond continuity and discreteness, the space-time uncertainty relations \eqref{eq:gnc_phys_Mup} implies that one cannot measure a \sT point as accurately as one wants. Therefore, at small scales, the \qST becomes blurry because it cannot be resolved with infinite precision. This is roughly the idea of the fuzziness of a \qST.

\section{Spectral triple}
\label{sec:gnc_st}
\paragraph{}
Given a space with topology, as depicted in \secref{gnc_found}, we further want to have a differential structure to generate essential elements of geometry, like the metric or the curvature. \namefont{Kac}'s drum (see \secref{gnc_found}) asked whether one could build the differential structure using only the (spectrum of the) Laplacian\footnote{
	The Laplacian corresponds to the d'Alembertian in a pseudo-Riemannian context, that is considering the geometry of a \sT.
}.
It turns out not to be possible, but, by considering the \enquote{square root} of the Laplacian, \ie the \Do, \namefont{Connes} was able to reconstruct the geometry. However, in order to define the \Do, one needs additionally a \Hsp on which the algebra is represented. Therefore, the geometry is gathered in three objects forming a \st: the algebra (of smooth functions), the \Do and the \Hsp.

The \st formalism is presented briefly here. A curious reader can find much more details in the following references. The main writing on \nCG through \st is \namefont{Connes}' book \cite{Connes_1994} which gives a full overview of the philosophy of this construction. Other physics oriented textbooks can be consulted like \cite{Landi_1997, Gracia-Bondia_2001, Varilly_2006}.

\paragraph{}
Here, we define the \st and gather a more extensive discussion on the motivations after.

\begin{Def}{Spectral triple}
	{gnc_st}
	A \st is a set of three data $(\algft{A}, \Hilbft{H}, \Dirop)$, where 
	\begin{nclist}
		\item $\algft{A}$ is a (unital) \Csalg,
		\item $\Hilbft{H}$ is a \Hsp, such that the algebra $\algft{A}$ represents faithfully on $\Hilbft{B}(\Hilbft{H})$, the set of bounded operators on $\Hilbft{H}$,
		\item $\Dirop$ is a self-adjoint operator on $\Hilbft{H}$, with compact resolvent and satisfying $[\Dirop, f] \in \Hilbft{B(H)}$ for any $f \in \algft{A}$.
	\end{nclist}
	The latter \st is said \emph{odd} as such and is further said \emph{even} given that there exists a $\Int_2$-grading on $\Hilbft{H}$, that is a map $\gamma: \Hilbft{H} \to \Hilbft{H}$ satisfying $\gamma^\dagger = \gamma$ and $\gamma^2 = 1$, such that
	\begin{subequations}
		\label{eq:gnc_st}
	\begin{align}
		[\gamma, f] = 0, &&
		\{\gamma, \Dirop\} = 0,
		\label{eq:gnc_st_z2}
	\end{align}
	for any $f \in \algft{A}$.
	
	Furthermore, we say that the \st is \emph{real} if there exists an anti-linear isometry $J: \Hilbft{H} \to \Hilbft{H}$ such that
	\begin{align}
		J^2 = \pm 1, &&
		J \Dirop = \pm \Dirop J, &&
		J \gamma = \pm \gamma J,
		\label{eq:gnc_st_real}
	\end{align}
	where the $+$ and $-$ are determined by the so-called KO-dimension of the triple.
	
	Finally, the \st should satisfy the zeroth-order condition
	\begin{align}
		[f, J g^\dagger J^{-1}] = 0,
		\label{eq:gnc_st_0th}
	\end{align}
	and the first-order condition
	\begin{align}
		\big[ [\Dirop, f], J g^\dagger J^{-1} \big] = 0,
		\label{eq:gnc_st_1st}
	\end{align}
	for any $f, g \in \algft{A}$.
	\end{subequations}
\end{Def}

\paragraph{}
There are many comments to be made about this definition, which are summarised below.
\begin{nclist}
	\item There is \aprio no need of further structure for $\algft{A}$ than being an \opalg, \ie a \Csalg or a \namefont{von Neumann} algebra. This requirement is made in order that $\algft{A}$ can be represented on a \Hsp $\Hilbft{H}$, and therefore for $\Hilbft{H}$ to exist. However, in some \nCST $\algft{A}$ is not a \Csalg itself but close to it, like a pre-\Csalg or a dense $*$-subalgebra of a \Csalg.
	\item The \st formalism was designed first for compact manifold, which corresponds to unital algebras. \nameref{thm:gnc_st_rt}, discussed below, was only proved in the compact case. Therefore, one can ask for $\algft{A}$ to have a unit, since unital algebras corresponds to compact spaces. The \st formulation above is the one used in the non-compact case.
	\item The \Hsp $\Hilbft{H}$ is necessary to define the \Do and so the differential structure. Therefore, it is defined as a representation of our \nCG $\algft{A}$ via $\pi: \algft{A} \to \Hilbft{B(H)}$. As already mentioned, $\pi$ is always implied in notations.
	\item The representation $\pi$ above is required to be faithful so that information stored in $\algft{A}$ are not lost when working with only operators on $\Hilbft{H}$.
	\item Self-adjointness (or Hermitianity) is required in a physical context as observables correspond to self-adjoint operators in \qM. As the \Do contains the gauge fields, it is required to be self-adjoint. For example, in particle physics, the \Do corresponds to the covariant derivative $\Dirop = i \gamma^\mu(\partial_\mu + A_\mu)$, where $A_\mu$ is the gauge field.
	\item The compact resolvent requirement states that the operator $(\Dirop - z)^{-1}$ is compact, for any $z \nin \Real$, in the case of a unital algebra $\algft{A}$ (and so a compact geometry). This condition is changed to $f(\Dirop - z)^{-1}$ been compact, for any $z \nin \Real$ and $f \in \algft{A}$, in the non-unital case.
	\item In the case of a spin manifold, \ie a space on which one can define fermonic (spinor) fields, one can write down
	\begin{align}
		[\Dirop, f] = - i\, c(\td f)
		\label{eq:gnc_st_Dborn}
	\end{align}
	where $c$ is the \namefont{Clifford} action. As the action is merely a left multiplication, equation \eqref{eq:gnc_st_Dborn} motivates both the requirement of boundedness for $[\Dirop, f]$ and the basic property of the quantised calculus that the differential $\td$ can be expressed through $[\Dirop, \cdot]$ in \eqref{eq:gnc_st_qc_diff}. Moreover, in the commutative case, the condition \eqref{eq:gnc_st_Dborn} is also linked to the smoothness of $f$.
	\item The grading map $\gamma$ is actually quite common for particle physicists since it plays the exact same role as $\gamma^5$. The chirality of a particle is actually a $\Int_2$-grading that splits the space into positive and negative chiralities. Here, $\gamma$ plays a similar role (and so is often called the chirality) as one can define $\Hilbft{H}^\pm$ depending if an element of $\Hilbft{H}$ will have eigenvalue $+1$ or $-1$ for $\gamma$. The grading map is here to ensure the orientability of underlying manifold.
	\item Similarly that bosons commute (with $\gamma^5$) and fermions anti-commute (with $\gamma^5$), the condition \eqref{eq:gnc_st_z2} states that space elements do not change the chirality but the \Do does.
	\item The element $J$, often called the \namefont{Tomita} involution, can be viewed as a generalisation of the complex $i$ to operators on a \Hsp. Therefore, it is used to disjoint \enquote{real} and \enquote{imaginary} quantities. It corresponds to the isometry part (in the polar decomposition) of the operator induced by the involution ${}^\dagger$ of $\algft{A}$. Note that there is a way to construct such a $J$ operator, satisfying \eqref{eq:gnc_st_0th}, through \namefont{Tomita}'s theorem.
	\item In the commutative case, one only requires that $g = Jg^\dagger J^{-1}$ so that the zeroth-order condition \eqref{eq:gnc_st_0th} is automatically satisfied.
	\item The first-order condition \eqref{eq:gnc_st_1st} states that the \Do is a first order differential operator\footnote{
		This can be compared to an equivalent relation with the Laplacian $\Delta$, which is of second order: $\big[[\Delta, f], g \big] = \metft{g}^{-1}(\td f, \td g)$, where $\metft{g}$ is the metric.	
	}.
	This expression also allows for the quantised calculus expressions like $f_1 [\Dirop, f_2] \cdots [\Dirop, f_{n+1}]$ to be coherent (as a representation of $\algft{A}^{\otimes (n+1)}$) and so to form a differential calculus over $\algft{A}$. 
\end{nclist}

\paragraph{}
The main motivation to study a \st as a \enquote{\nCG} is given by the following \cite{Connes_2013}
\begin{Thm}{\namefont{Connes}' reconstruction theorem}
	{gnc_st_rt}
	Let $(\algft{A}, \Hilbft{H}, \Dirop)$ be a \st satisfying
	\begin{clist}
		\item $\algft{A}$ is commutative.
		\item \label{it:gnc_st_rt_eig}
		Let $\lambda_0 \geqslant \lambda_1 \geqslant \cdots$ denotes the eigenvalues of $\vert \Dirop \vert^{-1}$, then it exists $n \in \NInt$, such that $\lambda_k = \BigO[k \to +\infty]{k^{-1/n}}$.
		\item \label{it:gnc_st_rt_dom}
		For any $f \in \algft{A}$, $f, [\Dirop, f] \in \Dom(\delta^j)$, for $j \in \nzNInt$ and $\delta(T) = [\vert \Dirop \vert, T]$.
		\item The triple is either odd or even (and if so $\gamma$ is defined).
		\item \label{it:gnc_st_rt_Hinf}
		The space of smooth vectors $\Hilbft{H}^\infty = \bigcap \limits_{j \in \nzNInt} \Dom(\Dirop^j)$ is a finitely generated projected $\algft{A}$-module with a hermitian pairing, given by the noncommutative integral.
	\end{clist}
	Then, there exists a compact oriented smooth Riemannian manifold $\manft{M}$ such that $\algft{A} = \func^\infty(\manft{M})$.
\end{Thm}
Beyond the very mathematical requirements that we try to clarify below, \nameref{thm:gnc_st_rt} states that a commutative \st corresponds to a classical geometry. Therefore, a \nCG would be given by a noncommutative \st.

In the above requirements, \ref{it:gnc_st_rt_eig} states that the eigenvalues of the inverse modulus of the \Do $\vert \Dirop \vert^{-1}$ are decreasing to zero as $k^{-1/n}$ for some $n$. As explained below with the distance formula \eqref{eq:gnc_st_dist}, $\vert \Dirop \vert^{-1}$ can be interpreted as a length element \enquote{$\td x$}. Therefore, $\vert \Dirop \vert^{-n}$, corresponding to a volume element in $n$-dimension $(\td x)^n$, has eigenvalues decreasing as $1/k$. The $n$ thus corresponds to the usual space dimension in commutative geometry\footnote{
	One notices that, in a more general context, this $n$ needs not be an integer. Part of the motivation for studying \nCG, was that it could depict non-integer dimensional space, like fractals.
}
and is called the metric dimension of the \st.

The condition \ref{it:gnc_st_rt_dom} requires that $f$ and $[\Dirop, f]$ are smooth, since one can derivate them an arbitrary number of times (corresponding to $j$ in the theorem) with the derivative $\delta$.

Finally, the requirement \ref{it:gnc_st_rt_Hinf} imposes that quantities such as $f \vert \Dirop \vert^{-n}$ are measurable and so that the noncommutative integral is well-defined. Indeed, $f \vert \Dirop \vert^{-n} = f (\td x)^n$ represents the volume element of the space on which we want to sum. But, considering \ref{it:gnc_st_rt_eig}, the sum of eigenvalues are logarithmically divergent $\sum \limits_{k < N} \lambda_k \sim \sum \limits_{k < N} \frac{1}{k} \sim \log(N)$, and so $\int f \vert \Dirop \vert^{-n}$ should also diverge logarithmically. Convergence of this quantity is ensured precisely on $\Hilbft{H}^\infty$, which plays a similar role that of the test functions for the distributions. 

Note that \nameref{thm:gnc_st_rt} can adapt to the case of a spin manifold.

\paragraph{}
Now that we have defined the \st and explained how it relates to the usual notion of geometry, we can go to the missing part of what makes this a \nCG, that is the differential structure. In order to do so, one needs a differential operator $\td$ that sends a $n$-form to a $n+1$-form, and the latter set of forms.

Inspired by the relation \eqref{eq:gnc_st_Dborn}, one can define the differential $\td$, by
\begin{align}
	\td f = [\Dirop, f]
	\label{eq:gnc_st_qc_diff}
\end{align}
for any $f \in \algft{A}$. The formula \eqref{eq:gnc_st_qc_diff} can be viewed as the equation of motion in classical Hamiltonian mechanics, where the \namefont{Poisson} bracket has been replaced by the commutator $[\cdot, \cdot]$. From there one defines a generic one-form $\omega \in \Omega^1(\algft{A})$ by
\begin{align}
	\omega = f [\Dirop, g]
	\label{eq:gnc_st_qc_1form}
\end{align}
for some $f, g \in \algft{A}$. By applying the differential $\td$ to this one-form several times, one obtains that a generic $n$-form $\omega \in \Omega^n(\algft{A})$ writes
\begin{align}
	\omega = f_1 [\Dirop, f_2] \cdots [\Dirop, f_{n+1}]
	\label{eq:gnc_st_qc_nform}
\end{align}
for any $f_1, \ldots, f_{n+1} \in \algft{A}$.

The fact that a one-form writes $\omega_{\ggft{u}} = \ggft{u} [\Dirop, \ggft{u}^\dagger]$, states that the failure of invariance of the \Do with respect to the gauge transformation $\ggft{u} \in \algft{A}$, satisfying $\ggft{u} \star \ggft{u}^\dagger = \ggft{u}^\dagger \star \ggft{u} = 1$, is a one-form:
\begin{align}
	\ggft{u} \Dirop \ggft{u}^\dagger
	= \Dirop + \omega_\ggft{u},
	\label{eq:gnc_st_gt}
\end{align}
which corresponds to the formalism of \gT.

\paragraph{}
From there, one can define the noncommutative integral (defined thanks to the \namefont{Dixmier} trace) and check that it matches the usual integration on space in the context of \nameref{thm:gnc_st_rt}. Using the noncommutative integral (or noncommutative trace) $\tr$, one can define the so-called spectral action for a fermionic field $\psi \in \Hilbft{H}$ as
\begin{align}
	S(\Dirop) 
	= \langle \psi, \Dirop \psi \rangle 
	+ \tr\left(\alpha\left(\frac{\Dirop^2}{\Lambda^2}\right)\right),
	\label{eq:gnc_st_sa}
\end{align}
where $\langle \cdot, \cdot \rangle$ is the inner product of $\Hilbft{H}$, $\alpha$ is a positive functional and $\Lambda$ is an energy scale. The tracial part of the action can be expanded by heat kernel methods (which generates an expansion over $1/\Lambda$), and gives rise to geometric invariants in the commutative setting. In other words, the spectral action displays all the gauge and diffeomorphic (\ie change of coordinate) invariant quantities that one can build within the geometry. Moreover, the expansion over the energy scale $\Lambda$ implies that one can treat \eqref{eq:gnc_st_sa} as an \eFT.

\paragraph{}
One of the key feature of the \st formalism, is that one can lift the notion of geodesic distance to the noncommutative case. Indeed, by dualising the geodesic distance formula, one obtains
\begin{align}
	\dist(\psi_1, \psi_2)
	&= \sup \left\{ |\psi_1(f) - \psi_2(f)|,\ f \in \algft{A},\ \left\Vert [\Dirop, f] \right\Vert \leqslant 1 \right\}
	\label{eq:gnc_st_dist}
\end{align}
which gives the distance between two arbitrary states $\psi_1, \psi_2 \in \Spst{\algft{A}}$. In the case of smooth functions, the expression $\left\Vert [\Dirop, f] \right\Vert$ corresponds to the maximal value of the (first) derivative of $f$. This is called the \namefont{Lipschitz} norm and can be defined for non-differentiable functions also. A direct consequence is that one can interpret the inverse modulus of $\Dirop$, noted $|\Dirop|^{-1}$, as a length element $\td x$.

\section{Quantum groups}
\label{sec:gnc_qg}
\paragraph{}
There are several definitions of \qg{s} coming from different approaches. \namefont{Drinfel'd} and \namefont{Jimbo} developed the construction of deformation of universal enveloping algebras of any semi-simple \Lieat. At about the same time, \namefont{Woronowicz} made his theory of compact matrix quantum groups. Moreover, an algebraic approach to quantised coordinate algebra was introduced by \namefont{Manin}. Even at this date, there is no global definition for what \qg{s} are, but as pointed out in \cite{Klimyk_1997}: \enquote{Instead of searching for a rigorous definition of a quantum group it seems to be more fruitful to look for classes of Hopf algebras that give rise to a rich theory with important applications and contain enough interesting examples.} Therefore, we will rather explain here the philosophy of \Hfat{s} and how they can be related to physics.

\paragraph{}
The usual textbooks on \Hfat and \qg{s} are \cite{Montgomery_1993, Majid_1995, Klimyk_1997}. The basic elements and examples of \Hfat needed for this manuscript are presented here. However, more mathematical details are postponed to \appref{ha}.

\subsection{The emergence of \namefont{Hopf} algebras}
\label{subsec:gnc_qg_emerg}
\paragraph{}
In the same spirit as in \secref{gnc_found}, we want to characterise a group $G$ through its dual, that is the space of functions over it. We denote the latter by $\dual{G}$. Physically, $G$ may be thought of as a group of symmetries. In a very similar fashion to the product \eqref{eq:gnc_found_cpdt}, one can define a product, and so an algebra structure, on $\dual{G}$ through \eqref{eq:gnc_gpf_alg_prod}. The group structure of $G$ is however not exported to $\dual{G}$ as such. As shown in \exref{gnc_gpf}, one needs to equip $\dual{G}$ with a \Hfat structure in order to read the group structure of $G$ in $\dual{G}$.

\begin{Ex}{The algebra of functions over a group}
	{gnc_gpf}
	Let $G$ be a group and let $\dual{G}$ be the algebra of complex-valued functions on $G$, \ie $f \in \dual{G}$ is a function from $G$ to $\Cpx$. $\dual{G}$ is called the dual of $G$.
	
	\paragraph{}
	The space $\dual{G}$ is canonically an algebra (see \defref{ha_as_alg}) if we consider the point-wise multiplication in $\Cpx$. Explicitly, for any $z_1, z_2 \in \Cpx$, $f_1, f_2 \in \dual{G}$ and $\ggft{u} \in G$,
	\begin{subequations}
		\label{eq:gnc_gpf_alg}
	\begin{align}
		\left. \begin{matrix}
			z_1 f_1 + z_2 f_2 : & G & \to & \Cpx \\
			& \ggft{u} & \mapsto & z_1\, f_1(\ggft{u}) + z_2\, f_2(\ggft{u})
		\end{matrix} \right]
		&&
		\text{\ie} \quad z_1\, f_1 + z_2\, f_2 \in \dual{G}, 
		\label{eq:gnc_gpf_alg_vs}
	\end{align}
	therefore, $\dual{G}$ is a vector space. Considering the product,
	\begin{align}
		(f_1 \cdot f_2)(\ggft{u})
		= f_1(\ggft{u})\, f_2(\ggft{u}),
		\label{eq:gnc_gpf_alg_prod}
	\end{align}
	\end{subequations}
	where the right-hand-side product corresponds to the product in $\Cpx$, $\dual{G}$ satisfies the definition of the algebra. The associativity property of the product is directly inherited from the associativity of the product of $\Cpx$. Note that this algebra is unital since $1(\ggft{u}) = 1$ belongs to $\dual{G}$. This algebra structure, however, does not allow one to characterise $G$ from $\dual{G}$ as it does not involve the product of $G$. In other words, only knowing the algebra structure of $\dual{G}$ one does not have all the information needed to reconstruct $G$.
	
	\paragraph{}
	Let $\Delta:\dual{G} \to \dual{(G \times G)}$ be the function defined as, for any $f \in \dual{G}$ and $\ggft{u}_1, \ggft{u}_2 \in G$,
	\begin{subequations}
		\label{eq:gnc_gpf_coa}
	\begin{align}
		\Delta(f)(\ggft{u}_1, \ggft{u}_2)
		= f(\ggft{u}_1 \ggft{u}_2).
		\label{eq:gnc_gpf_coa_cop}
	\end{align}
	This map is called the coproduct and allows one to get the missing information of the product of $G$. Indeed, it characterises how the function $f$ should split in $\dual{(G \times G)}$ to match the product of $G$ through \eqref{eq:gnc_gpf_coa_cop}. Considering $1$ to be the unit of $G$, one can also define a counit $\varepsilon: \dual{G} \to \Cpx$ through
	\begin{align}
		\varepsilon(f) 
		= f(1).
		\label{eq:gnc_gpf_coa_cou}
	\end{align}
	\end{subequations}
	Again, the unit of $\dual{G}$ only involves $1 \in \Cpx$ and does not resolve the unit of $G$, whereas the counit $\varepsilon$ is precisely defined to see how functions behave on $1 \in G$. As additional constraints, these two maps should satisfy, in some way, the associativity of the product in $G$ and the unit property $\ggft{u}1 = 1\ggft{u} = \ggft{u}$ of $1 \in G$, respectively. The way to do this is to see that the coproduct is coassociative \eqref{eq:gnc_gpf_coass} and the counit property \eqref{eq:gnc_gpf_cou} is satisfied. Those properties are shown in the proof below. This will form what is called the coalgebra sector of $\dual{G}$.
	
	\paragraph{}
	But every group element $\ggft{u} \in G$ has, by definition, an inverse $\ggft{u}^{-1}$. Therefore, we need another structure on $\dual{G}$ to account for the inverse. The latter map is denoted $S: \dual{G} \to \dual{G}$ and is called the antipode (or coinverse). It is defined as
	\begin{align}
		\big(S(f)\big)(\ggft{u})
		= f(\ggft{u}^{-1}),
		\label{eq:gnc_gpf_ant}
	\end{align}
	and once again it should match the inverse property $\ggft{u} \ggft{u}^{-1} = \ggft{u}^{-1} \ggft{u} = 1$ of $G$. This is implemented by requiring the $S$ satisfies the coinverse property \eqref{eq:gnc_gpf_coi}, as detailed in the proof. 
	
	Finally, one can check that the coalgebra sector is adapted to the algebra structure \eqref{eq:gnc_gpf_alg} of $\dual{G}$ through \eqref{eq:gnc_gpf_cohom}. This will be detailed in the proof below.
\end{Ex}

\begin{proof}
	We here show that some properties of \exref{gnc_gpf} are satisfied.
	
	First, let us check the coassociativity of the function $\Delta$ defined in \eqref{eq:gnc_gpf_coa_cop}. In algebraic terms it writes
	\begin{align}
		(\Delta \otimes \id) \circ \Delta
		= (\id \otimes \Delta) \circ \Delta.
		\label{eq:gnc_gpf_coass}
	\end{align}
	One computes, for any $f \in \dual{G}$, any $\ggft{u}_1, \ggft{u}_2, \ggft{u}_3 \in G$, 
	\begin{align*}
		\big(((\Delta \otimes \id) \circ \Delta)(f) \big)(\ggft{u}_1, \ggft{u}_2, \ggft{u}_3)
		&= f\big( (\ggft{u}_1\ggft{u}_2) \ggft{u}_3 \big) \\
		\big(((\id \otimes \Delta) \circ \Delta)(f) \big)(\ggft{u}_1, \ggft{u}_2, \ggft{u}_3)
		&= f\big( \ggft{u}_1 (\ggft{u}_2\ggft{u}_3) \big).
	\end{align*}
	Therefore, the equality is fulfilled thanks to the associativity of the product in $G$. Now, the counit property writes
	\begin{align}
		(\varepsilon \otimes \id) \circ \Delta
		= (\id \otimes \varepsilon) \circ \Delta
		= \id,
		\label{eq:gnc_gpf_cou}
	\end{align}
	and it is satisfied thanks to the unit property $1 \ggft{u} = \ggft{u} 1 = \ggft{u}$ of $1 \in G$. Indeed,
	\begin{align*}
		\big(((\varepsilon \otimes \id) \circ \Delta)(f) \big) (\ggft{u})
		&= f(1 \ggft{u}) \\
		\big(((\id \otimes \varepsilon) \circ \Delta)(f) \big) (\ggft{u})
		&= f(\ggft{u} 1) \\
		\big( \id(f) \big)(\ggft{u}) 
		&= f(\ggft{u}).
	\end{align*}
	In a similar fashion, the coinverse property writes
	\begin{align}
		m \circ (S \otimes \id) \circ \Delta
		= m \circ (\id \otimes S) \circ \Delta
		= \eta \circ \varepsilon
		\label{eq:gnc_gpf_coi}
	\end{align}
	where $m : \dual{(G \times G)} \to \dual{G}$ is the product of $\dual{G}$ \eqref{eq:gnc_gpf_alg_prod}, and $\eta : \Cpx \to \dual{G}$ is the unit of $\dual{G}$, with $\eta(z) = z1$ for any $z \in \Cpx$. It is satisfied, thanks to the inverse property $\ggft{u} \ggft{u}^{-1} = \ggft{u}^{-1}\ggft{u} = 1$ in $G$, through
	\begin{align*}
		\big( (m \circ (S \otimes \id) \circ \Delta) (f) \big) (\ggft{u})
		&= f(\ggft{u}^{-1} \ggft{u}) \\
		\big( (m \circ (\id \otimes S) \circ \Delta) (f) \big) (\ggft{u})
		&= f(\ggft{u} \ggft{u}^{-1}) \\
		\big( (\eta \circ \varepsilon) (f) \big) (\ggft{u})
		&= f(1).
	\end{align*}
	
	Second, the consistency relations for $\Delta$ and $\varepsilon$ with the algebra $\dual{G}$ state that $\Delta$ and $\varepsilon$ are algebra homomorphisms, that is, for any $f_1, f_2 \in \dual{G}$
	\begin{align}
		\Delta(f_1 f_2) = \Delta(f_1) \Delta(f_2), &&
		\varepsilon(f_1f_2) = \varepsilon(f_1) \varepsilon(f_2).
		\label{eq:gnc_gpf_cohom}
	\end{align}
	Considering any $\ggft{u}_1, \ggft{u}_2 \in G$, one has, using the product definition \eqref{eq:gnc_gpf_alg_prod},
	\begin{align*}
		\big( \Delta(f_1 f_2) \big) (\ggft{u}_1, \ggft{u}_2)
		= (f_1 f_2)(\ggft{u}_1 \ggft{u}_2)
		= f_1(\ggft{u}_1 \ggft{u}_2) &\, f_2(\ggft{u}_1 \ggft{u}_2)
		= \Delta(f_1)(\ggft{u}_1, \ggft{u}_2) \, \Delta(f_2)(\ggft{u}_1, \ggft{u}_2), \\
		\varepsilon(f_1 f_2)
		= (f_1 f_2)(1)
		= f_1(1) &\, f_2(1)
		= \varepsilon(f_1) \varepsilon(f_2).
		\qedhere
	\end{align*}
\end{proof}

\begin{Rmk}{}
	{gnc_ha}
	If one wants to have a fully algebraic version of a group $G$ as depicted in \exref{gnc_gpf}, one should have an abstract definition of $\dual{G}$ without the mention of $G$. This is done in the previous example, through the structure of \Hfat, to the exception of the coproduct. In \eqref{eq:gnc_gpf_coa_cop}, the coproduct $\Delta$ has an image in $\dual{(G \times G)}$, but from the point of view of the \Hfat, its image should be in $\dual{G} \otimes \dual{G}$. This distinction is also present for the product of $\dual{G}$ but it was not made explicit in the example.
	
	In general, one has the inclusion 
	\begin{align}
		\dual{G} \otimes \dual{G}
		\subset \dual{(G \times G)},
		\label{eq:gnc_ha_mis}
	\end{align}
	so that the \Hfat structure is more restrictive than the algebraic version we are looking for. Even if the amount of missing information is not quantified, it is not considered to be pathological in the literature.
	
	As a side remark, let us observe that the equality in \eqref{eq:gnc_ha_mis} stands when $G$ is finite. 
\end{Rmk}

\subsection{The structure of \namefont{Hopf} algebras}
\label{subsec:gnc_qg_struct}
\paragraph{}
Considering the construction of $\dual{G}$ above, the \qg formalism studies \Hfat and considers it mimics $\dual{G}$ as a space of functions over some, at this point undetermined, deformed version of some group $G$.  

\paragraph{}
In this context, a \qg is therefore a \Hfat $(\Hoft{H}, \cdot, 1, \Delta, \varepsilon, S)$, generically noted only $\Hoft{H}$. The product $\cdot$ makes $\Hoft{H}$ an algebra, with a unit $1$. The coproduct $\Delta : \Hoft{H} \to \Hoft{H} \otimes \Hoft{H}$ decomposes an element into its product constituents, as expressed by the \namefont{Sweedler} notations
\begin{align}
	\Delta(X)
	= \sum X_{(1)} \otimes X_{(2)}
	\label{eq:gnc_sweedn} 
\end{align}
for any $X \in \Hoft{H}$. In \eqref{eq:gnc_sweedn}, $X_{(1)}, X_{(2)} \in \Hoft{H}$ are the constituents of $X$. The sum symbol is here to express that several such decompositions may exist and the coproduct corresponds to the sum of all\footnote{
	This vague sentence is to be thought in a broad sense. We present below the multi-particle interpretation of the coproduct, because of this manuscript interest. However, other physical interpretations can be made of the coproduct, like in statistical physics. We refer to \cite{Majid_1995} for more insights on this topic.
}.
In the context of symmetries, one can interpret the coproduct of a symmetry as its action on a multi-particle state. For example, if one considers the angular momentum operator $J_z$, acting on a two-particle state $\ket{\psi_1} \otimes \ket{\psi_2}$, one has that
\begin{subequations}
	\label{eq:gnc_cop_amo}
\begin{align}
	J_z \big(\ket{\psi_1} \otimes \ket{\psi_2}\big)
	= J_z(\ket{\psi_1}) \otimes \ket{\psi_2}
	+ \ket{\psi_1} \otimes J_z(\ket{\psi_2}),
	\label{eq:gnc_cop_amo_phys}
\end{align}
which can be recovered from
\begin{align}
	\Delta (J_z) 
	= J_z \otimes 1 + 1 \otimes J_z.
	\label{eq:gnc_cop_amo_math}
\end{align}
\end{subequations}
The counit $\varepsilon : \Hoft{H} \to \Cpx$ is, in some sense, the \enquote{unit} of the coproduct. From the multi-particle state interpretation above, the counit is \enquote{suppressing} a particle in the sense that one goes from a symmetry in a $n$-particle state to a symmetry in a $(n-1)$-particle state by applying the counit.

The coassociativity property \eqref{eq:gnc_gpf_coass} states that, whenever $\Delta$ is applied to \eqref{eq:gnc_sweedn}, the choice of applying it on $X_{(1)}$ (\ie to the left) or on $X_{(2)}$ (\ie to the right) does not matter. In terms of multi-particle state interpretation, one can construct the action of a symmetry on a $n$-particle state via its action on a $(n-1)$-particle state. For example, from \eqref{eq:gnc_cop_amo_math}, one has
\begin{align*}
	\Delta^2(J_z)
	&= \Delta(J_z) \otimes 1 + \Delta(1) \otimes J_z
	= J_z \otimes \Delta(1) + 1 \otimes \Delta(J_z) \\
	&= J_z \otimes 1 \otimes 1 + 1 \otimes J_z \otimes 1 + 1 \otimes 1 \otimes J_z.
\end{align*}
The counit property \eqref{eq:gnc_gpf_cou} states also that applying $\varepsilon$ to the left or right of \eqref{eq:gnc_sweedn} gives the same result. When considering multi-particle states, this means that one can \enquote{suppress} any of the particle in a $n$-particle state.

Finally, the antipode $S : \Hoft{H} \to \Hoft{H}$ corresponds to a generalised inverse. The coinverse property \eqref{eq:gnc_gpf_coi} imposes that $S$ is actually a generalised version of the inverse, be it a right or left inverse. More precisely, with the \namefont{Sweedler} notation \eqref{eq:gnc_sweedn}, one has
\begin{align}
	\sum S(X_{(1)}) X_{(2)}
	= \sum X_{(1)} S(X_{(2)})
	= \varepsilon(X) 1.
	\label{eq:gnc_sweedn_coi}
\end{align}

We refer to \defref{ha_as_ha} for a more detailed definition of the \Hfat structure.

\paragraph{}
We now give a useful example of the construction of a \Hfat.

\begin{Ex}{The \namefont{Hopf} algebra of $U(\Lieft{g})$}
	{gnc_uealg}
	Consider $\Lieft{g}$ to be a semi-simple \Lieat. This construction is made for any semi-simple \Lieat, but one can consider $\Lieft{g}$ to be the \Lieat of \vf{s} in the context of \gR, or as infinitesimal symmetry transformations associated to some \Liegt.
	We will study the enveloping algebra $U(\Lieft{g})$ of $\Lieft{g}$. The latter corresponds to the algebra such that, for any $X, Y \in U(\Lieft{g})$,
	\begin{align*}
		[X, Y] = XY - YX
	\end{align*}
	where $[\cdot, \cdot]$ is the bracket product of $\Lieft{g}$. In general, $U(\Lieft{g})$ is wider then $\Lieft{g}$. As an example, the product of two \vf{s} is not a \vf because it does not obey the \Lru, only the commutator is. The enveloping algebra of \vf{s} then corresponds the set of all objects generated when multiplying \vf{s}.
	
	\paragraph{}
	From the algebra $U(\Lieft{g})$, one can construct a \Hfat by defining the following maps, for any $X \in U(\Lieft{g})$,
	\begin{align}
		\Delta(X)
		= X \otimes 1 + 1 \otimes X, &&
		\varepsilon(X)
		= 0, &&
		S(X)
		= - X.
		\label{eq:gnc_uealg_com}
	\end{align}
	From there, one can check that the \Hfat axioms are satisfied.
\end{Ex}

\subsection{The quantum space-time}
\label{subsec:gnc_qg_qst}
\paragraph{}
If the above \qg stands as the quantum symmetries, then what is the \qST having such symmetries? The answer lies again in the classical case and from which all the algebraic structures are exported. For a symmetry $\ggft{u} \in G$, one goes from a \sT point $x \in \manft{M}$ to its symmetric $y \in \manft{M}$ by acting with the transformation: $y = \ggft{u} \actl x$. Then, any \sT function $f \in \func^\infty(\manft{M})$ transform under $G$ via $(\ggft{u} \actl f)(x) = f (\ggft{u} \actl x)$. This corresponds to a module structure. We say that $\func^\infty(\manft{M})$ is a $G$-module. In addition, $\func^\infty(\manft{M})$ has an algebra structure thanks to the product \eqref{eq:gnc_found_cpdt}. This makes $\func^\infty(\manft{M})$ a $G$-module algebra.

\paragraph{}
More explicitly, given a \Hfat $\Hoft{H}$, a $\Hoft{H}$-module $\algft{A}$ is a vector space with a linear action $\actl : \Hoft{H} \otimes \algft{A} \to \algft{A}$ satisfying\footnote{
	Note that we define here a left action, but one can define similarly a right action. See \defref{ha_rt_mod}.	
}
\begin{align}
	(XY) \actl f
	&= X \actl ( Y \actl f), &
	1 \actl f &= f
	\label{eq:gnc_qg_qst_mod}
\end{align}
for any $X, Y \in \Hoft{H}$ and $f \in \algft{A}$. Physically, the action $\actl$ can be viewed as the action of a symmetry, like in \exref{gnc_qg_qst_modex}. In this picture, \eqref{eq:gnc_qg_qst_mod} states that applying the transformation $X$ and then $Y$ amounts to apply directly $XY$, and that applying the identity amounts not to transform $f$.

\begin{Ex}{The module of rotations on the Euclidean space}
	{gnc_qg_qst_modex}
	Let $\algft{A} = \func^\infty(\manft{M})$ with $\manft{M} = \Real^3$, the $3$-dimensional Euclidean space, and $X = R_z(\theta) \in \SOg{3}{\Real}$ be the rotation of angle $\theta$ around the $z$-axis. A space point, noted $x = \vec{x} = (x_1, x_2, x_3) \in \Mat{3,1}{\Real} \simeq \Real^3$, is transformed to $y = R_z(\theta) \vec{x}$ by matrix multiplication. Here, $y$ corresponds to $x$ rotated of angle $\theta$ with respect to the origin.	
	
	Let us consider observables of $\Real^3$, that is $f \in \func^\infty(\Real^3)$. $f$ could be any measurable quantity in space, like temperature, electric charge, \etc Considering a rotation of angle $\theta$ around the $z$-axis $X = R_z(\theta)$, we define the action $\big(R_z(\theta) \actl f \big)(x) = f \big( R_z(\theta)x \big)$, which corresponds to the evaluation of $f$ at the point $R_z(\theta)x$.  The algebraic structure linking the space $\func^\infty(\Real^3)$ and the symmetries $\SOg{3}{\Real}$ via the action $\actl$ is called a module, \ie $\func^\infty(\Real^3)$ is said to be a $\SOg{3}{\Real}$-module.
	
	In this example, left hand side of \eqref{eq:gnc_qg_qst_mod} states that rotating of angle $\theta$ and then of $\tilde{\theta}$ amounts to rotating of angle $\theta + \tilde{\theta}$, \ie 
	\begin{align*}
		R_z(\tilde{\theta}) \actl ( R_z(\theta) \actl f)
		&= \big( R_z(\tilde{\theta}) R_z(\theta) \big) \actl f
		= R_z(\tilde{\theta} + \theta) \actl f.
	\end{align*}
	In the same spirit, the right hand side of \eqref{eq:gnc_qg_qst_mod} imposes that a rotation of angle $0$ does not change the function $f$, \ie
	\begin{align*}
		R_z(0) \actl f
		&= 1 \actl f
		= f.
	\end{align*}
\end{Ex}

A $\Hoft{H}$-module algebra $\algft{A}$ is a $\Hoft{H}$-module and an algebra, such that both structures are compatible. More quantitatively, if $(\Hoft{H}, \cdot, 1_\Hoft{H}, \Delta, \varepsilon, S)$ is a \Hfat, then $(\algft{A}, \star, 1_\algft{A})$ is a $\Hoft{H}$-module algebra if and only if
\begin{align}
	X \actl (f \star g)
	= \sum \big( X_{(1)} \actl f \big) \star \big( X_{(2)} \actl g \big), &&
	X \actl 1_\algft{A}
	= \varepsilon(X) 1_\algft{A}
	\label{eq:gnc_qg_qst_modalg}
\end{align}
for any $f, g \in \algft{A}$ and $X \in \Hoft{H}$, where we use \namefont{Sweedler} notations \eqref{eq:gnc_sweedn}. 

\begin{Ex}{The module algebra of rotations on the Euclidean space}
	{gnc_qg_qst_modalgex}
	We consider the same situation as in \exref{gnc_qg_qst_modex}. In order to do so, one needs to complete $\SOg{3}{\Real}$ to have a \Hfat structure. This is done by considering $U(\SOa{3})$ as in \exref{gnc_uealg}. Therefore, an infinitesimal rotation $R_z$ has a coproduct and counit given by \eqref{eq:gnc_uealg_com}, and applying left hand side of \eqref{eq:gnc_qg_qst_modalg}, one obtains
	\begin{align*}
		R_z \actl (f g)
		&= (R_z \actl f) g + f (R_z \actl g).
	\end{align*}
	$R_z$ acts like a derivation, which is in agreement with the usual association of $R_z$ with a \vf in $\Real^3$: $R_z = x^1 \partial_2 - x^2 \partial_1$. The right hand side, states that
	\begin{align*}
		R_z \actl 1_\algft{A} = 0.
	\end{align*}
	This can be understood as rotating $1_\algft{A}(x) = 1$ does not change $1_\algft{A}$, \ie $1_\algft{A}(R_z(\theta)x) = 1_\algft{A}(x) = 1$, therefore any infinitesimal rotation $R_z$ of $1_\algft{A}$ vanishes.
\end{Ex}

\paragraph{}
Therefore, a \qST $\algft{A}$ having a \Hfat $\Hoft{H}$ as its space of symmetries is defined as a $\Hoft{H}$-module algebra. In order to make contact with \subsecref{gnc_qg_struct}, one can check that \eqref{eq:gnc_cop_amo_phys} is recovered from \eqref{eq:gnc_cop_amo_math} using \eqref{eq:gnc_qg_qst_modalg}.

Two main comments are in order.
\begin{nclist}
	\item One can construct a $\Hoft{H}$-module algebra canonically by considering the dual \Hfat of $\Hoft{H}$, see \eqref{eq:ha_rt_dcomod}. This is used in the context of the \kMt space-time in \secref{kM_kP}.
	\item In \eqref{eq:gnc_qg_qst_modalg}, the algebra sector of $\algft{A}$ (that is the product $\star$ and the unit $1_\algft{A}$) is linked to the coalgebra sector of $\Hoft{H}$ (that is the coproduct $\Delta$ and the counit $\varepsilon$). Therefore, the noncommutativity of $\algft{A}$ can be linked with the noncocommutativity of $\Hoft{H}$. The cocommutativity of a coproduct is defined as the interchangeability of the constituents: $\sum X_{(1)} \otimes X_{(2)} = \sum X_{(2)} \otimes X_{(1)}$.
\end{nclist}

\paragraph{}
Particular types of \Hfat have been studied in the physics literature, because of their link to physical systems or their relative simplicity.

\begin{Emph}{Braided geometry}
	The (quasi-)triangular \Hfat has an additional element called the \Rmat, generically denoted $\Rma \in \Hoft{H} \otimes \Hoft{H}$. It is invertible and must satisfy some consistency equations. When considering a \qST $\algft{A}$ with symmetries $\Hoft{H}$, \ie a $\Hoft{H}$-module algebra $\algft{A}$, one can consider braided commutativity \cite{Weber_2020b, Aschieri_2020}, that is when noncommutativity of $\algft{A}$ is controlled by the \Rmat through
	\begin{align}
		f \star g
		= \sum \big(\Rma_1^{-1} \actl g \big) \star \big(\Rma_2^{-1} \actl f \big),
		\label{eq:gnc_bnc}
	\end{align}
	where we noted $\Rma = \sum \Rma_1 \otimes \Rma_2 \in \Hoft{H} \otimes \Hoft{H}$ in the spirit of the \namefont{Sweedler} notation. The main point of this braided construction is that one can export many more commutative structures just by adapting them to be braided commutative. More details are given in \secref{ncft_Dt}.
\end{Emph}

\begin{Emph}{Bicrossproduct structure}
	One can export internal structures of groups to the \qg setting, especially the semi-direct product. Let $G$ be a group with semi-direct structure, \ie $G = G_1 \ltimes G_2$. This means that transformations with $G_2$ also involve elements of $G_1$. As an example, the \Pogt can be decomposed as $\Pog[1,3] = \Tran[1,3] \rtimes \SOg{1,3}{\Real}$, where $\Tran[1,3]$ denotes the group of translations and $\SOg{1,3}{\Real}$ the group of rotations and boosts. Using the \Hfat formalism, the semi-direct product can be generalised as the bicrossproduct \cite{Majid_1995} $\Hoft{H} = \Hoft{H}_1 \bicros \Hoft{H}_2$ via the introduction of an action $\actl$ of $\Hoft{H}_1$ on $\Hoft{H}_2$ and a coaction $\coactr$ of $\Hoft{H}_2$ on $\Hoft{H}_1$. As an example, the \kPt \Hfat is constructed as a quantum deformation of the \Pogt thanks to this bicrossproduct structure (see \secref{kM_kP} for more details).

	The \qST can emerge as the dual of $\Hoft{H}$. But \namefont{Majid} \cite{Majid_1995} has developed another idea on how geometry could emerge from this bicrossproduct structure, as the latter can be seen as the quantisation of homogeneous spaces. The semi-direct product actually appears when considering a \sT $\manft{M}$ with a \Liegt of symmetries $G$. As noted above, the symmetries $G$ acts on the \sT so that $\manft{M} \rtimes G$ and taking the dual point of view $\func^\infty(\manft{M}) \rtimes \Cpx G$, where $\Cpx G$ denotes the vector space of formal $\Cpx$-linear combinations of elements of $G$. At this point, $\func^\infty(\manft{M})$ can be turned into a noncommutative algebra $\algft{A}$ and $\Cpx G$ into a (noncocommutative) \Hfat $\Hoft{H}$, giving $\algft{A} \rtimes \Hoft{H}$. One further forms a bicrossproduct by requiring that the \sT \enquote{reacts} on the momentum space $\func^\infty(\manft{M}) \bicrosr \Cpx G$ and one exports that to the quantum setting $\algft{A} \bicrosr \Hoft{H}$.
\end{Emph}

\begin{Emph}{\namefont{Drinfel'd} twist}
	Starting from a \Hfat $\Hoft{H}$, one can construct another \Hfat $\tilde{\Hoft{H}}$ via a \Dt. The \Dt is an invertible element $\Hoft{F} = \sum \Hoft{F}_1 \otimes \Hoft{F}_2 \in \Hoft{H} \otimes \Hoft{H}$ satisfying the so-called $2$-cocycle condition \eqref{eq:gnc_dq_dt_2co} and the normalisation \eqref{eq:gnc_dq_dt_norm}. If we define 
	\begin{align}
		\Delta^{\Hoft{F}} 
		= \Hoft{F} \Delta \Hoft{F}^{-1}, &&
		S^{\Hoft{F}}
		= \chi S \chi^{-1}
		\label{eq:gnc_qg_Dt}
	\end{align}
	where $\chi = \Hoft{F}_1 S(\Hoft{F}_2) \in \Hoft{H}$, then $\tilde{\Hoft{H}} = (\Hoft{H}, \cdot, 1, \Delta^{\Hoft{F}}, \varepsilon, S^{\Hoft{F}})$ forms a new \Hfat. This is convenient in two ways. First, one can start with a very simple \Hfat structure, like the one in \exref{gnc_uealg}, and construct a more complex \Hfat using the \Dt. Second, if we consider a $\Hoft{H}$-module algebra $\algft{A}$, then by adapting its product it can be turned into a $\tilde{\Hoft{H}}$-module algebra. Explicitly, the new product of $\algft{A}$ is
	\begin{align}
		f \star_{\Hoft{F}} g
		= \sum \big( \Hoft{F}_1^{-1} \actl f \big) \star \big( \Hoft{F}_2^{-1} \actl g \big)
		\label{eq:gnc_qg_Dt_prod}
	\end{align}
where $\star$ was the previous product of $\algft{A}$. Thus, one can  consider a commutative algebra of functions over a classical \sT $\algft{A} = \func^\infty(\manft{M})$ with some symmetry group $\Hoft{H}$ (for example the \Lieat of \vf{s}) and construct a \qST with quantum symmetries by simply introducing a \Dt $\Hoft{F} \in \Hoft{H} \otimes \Hoft{H}$. See \subsecref{gnc_dq_dt} for more insights.
\end{Emph}

\section{Derivation based differential calculus}
\label{sec:gnc_dbdc}
\paragraph{}
When considering a \sT $\manft{M}$, one can define quantities through smooth functions $\func^\infty(\manft{M})$, like position. But, one needs also their derivatives, like in the computation of speed. When the \sT is curved, the speed vector does not \enquote{belong} to the \sT but to its tangent space. Moreover, comparing vectors attached to distant points is no longer straightforward. This lack of consistency generated by the geometry is in fact not a curse since it is precisely how we quantify and characterise the geometry. In other words, all the geometric information goes into \vf{s} from which we construct quantitive geometrical objects like the metric, the connection or the curvature.

Considering that we want to generalise the previous setting into a purely algebraic one, we should start by generalising the \vf{s}, from which all other quantities would follow.

\paragraph{}
The generalisation of vector fields to the non-commutative setting has been considered in several ways: via \Hfat (see section 14.1 of \cite{Klimyk_1997}), or via quantum principal fiber bundles \cite{Brzezinski_1993}, among others. The point of view taken here \cite{Dubois-Violette_2001} is to see that there is a one-to-one correspondence between the \vf{s} and the derivations over $\func^\infty(\manft{M})$, noted $\Der(\func^\infty(\manft{M}))$.

Given an algebra $\algft{A}$, a derivation $X \in \Der(\algft{A})$ over $\algft{A}$ is defined as a linear mapping satisfying the \Lru, \ie for any $f, g \in \algft{A}$,
\begin{align}
	X(f \star g)
	= X(f) \star g + f \star X(g)
	\label{eq:gnc_Lru}
\end{align}
where $\star$ denotes the product of $\algft{A}$. The derivations form a \Lieat, in the sense that given two derivations $X, Y \in \Der(\algft{A})$, then $[X, Y] = XY - YX$ is also a derivation.
\begin{proof}
	First, $[X, Y]$ is linear as a composition and sum of linear maps. One then need to check that it follows the \Lru \eqref{eq:gnc_Lru}. Let $f, g \in \algft{A}$,
	\begin{align*}
		[X, Y](f \star g)
		&= X \big( Y(f \star g) \big) - Y \big( X(f \star g) \big) \\
		&= X \big( Y(f) \star g + f \star Y(g) \big) - Y \big( X(f) \star g + f \star X(g) \\
		&\begin{aligned}
			\;=\;& XY(f) \star g + \cancel{Y(f) \star X(g)} + \bcancel{X(f) \star Y(g)} + f \star XY(g) \\
			&- YX(f) \star g - \bcancel{X(f) \star Y(g)} - \cancel{Y(f) \star X(g)} - f \star YX(g)
		\end{aligned} \\
		&= [X,Y](f) \star g + f \star [X,Y](g)
		\qedhere
	\end{align*}
\end{proof}
Furthermore, if we define the action of $\algft{A}$ on $\Der(\algft{A})$, by $f \actl X = f \star X$, then $\Der(\algft{A})$ is not an $\algft{A}$-module, since $f \star X$ does not follow the \Lru. But, as one can notice in performing this computation (done in the proof below), $f \star X$ would be a derivation if $f$ was commuting with any element of $\algft{A}$. Even if we work in a noncommutative context, it can exist elements of the algebra that commutes with all others. The set of those elements is called the center of the algebra and is denoted $\Cen{\algft{A}}$. Explicitly, $f$ is an element of the center $\Cen{\algft{A}}$ if and only if $f \star g = g \star f$ for any $g \in \algft{A}$. Thus, $\Der(\algft{A})$ is a $\Cen{\algft{A}}$-module.
\begin{proof}
	One needs to check that $f \star X \in \Der(\algft{A})$ for any $f \in \Cen{\algft{A}}$ and $X \in \Der(\algft{A})$. As above, $f \star X$ is a linear map because $X$ is. Now let us check the \Lru. Given any $g, h \in \algft{A}$,
	\begin{align*}
		(f \star X)(g \star h)
		&= f \star \big( X(g) \star h + g \star X(h) \big) \\
		&= f \star X(g) \star h + f \star g \star X(h) \\
		&= f \star X(g) \star h + g \star f \star X(h) \\
		&= (f \star X)(g) \star h + g \star (f \star X)(h)
		\qedhere
	\end{align*}
\end{proof}
One should note that, when considering the algebra of smooth functions $\algft{A} = \func^\infty(\manft{M})$, one has $\Cen{\algft{A}} = \func^\infty(\manft{M}) = \algft{A}$ since this algebra is commutative.

\paragraph{}
The \vf{s} are thus generalised, in the noncommutative context, as the derivations over the algebra. This observation is a cornerstone for the definition of some physical models developed in this manuscript. In order to build the \dbdc, one has to define the set of forms. Inspired by the commutative case, one defines a $n$-forms $\omega \in \Omega^n(\algft{A})$ as a $\Cen{\algft{A}}$-multilinear antisymmetric map from $\Der(\algft{A})^n \to \algft{A}$. This means that $\omega$ takes $n$ entries from $\Der(\algft{A})$, like $X_1, \ldots, X_n \in \Der(\algft{A})$, and sends it to $\algft{A}$:
\begin{subequations}
	\label{eq:gnc_dbdc_form}
\begin{align}
	\omega(X_1, \ldots, X_n) \in \algft{A}.
	\label{eq:gnc_dbdc_form_n}
\end{align}
The antisymmetry states that swapping two of the $n$ entries (here $X_j$ and $X_k$) generates a minus sign
\begin{align}
	\begin{aligned}
	&\omega(X_1, \ldots, X_{j-1}, X_j, X_{j+1}, \ldots, X_{k-1}, X_k, X_{k+1}, \ldots, X_n) \\
	& \quad = - \omega(X_1, \ldots, X_{j-1}, X_k, X_{j+1}, \ldots, X_{k-1}, X_j, X_{k+1}, \ldots, X_n)
	\end{aligned}
	\label{eq:gnc_dbdc_form_as}
\end{align}
for any $k, j = 1, \ldots, n$. Finally, $\Cen{\algft{A}}$-multilinearity states that any of the $n$ entries (here $X_j$) is linear for the scalars and for the action of $\Cen{\algft{A}}$:
\begin{align}
	\begin{aligned}
	&\omega(X_1, \ldots, X_{j-1}, f\star X_j + g \star Y_j, X_{j+1}, \ldots, X_n) \\
	& \quad = f \star \omega(X_1, \ldots, X_{j-1}, X_j, X_{j+1}, \ldots, X_n)
	+ g \star \omega(X_1, \ldots, X_{j-1}, Y_j, X_{j+1}, \ldots, X_n)
	\end{aligned}
\end{align}
for any $f, g \in \Cen{\algft{A}}$ and $Y_j \in \Der(\algft{A})$.
\end{subequations}

\paragraph{}
From there, we can define the (wedge) product between forms of different degrees (here $n$ and $m$) via
\begin{align}
\begin{aligned}
    (\omega \wedge \eta) & (X_1, \ldots, X_{n+m}) \\
    &= \frac{1}{n!m!} \sum_{\sigma\in \mathfrak{S}_{n+m}} (-1)^{{\sign}(\sigma)}
    \omega(X_{\sigma(1)}, \ldots, X_{\sigma(n)}) \star \eta(X_{\sigma(n+1)}, \ldots, X_{\sigma(n+m)}),
    \label{eq:gnc_dbdc_form_prod}
\end{aligned}
\end{align}
where $\omega \in \Omega^n(\algft{A})$, $\eta \in \Omega^m(\algft{A})$, $\mathfrak{S}_{n+m}$ denotes the set of permutations of $1, \ldots, n+m$ and $\sign$ stands for the signature of such a permutation. This definition is similar to the commutative one. One can notice however, the presence of the algebra product $\star$ in the expression \eqref{eq:gnc_dbdc_form_prod}. If $\star$ is noncommutative, then $\wedge$ will not be graded commutative, that is
\begin{align}
	\omega \wedge \eta
	\neq (-1)^{|\omega| \, |\eta|} \eta \wedge \omega
	\label{eq:gnc_dbdc_gdnc}
\end{align}
where $|\cdot|$ here denotes the degree of the form.

This observation is more general than the \dbdc formalism since it also occurs in other formalism of \nCG. One of the main consequence, like in any noncommutativity context, is that one has to pay attention to the order of the terms. Especially, when considering local coordinates in $(d+1)$-dimensional \sT $(x^0, \ldots, x^d)$, $d+1$-forms are expressed in the commutative case as
\begin{align}
	\omega = f \tdr{}{x^0} \wedge \cdots \wedge \td x^d
	\label{eq:gnc_dbdc_comm_form}
\end{align}
where $f$ is a smooth function. But since $\td x^j  \wedge \td x^k$ and $\td x^k \wedge \td x^j$ are a priori unrelated if $\algft{A}$ is noncommutative, then one has to take into account any permutation of the $\td x^j$ in the expression \eqref{eq:gnc_dbdc_comm_form}.

This implies that one can generate forms of arbitrary degrees. Indeed, in a commutative setting $\td x^j \wedge \td x^j = 0$ because of antisymmetry of $2$-forms due to graded commutativity. Therefore, if one wants to build a form of degrees higher then the \sT dimension, that is higher than $d+1$, one should add at least one $\td x^j$ in expression \eqref{eq:gnc_dbdc_comm_form}. However, this $\td x^j$ is already present, and from the previous argument, such a form should necessarily vanish. In the presence of a noncommutative product, the previous demonstration falls short because $\td x^j \wedge \td x^j$ can be non-zero due to \eqref{eq:gnc_dbdc_gdnc}. Thus, there is no upper bound on the degrees of non-vanishing forms.

\paragraph{}
Finally, one can jump to higher degrees by using the differential $\td: \Omega^n(\algft{A}) \to \Omega^{n+1}(\algft{A})$ which is defined, as in the commutative setting, via the \namefont{Koszul} formula
\begin{align}
\begin{aligned}
    \td \omega(X_1, \dots, X_{n+1}) 
    =& \sum_{j = 1}^{n+1} (-1)^{j+1} X_j \big( \omega( X_1, \ldots, \omitel{j}, \ldots, X_{n+1}) \big) \\
    &+ \sum_{1 \leqslant j < k \leqslant n+1} (-1)^{j+k} \omega( [X_j, X_k], X_1, \ldots, \omitel{j}, \ldots, \omitel{k}, \ldots, X_{n+1}),
    \label{eq:gnc_dbdc_koszul}
\end{aligned}
\end{align}
where $\omitel{j}$ denotes the omission of the element $X_j$. One can see that this definition does not involve the product of $\algft{A}$ and therefore, does not see the noncommutativity.

\paragraph{}
From the previous definition, two main properties arise
\begin{subequations}
	\label{eq:gnc_dbdc_prop}
\begin{align}
	\td^2 &= 0, 
	\label{eq:gnc_dbdc_prop_d2}\\
	\td(\omega \wedge \eta)
	&= \td\omega \wedge \eta + (-1)^{|\omega|}\, \omega \wedge \td\eta.
	\label{eq:gnc_dbdc_prop_der}
\end{align}
\end{subequations}
From \eqref{eq:gnc_dbdc_prop_der}, one reads that $\td$ is a graded derivation (satisfying a graded \Lru) and, from \eqref{eq:gnc_dbdc_prop_d2}, that it squares to zero. Therefore, $\td$ is a differential operator.

\paragraph{}
Finally, one defines the set of all forms as the sum of all sets of forms of any degrees
\begin{align}
	\Omega^\bullet(\algft{A})
	= \bigoplus_{n = 0}^\infty \Omega^n(\algft{A}),
	\label{eq:gnc_dbdc_allform}
\end{align}
where $\Omega^0(\algft{A}) = \algft{A}$. The triplet 
\begin{align}
	(\Omega^\bullet(\algft{A}), \wedge, \td)
	\label{eq:gnc_dbdc_diffalg}
\end{align}
is the differential algebra defining the \dbdc.

\section{Deformation quantization}
\label{sec:gnc_dq}
\paragraph{}
\GR and the \SM of particle physics are the two main models of theoretical physics that give very accurate predictions of the nature we observe. They are expressed as field theories on a classical \sT. If we were to generalise these two pictures to field theories on a \qST, then this new theory would need to account for the results given by the two former models, in some way. This is the precise point of \dq: considering a classical geometry, how can one deform it into a \nCG? 

The way the classical geometry is \enquote{kept track of} relies on a parameter, called the \dpt, that expresses if the geometry is to be considered classical or quantum. In most of our physical models, the \dpt corresponds to an energy scale (often associated to the \Pmast). Therefore, for energies comparable or above the \dpt, the noncommutative nature of the geometry gives relevant contributions, but for energies far below the \dpt, the geometry can be considered classical. The previous way of thinking is akin to \qM, where the \dpt is $\hbar$. \Dq was indeed first developed as a mathematical model of \qM \cite{Weyl_1927, von_Neumann_1931, Durhuus_2013}, as detailed in \secref{gnc_phys}.

We here give a brief introduction to \dq and refer to \cite{Sternheimer_1998, Waldmann_2016} for more historical and theoretical aspects.

\paragraph{}
After \namefont{Weyl} \cite{Weyl_1927} and \namefont{von Neumann} \cite{von_Neumann_1931} works, mathematicians have gone on and tried to deform more complex structures. The existence of a star-product on symplectic manifolds was proven by \namefont{de Wilde} and \namefont{Lecompte} \cite{De_Wilde_1983} and, in parallel, by \namefont{Bayen} and collaborators \cite{Bayen_1978}. The existence of the star-product on general \namefont{Poisson} manifold was established by \namefont{Kontsevich} \cite{Kontsevich_2003}. 

In general terms, let $\manft{M}$ be a \namefont{Poisson} manifold, that is a \sT with phase-space structure. A \spdtt $\star_{\kbar}$ \eqref{eq:gnc_spdt} is an associative product of function expressed formally as a power series expansion in the \dpt $\kbar$. That is for two functions $f$ and $g$
\begin{align}
	f \star_{\kbar} g
	= f \cdot g + \sum_{n = 1}^\infty \kbar^n C_n(f,g)
	\label{eq:gnc_dq_spdt}
\end{align}
where $\cdot$ is the classical product of functions \eqref{eq:gnc_found_cpdt}, $C_n$ is a bilinear differential operator and $\kbar$ is a constant. A priori $f$ and $g$ are taken to be smooth functions, \ie $f,g \in \func^\infty(\manft{M})$. But, since the operation $\star_{\kbar}$ \eqref{eq:gnc_dq_spdt} may not be convergent for any smooth functions, the set of considered functions might be smaller then $\func^\infty(\manft{M})$, and is generically called the multiplier space. In the context of formal deformation, we simply state that $f$ and $g$ can be written as formal power series in $\kbar$, that is
\begin{align}
	f = f_0 + \sum_{n=1}^\infty \kbar^n f_n
	\label{eq:gnc_dq_fps}
\end{align}
where $f_0$ is the classical smooth function. We usually note that $f$ is a formal power expansion in $\kbar$ of the form \eqref{eq:gnc_dq_fps} as $f \in \func^\infty(\manft{M})\llbracket \kbar \rrbracket$. Note that with expression \eqref{eq:gnc_dq_fps}, the operator $C_n$ in \eqref{eq:gnc_dq_spdt} writes
\begin{align}
	C_n(f,g)
	&= \sum_{m=0}^n f_m \, g_{n-m}.
\end{align}

The main feature of this star-product formalism is that the limit of a vanishing \dpt, corresponding here to $\kbar \to 0$, makes one recover (formally) all the classical structure of $\manft{M}$. More explicitly, $f \star_{\kbar} g \to f \cdot g$ and 
\begin{align}
	[f, g]_{\star_{\kbar}}
	&= f \star_{\kbar} g - g \star_{\kbar} f
	\to i \kbar \{f, g\},
	\label{eq:gnc_dq_cl_bra}
\end{align}
where $\{\cdot, \cdot\}$ correspond to the \namefont{Poisson} bracket of $\manft{M}$. 

In a physical context, the limit $\kbar \to 0$ can correspond to a low energy limit, if $1 / \kbar$ is an energy scale. Furthermore, the formula \eqref{eq:gnc_dq_cl_bra} corresponds mathematically to the usual sentence \enquote{quantising a physical theory is replacing the brackets with commutators}. It is the precise reason of the word quantisation in \enquote{\dq}. In this sense, the \spdtt formalism is a generalisation of canonical quantisation of \qM.

\paragraph{}
In order to perform computations, an explicit formula of $\star_{\kbar}$ is needed. In the mathematical literature, this field of research is called strict \dq. There are several ways to generate explicit \spdtt. In this manuscript, we follow the procedure of \namefont{Gutt} with deformation of the universal enveloping algebra \cite{Gutt_1983} and \namefont{Reiffel} through \calg techniques \cite{Reiffel_1990}. We also present the \Dt method.

\subsection{\tops{$C^*$}{C*}-algebras techniques}
\label{subsec:gnc_dq_ca}
\paragraph{}
The procedure presented here is the one of \cite{Gutt_1983, Reiffel_1990}. Let $\manft{M}$ be a \sT with a linear \namefont{Poisson} structure, that is, for any $f,g \in \func^\infty(\manft{M})$,
\begin{align}
	\{f, g\}(x)
	&= \tensor{\sC}{^{\nu\rho}_\mu} x^\mu \frac{\partial f}{\partial x^\nu} \frac{\partial g}{\partial x^\rho}
	\label{eq:gnc_dq_ca_LP}
\end{align}
where $x^\mu$ are (local) coordinates on $\manft{M}$ and $\tensor{\sC}{^{\nu\rho}_\mu}$ is a constant. Then, $\manft{M}$ naturally identifies to the dual $\dual{\Lieft{g}}$ of a \Lieat $\Lieft{g}$ with \namefont{Kirillov-Kostant-Souriau} structure, where $\sC$ is the structure constant of $\Lieft{g}$. Those are called linear \namefont{Poisson} or \namefont{Lie-Poisson} structures. In this context, $\{x^\mu\}_{\mu = 0, \ldots, d}$ corresponds to coordinate functions of $\dual{\Lieft{g}}$ that satisfies
\begin{align}
	\{x^\mu, x^\nu\} = \tensor{\sC}{^{\mu\nu}_\rho} x^\rho.
	\label{eq:gnc_dq_ca_Lb}
\end{align}
The \Lieat $\Lieft{g}$ is associated to a \Liegt $\manft{G}$, which can be interpreted as the momentum space. The first step of this construction is to build the \calg of $\manft{G}$. We refer to textbooks like \cite{Deitmar_2014} for more mathematical details on harmonic analysis and group theory. This introduction will also be repeated in the more explicit case of the deformed group of momentum in \subsecref{ncft_p4_dms}.

\begin{Emph}{\Calg}
In order to define a convolution, we need a notion of \enquote{sum} (or \enquote{integral}) over the group $\manft{G}$. In other words, we need a measure $\lHm$ over $\manft{G}$. In the case of locally compact groups (which we suppose always to be the case), there exists a unique left-invariant (resp.~right-invariant) measure called the left (resp.~right) \Haarm, denoted $\lHm$ (resp.~$\rHm$). The invariance states that
\begin{align}
	\lHm(\ggft{u}_1 \ggft{u}_2)
	&= \lHm(\ggft{u}_2), &
	\rHm(\ggft{u}_1 \ggft{u}_2)
	&= \rHm(\ggft{u}_1),
	\label{eq:gnc_dq_ca_Hm}
\end{align}
for any $\ggft{u}_1, \ggft{u}_2 \in \manft{G}$. The uniqueness of both measures imposes that there exists a positive function $\Delta : \manft{G} \to \piReal$ linking the two, \ie
\begin{align}
	\rHm(\ggft{u}) = \Delta(\ggft{u}^{-1}) \, \lHm(\ggft{u})
	\label{eq:gnc_dq_ca_mf}
\end{align}
for any $\ggft{u} \in \manft{G}$. The function $\Delta$ is called the \mft. It is a group homomorphism, meaning it satisfies
\begin{align}
	\Delta(\ggft{u}_1 \ggft{u}_2) 
	&= \Delta(\ggft{u}_1) \Delta(\ggft{u}_2), &
	\Delta(\ggft{u}^{-1})
	&= \Delta(\ggft{u})^{-1}, &
	\Delta(1)
	&= 1.
	\label{eq:gnc_dq_ca_mf_hom}
\end{align}
In the case where $\Delta = 1$, the group $\manft{G}$ is said to be unimodular and one has $\lHm = \rHm$.

We can now integrate functions on the group like $F : \manft{G} \to \Real$ as \begin{align}
	\int_\manft{G} \lHm(\ggft{u})\ F(\ggft{u}).
	\label{eq:gnc_dq_ca_int}
\end{align}
The set of functions for which \eqref{eq:gnc_dq_ca_int} does not diverge is noted $L^1(\manft{G})$ and corresponds to integrable functions on the group. The choice of left or right \Haarm in \eqref{eq:gnc_dq_ca_int} has no impact, since one can go from one to the other via a change of variable.

Finally, given two functions $F_1, F_2 \in L^1(\manft{G})$, one can define the convolution product $\cpdt$ through
\begin{align}
	(F_1 \cpdt F_2)(\ggft{u})
	&= \int_\manft{G} \lHm(\ggft{u}_0)\ F_1(\ggft{u}_0) \, F_2(\ggft{u} \ggft{u}_0^{-1}).
	\label{eq:gnc_dq_ca_cpdt}
\end{align}
The space $L^1(\manft{G})$ together with the product $\cpdt$ forms an algebra that is called the \calg of $\manft{G}$. One can show that $\cpdt$ is commutative if and only if the group $\manft{G}$ is Abelian.
\end{Emph} 

\paragraph{}
This new product $\cpdt$ is a key ingredient to define our \spdtt. One then just needs to make the link between the function over $\dual{\Lieft{g}}$, that are functions on the \sT $\manft{M}$, and functions of $\manft{G}$, that are functions on the momentum space. This link is made, as in \qM, through the \Ftt:
\begin{align}
	\Ft(f)(\ggft{u}) 
	&= \int_{\dual{\Lieft{g}}} e^{i \langle \log(\ggft{u}),\, x \rangle} f(x) \tdr{}{x}, &
	\Ft^{-1}(F)(x)
	&= \int_{\manft{G}} e^{-i \langle \log(\ggft{u}),\, x \rangle} F(\ggft{u}) \, \lHm(\ggft{u}),
	\label{eq:gnc_dq_ca_Four}
\end{align}
where $\td x$ is the \namefont{Lebesgue} measure on $\dual{\Lieft{g}}$, $\langle \cdot, \cdot \rangle : \Lieft{g} \times \dual{\Lieft{g}} \to \Real$ is the dual pairing between $\Lieft{g}$ and $\dual{\Lieft{g}}$, and $\log : \manft{G} \to \Lieft{g}$ is simply here to make sense of the expression $\langle \log(\ggft{u}),\, x \rangle$ via the correspondence of the \Lieat $\Lieft{g}$ and the \Liegt $\manft{G}$. Finally, we define the \spdtt on $\manft{M}$ through
\begin{align}
	f \star g
	&= \Ft^{-1} \big( \Ft(f) \cpdt \Ft(g) \big).
	\label{eq:gnc_dq_ca_spdt}
\end{align}

\paragraph{}
One can make several comments of the expression \eqref{eq:gnc_dq_ca_spdt}.

An explicit expression of the \spdtt can be obtained from \eqref{eq:gnc_dq_ca_Four}, \eqref{eq:gnc_dq_ca_spdt} together with the expression of the \Haarm. The latter is derived via the group law, which itself comes from the \Lieat structure \eqref{eq:gnc_dq_ca_Lb} via the \BCH \cite{Van-Brunt_2015, Van-Brunt_2018}. A derivation of the \spdtt \eqref{eq:gnc_dq_ca_spdt} is done for a deformation of \Minkt space-time in \secref{kM_kP}.

This \spdtt is non-local because of the convolution product \eqref{eq:gnc_dq_ca_cpdt}. Indeed, if one interprets $\ggft{u}_0$ as a momentum, then the integration over $\ggft{u}_0$ imposes that $F_1 \cpdt F_2$ at momentum $\ggft{u}$ is constituted of the (convoluted) sum of $F_1$ and $F_2$ for all momenta. Therefore, given an energy, the value of $F_1 \cpdt F_2$ at this energy depends of the value of $F_1$ and $F_2$ at all energies. Going back to position space $x$, this will impose that the value of $f \star g$ at $x$ depends of the value of $f$ and $g$ everywhere in space and time.

The \dpt is not explicitly shown here, but is present in the structure constant $\sC$ of \eqref{eq:gnc_dq_ca_Lb} and so in the group law via the \BCH. The commutative limit thus corresponds to $\sC \to 0$ and so to an Abelian group law. In this case, one can show that the successive integrations of \eqref{eq:gnc_dq_ca_spdt} involve \Ddf{s} and gives $f \star g = f \cdot g$, where $\cdot$ is the commutative product of functions \eqref{eq:gnc_found_cpdt}.

Finally, to make contact with the quantisation of mechanics presented in \secref{gnc_phys}, one can consider a (faithful $*$-) representation of the group algebra on a \Hsp $\pi: L^1(\manft{G}) \to \Hilbft{B}(\Hilbft{H})$. Thus, the \spdtt \eqref{eq:gnc_dq_ca_spdt} can be expressed via a quantisation map, similarly to \eqref{eq:gnc_qm_spdt_gen}, with $\Qmap = \pi \circ \Ft$.

\subsection{Deformations through \namefont{Drinfel'd} twist}
\label{subsec:gnc_dq_dt}
\paragraph{}
The other mainly used approach to an explicit \spdtt construction is given by \Dt deformation of product \cite{Drinfeld_1983, Drinfeld_1990}. The main idea of this construction is to use \exref{gnc_uealg} in order to form a commutative \Hfat out of a \Lieat $\Lieft{g}$ and then to deform it to a noncommutative algebra via the \Dt through \eqref{eq:gnc_qg_Dt_prod}. We explicit this below. Note that this \subsecref{gnc_dq_dt} uses \qg{s} notions, so we refer the reader to \secref{gnc_qg} or \appref{ha} for more details on \Hfat{s}.

\paragraph{}
Let $\Lieft{g}$ be a \Lieat, that can be the \Lieat of \vf{s} in the context of \gR, or as infinitesimal symmetry transformations associated to some \Liegt. We consider the universal enveloping algebra $U(\Lieft{g})$ which corresponds to the algebra such that the commutator of the product of $U(\Lieft{g})$ corresponds to the bracket $[\cdot, \cdot]$ of $\Lieft{g}$. This algebra can be endowed with a trivial \Hfat structure, as developed in \exref{gnc_uealg}.

Let $\Hoft{F} = \sum \Hoft{F}_1 \otimes \Hoft{F}_2 \in U(\Lieft{g}) \otimes U(\Lieft{g})$ be an invertible element. We say that $\Hoft{F}$ is a \Dt if it further satisfies
\begin{subequations}
	\label{eq:gnc_dq_dt}
\begin{align}
	(\Hoft{F} \otimes 1) (\Delta \otimes \id) (\Hoft{F})
	&= (1 \otimes \Hoft{F}) (\id \otimes \Delta) (\Hoft{F}), &
	\text{($2$-cocycle condition)}&
	\label{eq:gnc_dq_dt_2co} \\
	(\id \otimes \varepsilon) (\Hoft{F})
	&= (\varepsilon \otimes \id) (\Hoft{F})
	= 1, &
	\text{(normalisation)}&
	\label{eq:gnc_dq_dt_norm} \\
	\Hoft{F} &= 1 \otimes 1 + \BigO{\kbar}, &
	\text{(semi-calssical limit)}&
	\label{eq:gnc_dq_dt_scl}
\end{align}
\end{subequations}
where $\Delta$ and $\varepsilon$ are the coproduct and counit of $U(\Lieft{g})$ respectively and $\kbar$ is the deformation parameter. The \Dt $\Hoft{F}$ can be viewed as a function of the deformation parameter $\kbar$, and the condition \eqref{eq:gnc_dq_dt_scl} ensures that the twist vanish at the commutative limit $\kbar \to 0$.

Let us define 
\begin{align}
	\Delta^\Hoft{F}
	&= \Hoft{F} \Delta \Hoft{F}^{-1}, &
	S^{\Hoft{F}}
	&= \chi S \chi^{-1},
	\label{eq:gnc_dq_dt_tHalg}
\end{align}
where $S$ is the antipode of $U(\Lieft{g})$ and we noted $\chi = \Hoft{F}_1 S(\Hoft{F}_2)$. Then, the set $U(\Lieft{g})^\Hoft{F} = \big( U(\Lieft{g}), \cdot, 1, \Delta^\Hoft{F}, \varepsilon, S^\Hoft{F} \big)$ is a \Hfat (see \thmref{ha_as_Dt_tHalg}). Note that it is often denoted $U(\Lieft{g})_\star$ in the physics literature, in reference to the \spdtt $\star$ \eqref{eq:gnc_dq_dt_spdt}.

\paragraph{}
Now, let $\manft{M}$ be a \sT that we want to quantise and $\manft{G}$ a \Liegt of symmetries that acts on $\manft{M}$. We apply the previous procedure to $\Lieft{g}$ the \Lieat of $\manft{G}$. In this context, $U(\Lieft{g})$ corresponds to the (infinitesimal) symmetries of $\func^\infty(\manft{M})$ because $\func^\infty(\manft{M})$ is a $U(\Lieft{g})$-module algebra (see details in \subsecref{gnc_qg_qst}).

We want to deform this picture by starting with the symmetries. As detailed above, one can consider a \Dt $\Hoft{F}$, to deform $U(\Lieft{g})$ in a \enquote{non-trivial} quantum group $U(\Lieft{g})^\Hoft{F}$. The \qST $\algft{A}$, corresponding to a deformation of $\func^\infty(\manft{M})$, which has $U(\Lieft{g})^\Hoft{F}$ as its algebra of quantum symmetries, is determined by the new product
\begin{align}
	f \star g
	&= \sum \big( \Hoft{F}_1^{-1} \actl f \big)
	\big( \Hoft{F}_2^{-1} \actl g \big)
	\label{eq:gnc_dq_dt_spdt}
\end{align}
for any $f, g \in \algft{A}$.

\paragraph{}
At this point several comments are in order.

First, the noncommutativity of \eqref{eq:gnc_dq_dt_spdt}, which is linked to the \enquote{quantum} trait of $\algft{A}$, is linked to the noncocommutativity of $\Delta^\Hoft{F}$. The undeformed coproduct \eqref{eq:gnc_uealg_com} is cocommutative, and it is linked to the commutativity of $\func^\infty(\manft{M})$. Therefore, one pictures that the noncommutativity of $\algft{A}$ comes entirely from the \Dt $\Hoft{F}$.

The \dpt is not shown explicitly here. However, it is usually contained in the \Dt expression, so that the \spdtt \eqref{eq:gnc_dq_dt_spdt} is indeed parametrized by the \dpt. The commutative limit of $\Hoft{F}$ is given by the requirement \eqref{eq:gnc_dq_dt_scl}, and combined with \eqref{eq:gnc_dq_dt_spdt} one can verify that the \spdtt of $\algft{A}$ corresponds to the commutative product of $\func^\infty(\manft{M})$ in the commutative limit.

\begin{Ex}{Abelian \namefont{Drinfel'd} twist}
	{gnc_dq_dt_abe}
	Let $X, Y$ be two commuting elements of $U(\Lieft{g})$. For example, one could consider $X$ and $Y$ to be some generators of translations in the case where $\Lieft{g}$ is the \Lieat of the \Pogt. Let
	\begin{align*}
		\Hoft{F} 
		&= \exp \big( i \kbar\, X \otimes Y \big)
		\in U(\Lieft{g}) \otimes U(\Lieft{g}),
	\end{align*}
	then one can show that $\Hoft{F}$ is a \Dt (see proof below). From there, the \spdtt \eqref{eq:gnc_dq_dt_spdt} writes
	\begin{align*}
		f \star g
		&= \sum_{n=0}^\infty \frac{(i \kbar)^n}{n!} \, (X^n \actl f) \, (Y^n \actl g).
 	\end{align*}
	
	Note that, one could have equivalently considered, instead of $X$ and $Y$, a family of commuting elements $\{X_\mu\}_{\mu=0, \ldots, d}$ of $U(\Lieft{g})$. In this case, a \Dt could write
	\begin{align*}
		\Hoft{F} 
		&= \exp \big( i \kbar\, \Theta^{\mu\nu}\, X_\mu \otimes X_\nu \big)
		\in U(\Lieft{g}) \otimes U(\Lieft{g})
	\end{align*}
	where $\Theta$ is matrix of constant coefficients. If we further consider that $X_\mu$ acts on functions as a derivation on the $\mu$-th coordinate, one has $X_\mu \actl f = \partial_\mu f$. Therefore, the \spdtt \eqref{eq:gnc_dq_dt_spdt} writes
	\begin{align*}
		f \star g
		&= \sum_{n=0}^\infty \frac{(i \kbar)^n}{n!} \Theta^{\mu_1 \nu_1} \cdots \Theta^{\mu_n \nu_n} \, \partial_{\mu_1} \cdots \partial_{\mu_n} f \, \partial_{\nu_1} \cdots \partial_{\nu_n} g.
	\end{align*}
\end{Ex}

\begin{proof}
	Let $\Hoft{F} = \exp( i\kbar\, X \otimes Y)$ as in \exref{gnc_dq_dt_abe}. It is invertible straightforwardly as it is expressed via an exponential. Thus, in order to show that $\Hoft{F}$ is a \Dt, one needs to verify \eqref{eq:gnc_dq_dt}.
	
	Considering first the $2$-cocycle condition, one uses the fact that $\Delta$ is an algebra homomorphism to obtain that $\Delta(\exp(X)) = \exp(\Delta(X))$ for any $X \in U(\Lieft{g})$. Then, using the coproduct expression \eqref{eq:gnc_uealg_com}, one computes
	\begin{align*}
		(\Hoft{F} \otimes 1) (\Delta \otimes \id)(\Hoft{F})
		&= \exp(i\kbar \, X \otimes Y \otimes 1) \exp\big(i\kbar \, (X \otimes 1 + 1 \otimes X) \otimes Y \big) \\
		&= \exp\big( i\kbar \, (X \otimes Y \otimes 1 + X \otimes 1 \otimes Y + 1 \otimes X \otimes Y) \big), \\
		(1 \otimes \Hoft{F}) (\id \otimes \Delta)(\Hoft{F})
		&= \exp(i\kbar \, 1 \otimes X \otimes Y) \exp\big(i\kbar \, X \otimes (Y \otimes 1 + 1 \otimes Y) \big) \\
		&= \exp\big( i\kbar \, (1 \otimes X \otimes Y + X \otimes Y \otimes 1 + X \otimes 1 \otimes Y ) \big),
	\end{align*}
	so that \eqref{eq:gnc_dq_dt_2co} is satisfied.
	
	Again, using the fact that the counit is an algebra homomorphism, one obtains that $\varepsilon(\exp(X)) = \exp(\varepsilon(X))$. From the expression \eqref{eq:gnc_uealg_com}, one computes
	\begin{align*}
		(\id \otimes \varepsilon)(\Hoft{F})
		&= \exp( i\kbar \, X \otimes \varepsilon(Y))
		= \exp(0) = 1 \\
		(\varepsilon \otimes \id)(\Hoft{F})
		&= \exp( i\kbar \, \varepsilon(X) \otimes Y)
		= \exp(0) = 1,
	\end{align*}
	so that \eqref{eq:gnc_dq_dt_norm} is satisfied.
	
	Finally, one obtains \eqref{eq:gnc_dq_dt_scl} by simply expanding the exponential into an infinite sum
	\begin{align*}
		\Hoft{F}
		&= 1 \otimes 1 + \kbar \, \sum_{n=1}^\infty \frac{i^n \kbar^{n-1}}{n!} X^n \otimes Y^n
		= 1 \otimes 1 + \BigO{\kbar}.
		\qedhere
	\end{align*}
\end{proof}

\paragraph{}
We have gathered all the ingredients to construct a \qST out of a classical \sT. The next step is to look at the behaviour of fields on this \qST. Therefore, the next Chapters look at a toy model of \phif and try to capture the deformation that noncommutativity induces on $\Ug{1}$ \gT.
\chapter{Noncommutative field and gauge theories}
\label{ch:ncft}
\paragraph{}
Over classical \sT{s}, \gT[ies] are expressed through a (principal) fiber bundle, on which a gauge group acts. This fiber bundle corresponds to the geometric space where the connection (or the gauge fields) and the curvature (or the field strength) live.

For example, in the context of \gR, the considered bundle is the tangent bundle. A \vf is expressed as a section of the tangent bundle, \ie to any point in \sT $x \in \manft{M}$, a \vf $X \in \Gamma(\manft{M})$, associate a vector in the tangent space at $x$, $X(x) \in T_x\manft{M}$. Given two distant points $x, y \in \manft{M}$, the vectors $X(x)$ and $X(y)$ cannot be compared as they live in different vector spaces. Therefore, one needs the notion of a parallel transport, or equivalently a connection (or even a covariant derivative), to transport $X(x)$ in the tangent space at $y$, \ie in $T_y\manft{M}$. The covariant derivative is defined, in the way of \namefont{Koszul} \cite{Koszul_1960}, as $\nabla : \Gamma(\manft{M}) \times \Gamma(\manft{M}) \to \Gamma(\manft{M})$ and writes in local coordinates 
\begin{align}
	\nabla_\mu(\partial_\nu)
	&= \Gamma_{\mu\nu}^\rho \partial_\rho
	\label{eq:ncft_con_GR}
\end{align}
where $\Gamma_{\mu\nu}^\rho$ is the connection. Finally, the curvature $R: \Gamma(\manft{M})^2 \times \Gamma(\manft{M}) \to \Gamma(\manft{M})$ is defined as the mismatch of the starting and ending point when performing a loop of parallel transports (see \figref{curv}). In local coordinates, this writes
\begin{align}
	R_{\mu\nu}(\partial_\rho)
	&= \nabla_\mu(\nabla_\nu(\partial_\rho)) - \nabla_\nu(\nabla_\mu(\partial_\rho))
	= \tensor{R}{_{\mu\nu\rho}^\sigma} \partial_\sigma.
	\label{eq:ncft_cur_GR}
\end{align}

One can generalise this picture to the case of a principal fiber bundle $\manft{P}$ over a \sT $\manft{M}$, with a structure group given by a \Liegt $\manft{G}$. In such a bundle $\manft{G}$ acts on $\manft{P}$ via a right action. One then considers $(V, \rho)$ a representation of $\manft{G}$ and considers the associated vector bundle $\manft{X} = \manft{P} \times_\rho V$. The notation $\times_\rho$ denotes here the fact that $\manft{G}$ acts on the $V$ part of $\manft{X}$ via $\rho$. The sections $s : \manft{M} \to \manft{X}$, noted $s \in \Gamma(\manft{X})$, on such a bundle form a $\funcs(\manft{M})$-module for the point-wise product. As above, one can compare distant regions of this bundle thanks to a covariant derivative $\nabla : \Gamma(\manft{M}) \times \Gamma(\manft{X}) \to \Gamma(\manft{X})$, and can asses the loop mismatch via the curvature $R : \Gamma(\manft{M})^2 \times \Gamma(\manft{X}) \to \Gamma(\manft{X})$ (see \figref{curv}).

\paragraph{}
In this context, the other main example is the case where $\manft{G} = \Ug{1}$ and $V = \Real$, which corresponds to $\Gamma(\manft{X}) = \funcs(\manft{M}) \otimes \mathbb{R}$. From the definition of $\nabla$, one has in local coordinates (see \eqref{eq:ncft_db_U1_gf})
\begin{align}
	\nabla_\mu f
	&= \partial_\mu f - \iCpx A_\mu f
	\label{eq:ncft_con_QFT}
\end{align}
for any $f \in \funcs(\manft{M})$. Note that for simplicity we have abbreviated $f = f \otimes 1 \in \funcs(\manft{M}) \otimes \Real$, by using $\funcs(\manft{M}) \otimes \Real \simeq \funcs(\manft{M})$. One recognizes in \eqref{eq:ncft_con_QFT} the particle physics expression of the covariant derivative together with the gauge field $A_\mu$. The curvature then writes
\begin{align}
	R_{\mu\nu}(f)
	&= \iCpx \big(\partial_\mu A_\nu - \partial_\nu A_\mu - \iCpx [A_\mu, A_\nu] \big) f
	= \iCpx F_{\mu\nu} f
	\label{eq:ncft_cur_QFT}
\end{align}
where one recognize the field strength $F$. Therefore, (classical) $\SUg{n}$ \gT[ies] can also be written in this general framework.

\begin{Figure}
	[label={fig:curv}]%
	{
		Illustration of the formula \eqref{eq:ncft_db_cur}, consisting in a loop of parallel transports, for $s$ a \vf.
	}%
	\begin{tikzpicture}[scale = 2.7]
	\draw[darkblue, thin] (0,0) to[out=15, in=160] (2,0)
		to[out=40, in=180] (5.4,1)
		to[out=120, in=-30] (4.4,2.2)
		to[out=180 , in=40] (0,0);
	\node[darkblue] at (3.3,.6) {$\manft{M}$};	
	
	\draw[black, -To, thin] (1,.25) to (2,1.4);
	\draw[black, -To, thin] (2,1.4) to (4,1.7);
	\draw[black, -To, thin] (4,1.7) to (4.2, 1.65);
	\draw[black, -To, thin] (4.2,1.65) to (3.9,1);
	\draw[black, -To, thin] (3.9,1) to (1,.25);
	\node at (1.5, 1) {$X$};
	\node at (3, 1.65) {$Y$};
	\node at (3.85, 1.55) {$-[X,Y]$};
	\node at (3.8, 1.2) {$-X$};
	\node at (2.4, .5) {$-Y$};

	\draw[darkred, -To] (1,.25) to (1.5, .5) node[anchor=west] {$s$};
	\draw[darkred, -To] (1,.25) to (1.4, .6);
	\draw[darkred, -To] (2, 1.4) to (2.4, 1.6) node[anchor = west] {$s$};
	\draw[darkred, -To] (4,1.7) to (4.4, 2) node[anchor = south] {$s$};
	\draw[darkred, -To] (4.2, 1.65) to (4.55, 1.95) node[anchor = north west] {$s$};
	\draw[darkred, -To] (3.9,1) to (4.25, 1.4) node[anchor = west] {$s$};
	
	\draw[darkgreen, -To] (1.55, .55) to (1.45, .65) node[anchor =south west, yshift = -5pt] {$R_{X,Y}(s)$};
\end{tikzpicture}%
\end{Figure}

\paragraph{}
To generalise \gT[ies] to the algebraic context of \nCG, we first need a generalisation of the notion of (principal) fiber bundle. The algebraic analogue of sections of the bundle is, in fact, directly given by the \namefont{Serre-Swan} theorem, which states that fiber bundles are in one-to-one correspondence with the notion of (projective) module (see the \defref{ha_rt_mod} and the \exref{gnc_qg_qst_modex} for an explicit example). Therefore, the first element to build a \gT is a module, generically denoted by $\modft{X}$ here, that is supposed to stand as the noncommutative counterpart of $\Gamma(\manft{X})$. The covariant derivative and the curvature definitions are then exported from the commutative case as such. Finally, one can also implement the gauge transformations in this setting.

\paragraph{}
To summarise, we consider here a \qST $\algft{A}$ as introduced in \chapref{gnc}. The \gT over this \qST is defined on a module $\modft{X}$ over the algebra $\algft{A}$ in which we implement the notions of covariant derivative, associated curvature and gauge transformations. From there, one recovers the usual physical quantities (gauge fields, field strength, \etc) on the \qST.

Note that these notions depend on how one generalises the \vf{s} $\Gamma(\manft{M})$ and so how one generalises the \dc. As already expressed in \chapref{gnc}, there are three main ways of doing so, leading thus to different formulations of \gT on \qST{s}. Here, we detail two of such constructions: \gT[ies] coming from \dbdc and the one coming from \Dt construction. The way \gT[ies] are implemented in the \st formalism was roughly introduced by \eqref{eq:gnc_phys_SM_g} and \eqref{eq:gnc_st_gt} and is not treated further here.

The previous scheme of \gT has been applied to several \qST{s}, like to \kMt (see \secref{kM_gt}). For a complete review of \gT[ies] on \qST, see \cite{Hersent_2023a}. It should also be noted that the following constructions are made for right modules, but could be equally well made for left modules or bimodules \cite{Hersent_2023a}.

\paragraph{}
Note that building a \gT on a noncommutative space is \aprio not straightforward. Indeed, consider  $\Lieft{g}$ to be the \Lieat of infinitesimal gauge transformations, with associated gauge (\namefont{Lie}) group $G$, with a matrix representation. The gauge field $A$ is a $\Lieft{g}$-valued connection. Then, the noncommutative analogue of $A_\mu$ is an element of $\Lieft{g} \otimes \algft{A}$, which, in the matrix representation, corresponds to a matrix with coefficients in $\algft{A}$. However, since $\algft{A}$ is noncommutative, the \Lieat closure rules are likely to be broken. Explicitly, for $\alpha \otimes f$, $\beta \otimes g \in \Lieft{g} \otimes \algft{A}$, one has
\begin{align}
	[\alpha \otimes f, \beta \otimes g]
	&= [\alpha, \beta] \otimes (f \star g) + \alpha \beta \otimes [f, g]_\star,
	\label{eq:ncft_nclo}
\end{align}
which would corresponds to a term of the form $[A_\mu, A_\nu]$ in the field strength. The first term of \eqref{eq:ncft_nclo} is stable in $\Lieft{g} \otimes \algft{A}$ but the second is not, since in general $\alpha \beta \nin \Lieft{g}$, and does not vanish as the \spdtt is not commutative. The usual solution to this issue is to consider that the connection takes values in $U(\Lieft{g}) \otimes \algft{A}$, where $U(\Lieft{g})$ is the universal enveloping algebra of $\Lieft{g}$. Then, some conditions may be imposed on $A$ to recover a connection that takes values in $\Lieft{g} \otimes \algft{A}$. For the most used case of $G = \Ug{n}$, which corresponds to $\Lieft{g} = \Ua{n}$, one has that $U(\Ua{n}) \simeq \Mat{n}{\Cpx}$, so that $A$ takes values in $\Mat{n}{\Cpx} \otimes \algft{A}$. To recover $\Ua{n}$ from $\Mat{n}{\Cpx}$, a hermiticity condition is imposed on $A$ using the involution of $\algft{A}$.

Each of the following proposals eliminates the previous problem, mainly by considering different types of deformation. The \gT schemes, based on derivations, developed in \secref{ncft_db} bypass this problem by considering a deformed gauge group. In the case of the \Dt formulation of \secref{ncft_Dt}, another solution was pointed out. One can deform the gauge transformation thanks to the \Dt in such a way that all the deformations go into the \Lru which arises when one gauge transforms a product. The \Lru of the gauge transformation is seen, in this formalism, in its coproduct. For its part, the \namefont{Seiberg-Witten} map of \secref{ncft_SW} requires the gauge parameter to depend on the undeformed gauge field, leading to a modified gauge transformation.

\section{Derivation based theories}
\label{sec:ncft_db}
\paragraph{}
In this \secref{ncft_db}, we introduce the \gT built on \dbdc, as it was set up in \cite{Dubois-Violette_1988, Dubois-Violette_1996}. For a review, see \cite{Dubois-Violette_2001}. We consider $\algft{A}$ to be a \qST and $\modft{X}$ to be a $\algft{A}$-module. The \vf{s} of $\algft{A}$ are considered to be the derivations of the algebra $\Der(\algft{A})$, following the construction of \secref{gnc_dbdc}.

\paragraph{}
Given a derivation $X \in \Der(\algft{A})$, we define a connection \latint{\`{a} la} \namefont{Koszul} \cite{Koszul_1960} $\nabla_X : \modft{X} \to \modft{X}$ as
\begin{subequations}
	\label{eq:ncft_db_con}
\begin{align}
	\nabla_X(s \actr f)
	&= \nabla_X(s) \actr f + s \actr X(f),
	& \text{(\namefont{Leibniz} rule)}
	\label{eq:ncft_db_con_Lr} \\
	\nabla_{X + zY}(s)
	&= \nabla_X(s) + \nabla_Y(s) \actr z,
	& \text{($\Cen{\algft{A}}$-linearity)}
	\label{eq:ncft_db_con_lin}
\end{align}
\end{subequations}
for any $Y \in \Der(\algft{A})$, $s \in \modft{X}$, $f \in \algft{A}$ and $z \in \Cen{\algft{A}}$. Considering the \dc constructed in \secref{gnc_dbdc}, the definition \eqref{eq:ncft_db_con} is equivalent to
\begin{align}
	\nabla &: \modft{X} \to \modft{X} \otimes_\algft{A} \Omega^1(\algft{A}), &
	\nabla(s \actr f) 
	&= \nabla(s) \actr f + s \otimes \td f,
\end{align}
where $\td : \algft{A} \to \Omega^1(\algft{A})$ is the differential. It straightforwardly extends to the covariant derivative $\nabla: \modft{X} \to \modft{X} \otimes_\algft{A} \Omega^\bullet(\algft{A})$.

\paragraph{}
Performing a loop of parallel transports $0 \to X \to XY \to XY - Y \to XY - YX$ (see \figref{curv}), one does not necessarily come back to the same point. This mismatch is measured by $R_{X,Y} : \modft{X} \to \modft{X}$, the curvature associated to $\nabla$, thus defined as
\begin{align}
	\begin{aligned}
	R_{X,Y}(s)
	&= [\nabla_X, \nabla_Y](s) - \nabla_{[X,Y]}(s) \\
	&= \nabla_X( \nabla_Y(s) ) - \nabla_Y( \nabla_X(s) ) - \nabla_{[X,Y]}(s).
	\end{aligned}
	\label{eq:ncft_db_cur}
\end{align}

One can show, using \eqref{eq:ncft_db_con_Lr}, that the curvature is a module homomorphism, that is $R_{X,Y}(s \actr f) = R_{X,Y}(s) \actr f$. This condition actually justifies that one can only consider the components $F_{\mu\nu}$ in \eqref{eq:ncft_cur_QFT} or $\tensor{R}{_{\mu\nu\rho}^{\sigma}}$ in \eqref{eq:ncft_cur_GR} instead of the full curvature.

\begin{proof}
	\begin{align*}
		R_{X,Y}(s \actr f) 
		&= \nabla_X \nabla_Y(s \actr f) - \nabla_Y \nabla_X(s \actr f) - \nabla_{[X,Y]}(s \actr f) \\
		&\begin{aligned}
			&= \nabla_X\big( \nabla_Y(s) \actr f + s \actr Y(f) \big) 
			- \nabla_Y\big( \nabla_X(s) \actr f + s \actr X(f) \big) \\
			&\pe - \nabla_{[X,Y]}(s) \actr f - s \actr [X,Y](f)
		\end{aligned} \\
		&\begin{aligned}
			&= \nabla_X \nabla_Y(s) \actr f + \cancel{\nabla_Y(s) \actr X(f)} + \bcancel{\nabla_X(s) \actr Y(f)} + \xcancel{s \actr XY(f)} \\
			&\pe - \nabla_Y \nabla_X(s) \actr f - \bcancel{\nabla_X(s) \actr Y(f)} - \cancel{\nabla_Y(s) \actr X(f)} - \xcancel{s \actr YX(f)} \\
			&\pe - \nabla_{[X,Y]}(s) \actr f - \xcancel{s \actr [X,Y](f)}
		\end{aligned} \\
		&= R_{X,Y}(s) \actr f
		\qedhere
	\end{align*}
\end{proof}

\paragraph{}
The gauge group is defined as the group of automorphisms of $\modft{X}$, that is the invertible linear maps $\varphi : \modft{X} \to \modft{X}$ such that $\varphi(s \actr f) = \varphi(s) \actr f$. The gauge transformation of $\nabla$ and consequently of the curvature are given by
\begin{align}
	\nabla^\varphi_X
	&= \varphi^{-1} \circ \nabla_X \circ \varphi, &
	R^\varphi_{X,Y}
	&= \varphi^{-1} \circ R_{X,Y} \circ \varphi.
	\label{eq:ncft_db_gt}
\end{align}
One can actually check that the gauge transformed connection $\nabla^\varphi$ is indeed a connection as it satisfies \eqref{eq:ncft_db_con}.

\paragraph{}
In the context of unitary gauge groups, as in \namefont{Yang-Mills} theory, one further needs a Hermitian structure $(\cdot, \cdot) : \modft{X} \times \modft{X} \to \algft{A}$, \ie a sesquilinear map (see \eqref{eq:oa_tas_Hsf_lin} and \eqref{eq:oa_tas_Hsf_slin}) which satisfies
\begin{align}
	(s_1, s_2)^\dagger
	&= (s_2, s_1), &
	(s_1 \actr f_1, s_2 \actr f_2)
	&= f_1^\dagger \star (s_1, s_2) \star f_2
	\label{eq:ncft_db_Hf}
\end{align}
for any $s_1, s_2 \in \modft{X}$ and $f_1, f_2 \in \algft{A}$. In this context, the connection is Hermitian if and only if
\begin{align}
	X \big( (s_1, s_2) \big)
	&= (\nabla_X(s_1), s_2) + (s_1, \nabla_X(s_2) )
	\label{eq:ncft_db_Hcon}
\end{align}
for any $X \in \Der(\algft{A})$ such that $X(f^\dagger) = X(f)^\dagger$. One says that $X$ is a real derivative when the latter condition holds. When considering deformed \namefont{Yang-Mills} theories, the condition \eqref{eq:ncft_db_Hcon} is equivalent to $A_\mu^\dagger = A_\mu$, which corresponds to $A_\mu$ being real-valued in the commutative limit.

Finally, a gauge transformation $\varphi$ is said to be unitary if
\begin{align}
	( \varphi(s_1), \varphi(s_2) )
	&= (s_1, s_2).
	\label{eq:ncft_db_ugt}
\end{align}
We denote the set of unitary gauge transformations as $\ugt{\modft{X}}$.

\paragraph{}
All the previous definitions make more sense when applied to the specific case of $\modft{X} = \algft{A}$, as in \exref{ha_rt_ncop}.

\begin{Emph}{Noncommutative electrodynamics}
	Consider\footnote{
		\label{fn:ncft_db_U1}%
		In accordance with the commutative case described in \eqref{eq:ncft_con_QFT}, it would be more appropriate to consider that $\modft{X} = \algft{A} \otimes \Cpx$. In this decomposition, $\Cpx$ is as the $1$-dimensional vector space to which $\algft{A}$ is associated to form a vector bundle. However, one can use that $\algft{A} \otimes \Cpx \simeq \algft{A}$ to simplify the notations.
	}
	$\modft{X} = \algft{A}$, with action $\actr = \star$ and Hermitian structure $(f, g)_\algft{A} = f^\dagger \star g$. One can start by actually checking that $\star$ is indeed an action and that $(\cdot, \cdot)_\algft{A}$ satisfies \eqref{eq:ncft_db_Hf}. The equation \eqref{eq:ncft_db_con} now states how the connection $\nabla$ behaves on products, \ie
	\begin{align}
		\nabla_X(g \star f)
		&= \nabla_X(g) \star f + g \star X(f)
		\label{eq:ncft_db_U1_con}
	\end{align}
	and considering $g = 1$, this gives
	\begin{align}
		\nabla_X(f) = X(f) - \iCpx A_X \star f,
		\label{eq:ncft_db_U1_gf}
	\end{align}
	where we noted $\nabla_X(1) =  - \iCpx A_X$. If one considers local coordinates, then \eqref{eq:ncft_db_U1_gf} is akin to \eqref{eq:ncft_con_QFT} for $X = \partial_\mu$. Our $\nabla$ thus corresponds to the deformed covariant derivative associated to some gauge field $A$. Using the curvature definition \eqref{eq:ncft_db_cur}, one computes with \eqref{eq:ncft_db_U1_gf} that
	\begin{align}
		\begin{aligned}
		R_{X,Y}(f)
		&= \iCpx \big( X(A_Y) - Y(A_Y) - \iCpx [A_X, A_Y]_\star + A_{[X,Y]} \big) \star f \\
		&= F_{X,Y} \star f,
		\end{aligned}
		\label{eq:ncft_db_U1_cur}
	\end{align}
	which again gives the same expression as \eqref{eq:ncft_cur_QFT} if $X = \partial_\mu$ and $Y = \partial_\nu$. Furthermore, if one requires that $\nabla$ is Hermitian, then, using \eqref{eq:ncft_db_Hcon} and \eqref{eq:ncft_db_U1_gf}, one computes that
	\begin{align}
		A_X^\dagger = A_X
		\label{eq:ncft_db_U1_Hgf}
	\end{align}
	for $X$ a real derivation.
	
	A gauge transformation $\varphi$ is fully determined by its value at $1$ since $\varphi(f) = \varphi(1) \star f$. We denote $\ggft{u} = \varphi(1)$ in the following. Through \eqref{eq:ncft_db_gt}, the gauge transform of the gauge field and the field strength are calculated to be
	\begin{subequations}
		\label{eq:ncft_db_U1_gt}
	\begin{align}
		A^{\ggft{u}}_X
		&= \ggft{u}^\dagger \star A_X \star \ggft{u} - \iCpx \ggft{u}^\dagger \star X(\ggft{u}),
		\label{eq:ncft_db_U1_gt_gf} \\
		F_{X,Y}^{\ggft{u}}
		&= \ggft{u}^\dagger \star F_{X,Y} \star \ggft{u}.
		\label{eq:ncft_db_U1_gt_fs}
	\end{align}
	\end{subequations}
	The unitary gauge group, given by \eqref{eq:ncft_db_ugt}, writes
	\begin{align}
		\ugt{1}
		&= \big\{ \ggft{u} \in \algft{A},\ \ggft{u}^\dagger \star \ggft{u} = \ggft{u} \star \ggft{u}^\dagger = 1 \big\}.
		\label{eq:ncft_db_U1_ugt}
	\end{align}
\end{Emph}

Several comments are in order.

\paragraph{}
The deformed \enquote{Abelian} \gT developed above actually is in fact more like a non-Abelian one, because the product $\star$ is noncommutative. Indeed, the bracket term $[A_X, A_Y]_\star$ is present in \eqref{eq:ncft_db_U1_cur} even if electrodynamics has been considered. However, in Abelian \gT[ies], the field strength contains directly measurable quantities, that are the electric and magnetic fields. In this sense, it must be gauge invariant, as the measured fields should not depend on the chosen gauge. From \eqref{eq:ncft_db_U1_gt_fs} it follows that, due to the noncommutativity of $\star$, the deformed field strength $F$ is not gauge invariant, but rather gauge covariant.

Accordingly, the field strength may not be a physical quantity in the noncommutative theory and one should build the relevant fields out of $F$. This is already done in non-Abelian \gT when one considers the gauge bosons of the electroweak interaction not to be the hypercharge and isospin fields $B$, $W^1$, $W^2$ and $W^3$ (which would all correspond to different copies of $A$ here) but rather $W^{\pm} = \frac{W^1 \mp i W^2}{\sqrt{2}}$, $\gamma = \cos(\theta_W) B + \sin(\theta_W) W^3$ and $Z^0 = \cos(\theta_W) W^3 - \sin(\theta_W) B$. In the case of \Moy space, it is actually more convenient to work with the so-called covariant coordinate or invariant connection, $\mathcal{A}_\mu = \iCpx (A_\mu + \Theta^{-1}_{\mu\nu} x^\nu)$, rather than $A_\mu$. However, there are \aprio an infinite number of possible candidate that would have the suitable properties to be a gauge field and the good commutative limit. The question of which one is the physically relevant quantity has not been settled yet.

\paragraph{}
The \gT considered here already has some phenomenological consequences on the physical model under study. Indeed, if the noncommutative space $\algft{A}$ is constructed as the deformation of some commutative space, as in \secref{gnc_dq}, one can consider the expansion of $F$ up to a given order $m$ in the \dpt. This would give an effective \namefont{Yang-Mills} action with supplementary operators of (mass) dimension $5$ to $4 + m$, for which the \dpt is the scale of \enquote{new physics}. From another perspective, one could also look into the deformation induced by the symmetries. Indeed, the gauge group $\ugt{1}$ is a deformed version of the commutative $\Ug{1}$ gauge group but is not $\Ug{1}$. Therefore, the usual gauge group should be broken at scales close to the \dpt. A more extended discussion on phenomenology of \qST{s} is given in \secref{qg_ph}.

\paragraph{}
One can construct similarly $\ugt{n}$ \gT[ies], generically called \nCYM[ies].

\begin{Emph}{\NCYM}
	Consider\footnote{
		In accordance with the footnote \ref{fn:ncft_db_U1} in the example of noncommutative electrodynamics, one should rather consider $\modft{X} = \algft{A} \otimes \Cpx^n$. The notations are simplified by using $\algft{A} \otimes \Cpx^n \simeq \algft{A}^{\otimes n}$.
	}
	$\modft{X} = \algft{A}^{\otimes n}$, for some $n \in \nzNInt$, with action $\actr = \star \otimes \cdots \otimes \star$ (see \exref{ha_rt_ncop}). For convenience, one writes
	\begin{align}
		\bel_j 
		= 0 \otimes \cdots \otimes 0 
		\otimes \overset{(j)}{1} \otimes 
		0 \otimes \cdots \otimes 0,
	\end{align}
	where the only non-zero entry is at the $j$-th place, for any $j = 1, \ldots, n$. Thus, for any $s \in \modft{X}$, one can decompose $s = \sum \limits_{j=1}^n \bel_j \star s^j$, where $s^j \in \algft{A}$. From there, the action writes $s \actr f = \sum \limits_{j=1}^n \bel_j \star s^j \star f$, for any $f \in \algft{A}$. The Hermitian structure considered is
	\begin{align}
		(s_1, s_2) 
		&= \sum_{j=1}^n (s_1^j)^\dagger \star s_2^j.
	\end{align}
	
	Thanks to the \Lru \eqref{eq:ncft_db_con_Lr}, one has
	\begin{align}
		\nabla_X(s)
		&= \sum_{j = 1}^n \nabla_X(\bel_j \star s^j)
		= \sum_{j=1}^n \bel_j \star X(s^j) + \nabla_X(\bel_j) \star s^j.
		\label{eq:ncft_db_Un_Hf}
	\end{align}
	Therefore, the connection is fully determined by its values on the basis $\{\bel_j\}_j$, which are written in components as $\nabla_X(\bel_j) = -i \sum \limits_{k = 1}^n \bel_k \star (A_X)^k_j$. If one associates the module with column vectors of elements in $\algft{A}$, that is $\modft{X} = \Mat{1,n}{\algft{A}} = \Mat{1,n}{\Cpx} \otimes \algft{A}$, one can write the connection $A_X$ under a matrix form
	\begin{align}
		A_X
		= \sum_{j,k = 1}^n \bel^j \star \bel_k \star (A_X)^k_j
		= \begin{pmatrix}
			(A_X)^1_1 & \cdots & (A_X)^1_n \\
			\vdots    & \ddots & \vdots    \\
			(A_X)^n_1 & \cdots & (A_X)^n_n
		\end{pmatrix}
		\in \Mat{n}{\algft{A}}
		= \Mat{n}{\Cpx} \otimes \algft{A}
	\end{align}
	where $\{\bel^j\}_j$ are the basis elements of $\Mat{n,1}{\algft{A}}$. It is also quite convenient to combine the matrix transpose $\tp{}$ and the involution ${}^\dagger$ of $\algft{A}$ to get a generalisation of the adjoint matrix ${}^\ddagger = \tp{} \otimes {}^\dagger$. In these notations $\bel_j^\ddagger = \bel^j$, and \eqref{eq:ncft_db_Un_Hf} writes $(s_1, s_2) = s_1^\ddagger \star s_2$, where $\star$ here denotes the matrix product with $\star$.
	
	Equipped with these notations, the requirement that $\nabla$ is Hermitian \eqref{eq:ncft_db_Hcon} now writes
	\begin{align}
		A_X^\ddagger
		= A_X.
	\end{align}
	Furthermore, the gauge transformations are fully determined by their action on the basis $\{\bel_j\}_j$, since $\varphi(s) = \sum \limits_{j=1}^n \varphi(\bel_j) \star s^j$. Indeed, one computes that $\varphi(\bel_j) = \sum \limits_{k=1}^n \ggft{u}_j^k \star \bel_k$ with $\ggft{u}_j^k \in \algft{A}$. The latter element can be written in matrix notations as $\ggft{u} = (\ggft{u}_j^k)_{j,k} \in \Mat{n}{\algft{A}}$. Such a gauge transformation is said unitary if $\ggft{u}^\ddagger \star \ggft{u} = \Matid{n}$, as one computes from \eqref{eq:ncft_db_ugt}. Therefore, the (unitary) gauge group is
	\begin{align}
		\ugt{n}
		&= \big\{ \ggft{u} \in \Mat{1,n}{\algft{A}},\ \ggft{u}^\ddagger \star \ggft{u} = \ggft{u} \star \ggft{u}^\ddagger = \Matid{n} \big\}.
	\end{align}
	Finally, the connection $A$ and the curvature $F$ are computed to gauge transform as
	\begin{subequations}
		\label{eq:ncft_db_Un_gt}
	\begin{align}
		A^{\ggft{u}}_X
		&= \ggft{u}^\ddagger \star A_X \star \ggft{u} - \iCpx \ggft{u}^\ddagger \star X(\ggft{u}),
		\label{eq:ncft_db_Un_gt_gf} \\
		F_{X,Y}^{\ggft{u}}
		&= \ggft{u}^\ddagger \star F_{X,Y} \star \ggft{u}.
		\label{eq:ncft_db_Un_gt_fs}
	\end{align}
	\end{subequations}
\end{Emph}

\begin{proof}
	We here derive some equations stated above, in the case of noncommutative electrodynamics $\ugt{1}$. The $\ugt{n}$ case can be computed similarly.
	
	First let us prove \eqref{eq:ncft_db_U1_cur}. To do so, one mainly uses \eqref{eq:ncft_db_U1_con} and \eqref{eq:ncft_db_U1_gf}.
	\begin{align*}
		R_{X,Y}(f)
		&= \nabla_X( \nabla_Y(f) ) - \nabla_Y( \nabla_X(f) ) - \nabla_{[X,Y]}(f) \\
		&= \nabla_X \big( Y(f) - \iCpx A_Y \star f \big)
		- \nabla_Y \big( X(f) - \iCpx A_X \star f \big)
		- [X,Y](f) + \iCpx A_{[X,Y]} \star f \\
		&= \xcancel{XY(f)} - \cancel{\iCpx A_X \star Y(f)} - \iCpx \nabla_X(A_Y) \star f - \bcancel{\iCpx A_Y \star X(f)} \\
		&\pe - \xcancel{YX(f)} + \bcancel{\iCpx A_Y \star X(f)} + \iCpx \nabla_Y(A_X) \star f + \cancel{\iCpx A_X \star Y(f)} \\
		&\pe - \xcancel{[X,Y](f)} + \iCpx A_{[X,Y]} \star f \\
		&= - \iCpx \big( X(A_Y) - Y(A_X) - \iCpx [A_X, A_Y] - A_{[X,Y]} \big) \star f.
	\end{align*}
	
	The hermitian structure considered together with \eqref{eq:ncft_db_Hcon}, imposes that for any real derivation $X$, one has
	\begin{align*}
		X\big( (f, g) \big)
		&= X( f^\dagger \star g)
		= X(f)^\dagger \star g + f^\dagger \star X(g) \\
		= (\nabla_X(f), g) + (f, \nabla_X(g))
		&= \big( X(f) - i A_X \star f \big)^\dagger \star g
		+ f^\dagger \star \big( X(g) - i A_X \star g \big) \\
		&= X(f)^\dagger \star g + f^\dagger \star X(g) + i f^\dagger \star( A_X^\dagger - A_X ) \star g,
	\end{align*}
	from which one obtains \eqref{eq:ncft_db_U1_Hgf}.
	
	The unitary elements of the gauge group given by \eqref{eq:ncft_db_ugt} reads
	\begin{align*}
		(\varphi(f), \varphi(g))
		&= ( \ggft{u} \star f, \ggft{u} \star g)
		= f^\dagger \star \ggft{u}^\dagger \star \ggft{u} \star g \\
		= (f, g)
		&= f^\dagger \star g,
	\end{align*}
	which implies \eqref{eq:ncft_db_U1_ugt}.
	
	Finally, given the expressions for the gauge transformations \eqref{eq:ncft_db_gt}, one computes
	\begin{align*}
		\nabla^\varphi_X(f)
		&= \varphi^{-1} \big( \nabla_X(\varphi(f)) \big)
		= \ggft{u}^\dagger \star \big( \nabla_X( \ggft{u} \star f) \big) \\
		&= \ggft{u}^\dagger \star \big( \nabla_X( \ggft{u} ) \star f + \ggft{u} \star X(f) \big) \\
		&= \ggft{u}^\dagger \star \big( X(\ggft{u}) - \iCpx A_X \star \ggft{u} \big) \star f + X(f) \\
		&= X(f) - \iCpx A_X^{\ggft{u}} \star f
	\end{align*}
	where $\varphi^{-1}(f) = \ggft{u}^{-1} \star f = \ggft{u}^\dagger \star f$, for $\ggft{u} \in \ugt{1}$. One directly reads \eqref{eq:ncft_db_U1_gt_gf} and deduces \eqref{eq:ncft_db_U1_gt_fs} from standard computations.
\end{proof}

Finally, suppose that there exists an integral over $\algft{A}$, potentially inherited from the integral over the \sT $\manft{M}$, for which $\algft{A}$ is the deformed space of smooth functions. In the context of \nCYM, let us consider the action
\begin{align}
	S
	&= \int \tdl{d+1}{x} \tr \big( F^{\mu\nu} \star F_{\mu\nu}^\ddagger \big)
	\label{eq:ncft_db_YMa}
\end{align}
where $\tr$ stands for the matrix trace. If one considers that the integral is cyclic for $\star$, that is
\begin{align}
	\int \tdl{d+1}{x} f \star g
	&= \int \tdl{d+1}{x} g \star f
	\label{eq:ncft_db_cyc}
\end{align}
then, the action \eqref{eq:ncft_db_YMa} is gauge invariant for the $\ugt{n}$ gauge group. Indeed, for any $\ggft{u} \in \ugt{n}$, one has
\begin{align}
\begin{aligned}
	S^{\ggft{u}}
	&= \int \tdl{d+1}{x} \tr \big( (F^\ggft{u})^{\mu\nu} \star (F^\ggft{u})_{\mu\nu}^\ddagger \big) \\
	&= \int \tdl{d+1}{x} \tr \big( \ggft{u}^\ddagger \star F^{\mu\nu} \star \ggft{u} \star (\ggft{u} \star F_{\mu\nu} \star \ggft{u}^\ddagger )^\ddagger \big) \\
	&= \int \tdl{d+1}{x} \tr \big( \ggft{u}^\ddagger \star F^{\mu\nu} \star \cancel{\ggft{u}} \star \cancel{\ggft{u}^\ddagger} \star F_{\mu\nu}^\ddagger \star \ggft{u} \big) \\
	&= \int \tdl{d+1}{x} \tr \big( \cancel{\ggft{u}} \star \cancel{\ggft{u}^\ddagger} \star F^{\mu\nu} \star F_{\mu\nu}^\ddagger \big)
	= S
\end{aligned}
	\label{eq:ncft_db_YMgi}
\end{align}
By construction, the action \eqref{eq:ncft_db_YMa} gives the \namefont{Yang-Mills} action at the commutative limit, and thus consists of a coherent deformation of the \namefont{Yang-Mills} theory. In the context of the \Moy space, this model is discussed in \secref{ncft_Moy}. When one builds a \gT similar to \eqref{eq:ncft_db_YMa} on the \kMt space-time, the lack of cyclicity \eqref{eq:ncft_db_cyc} prevents the action \eqref{eq:ncft_db_YMa} to be straightforwardly gauge invariant. The latter obstacle and the possible ways around it are discussed in \secref{kM_gt}.

\section{\namefont{Drinfel'd} twist based theories}
\label{sec:ncft_Dt}
\paragraph{}
Another way of defining the differential calculus in a consistent manner consists of using the twist deformation of a classical differential calculus \cite{Majid_1999, Sitarz_2001}. As detailed in \subsecref{gnc_dq_dt}, one starts with a \sT $\manft{M}$ and its \Lieat of \vf{s} $\Gamma(\manft{M})$. Then, a \Dt, defined on the enveloping algebra, is used to deform the vector fields and derive the corresponding \qST $\algft{A}$. Finally, one defines the bundle $\modft{X}$ as a $\algft{A}$-module.

Within this scheme, deformed $\Ug{n}$ \gT[ies] have first been considered in \cite{Aschieri_2006a, Vassilevich_2006, Chaichian_2006} and adapted to gravity in \cite{Aschieri_2005, Aschieri_2006b}.

\paragraph{}
Given a \sT $\manft{M}$, its set of \vf{s} forms a \Lieat. Thus, one can deform its universal enveloping algebra, that we note\footnote{
	In \cite{Aschieri_2006b}, $\Xi = \Gamma(\manft{M})$ which makes our notations differ a bit.
}
$\Xi = U(\Gamma(\manft{M}))$ for simplicity, via a \Dt (see \subsecref{gnc_dq_dt} for more details). Considering a \Dt $\Hoft{F} \in \Xi \otimes \Xi$, one defines the \Hfat $\Xi^{\Hoft{F}}$ of deformed \vf{s}. The associated \qST $\algft{A}$ is a $\Xi^{\Hoft{F}}$-module algebra and corresponds to a deformation of the smooth function algebra $\func^\infty(\manft{M})$. The new (noncommutative) product on $\algft{A}$ is given by
\begin{align}
	f \star g
	&= \sum \big( \Hoft{F}_1^{-1} \actl f \big) \cdot \big( \Hoft{F}_2^{-1} \actl g \big)
	\label{eq:ncft_Dt_spdt}
\end{align}
for any $f, g \in \algft{A}$, where we noted $\Hoft{F} = \sum \Hoft{F}_1 \otimes \Hoft{F}_2$. This product corresponds to \eqref{eq:gnc_dq_dt_spdt}.

One can then take the usual differential structure of $\manft{M}$ and twist it all the way to the end. Let us begin by the tensor product $\otimes$, which can be twisted through
\begin{align}
	X \otimes_{\Hoft{F}} Y
	&= \sum \big( \Hoft{F}_1^{-1} \actl X \big) \otimes \big( \Hoft{F}_2^{-1} \actl Y \big)
	\label{eq:ncft_Dt_ttp}
\end{align}
for any $X, Y \in \Xi^{\Hoft{F}}$, where $\Hoft{F}_j^{-1} \actl X = [\Hoft{F}_j^{-1}, X]$ stands for the \namefont{Lie} derivative. The product $\otimes_{\Hoft{F}}$ is often called the star tensor product and denoted $\otimes_\star$. The definition \eqref{eq:ncft_Dt_ttp} can be exported to tensors of any rank. Then, the wedge product of forms is twisted in a similar way to
\begin{align}
	\omega \wedge_{\Hoft{F}} \eta
	&= \sum \big( \Hoft{F}^{-1}_1 \actl \omega \big) \wedge \big( \Hoft{F}_2^{-1} \actl \eta \big)
	\label{eq:ncft_Dt_swed}
\end{align}
for any $\omega, \eta \in \Omega^1(\algft{A})$, \ie they are linear functional from $\Xi^{\Hoft{F}}$ to $\algft{A}$. The algebra of forms $\Omega^\bullet(\algft{A})$ is built from the star wedge product \eqref{eq:ncft_Dt_swed} and actually corresponds to multilinear braided antisymmetric maps from $(\Xi^{\Hoft{F}})^n$ to $\algft{A}$. 

\paragraph{}
The fact that the star wedge is braided antisymmetric, that is it satisfies \eqref{eq:ncft_Dt_basym}, is not innocuous. The \Hfat $\Xi$ is a triangular \Hfat (see \defref{ha_as_qtHalg}) with a \Rmat being trivial, \ie $\Rma = 1 \otimes 1$. The \Dt deformation actually conserves this triangularity property (see \thmref{ha_as_Dt_tqtHalg}) with a new \Rmat given by $\Rma = \Hoft{F}_{21} \Hoft{F}^{-1}$. The triangularity property thus transforms commutativity into braided commutativity 
\begin{align}
	f \star g
	&= \sum \big( \Rma^{-1}_1 \actl g \big) \star \big( \Rma^{-1}_2 \actl f \big),
	\label{eq:ncft_Dt_bcom}
\end{align}
and antisymmetry into braided antisymmetry
\begin{align}
	\omega \wedge_{\Hoft{F}} \eta
	&= - \sum \big(\Rma_1^{-1} \actl \eta \big) \wedge_{\Hoft{F}} \big( \Rma_2^{-1} \actl \omega \big).
	\label{eq:ncft_Dt_basym}
\end{align}
Beyond the invocation of the triangular structure, one can check that \eqref{eq:ncft_Dt_bcom} and \eqref{eq:ncft_Dt_basym} are satisfied, as done in the proof below.

The main remark to make here is that the \Rmat parametrises the noncommutativity, which could be expected since it parametrizes the noncocommutativity of the coproduct of $\Xi^{\Hoft{F}}$. The triangularity condition $\Rma_{21} \Rma = 1 \otimes 1$ allows the braiding to be a symmetry\footnote{
	At this point, one should stop calling $\Rma$ a braiding since it is a symmetry through the triangular condition. In other words, a braiding, contrary to a symmetry, never brings one back to the original position, as when one braids hairs. However, the triangular condition specifically states that \enquote{braiding} two times is akin to doing nothing and thus does not correspond to a braiding properly speaking. Here, we stick to the denomination of \cite{Aschieri_2020} in which everything is called \enquote{braided} even in the triangular case.
}.
Indeed, if one applies \eqref{eq:ncft_Dt_bcom} two times, one gets $f \star g = f \star g$. Finally, in the context of \dq, the semi-classical condition on the \Dt \eqref{eq:gnc_dq_dt_scl}, explicitly $\Hoft{F} = 1 \otimes 1 + \BigO{\kbar}$, imposes a similar condition on the \Rmat, \ie $\Rma = 1 \otimes 1 + \BigO{\kbar}$. Therefore, in the commutative limit $\kbar \to 0$, the \Rmat becomes trivial and the product $\star$ equals the commutative product of functions.

\begin{proof}
	We verify here the equality \eqref{eq:ncft_Dt_bcom}. The computation leading to \eqref{eq:ncft_Dt_basym} is very similar. The main thing we need is $\Rma = \Hoft{F}_{21} \Hoft{F}^{-1}$, \ie $\Rma^{-1} = \Hoft{F} \Hoft{F}^{-1}_{21}$.
	\begin{align*}
		\sum \big( \Rma^{-1}_1 \actl g \big) \star \big( \Rma^{-1}_2 \actl f \big)
		&= \sum \big( (\Hoft{F}_1^{-1} \Rma^{-1}_1) \actl g \big) \big( (\Hoft{F}_2^{-1} \Rma^{-1}_2) \actl f \big) \\
		&= \sum \big( (\Hoft{F}_1^{-1} \Hoft{F}_1 \Hoft{F}_2^{-1}) \actl g \big) \big( (\Hoft{F}_2^{-1} \Hoft{F}_2 \Hoft{F}_1^{-1}) \actl f \big) \\
		&= \sum \big( \Hoft{F}_2^{-1} \actl g \big) \big( \Hoft{F}_1^{-1} \actl f \big)
		= f \star g
		\qedhere
	\end{align*}
\end{proof}

\paragraph{}
The construction of a \gT on the \qST $\algft{A}$ can follow the same steps as in \secref{ncft_db} by replacing $\Der(\algft{A})$ by $\Xi^\Hoft{F}$. One has to be careful about the actions though. Explicitly, consider a $\algft{A}$-module $\modft{X}$, as the generalised fiber bundle. A connection $\nabla_X : \modft{X} \to \modft{X}$, for any $X \in \Xi^{\Hoft{F}}$, is defined as
\begin{subequations}
	\label{eq:ncft_Dt_con}
\begin{align}
	\nabla_X(s \actr_{\modft{X}} f)
	&= \nabla_X(s) \actr_{\modft{X}} f + s \actr_{\modft{X}} (X \actl f),
	& \text{(\namefont{Leibniz} rule)}
	\label{eq:ncft_Dt_con_Lr} \\
	\nabla_{X + zY}(s)
	&= \nabla_X(s) + \nabla_Y(s) \actr_{\modft{X}} z,
	& \text{($\Cen{\algft{A}}$-linearity)}
	\label{eq:ncft_Dt_con_lin}
\end{align}
\end{subequations}
for any $Y \in \Xi^{\Hoft{F}}$, $s \in \modft{X}$ and $z \in \algft{A}$. In the previous expression, $\actr_{\modft{X}}$ corresponds to the action of $\algft{A}$ on $\modft{X}$, which is to be thought as a \enquote{scalar} product on the bundle. Still, it is actually very different from $\actl$ introduced in \eqref{eq:ncft_Dt_spdt}, which corresponds here to the action of $\Xi^{\Hoft{F}}$ on $\algft{A}$ and should be thought as an action of derivation. Explicitly, $X \in \Xi^{\Hoft{F}}$ is a generalised \vf and $f \in \algft{A}$ a generalised function, so that $X \actl f$ corresponds to the derivative of $f$ \enquote{along} $X$.

Beyond this subtlety, one can construct the \nCYM with $\ugt{n}$ gauge group, for any $n$, as in \secref{ncft_db}. Rather than repeating this construction, we make two important remarks.

\paragraph{}
In the \gT as originally developed in \cite{Aschieri_2006a, Vassilevich_2006, Chaichian_2006}, the gauge group is undeformed, and the papers mainly evolve around matching the deformed field theory with the undeformed gauge. Besides, these studies were mainly done on the \Moy space.

The formulation of the latter authors may look different from ours as they work with infinitesimal gauge transformations, but one can relate the two. Explicitly, considering a $\Ug{n}$ \gT, a field $\phi$ transforms under the gauge $\ggft{u} \in \Ug{n}$ as $\phi^\ggft{u} = \ggft{u} \phi$ and the gauge field $A$ as in \eqref{eq:ncft_db_Un_gt_gf}. If now one considers $\ggft{u} = \exp(\iCpx \alpha_a T^a)$, where $T^a$ are the generators of the \Lieat $\Ua{n}$, then $\ggft{u}$ stands for the full transformation and $\alpha = \alpha_a T^a$ for the infinitesimal one. Therefore, taking the expansion to first order in $\alpha$, one computes the infinitesimal gauge transformations of $\phi$ and $A$ to be
\begin{align}
	\delta_\alpha \phi
	&= \phi^{\alpha} - \phi
	= \iCpx \alpha \phi, &
	\delta_\alpha A_\mu
	&= A^{\alpha}_\mu - A_\mu
	= \partial_\mu \alpha + \iCpx [\alpha, A_\mu].
	\label{eq:ncft_Dt_undef_gt}
\end{align}
Note that here, we can consider gauge transformation to be either local or global, by considering $\ggft{u}$ and $\alpha$ to be $x$-dependant or not. This can be enlarged to any \Lieat $\Lieft{g}$ of any gauge group $G$.

In their early works, the latter authors considered deformed transformations $\delta^\star_\alpha$ to act as in the commutative case, that is through \eqref{eq:ncft_Dt_undef_gt}. The deformation appears when one makes $\delta^\star_\alpha$ act on a product, like $\delta^\star_\alpha(\phi_1 \star \phi_2)$. Equivalently, the deformation is contained in the coproduct of the gauge differential, that is in $\Delta(\delta^\star_\alpha)$, which can be computed with the \Dt $\Hoft{F}$. One can check that the usual gauge transformation of the field strength $F_{\mu\nu}$ is recovered and that the algebra of gauge transformation indeed closes through $\delta^\star_\alpha \delta^\star_\beta - \delta^\star_\beta \delta^\star_\alpha = \delta^\star_{- \iCpx [\alpha, \beta]}$. Finally, one obtains that an action of the form
\begin{align}
	S
	&= \int \tdl{d+1}{x} \tr \big( F^{\mu\nu} \star F_{\mu\nu} \big),
	\label{eq:ncft_Dt_YMa}
\end{align}
is gauge invariant, upon cyclicity of the integral with respect to $\star$ \eqref{eq:ncft_db_cyc}.

\paragraph{}
The second remark concerns the notion of right or left modules. In the commutative theory, there is no ordering problem because every function commutes. When constructing a noncommutative theory, the ordering has a primordial importance, all the more that the ordering choice is not seen in the commutative limit. In a sense, the ordering is a new symmetry of the noncommutative theories as discussed in \secref{ncft_p4}.

Within the context of braided geometry \cite{Weber_2020b, Aschieri_2020} discussed above, the braided commutativity property \eqref{eq:ncft_Dt_bcom} is reducing the importance of the ordering. For example, the braided commutativity implies that left module $\modft{X}$ is also a braided right module, that is, given the left action $\actl_{\modft{X}}$, one can define a right action $\actr_{\modft{X}}$ through\footnote{
	One could equivalently start with a bimodule structure and require that the two actions are braided symmetric in the sense that they satisfy \eqref{eq:ncft_Dt_bsyma}.
}
\begin{align}
	f \actl_{\modft{X}} s
	&= \sum (\Rma_1^{-1} \actl s) \actr_{\modft{X}} (\Rma_2^{-1} \actl f).
	\label{eq:ncft_Dt_bsyma}
\end{align}
Many other structures can be braided in a similar fashion \cite{Weber_2020a}, like derivations or connections. The braided derivations $\Der_\Rma(\algft{A})$, defined as the linear functional satisfying a braided \Lru
\begin{align}
	X(f \star g)
	&= X(f) \star g + \sum \big(\Rma_1^{-1} \actl f \big) \star \big( (\Rma_2^{-1} \actl X)(g) \big),
	\label{eq:ncft_Dt_bLru}
\end{align}
form a $\algft{A}$-bimodule for the braided symmetric actions of \eqref{eq:ncft_Dt_bsyma}, whereas the derivations $\Der(\algft{A})$ only form a $\Cen{\algft{A}}$-module.

\section{\namefont{Seiberg-Witten} map}
\label{sec:ncft_SW}
\paragraph{}
In the context of the \Moy space, for deformed $\Ug{n}$ \gT[ies], \namefont{Seiberg} and \namefont{Witten} \cite{Seiberg_1999} (see also \cite{Schomerus_1999}) found a correspondence between noncommutative gauge fields, noted $\hat{A}$ in this \secref{ncft_SW}, and the ordinary (commutative) one, noted $A$. The relation \eqref{eq:ncft_SW_gfgt} found in \cite{Seiberg_1999} was first considered for open strings with a magnetic field, but was also applied in fully noncommutative contexts thanks to its fairly general form.

The \namefont{Seiberg-Witten} map is defined by
\begin{align}
    \hat{A}_\mu (A_\nu) + \hat{\delta}_{\hat{\alpha}} \hat{A}_\mu (A_\nu)
    = \hat{A}_\mu (A_\nu + \delta_\alpha A_\nu)
    \label{eq:ncft_SW_SWm}
\end{align}
where $\hat{\delta}_{\hat{\alpha}}$ is the infinitesimal noncommutative gauge transformation with parameter $\hat{\alpha}$ given by
\begin{align}
   \hat{\delta}_{\hat{\alpha}} \hat{A}_\mu
   = \partial_\mu \hat{\alpha} + i [\hat{\alpha}, \hat{A_\mu}]_\theta.
   \label{eq:ncft_SW_gfgt}
\end{align}
A strong hypothesis of this construction is the so-called \enquote{gauge equivalence}. The main idea being that if $A$ and $A'$ are related by a gauge transformation $\alpha$, \ie $A' = A + \delta_\alpha A$, then $\hat{A}$ and $\hat{A}'$ should relate by a deformed gauge transformation $\hat{\alpha}$, \ie $\hat{A}' = \hat{A} + \hat{\delta}_{\hat{\alpha}} \hat{A}$. This implies that the deformed gauge transformation depends on the undeformed one and on the gauge field. In other words, the quantity $\hat{\alpha}$ depends on $\alpha$ and $A$, namely $\hat{\alpha} = \hat{\alpha}(A,\alpha)$. In this sense, \eqref{eq:ncft_SW_SWm} is a mathematical formulation of the gauge equivalence.

\paragraph{}
The full expression of \eqref{eq:ncft_SW_gfgt} is obtained by expansion in powers of $\dpMoy$. For instance, up to the second order, one obtains
\begin{align}
     \hat{\delta}_{\hat{\alpha}} \hat{A}_\mu
     = \partial_\mu \hat{\alpha} 
     - \Theta^{\rho\sigma} \, \partial_\rho \alpha \, \partial_\sigma A_\mu 
     + \BigO{\dpMoy^2},
\end{align}
which, combined with \eqref{eq:ncft_SW_SWm}, yields
\begin{align}
    \hat{A}_\mu(A_\nu)
    &= A_\mu  - \frac{1}{2} \Theta^{\rho\sigma} A_\rho(\partial_\sigma A_\mu + F_{\sigma\mu}) + \BigO{\dpMoy^2},
    \label{eq:ncft_SW_gfgt2}
\end{align}
where $F_{\mu\nu}$ is the ordinary field strength. Accordingly, the noncommutative field strength takes the form
\begin{equation}
    \hat{F}_{\mu\nu}
    = F_{\mu\nu} + \Theta^{\rho\sigma} (F_{\mu\rho} F_{\nu\sigma} - A_\rho \partial_\sigma F_{\mu\nu}) + \BigO{\dpMoy^2}.
    \label{eq:ncft_SW_fs}
\end{equation}
One would proceed similarly whenever a (fermionic) matter field is included so as to obtain a $\dpMoy$-expanded \nCFT.

\paragraph{}
This construction suffers from two main caveats despite its explicit formulation. First, the map \eqref{eq:ncft_SW_SWm} can only be computed as an infinite power expansion over $\dpMoy$. This is highly sufficient when one wants to characterise the semi-classical behaviour of the theory, but is hampering a full treatment. Moreover, it was shown \cite{Martin_2007} that the matter field part of the theory was not renormalisable, at least in the standard way.

The \namefont{Seiberg-Witten} map has attracted lot of attention concerning the study of quantum properties of these \gT[ies], or their phenomenological traits. One can find the relevant references in \cite{Hersent_2023a}.

\section{New approaches}
\label{sec:ncft_na}
\paragraph{}
More recent approaches to \gT[ies] on \qST have been proposed and are gathered here.

\subsection{Gauge theories with \tops{$L_\infty$}{L-infinity}-algebras}
\label{subsec:ncft_na_Li}
\paragraph{}
The algebraic framework of $L_\infty$-algebras was shown to be able to render field theory dynamics and \gT[ies], at least at the classical level \cite{Hohm_2017}. An $L_\infty$-algebra is a sort of infinite extension of a \Lieat, where the bracket $[\cdot, \cdot]$ for two elements has counterpart brackets for $3$, $4$, ... up to infinity elements. The \namefont{Jacobi} identity of the bracket $[\cdot, \cdot]$ has now also counterparts for the other brackets. Explicitly, a $L_\infty$-algebra is a graded vector space $V = \bigoplus \limits_{k \in \Int} V_k$ with graded antisymmetric multilinear maps $\ell_n : \bigotimes \limits_{n \in \NInt} V \to V$ called $n$-brackets. Thus, the $2$-bracket $\ell_2$ is the analogue of $[\cdot, \cdot]$. However, classical gauge theories can be recovered by only using 4 degrees, \ie $V = V_0 \oplus \cdots \oplus V_3$, where degree $0$ fields are gauge parameters, degree $1$ fields are gauge fields, degree $2$ fields encode the equations of motion and degree $3$ fields encode the \namefont{Noether} identities. 

\paragraph{}
The authors of \cite{Blumenhagen_2018, Kupriyanov_2020b} advocate that the $L_\infty$-algebra framework is natural in the context of \dq. Furthermore, the $L_\infty$-algebras are known to encode both noncommutative and non-associative algebras, a case that may occur when deforming the so-called \enquote{quasi-\namefont{Poisson} structures}, according to \cite{Blumenhagen_2018}. Therefore, the latter authors generalised the $L_\infty$-algebra gauge theory to noncommutative gauge theory in the context of deformed \namefont{Chern-Simons} and \namefont{Yang-Mills} actions.

As a brief summary, we detail how the \nCYM fits into the $L_\infty$-algebra setting. If one requires the general gauge transformation of a vector field $A \in V_{1}$, defined by 
\begin{align}
	\delta_\alpha A
	&= \sum_{n = 0}^{+\infty} \frac{1}{n!} (-1)^{\frac{n(n-1)}{2}} \ell_{n+1}(\alpha, A, \ldots, A),
	\label{eq:ncft_na_Ligt}
\end{align}
to correspond to the straightforward deformed gauge transformation $\delta_\alpha A = \partial \alpha + \iCpx [\alpha, A]_\star$, then one has
\begin{align}
	\ell_1(f)
	&= \partial f, &
	\ell_2(f, A)
	&= \iCpx [f, A]_\star.
	\label{eq:ncft_na_Linb}
\end{align}
The higher $n$-brackets are determined by the equations of motion for $A$. Moreover, one can check that the \namefont{Jacobi} identities are satisfied. For example, the first one states that $\tensor{\epsilon}{_\mu^{\nu\rho}} \partial_\nu \partial_\rho f = 0$, which is always true and the second one requires that the bracket $[f, \cdot]_\star$
satisfies the \Lru.

\paragraph{}
Another version of $L_\infty$-algebras encoding \nCG was constructed in \cite{Dimitrijevic_2022} and presented below. Given a classical \sT $\manft{M}$, one considers the universal enveloping algebra of vector fields $\Xi = U(\Gamma(\manft{M}))$. The classical $L_\infty$-algebra associated to $\manft{M}$ corresponds to a $\Int$-graded $\Xi$-module for which the $n$-bracket $\ell_n$ commutes with the action of $\Xi$. Since the elements of $\Xi$ are polynomials of derivatives, it simply means that the derivatives act straightforwardly on the $n$-bracket, as the generalised analogue of the relation $\partial [f,g] = [\partial f, g] + [f, \partial g]$.

The noncommutative version of the latter classical picture is obtained via a \Dt deformation, similarly as described in \secref{ncft_Dt}. The noncommutative $L_\infty$-algebra is defined to be a $\Xi^\Hoft{F}$-module, with $\Hoft{F} \in \Xi \otimes \Xi$ the \Dt. The deformed $n$-bracket $\ell_n^\Hoft{F}$ are defined as the twisted versions of $\ell_n$, that is
\begin{align}
	\ell_n^\Hoft{F}(f_1, \ldots, f_n)
	&= \ell_n( f_1 \otimes_{\Hoft{F}} \cdots \otimes_{\Hoft{F}} f_n)
\end{align}
with $\ell_1^\Hoft{F} = \ell_1$, where $\otimes_\Hoft{F}$ is defined in \eqref{eq:ncft_Dt_ttp}. One can show that the deformed $n$-bracket is braided graded antisymmetric, \ie
\begin{align}
	\ell_n^{\Hoft{F}}(f_1, \ldots, f_j, \ldots, f_k, \ldots, f_n)
	&= (-1)^{\vert f_j \vert \, \vert f_k \vert}
	\ell_n^{\Hoft{F}}(f_1, \ldots, \Rma_1^{-1} \actl f_k, \ldots, \Rma_2^{-1} \actl f_j, \ldots, f_n)
\end{align}
where $\vert f_j\vert$ is the degree of $f_j$, and $\Rma = \Hoft{F}_{21} \Hoft{F}^{-1}$ the \Rmat associated to $\Hoft{F}$. The gauge theory is then defined by twisting the usual (undeformed) gauge theory on $L_\infty$-algebra. For example, the deformed gauge transformation of $A$ \eqref{eq:ncft_na_Ligt} becomes
\begin{align}
	\delta^\star_\alpha A
	&= \sum_{n = 0}^{+ \infty} \frac{1}{n!} (-1)^{\frac{n(n-1)}{2}} \ell^\Hoft{F}_{n+1}(\alpha, A, \ldots, A).
	\label{eq:ncft_na_tLigt}
\end{align}
The deformed action, defined as the twisted classical action, can be shown to be gauge invariant under \eqref{eq:ncft_na_tLigt}.

\paragraph{}
For the case of the \Moy space, deformed with the twist \eqref{eq:ncft_Moy_Dt}, the study of the deformed $\Ug{1}$ gauge theory is undertaken with $L_\infty$-algebra \cite{Dimitrijevic_2022}. The quantisation method uses the \namefont{Batalin–Vilkovisky} formalism, which had previously been formulated in the $L_\infty$-algebra setting. The photon propagator is unchanged compared to the undeformed case and there are no three-photon or four-photon interactions, contrary to an action of the form \eqref{eq:ncft_db_YMa}. However, the fermion-photon vertex is non-trivially deformed. Therefore, the first deformed diagram would be the photon self-energy one. It appears that this diagram is UV-divergent, triggering a \UVIR in the theory. We refer to \secref{ncft_p4} for more details on \nCFT and how the \UVIR arises. Contrary to \eqref{eq:ncft_p4_nc2p}, the \npf{2}, at one loop, has no non-planar diagrams and still triggers a mixing. The authors of \cite{Dimitrijevic_2022} advocate that this \UVIR comes from an unadapted quantisation procedure for braided theories.

\subsection{\namefont{Poisson} gauge theories}
\label{subsec:ncft_na_Pgt}
\paragraph{}
The \namefont{Poisson} \gT[ies] correspond to noncommutative \gT[ies] in which the gauge transformations of the gauge field $A$ and the field strength $F$ are tuned so as to gauge transform classically and therefore as to obtain a gauge invariant action in the usual manner.

\paragraph{}
One starts with an \emph{ansatz} for the gauge transformation of the gauge field $A$ and the expression of the curvature $F$ of the form
\begin{align}
    \delta_\alpha A_\mu 
    &= \gamma_\mu^\nu(A) \, \partial_\nu(f) + \{ A_\mu, \alpha \}, &
    F_{\mu\nu}
    &= \tensor{P}{_{\mu\nu}^{\rho\sigma}}(A) \, \partial_\rho A_\sigma + \tensor{R}{_{\mu\nu}^{\rho\sigma}}(A) \, \{A_\rho, A_\sigma\},
    \label{eq:ncft_na_Pgt}
\end{align}
where $\gamma$, $P$ and $R$ are to be determined. To match with the usual commutative limit, these new fields have to satisfy the following condition
\begin{align}
    \gamma_\mu^\nu 
    &= \delta_\mu^\nu + \BigO{\kbar}, &
    \tensor{P}{_{\mu\nu}^{\rho\sigma}}
    &= \delta_\mu^\rho \delta_\nu^\sigma - \delta_\mu^\sigma \delta_\nu^\rho + \BigO{\kbar}, &
    \tensor{R}{_{\mu\nu}^{\rho\sigma}}
    &= \frac{1}{2} \big( \delta_\mu^\rho \delta_\nu^\sigma - \delta_\mu^\sigma \delta_\nu^\rho \big) + \BigO{\kbar}.
    \label{eq:ncft_na_Pgscl}
\end{align}
One imposes additional conditions corresponding to the closure of the gauge transformation and the covariance of the field strength, which writes respectively
\begin{align}
	[\delta_\alpha, \delta_\beta] A 
	&= \delta_{\{\alpha, \beta \}} A, &
	\delta_\alpha F 
	&= \{F, \alpha \}.
	\label{eq:ncft_na_Pggt}
\end{align}
This whole set of conditions imposes constraints on $\gamma$, $P$, $R$, to which a solution, if it exists, can give rise to a gauge invariant action. Indeed, considering a \namefont{Yang-Mills}-like action with integrand $F^{\mu\nu} \star F_{\mu\nu}$, the gauge transformation \eqref{eq:ncft_na_Pggt} implies that the Lagrangian transforms covariantly, which leads to the gauge invariant of this action upon integral cyclicity \eqref{eq:ncft_db_cyc}.

Note that this \emph{ansatz} was made in the semi-classical limit, \ie approximating the $\star$-commutator of functions by the \namefont{Poisson} bracket $\{\cdot, \cdot\}$. This is the reason why the latter \namefont{Poisson} bracket appears in \eqref{eq:ncft_na_Pggt}.

\paragraph{}
The fact that the expression for the noncommutative field strength considered here in \eqref{eq:ncft_na_Pgt} is different from other previous proposals, say \eqref{eq:ncft_db_U1_cur}, can be related to the discussion of \secref{ncft_db}. Indeed, the noncommutative formulations of \gT[ies] work through analogues and lack of physical intuition when defining the \enquote{physical} fields. Therefore, there is no argument for the \enquote{noncommutative electromagnetic field} (if such a thing makes sense) to be \eqref{eq:ncft_db_U1_cur} rather than \eqref{eq:ncft_na_Pgt}.

\paragraph{}
This framework was first developed in \cite{Kupriyanov_2020a} for the \Moy space and extended to \Lieat-type noncommutativity in \cite{Kupriyanov_2022}. A \Lieat-type noncommutativity corresponds to a \qST in which the bracket coordinates is linear, that is
\begin{align}
	[x^\mu, x^\nu]_\star
	&= \tensor{\sC}{^{\mu\nu}_\rho} x^\rho
	\label{eq:ncft_na_Latnc}
\end{align}
with $\sC$ a constant. As examples, $\Rcl$ \eqref{eq:gnc_R3l_coord}, \kMt \eqref{eq:kM_kP_kM_alg} and \rMt \eqref{eq:kM_rM_co} are of this type.

\paragraph{}
Furthermore, it was shown \cite{Kurkov_2022} that the \namefont{Poisson} gauge theory can be cast into a $L_\infty$-algebra formalism. In the slowly varying field approximation, the starred bracket $[\cdot, \cdot]_\star$, corresponding to $\ell_2$ as in \eqref{eq:ncft_na_Linb}, is replaced by the \namefont{Poisson} bracket $\{\cdot, \cdot\}$. The higher degree $n$-brackets are determined by the so-called $L_\infty$-bootstrap, using the generalised \namefont{Jacobi} identities.

\section{The example of the \namefont{Moyal} space}
\label{sec:ncft_Moy}
\paragraph{}
In this \secref{ncft_Moy}, the \Moy space is briefly introduced together with its noncommutative \gT. This \gT relies on methods introduced in \secref{ncft_db} and \secref{ncft_Dt}. The \UVIR popping out of the photon propagator correction (\npf{2}), at one-loop, is discussed. Some other \gT[ies], mainly developed to cure the mixing, are slightly discussed. We refer to \cite{Hersent_2023a} for more details and an extensive list of references.

\paragraph{}
The \Moy space, generically denoted $\MoyR{4}$, corresponds to a \dq of the (symplectic) space $\Real^4$. One can deform in a similar fashion the space $\Real^{2n}$ for any $n \in \nzNInt$, but we stick to $4$ dimensions here. Its coordinates satisfy 
\begin{align}
	[x^\mu, x^\nu]_{\star_\dpMoy}
	&= \iCpx \Theta^{\mu\nu}, &
	\text{with }
	\Theta = \dpMoy
	\begin{pmatrix}
		0 & 1 & 0 & 0 \\
		-1 & 0 & 0 & 0 \\
		0 & 0 & 0 & 1 \\
		0 & 0 & -1 & 0
	\end{pmatrix},
	\label{eq:ncft_Moy_coord}
\end{align}
where $\dpMoy$ is the \dpt and $\star_\dpMoy$ is the \spdtt of $\MoyR{4}$.

The expression of the \spdtt $\star_\dpMoy$ can be obtained either through \calg techniques (see \subsecref{gnc_dq_ca}), or by \Dt deformation (see \subsecref{gnc_dq_dt}). 

In the first case, one considers the \Lieat $\Lieft{g}$ to be the \namefont{Heisenberg} algebra (that is the algebra of \eqref{eq:ncft_Moy_coord}) in $5$ dimensions (or $2n+1$ in the general case). This has to do with the fact that one needs to add $1$ as a generator in order for the \namefont{Lie} bracket \eqref{eq:ncft_Moy_coord} to close. The \Haarm of the \namefont{Heisenberg} group is the usual \namefont{Lebesgue} measure. One still needs to go down from $5$ dimensions to $4$ and this is done via the map
\begin{align}
	f^{\#}(x_1, \ldots, x_4)
	&= \int_{\Real} \tdl{}{z} f(z, x_1, \ldots, x_4) \ e^{- 2\pi \iCpx \dpMoy z}
	\label{eq:ncft_Moy_5to4}
\end{align}
The expression of the \Ftt is then given by the usual expression of the \Ftt on $\Real^4$, and one computes from expression \eqref{eq:gnc_dq_ca_spdt} that
\begin{align}
	(f \star_\dpMoy g)(x)
	&= \frac{1}{(\pi \dpMoy)^4} \int_{\Real^4 \times \Real^4} \td^{4}{y} \, \tdl{4}{z} 
	f(x + y) \ g(x + z) \ e^{-2 \iCpx \, y^\mu \Theta^{-1}_{\mu\nu} z^\nu}.
	\label{eq:ncft_Moy_spdt}
\end{align}

The second case relies on the \Dt
\begin{align}
	\Hoft{F}
	&= \exp \left( - \frac{\iCpx}{2} \Theta^{\mu\nu} \partial_\mu \otimes \partial_\nu \right)
	\label{eq:ncft_Moy_Dt}
\end{align}
which corresponds to an Abelian \Dt (see \exref{gnc_dq_dt_abe}), for which the considered vector fields correspond to the derivative $\{\partial_\mu = \frac{\partial}{\partial x^\mu}\}_\mu$. The corresponding \spdtt expression is
\begin{align}
	(f \star_\dpMoy g)(x)
	&= \sum_{n = 0}^{+ \infty} \frac{1}{n!} \left(\frac{\iCpx}{2}\right)^{\!\! n} \Theta^{\mu_1 \nu_1} \cdots \Theta^{\mu_n \nu_n} \
	\partial_{\mu_1} \cdots \partial_{\mu_n} f(x) \ 
	\partial_{\nu_1} \cdots \partial_{\nu_n} g(x).
	 \label{eq:ncft_Moy_spdt_Dt}
\end{align}
The expression \eqref{eq:ncft_Moy_spdt_Dt} can be made equal to \eqref{eq:ncft_Moy_spdt} by writing the (infinite) \namefont{Taylor} expansion of $f$ and $g$ in \eqref{eq:ncft_Moy_spdt}. One can also check that both previous expression of $\star_\dpMoy$ satisfies \eqref{eq:ncft_Moy_coord}. The corresponding involution ${}^\dagger$ is the complex conjugation. 

Note that the integral of $\Real^4$ defines an integral over $\MoyR{4}$ and satisfies
\begin{align}
	\int_{\Real^4} \tdl{4}{x} (f \star_\dpMoy g)(x)
	&= \int_{\Real^4} \tdl{4}{x} (g \star_\dpMoy f)(x)
	= \int_{\Real^4} \tdl{4}{x} f(x) \, g(x).
	\label{eq:ncft_Moy_int}
\end{align}
The first equality states that the integral is cyclic and the second one that it is closed.

\paragraph{}
The deformation of the space $\Real^4$ induces deformation in the symmetries of the space \cite{Chaichain_2004, Aschieri_2005, Balachandran_2005}. These deformations are more easily obtained in the \Dt formalism \eqref{eq:ncft_Moy_Dt}. One interprets the derivations $\partial_\mu$ as the action of the usual translation $P_\mu$ on functions, so that \eqref{eq:ncft_Moy_Dt} is actually a \Dt for the \Palgt $\Palg[4]$ of $\Real^4$. The twist thus deforms $U(\Palg[4])$ into $\Palg[4]_\dpMoy = U(\Palg[4])^{\Hoft{F}}$, the latter being the new symmetries of the \Moy space.

The symmetries of the field theory one builds on the \Moy space are therefore deformed. It induces phenomenological constraints on the theory as discussed in \secref{qg_ph}. Through \Pdef, the field theory may exhibit a \LIV.

\paragraph{}
In order to build the differential calculus on the \Moy space, one can consider the Abelian \Lieat $\Lieft{D}_1$ of derivations generated by the $\partial_\mu$'s. It is a \namefont{Lie} subalgebra of the full set of derivations, $\Lieft{D}_1 \subset \Der(\MoyR{4})$, so that one can trade it for $\Der(\MoyR{4})$ in the construction of \secref{ncft_db}. Therefore, the gauge invariant action of \nCYM \eqref{eq:ncft_db_YMa} can be built through the connection and its curvature defined on $\Lieft{D}_1$.

Note that if one were to consider the full set of derivations $\Der(\MoyR{4})$, then there would be an infinite number of components in $A$, instead of $4$, thus leading to an infinite number of degrees of freedom in the theory. Therefore, one usually chooses to work with a restricted amount of derivations to avoid this inconvenience. One should remark, however, that both cases lead to the same commutative limit with \aprio $4$ degrees of freedom, since in this case any (real) derivation $X \in \Der(\Real^4)$ can be written as a linear combination of the $\partial_\mu$'s, \ie $X = X^\mu \partial_\mu$, with $X^\mu \in \Real$.

%\begin{proof}
%	Let $f \in \func^\infty{\Real^4}$ and $X \in \Der(\Real^4)$. Given $x_0 \in \Real^4$, one can \namefont{Taylor} expand $f$ around $x_0$ as
%	\begin{align*}
%		f(x) 
%		&= f(x_0) + (x-x_0)^\mu \partial_\mu f(x_0)
%		+ (x-x_0)^\mu (x-x_0)^\nu \tilde{f}_{\mu\nu}(x_0).
%	\end{align*}
%	Applying $X$ on both sides of the previous expression gives
%	\begin{align*}
%		X(f)(x)
%		&= X(x^\mu) \partial_\mu f(x_0)
%		+ \tilde{f}_{\mu\nu}(x_0) \big( X(x^\mu x^\nu) - x^\mu_0 X(x^\nu) - X(x^\mu) x^\nu_0 \big),
%	\end{align*}
%	where we used that $X(c) = 0$ for any constant $c$. Then, evaluating at $x = x_0$, gives the wanted result with $X^\mu = X \circ \varphi^\mu$, where $\varphi^\mu(x) = x^\mu$.
%\end{proof}

\paragraph{}
The simplest action for a \nCYM corresponds to \eqref{eq:ncft_db_YMa}. Even in the deformed $\Ug{1}$ case, this action has $3$-vertex and $4$-vertex interactions for $A$. The vertices are quite similar to the commutative (non-Abelian) one with extra phase factors of the form $\sin\left( \frac{p \Theta q}{2} \right)$ with external momenta $p$ and $q$. Whenever one uses a \BRST procedure on this action, one obtains a vanishing \npf{1} (contrary to \kMt, see \secref{kM_gt}) and a vacuum polarisation tensor (\npf{2}) which, at one-loop, triggers a \UVIR due to the appearance of a new infra-red singularity.

The latter behaviour, described in \secref{ncft_p4}, distinguishes some diagrams, called planar, that diverge at large momenta (UV) and some others, called non-planar, that are singular at vanishing momenta (IR). In the mainstream meaning of renormalisability, a theory with a \UVIR is not perturbatively renormalisable due to the presence of quadratic and linear IR divergences. Indeed, this IR singularity induces an uncontrolled UV divergence for higher loop orders for which one cannot predicts the counter term expression. Note that in this case, the IR limit expression of the polarisation tensor is gauge invariant, so that the \UVIR of this gauge theory is not a gauge artefact.

Some methods have succeeded to get rid of the \UVIR and have shown to be fully renormalisable to all orders, as discussed in \secref{ncft_p4}. However, these methods are bound to the \Moy space and involves a modification of the action \eqref{eq:ncft_db_YMa}.

\paragraph{}
Another heavily studied gauge theory on the \Moy space is the so-called induced gauge theory \cite{deGoursac_2007}. The term \enquote{induced} stems from the fact that the study of counter-term suggested to work with powers of the covariant coordinate $\mathcal{A}$ instead of $A$. 

\begin{Emph}{The gauge invariant connection}
	In the case of the \Moy space, the derivations $\partial_\mu$ are inner derivations, meaning that it exists $\xi_\mu \in \MoyR{4}$ such that
	\begin{align}
		\partial_\mu f
		&= [\xi_\mu , f]_{\star_\dpMoy}
		\label{eq:ncft_Moy_inder}
	\end{align}
	for any $f \in \MoyR{4}$. One can even explicitly compute $\xi_\mu = - \iCpx \Theta^{-1}_{\mu\nu} x^\nu$. In this case, one can build a so-called gauge invariant connection $\nabla^{\text{inv}}$ on $\MoyR{4}$. Indeed, if one considers the expression \eqref{eq:ncft_db_U1_gf} where the gauge field $A$ is replaced by $\xi$ above, then one has 
	\begin{align}
		\nabla^{\text{inv}}_\mu(f)
		&= \partial_\mu(f) - \iCpx \, \Theta^{-1}_{\mu\nu} x^\nu \star_\dpMoy f 
		= \iCpx \, f \star_\dpMoy \Theta^{-1}_{\mu\nu} x^\nu,
		\label{eq:ncft_Moy_gic}
	\end{align}
	where we used \eqref{eq:ncft_Moy_inder}. Given a connection $\nabla$ on $\MoyR{4}$, it follows that $\iCpx(\nabla - \nabla^{\text{inv}})$ is a $1$-form tensor given by $\mathcal{A} = A + \xi$, which gauge transforms as $\mathcal{A}^{\ggft{u}} = \ggft{u}^\dagger \star_\dpMoy \mathcal{A} \star_\dpMoy \ggft{u}$, for any $\ggft{u} \in \ugt{1}$. Finally, one computes from \eqref{eq:ncft_db_U1_cur}, that the field strength writes
	\begin{align}
		F_{\mu\nu}
		&= [\mathcal{A}_\mu, \mathcal{A}_\nu]_{\star_\dpMoy} - \iCpx \Theta^{-1}_{\mu\nu}.
		\label{eq:ncft_Moy_gicur}
	\end{align}
	One can note that the curvature associated to $\nabla^{\text{inv}}$ is $- \iCpx \Theta^{-1}_{\mu\nu}$, which corresponds to the last term of \eqref{eq:ncft_Moy_gicur}.
	
	$\mathcal{A}_\mu = A_\mu - \iCpx \Theta^{-1}_{\mu\nu} x^\nu$ is sometimes called the covariant coordinate. It can be straightforwardly generalised to the case of $\ugt{n}$ gauge theory. 
\end{Emph}

In view of the gauge transformation of $\mathcal{A}$, a term of the form $\int \mathcal{A}_\mu \star_\dpMoy \mathcal{A}^\mu$ is gauge invariant, so that a general gauge invariant action writes
\begin{align}
	S(\mathcal{A})
	&= \int \tdl{4}{x} \left( \, \frac{1}{4} F_{\mu\nu} \star_\dpMoy F^{\mu\nu} + \frac{\Omega^2}{4} \big\{ \mathcal{A}_\mu, \mathcal{A}_\nu \big\}^2_\dpMoy + \kappa \, \mathcal{A}_\mu \star_\dpMoy \mathcal{A}^\mu \right).
\end{align}
In the previous expression, $F$ is defined in terms of $\mathcal{A}$ through \eqref{eq:ncft_Moy_gicur}, while $\Omega$ and $\kappa$ are constants. The latter action actually corresponds to a type IIB (IKKT) matrix model \cite{Ishibashi_1997}. It was also shown to have a \st formulation \cite{Gayral_2013}.

This theory exhibits a very complex vacuum configuration \cite{deGoursac_2008}, which heavily depends on $\Omega$ and $\kappa$. For specific values of the latter constants, tedious expressions were found. This consists in the main obstacle in studying the renormalisability of such models. Note also that the previous model \aprio suffers from a non-vanishing tadpole (\npf{1}).

Note that the previous computations are done in the matrix basis of \Moy, which consists of a convenient tool for computation.

\begin{Emph}{\namefont{Moyal} matrix basis}
	The matrix basis \cite{Gracia-Bondia_1988} $\{f_{mn}\}_{m,n \in \NInt}$ is an orthonormal basis of $\Schw{\Real^4} \subset \MoyR{4}$, the set of \namefont{Schwarz} function of $\Real^4$, \ie
	 \begin{subequations}
	 	\label{eq:ncft_Moy_mb}
    \begin{align}
        f_{mn} \star_\dpMoy f_{kl} 
        &= \delta_{nk} f_{ml}, &
        f_{mn}^\dagger 
        &= f_{nm}, 
        \label{eq:ncft_Moy_mb_norm}
    \end{align}%
    \vspace{\dimexpr-\abovedisplayskip-\belowdisplayskip-\baselineskip+\jot + 4pt}%
    \begin{align}
        \int \tdl{4}{x} \big(f_{mn}^\dagger \star_\dpMoy f_{kl} \big) (x)
        = 2\pi \, \dpMoy \, \delta_{mk} \delta_{nl}.
        \label{eq:ncft_Moy_mb_orth}
    \end{align}
    \end{subequations}
    Thus any elements $g, h \in \MoyR{4}$ can be decomposed as
	\begin{align}
		g
		&= \sum_{m,n = 0}^{+\infty} g_{mn} f_{mn}, &
		h
		&= \sum_{m,n = 0}^{+\infty} h_{mn} f_{mn},
		\label{eq:ncft_Moy_mbd}
	\end{align}	    
    where $g_{mn}, h_{mn} \in \Cpx$. It is called the \enquote{matrix} basis since the decomposition \eqref{eq:ncft_Moy_mbd} implies that $g$ is fully determined by the infinite size matrix $\{g_{mn}\}_{mn}$. Furthermore, the product of functions looks similar to a matrix product as one can compute
    \begin{align}
    	(g \star_\dpMoy h)(x)
    	&= \sum_{m,n = 0}^{+ \infty} \left( \sum_{k = 0}^{+\infty} g_{mk} h_{kn} \right) \, f_{mn}(x), &
    	\text{\ie} \
    	(g \star_\dpMoy h)_{mn}
    	&= \sum_{k=0}^{+\infty} g_{mk} h_{kn}.
    	\label{ncft_Moy_mbp}
    \end{align}
\end{Emph}

\section{\namefont{Lie} algebra-type noncommutative \tops{$\phi^4$}{phi\^4}-theory}
\label{sec:ncft_p4}
\paragraph{}
In this section, we consider $\algft{A}$ to be a \qST corresponding to the deformation of $\manft{M}$ a $d+1$-dimensional \sT with \spdtt $\star$ and involution ${}^\dagger$. The (local) coordinate functions of $\manft{M}$, $\{x^\mu\}_\mu$ are assumed to satisfy 
\begin{align}
	[x^\mu, x^\nu]_\star
	&= \tensor{\sC}{^{\mu\nu}_\rho} x^\rho
	\label{eq:ncft_p4_Latnc}
\end{align}
with $\tensor{\sC}{^{\mu\nu}_\rho} \in \Cpx$. The relation \eqref{eq:ncft_p4_Latnc} is sometimes called a \Lieat-type noncommutativity. As examples, $\Rcl$ \eqref{eq:gnc_R3l_coord}, \kMt \eqref{eq:kM_kP_kM_alg} and \rMt \eqref{eq:kM_rM_co} are of this type. One can also write the \Moy space \eqref{eq:ncft_Moy_coord} under this form, as detailed in \subsecref{ncft_p4_ex}.

We want to study a \phif on such a \qST $\algft{A}$ of the form
\begin{align}
	S(\phi)
	&= \int \tdl{d+1}{x} K[\phi](x) + \frac{\cC^2}{4!} (\phi \star \phi \star \phi \star \phi)(x)
	\label{eq:ncft_p4_act}
\end{align}
where $\cC$ is the coupling constant and $K$ is a kinetic term (differential operator). Note that the integral in \eqref{eq:ncft_p4_act} is possibly inherited from the integral over $\manft{M}$. A usual expression for $K$, in the context of noncommutative \phif, is
\begin{align}
	K[\phi]
	&= \metft{g}^{\mu\nu} \partial_\mu \phi \star \partial_\nu \phi + m^2 \phi \star \phi
	\label{eq:ncft_p4_kin}
\end{align}
where $m$ is a mass term and $\metft{g}$ is the metric of $\manft{M}$.

\paragraph{}
We want here to capture the simplest quantum properties of the action \eqref{eq:ncft_p4_act} by computing the \npf{2} at one-loop thanks to the generating functional of the connected \namefont{Green} functions. The latter study was performed in \cite{Hersent_2024a}.

\subsection{Deformed momentum space}
\label{subsec:ncft_p4_dms}
\paragraph{}
Similarly to most textbook detailing the computation of a quantum \phif, one handles more easily the action \eqref{eq:ncft_p4_act} by first going to momentum space. The usual way of describing a momentum in noncommutative field theories is by considering wave packets, that are functions of the form $e^{\iCpx p_\mu x^\mu}$ where $p$ is the momentum. When one multiplies two wave packets in the commutative theory, the momenta add up through
\begin{align}
	e^{\iCpx p_\mu x^\mu} \cdot e^{\iCpx q_\mu x^\mu} 
	&= e^{\iCpx (p_\mu + q_\mu) x^\mu}.
	\label{eq:ncft_p4_+}
\end{align}	
However, in the noncommutative case, this no longer holds in general. The main reason to assume \Lieat-type noncommutativity \eqref{eq:ncft_p4_Latnc} is that a similar equation to \eqref{eq:ncft_p4_+} can be written thanks to the \BCH. In other words, one has
\begin{align}
	e^{\iCpx p_\mu x^\mu} \star e^{\iCpx q_\mu x^\mu} 
	&= e^{\iCpx (p_\mu \dplus q_\mu) x^\mu}.
	\label{eq:ncft_p4_d+}
\end{align}
where $\dplus$ is a deformed version of the addition of momenta\footnote{
	The latter is sometimes noted $+_\star$, $\oplus$ or $\oplus_\star$ in the literature.
}
(see \subsecref{ncft_p4_ex} for examples). Its explicit expression, given by the \BCH, is in general very cumbersome. Therefore, we consider a general $\dplus$ law in the following. Note, however, that the commutative limit of $\dplus$ is straightforwardly $+$ since $\star$ goes to $\cdot$ and so \eqref{eq:ncft_p4_d+} boils down to \eqref{eq:ncft_p4_+}. From \eqref{eq:ncft_p4_d+}, one can show that $\dplus$ is associative, because $\star$ is, and that it is commutative if and only if $\star$ is. In our case, we consider that $\dplus$ is noncommutative, despite its additive notation.

The involution ${}^\dagger$ allows one to define the deformed inverse of momenta $\dminus$ through $(e^{\iCpx p_\mu x^\mu})^\dagger = e^{\iCpx (\dminus p_\mu) x^\mu}$. One then can check that the usual group rules stand
\begin{align}
	p_\mu \dplus (\dminus p_\mu)
	&= 0, &
	\dminus ( \dminus p_\mu)
	&= p_\mu, &
	\dminus (p_\mu \dplus q_\mu)
	&= (\dminus q_\mu) \dplus (\dminus p_\mu).
\end{align}

\paragraph{}
The wave packet is not the only way to define a particle. One could equivalently go through particle state definition by using the deformed symmetries, \ie deformed \namefont{Poincar\'{e}}. Yet, in this formalism, the multi-particle state seems rather complex to define, specifically due the deformed $+$ law for momenta \cite{Arzano_2023}. Note that a definition of a multi-particle state in (light-like\footnote{
	The light-like \kMt corresponds to a deformation of the \Minkt space-time satisfying \eqref{eq:kM_kP_gkM} with $a^\mu$ being light-like, \ie $a^\mu a_\mu = 0$.
})
\kMt, based on wave packet study, was proposed \cite{Fabiano_2023}. However, the deformed $+$ law is unbraided in this precise context.

\paragraph{}
It is of major importance to note that the wave packet definition  suffers from an ordering ambiguity.

\begin{Emph}{The ordering ambiguity of noncommutative wave packets}
In noncommutative field theories, wave packet expressions of the form $e^{\iCpx p_\mu x^\mu}$ may be written in many distinct ways. A non-exahaustive list of possibilities could be
\begin{align}
	e^{\iCpx p_\mu x^\mu}
	&= e^{\iCpx p_0 x^0} \cdots e^{\iCpx p_d x^d}, &
	e^{\iCpx p_\mu x^\mu}
	&= e^{\iCpx (p_0 x^0 + \cdots + p_d x^d)}, &
	e^{\iCpx p_\mu x^\mu}
	&= e^{\iCpx p_d x^d} \cdots e^{\iCpx p_0 x^0}.
	\label{eq:ncft_p4_ord}
\end{align}
All the different expressions (mentioned above or not) have the same commutative limit $e^{\iCpx p_\mu x^\mu}$. The major issue at stake here is that the $\dplus$ law expression depends on the chosen ordering so that the ordering prescription affects the rest of the computation. This would mean either that the theory \emph{needs} a specific \enquote{physical} ordering to match observations, or that the physical observables need to be invariant under the ordering choice. However, it is not clear that \enquote{physical observables} computed from \eqref{eq:ncft_p4_act} are not changed when considering a different ordering.

The latter assumption of invariance has however a nice interpretation in view of the \kMt case. Indeed, when studying the momentum space of \kMt, it has been shown \cite{Mercati_2018} that all orderings of \eqref{eq:ncft_p4_ord} are linked by a coordinate transformation of the momenta. In this view, requiring that the theory is ordering independent would mean that one imposes general covariance on the (curved) momentum space. To summarise, the ordering ambiguity \eqref{eq:ncft_p4_ord} only arises at the noncommutative level and therefore could stand as a new physical symmetry of the model.
\end{Emph}

The latter question is still open in the noncommutative literature and will not affect much the computation, since it is performed for a generic $\dplus$ law.

\paragraph{}
In order to write \eqref{eq:ncft_p4_act} in momentum space, one first needs to define the (deformed) \Ftt of $\phi$, and its inverse. The latter requires to know how to integrate (or sum) over momenta, which is no more straightforward. Through the wave packet formalism, the momentum space is defined by exponentiating the \Lieat of coordinates \eqref{eq:ncft_p4_Latnc}, and in this sense forms a \Liegt. For the sake of computability, we assume that this \Liegt is locally compact. This allows us to define a \Haarm on the \Liegt. See \cite{Deitmar_2014} for more details on harmonic analysis and group theory.

\begin{Emph}{Integrations on locally compact groups}
	We consider here our (noncommutative) group of momenta, which is supposed to be locally compact. Then, there exists a unique left-invariant (resp.~right-invariant) measure called the left (resp.~right) \Haarm, denoted $\lHm$ (resp.~$\rHm$), \ie it satisfies
	\begin{align}
		\lHm(p \dplus q)
		&= \lHm(q), &
		\big( \text{resp.~} \rHm(p \dplus q)
		&= \rHm(p) \big)
		\label{eq:ncft_p4_Hm}
	\end{align}
	for any momenta $p$ and $q$. If one further defines $\lHm_p(q) = \lHm(p \dplus q)$, then it is itself left-invariant because
	\begin{align*}
		\lHm_p(q^1 \dplus q^2)
		&= \lHm(q^1 \dplus q^2 \dplus p)
		= \lHm(q^2 \dplus p)
		= \lHm_p(q^2).
	\end{align*}
	By uniqueness, $\lHm$ and $\lHm_p$ must be proportional to each other, so that one can write $\lHm_p = \Delta(p) \, \lHm$, where the map $\Delta$ is called the modular function of the group. One can show that $\Delta$ is a continuous group homomorphism, that is that
	\begin{align}
		\Delta(p \dplus q)
		&= \Delta(p) \Delta(q), &
		\Delta(\dminus p)
		&= \Delta(p)^{-1}, &
		\Delta(0)
		&= 1
	\end{align}
	with $\Delta(p) \in \piReal$.
	
	The modular function actually quantifies the difference between the left and the right \Haarm. Indeed, if one considers $\lHm_{\dminus}(p) = \lHm( \dminus p) = \Delta(\dminus p) \lHm(p)$ then it is right invariant through
	\begin{align*}
		\lHm_{\dminus}(p \dplus q)
		&= \lHm \big( \dminus( p \dplus q ) \big)
		= \lHm \big( (\dminus q) \dplus (\dminus p) \big)
		= \lHm( \dminus p)
		= \lHm_{\dminus}(p).
	\end{align*}
	Therefore, up to an irrelevant positive constant, $\rHm(p) = \Delta( \dminus p) \lHm(p)$. The case of a unimodular group corresponds to $\Delta = 1$, thus implying that the left and right \Haarm are equal.
\end{Emph}

In the computation below, we consider the left \Haarm, but one could equivalently consider the right \Haarm as they are related by the modular function.

\paragraph{}
In the context of group integration, we require the analogue of the \Ddf $\delta$ to be defined as
\begin{subequations}
	\label{eq:ncft_p4_Ddf}
\begin{align}
	\int \lHm(p)\ f(p)\, \delta(p \dplus q)
	&= f(\dminus q), 
	\label{eq:ncft_p4_Ddf_l} \\
	\int \lHm(p)\ f(p)\, \delta(q \dplus p)
	&= \Delta(\dminus q) \, f( \dminus q)
	\label{eq:ncft_p4_Ddf_r}
\end{align}
\end{subequations}
for any function $f$. Note that this definition \eqref{eq:ncft_p4_Ddf} should be slightly changed in the case of the right \Haarm. The latter definition is purely algebraic in flavour since it solves a possible mismatch when computing $\int \lHm(p)\, f(p) \delta(p \dplus q)$ by integrating the $\delta$ directly, or by performing the change of variable $p \to p \dminus q$ first. From the definition \eqref{eq:ncft_p4_Ddf}, one can show the following rules hold
\begin{subequations}
	\label{eq:ncft_p4_dprop}
\begin{align}
	\delta(p \dplus q)
	&= \Delta(\dminus q) \, \delta(q \dplus p),
	\label{eq:ncft_p4_dprop_cyc} \\
	\delta( \dminus p)
	&= \delta(p),
	\label{eq:ncft_p4_dprop_min} \\
	\int \lHm(p)\ f(p) \, \delta(p \dminus q)
	&= \int \lHm(p)\ f(p) \, \delta(q \dminus p)
	= f(q).
	\label{eq:ncft_p4_dprop_app}
\end{align}
\end{subequations}
While \eqref{eq:ncft_p4_dprop_min} and \eqref{eq:ncft_p4_dprop_app} are expected noncommutative versions of the usual \Ddf properties, the deformed cyclicity \eqref{eq:ncft_p4_dprop_cyc} is remarkable and can actually be linked with the cyclicity of the integral through
\begin{align}
\begin{aligned}
	\int \tdl{d+1}{x} (f \star g)(x)
	&= \int \lHm(p) \lHm(q)\ f(p) \, g(q) \, \delta(p \dplus q) \\
	&= \int \lHm(p) \lHm(q)\ \Delta(\dminus q) \, f(p) \, g(q) \, \delta(q \dplus p) \\
	&= \int \tdl{d+1}{x} (\Delta(g) \star f)(x)
\end{aligned}
	\label{eq:ncft_p4_cyc}
\end{align}
where $\Delta$, in the last line, denotes the \Ftt operator of the (inverse) modular function. Therefore, the (non-)cyclicity property of the integral over $\algft{A}$ is governed by the modular function of the momentum space. In the case of a unimodular space, like the \Moy space, the integral is necessarily cyclic, while in a non-unimodular case, like \kMt, the integral is not cyclic. In the latter case, the loss of cyclicity can be shown to correspond to the \kPt generator $\kPE$, which indeed corresponds to the \Ftt operator of the (inverse) modular function.

Note that the previous analysis was made regardless of any \spdtt explicit expression. The importance of the cyclicity of the integral in the construction of noncommutative \gT was made clear in \eqref{eq:ncft_db_YMgi}. This is why the loss of cyclicity in some noncommutative theories has been seen as the main obstacle to the construction of a \gT. From \eqref{eq:ncft_p4_cyc}, one can check if the integral is cyclic by deriving the modular function of the momentum space, for any (\Lieat-type) \qST. Two main paths have been followed to construct gauge invariant actions without the cyclicity of the integral, either by restoring the cyclicity through a non-trivial measure on the position space, or by trying to deform the gauge transformation of the field strength $F$. This will be discussed in more details in \secref{kM_gt}.

\paragraph{}
With the previous tools in hand, we define the \Ftt and its inverse\footnote{
	The \Ftt of any functions $f$ is also denoted $f$ in accordance with the usual \qFT notations.
}
as
\begin{align}
	\phi(p)
	&= \int \tdl{d+1}{x} e^{\iCpx (\dminus p)_\mu x^\mu} \star \phi(x), &
	\phi(x)
	&= \frac{1}{(2 \pi)^{d+1}} \int \lHm(p)\ \phi(p) \, e^{\iCpx p_\mu x^\mu}.
	\label{eq:ncft_p4_Ft}
\end{align}
The fact that the field $\phi$ is left $\star$-multiplied by the exponential in \eqref{eq:ncft_p4_Ft} is linked to the choice of the left \Haarm. A right-invariant measure choice would require a right $\star$-multiplication.

Using this \Ftt, one writes the action \eqref{eq:ncft_p4_act} as
\begin{align}
	S(\phi) 
	&= \int \lHm(k)\ \phi(k) K(k) \phi(\dminus k)
	+ \frac{\cC^2}{4!} \int \lHm(k^1) \cdots \lHm(k^4)\ \delta\big(k^1 \dplus \cdots \dplus k^4\big) \phi(k^1) \cdots \phi(k^4)
	\label{eq:ncft_p4_actmom}
\end{align}
where $K$ is the \Ftt operator of the kinetic operator $K$ in \eqref{eq:ncft_p4_act}. For example, the kinetic term \eqref{eq:ncft_p4_kin} has a \Ftt operator given by
\begin{align}
	K(k) 
	= \metft{g}^{\mu\nu} k_\mu ( \dminus k )_\nu + m^2
	\label{eq:ncft_p4_kinmom}
\end{align}
In order for $K$ to only depend on $k$, the kinetic term needs to be translation invariant, a property which is assumed in the following. The expression \eqref{eq:ncft_p4_kin} generically is translation invariant. Moreover, we require that the kinetic operator satisfies
\begin{align}
	K(\dminus k)
	= K(k)
	\label{eq:ncft_p4_kincond}
\end{align}
which is satisfied by \eqref{eq:ncft_p4_kinmom}.

\subsection{UV/IR mixing}
\label{subsec:ncft_p4_UVIR}
\paragraph{}
In order to quantise \eqref{eq:ncft_p4_actmom}, we consider the generating functional of the connected \namefont{Green} functions
\begin{align}
	Z(J)
	&= \int \tdl{}{\phi} \exp\left( - S(\phi) 
	+ \frac{1}{2} \int \tdl{d+1}{x} (J \star \phi)(x)
	+ \frac{1}{2} \int \tdl{d+1}{x} (\phi \star J)(x) \right)
	\label{eq:ncft_p4_gfcgf}
\end{align}
where $J$ is a source term. The symmetrized\footnote{
	In the unimodular case, the integral is cyclic, thanks to \eqref{eq:ncft_p4_cyc}, and the symmetrize \spdtt corresponds to the \spdtt.
}
\spdtt $\frac{1}{2}( J \star \phi + \phi \star J)$ is here to ensure that, when considering the change of variable $\phi \to \phi + K^{-1} J$, the usual simplification
\begin{align*}
	&- \frac{1}{2} \int \lHm(k)\ \phi(k) K(k) \phi(\dminus k) + \frac{1}{2} \int \tdl{d+1}{x} (J \star \phi)(x) + (\phi \star J)(x) \\
	=  &- \frac{1}{2} \int \lHm(k)\ \phi(k) K(k) \phi(\dminus k) - \frac{1}{2} \int \lHm(k)\ J(k) K^{-1}(k) J(\dminus k)
\end{align*}
occurs, where \eqref{eq:ncft_p4_kincond} has been used.

Then, the perturbative expansion of the generating function writes
\begin{align}
\begin{aligned}
	Z(J)
	&= Z(0)\ \mathrm{exp}\left( - S_{\text{int}}\left(\frac{\partial}{\partial J} \right) \right) \mathrm{exp} \left( - \frac{1}{2} \int \lHm(k)\ J(k) K^{-1}(k) J(\dminus k) \right) \\
	&= Z(0) \sum_{n=0}^\infty \frac{(-1)^n}{n!} S_{\text{int}}^n\left(\frac{\partial}{\partial J}\right) \mathrm{exp} \left( - \frac{1}{2} \int \lHm(k)\ J(k) K^{-1}(k) J(\dminus k) \right).
	\label{eq:ncft_p4_gfexp}
\end{aligned}
\end{align}

\paragraph{}
The \npf{2} at one-loop order corresponds to
\begin{align}
	\big\langle \phi(p) \phi(q) \big\rangle_{\text{1-loop}}
	= - \left. \frac{\partial}{\partial J(p)} \frac{\partial}{\partial J(q)}   S_{\text{int}} \left( \frac{\partial}{\partial J} \right) \mathrm{exp} \left( -  \frac{1}{2} \int \lHm(k)\ J(k) K^{-1}(k) J(\dminus k) \right) \right|_{J = 0}.
	\label{eq:ncft_p4_2p1l}
\end{align}
where one-loop means that only the second term ($n=1$) in the exponential expansion of \eqref{eq:ncft_p4_gfexp} is considered. Note that we have $\frac{\partial J(p)}{\partial J(q)} = \delta(p \dminus q)$. After the calculation and the removal of the disconnected components, one derives the relevant diagrams. Eight of these diagrams are planar, as pictured in \figref{ncft_p}, and four are non-planar, see \figref{ncft_np}.

\begin{Figure}%
	[label={fig:ncft_p}]%
	{
		Planar contributions to the \npf{2} \eqref{eq:ncft_p4_2p1l}.
	}%
	\foreach \x in {1,...,4}{%
	\begin{Subfigure}
		[label={fig:ncft_p\x}]
		{4}{}
		\input{Figures/NCFT_p4_p\x}
	\end{Subfigure}%
	}
	\foreach \x in {5,...,8}{%
	\begin{Subfigure}
		[label={fig:ncft_p\x}]
		{4}{}
		\input{Figures/NCFT_p4_p\x}
	\end{Subfigure}%
	}%
\end{Figure}

\begin{Figure}%
	[label={fig:ncft_np}]%
	{%
		Non-planar contributions to the \npf{2} \eqref{eq:ncft_p4_2p1l}.
	}%
	\foreach \x in {1,...,4}{%
	\begin{Subfigure}%
		[label={fig:ncft_np\x}]%
		{4}{}%
		\input{Figures/NCFT_p4_np\x}%
	\end{Subfigure}%
	}%
\end{Figure}

All the diagrams under study are not equal, but they factorise so that the \npf{2} has two contributions, one planar (first term of \eqref{eq:ncft_p4_nc2p}) and one non-planar (second term of \eqref{eq:ncft_p4_nc2p}). Thus, it writes
\begin{align}
\begin{aligned}
	\big\langle \phi(p) \phi(q) \big\rangle_{\text{1-loop}}
	&= \frac{\cC^2}{4!}\ \delta(p \dplus q)\ (1 + \Delta(q)) \int \lHm(k)\ K^{-1}(k)\ (3 + \Delta(k)) \\
	&+ \frac{\cC^2}{4!} \int \lHm(k)\ K^{-1}(k)\ \Big(1 + \Delta(k)^{-1} \Big) \Big(1 + \Delta(q) \Delta(k)^{-2} \Big)\ \delta(p \dplus k \dplus q \dminus k).
\end{aligned}
	\label{eq:ncft_p4_nc2p}
\end{align}

The commutative limit of this \npf{2} is done by considering $\dplus = +$, $\dminus = -$, $\Delta = 1$ and $\lHm(k) = \tdl{d+1}{k}$ (the \namefont{Lebesgue} measure). Therefore, it yields
\begin{align}
	\big\langle \phi(p) \phi(q) \big\rangle_{\text{1-loop}}
	&= \frac{\cC^2}{2} \, \delta(p + q) \int \tdl{d+1}{k} K^{-1}(k),
	\label{eq:ncft_p4_com2p}
\end{align}
which corresponds to the usual formula for a \phif.

\paragraph{}
The planar contribution of \eqref{eq:ncft_p4_nc2p}, which consists of the first term, is very similar to a deformed version of the commutative \npf{2} \eqref{eq:ncft_p4_com2p}. The conservation of momenta, stored in the \Ddf, is now deformed and the measure is the noncommutative one. However, the non-planar contribution is far less casual, and possesses a deformed conservation of momenta that involves the loop momentum $k$.

\paragraph{}
By studying the \UVIR on the \Moy space, as in \eqref{eq:gnc_mix}, or in other \qST (see references in \cite{Hersent_2024a}), one realises that this phenomenon, even if not well defined, always has the same structure. One of the goals of \cite{Hersent_2024a} was to capture this structure in an unambiguous description of the \UVIR. The first criterion is given by the ultra-violet divergence of the planar diagram. The second one appears already in \Moy and corresponds to the divergence of the non-planar contribution when the external momenta vanishes (infra-red), the latter divergence being due to the ultra-violet divergent behaviour of the planar contribution. Finally, if one only includes this two criterion then the commutative contribution \eqref{eq:ncft_p4_com2p} would satisfy the \UVIR. Therefore, one needs to add a third point that requires the non-planar contribution to be ultra-violet finite for non-vanishing external momenta. In other words, \cite{Hersent_2024a} defines the \UVIR as
\begin{clist}
	\item \label{it:ncft_UV}
	The planar contribution is diverging in the UV.
	\item \label{it:ncft_IR}
	The non-planar contribution is singular in the IR, due to the UV divergence of \ref{it:ncft_UV}.
	\item \label{it:ncft_fin}
	The non-planar contribution is UV finite.
\end{clist}

The analysis carried out in \cite{Hersent_2024a} showed that a relevant criterion for the latter definition is the (UV) divergence of
\begin{align}
	\int \lHm(k)\ K^{-1}(k).
	\label{eq:ncft_p4_intprop}
\end{align}
The modular function (or its inverse) can be accommodated to be bounded so that \eqref{eq:ncft_p4_nc2p} is mainly governed by a contribution akin to \eqref{eq:ncft_p4_intprop}. In this sense, the divergence of \eqref{eq:ncft_p4_intprop} can be shown to be equivalent to both \ref{it:ncft_UV} and \ref{it:ncft_IR}.

Its link with the criterion \ref{it:ncft_fin} is more tricky since it necessitates to integrate out the $\delta(p \dplus k \dplus q \dminus k)$ contribution. The equation $p \dplus k \dplus q \dminus k = 0$ has no trivial solution for $k$ when $p, q \neq 0$. Besides, one has to distinguish between \enquote{commutative} components of $\dplus$, that is the set of index $\mu$ such that $p_\mu \dplus q_\mu = p_\mu + q_\mu$, and the \enquote{purely noncommutative} components, which consists of the rest, \ie $k = (k_{\text{nc}}, k_{\text{c}})$. By doing so, the $\delta$s with commutative components can be taken out of the integral, and the noncommutative ones give rise to a solution $k^*_{\text{nc}} \neq 0$ such that $p_{\text{nc}} \dplus k^*_{\text{nc}} \dplus q_{\text{nc}} \dminus k^*_{\text{nc}} = 0$. What remains is the integral of $K^{-1}(k^*_{\text{nc}}, k_{\text{c}})$ over the commutative components $k_{\text{c}}$. Since the explicit formula for $k^*_{\text{nc}}(k_{\text{c}}, p, q)$ and the behaviour of the propagator $K^{-1}$ with respect to $k_{\text{c}}$ are not known, it is quite hard to quantify the would-be UV finiteness of \eqref{eq:ncft_p4_intprop}.

\paragraph{}
The \UVIR is thought to spoil the perturbative renormalisability of the theory. The renormalisation is not present in this discussion. It appears that in the case where \eqref{eq:ncft_p4_intprop} is finite, no perturbative renormalisation is needed, and the \UVIR is not present. However, the link between the renormalisability of the theory and the badness of the divergence of \eqref{eq:ncft_p4_intprop} have not been carried out yet.

\paragraph{}
Most studies on the \UVIR has been done on the \Moy space. There were two main outcome of such studies: either the author wanted to get rid of it, or they wanted to use its properties for physical purpose.

Two main models are curing the field theory of the mixing: the \namefont{Grosse-Wulkenhaar} model \cite{Grosse_2005a, Grosse_2003, Grosse_2005b} and the IR damping model \cite{Gurau_2009}. The two are considering the action \eqref{eq:ncft_p4_act} with a kinetic term of the form \eqref{eq:ncft_p4_kin}, and an extra-term. In the \namefont{Grosse-Wulkenhaar} model, the extra-term correspond to a harmonic oscillator of the form $x^2 \phi \star_\dpMoy \phi$. The latter term introduces an exponential decrease of the propagator in the IR, thus getting rid of the singularity. It was shown to be renormalisable to all orders, but breaks translation invariance due to the presence of $x^2$ in the action. The IR damping model has an extra term of the form $1 / k^2$ in the kinetic term. This imposes a better convergence of the integral of the propagator in the IR and thus get rid of the IR singularity. This model was also shown to be renormalisable to all orders, but exhibits a propagator with unusual behaviour for small momenta. Note that other solutions were proposed in the literature \cite{Mirza_2006, Salim_2006, Schenkel_2011}.

Still, some authors \cite{Craig_2020, Steinacker_2007} have used the \UVIR as a physical trait of \nCG and studied some insight it could bring on new physics. See \subsecref{gnc_phys_pp} for an extended introduction to these models.

\subsection{Examples on known quantum space-times}
\label{subsec:ncft_p4_ex}
\paragraph{}
Finally, we apply the previous results to the \Moy space and the \kMt space-time.

\paragraph{}
The \Moy space\footnote{
	This paragraph is written for $4$-dimensional \Moy space, but can be straightforwardly generalised to any dimension, see \cite{Hersent_2024a}.
},
described more specifically in \secref{ncft_Moy}, has a \Lieat of coordinates given by \eqref{eq:ncft_Moy_coord}, which is not, strictly speaking, of \Lieat-type. However, if one adds the coordinate $x^{5} = 1$, one obtains that
\begin{align}
	[x^\mu, x^\nu]_{\star_\dpMoy}
	&= \iCpx \Theta^{\mu\nu} x^5, &
	[x^\mu, x^5]_{\star_\dpMoy}
	&= 0
	\label{eq:ncft_p4_Mc}
\end{align}
which is of \Lieat-type \eqref{eq:ncft_p4_Latnc} with $\tensor{\sC}{^{ab}_c} = \iCpx \Theta^{ab} \delta_c^5 (1 - \delta^a_c) (1 - \delta^b_c)$, where $a, b, c = 0, \ldots, 5$. The corresponding momentum space can be computing to be
\begin{subequations}
	\label{eq:ncft_p4_Mm}
\begin{align}
	p_\mu \dplus q_\mu 
	&= p_\mu + q_\mu, &
	p_5 \dplus q_5
	&= p_5 + q_5 + \iCpx \, p_\mu \Theta^{\mu\nu} q_\nu,
	\label{eq:ncft_p4_Mm_dp} \\
	\dminus p_\mu 
	&= - p_\mu, &
	\dminus p_5
	&= - p_5
	\label{eq:ncft_p4_Mm_dm}
\end{align}
\end{subequations}
As it is associated to $x^5 = 1$, the $p_5$ \enquote{momenta} actually corresponds to a pure phase term $e^{\iCpx p_5 x^5} = e^{\iCpx p_5}$. This group is further unimodular, so that $\Delta = 1$ and has the \namefont{Lebesgue} measure as its \Haarm, \ie $\lHm(p) = \td^{4}p$.

For the \npf{2} at one-loop, one computes then that the planar diagrams are equal to $\delta(p_\mu + q_\mu) \, \delta(p_5 + q_5)$ and the non-planar diagrams to $\delta(p_\mu + q_\mu) \, \delta(p_5 + q_5 + 2 p_\mu \Theta^{\mu\nu} k_\nu)$, where $p$ and $q$ are external moment and $k$ is the internal one. The \npf{2} then writes
\begin{align}
	\big\langle \phi(p) \phi(q) \big\rangle_{\text{1-loop}}
	&= \frac{\cC^2}{6} \delta(p + q) \int \tdl{4}{k} K^{-1}(k) \left( 2 + e^{2 \iCpx \, p_\mu \Theta^{\mu\nu} k_\nu} \right)
	\label{eq:ncft_p4_Mnc2p}
\end{align}
which exactly corresponds to the expression \eqref{eq:gnc_mix} with the kinetic operator \eqref{eq:ncft_p4_kinmom}. By taking the latter kinetic operator, the same analysis of the \UVIR as in \subsecref{gnc_phys_pp} can be done. Therefore, \ref{it:ncft_UV}, \ref{it:ncft_IR} and \ref{it:ncft_fin} are fulfilled. On the other hand, the integral of the propagator \eqref{eq:ncft_p4_intprop} for the kinetic operator \eqref{eq:ncft_p4_kinmom} with the laws \eqref{eq:ncft_p4_Mm} simplifies to
\begin{align}
	\int \lHm(k)\ K^{-1}(k)
	&= \int \tdl{4}{k} \frac{1}{- k^2 + m^2}
\end{align}
which is indeed (UV) divergent. Note that as \Moy is a deformation of $\Real^4$, we took the Euclidean metric $\metft{g} = (+ \cdots +)$ in \eqref{eq:ncft_p4_kinmom}.

\paragraph{}
The \kMt space-time $\kM[d]$ corresponds to a deformation of the \Minkt \sT $\Mink$, with a \Lieat of coordinates satisfying
\begin{align}
	[x^0, x^j]_{\star_\dpkM}
	&= \frac{\iCpx}{\dpkM} x^j, &
	[x^j, x^k]_{\star_\dpkM}
	&= 0,
	\label{eq:ncft_p4_kMc}
\end{align}
where $j, k = 1, \ldots, d$ and $\dpkM$ is the \dpt. How this quantum space-time is constructed and how \eqref{eq:ncft_p4_kMc} arises is explained in more detail in the \chapref{kM}. One can write \eqref{eq:ncft_p4_kMc} under the form \eqref{eq:ncft_p4_Latnc} with $\tensor{\sC}{^{\mu\nu}_\rho} = \frac{\iCpx}{\dpkM} (\delta^\mu_0 \delta^\nu_\rho - \delta^\nu_0 \delta^\mu_\rho)$. One computes the corresponding momentum space to be
\begin{subequations}
	\label{eq:ncft_p4_kMm}
\begin{align}
	p_0 \dplus q_0
	&= p_0 + q_0, &
	p_j \dplus q_j
	&= p_j + e^{-p_0/\dpkM} \, q_j,
	\label{eq:ncft_p4_kMm_dp} \\
	\dminus p_0
	&= - p_0, &
	\dminus p_j
	&= - e^{p_0 / \dpkM} \, p_j
	\label{eq:ncft_p4_kMm_dm}
\end{align}
\end{subequations}
This group is not unimodular with a left \Haarm $\lHm(p) = e^{d p_0 / \dpkM} \td^{d+1}p$ and a right \Haarm $\rHm(p) = \td^{d+1}p$. Therefore, the modular function is $\Delta(p) = e^{d p_0 / \dpkM}$.

The \npf{2} at one-loop writes 
\begin{align}
\begin{aligned}
	\big\langle \phi(p) \phi(q) \big\rangle_{\text{1-loop}}
	&= \frac{\cC^2}{4!} \delta(p_\mu \dplus q_\mu) \big( 1 + e^{d q_0 / \dpkM} \big) \int \tdl{d+1}{k} K^{-1}(k) e^{d k_0 / \dpkM} \left( 3 + e^{d k_0 / \dpkM} \right) \\
	&+ \frac{\cC^2}{4!} \delta(p_0 + q_0) \int \tdl{d+1}{k} K^{-1}(k) \left( 1 + e^{d k_0 / \dpkM} \right) \left( 1 + e^{- d (p_0 + 2 k_0) / \dpkM} \right) \\
	&\times \delta \left( p_j + e^{-(p_0 + k_0) / \dpkM} q_j - \big( 1 - e^{-p_0 / \dpkM} \big) k_j \right).
\end{aligned}
	\label{eq:ncft_p4_kMnc2p}
\end{align}
It was shown in \cite{Hersent_2024a}, that for $d > 1$ and a kinetic term of the form \eqref{eq:ncft_p4_kinmom}, the \npf{2} \eqref{eq:ncft_p4_kMnc2p} is neither diverging in the UV nor IR singular. Therefore, \ref{it:ncft_UV} and \ref{it:ncft_IR} are not fulfilled. The requirement \ref{it:ncft_fin} is satisfied because the \npf{2} at one-loop is finite. On the other hand, if one studies the integral of the propagator, one has
\begin{align}
	\int \lHm(k)\ K^{-1}(k)
	&= 4\pi \left( \frac{4 \pi \dpkM m}{d} \right)^{\!\!\frac{d-1}{2}} \BessK_{\frac{d-1}{2}} \left( \frac{m d}{2 \dpkM} \right)
	\label{eq:ncft_p4_kMintprop}
\end{align}
where $\BessK$ is a \namefont{Bessel} function and $m$ is the mass of the scalar field $\phi$. Therefore, \eqref{eq:ncft_p4_kMintprop} is finite, even in the massless case (see \cite{Hersent_2024a}), which is in accordance with the fact that \ref{it:ncft_UV} and \ref{it:ncft_IR} are not fulfilled. Note that, in \eqref{eq:ncft_p4_kMintprop}, $\dpkM$ is playing the role of a UV cut-off, as expected by \namefont{Snyder} (see \subsecref{gnc_phys_pp}).

Note that other scalar field theory on \kMt suffers from a \UVIR \cite{Grosse_2006}. This enlightens the fact that the choice of the \qST is not linked to the presence of the mixing, as this study was made on a generic space. The source of the \UVIR seems rather to come from the (integral of the) propagator.

\paragraph{}
Finally, let us see the impact of the ordering ambiguity \eqref{eq:ncft_p4_ord} on the \npf{2} computation. Due to cumbersomeness, the \npf{2} could not be explicitly computed for two different orderings. However, one can already compare the expressions of the integral of the propagators.

The deformed additive laws \eqref{eq:ncft_p4_kMm} were obtained for time-to-the-right ordering, corresponding to the right term of \eqref{eq:ncft_p4_ord}. Therefore, we denote by $\dplus_{\text{right}}$ the law \eqref{eq:ncft_p4_kMm} in the following paragraph. One could consider the \enquote{sum} ordering, corresponding to the middle term of \eqref{eq:ncft_p4_ord} and find that
\begin{align}
	p_\mu \dplus_{\text{sum}} q_\mu 
	&= \frac{p_0 + q_0}{e^{-q_0 / \dpkM} - e^{p_0 /\dpkM}} \left( \frac{1 - e^{p_0 / \dpkM}}{p_0} p_\mu + \frac{e^{-q_0/\dpkM} - 1}{q_0} q_\mu \right), &
	\dminus_{\text{sum}} p_\mu 
	&= - p_\mu.
	\label{eq:ncft_p4_kMms}
\end{align}
Note that the previous expression is finite in the limit $p_0,q_0 \to 0$. The modular function is unchanged $\Delta(p) = e^{d p_0/\dpkM}$, and the left \Haarm writes $\lHm_{\text{sum}}(p) = \left( \frac{1 - e^{p_0 / \dpkM}}{p_0} \right)^d \td^{d+1}p$. Expressions \eqref{eq:ncft_p4_kMm} and \eqref{eq:ncft_p4_kMms} can be linked through
\begin{align}
	p_0 \dplus_{\text{right}} q_0
	&= p_0 \dplus_{\text{sum}} q_0, &
	p_j \dplus_{\text{right}} q_j
	&= \frac{1}{g\left( \frac{p_0 + q_0}{\dpkM} \right)} \left( g\left( \frac{p_0}{\dpkM} \right) p_j \dplus_{\text{sum}} g\left( \frac{q_0}{\dpkM} \right) q_j \right)
	\label{eq:ncft_p4_kMm_r2s}
\end{align}
where $g(x) = \frac{x}{1 - e^{-x}}$.

Considering the kinetic term \eqref{eq:ncft_p4_kinmom}, with the \Minkt metric $\metft{g} = (+ - \cdots -)$, one obtains that
\begin{subequations}
	\label{eq:ncft_p4_kMipcc}
\begin{align}
	& \text{Right ordering:} &
	\int \lHm_{\text{right}}(k) K_{\text{right}}^{-1}(k) 
	&= \int \tdl{d+1}{k} \frac{e^{d k_0 /2 \dpkM}}{- k_0^2 + k_j^2 + m^2},
	\label{eq:ncft_p4_kMipcc_r} \\	
	& \text{Sum ordering:} &
	\int \lHm_{\text{sum}}(k) K_{\text{sum}}^{-1}(k) 
	&= \int \tdl{d+1}{k} \frac{e^{d k_0 /2 \dpkM}}{- k_0^2 + k_j^2 + m^2} \left( - \frac{\sinh\left( \frac{k_0}{2 \dpkM} \right)}{k_0} \right)^{\!\!d}.
	\label{eq:ncft_p4_kMipcc_s}
\end{align}
\end{subequations}

The integral \eqref{eq:ncft_p4_kMipcc_r} was computed in \eqref{eq:ncft_p4_kMintprop}, however, the integral \eqref{eq:ncft_p4_kMipcc_s} could not be computed. Their distinct expressions might be a computational artefact, but suggests that these two ordering have different \npf{2}{s}. If the \npf{2} is not ordering independent and one considers that the \npf{2} is an observable \cite{Buscemi_2013} then, either there is a single physical ordering, or the result should be ordering invariant by a specific choice of propagator. This point needs to be more carefully analysed.
\chapter{The \tops{$\dpkM$}{kappa}-\namefont{Minkowski} space-time}
\label{ch:kM}

\paragraph{}
The (quantum) deformations of the \Minkt \sT have been studied extensively for the insights they might provide on the physics of \qST{s}. Indeed, the \kMt space-time consists of the first (non-trivial) deformation of a space with time. Furthermore, its space of symmetries, the \kPt algebra, has a flourishing phenomenology, principally linked to \qG (see \secref{qg_ph}). The \dpt $\dpkM$ has a mass dimension and is therefore sometimes associated with the \Pmast. \kMt is considered as a good candidate to shed some light on possible \qG effects, at least in some regime.

We here depict some recent topics concerning field and \gT[ies] on deformed \Minkt space-times. Some older models are evoked.

\section{From the deformed symmetries of \tops{$\dpkM$}{kappa}-\namefont{Poincar\'{e}} to the \tops{$\dpkM$}{kappa}-\namefont{Minkowski} space-time}
\label{sec:kM_kP}
\paragraph{}
The \kMt space-time, noted $\kM$, was first define thirty years ago by \namefont{Majid} and \namefont{Ruegg} \cite{Majid_1994} as the space having the \kPt \Hfat $\kP[1,d]$ as its space of symmetries. In other words, \kMt needed to be a $\kP[1,d]$-module algebra, as explained in \subsecref{gnc_qg_qst}. The latter deformation of the \Palgt was introduced three years earlier by \namefont{Lukierski}, \namefont{Nowicki}, \namefont{Ruegg} and \namefont{Tolstoy} \cite{Lukierski_1991, Lukierski_1992}. For an historical review of the construction of \kMt and \kPt see \cite{Lukierski_2015}.

\paragraph{}
Starting with the \Palgt, one cannot perform a \namefont{Drinfel'd-Jimbo} quantisation procedure, since the latter \Lieat is not semi-simple. This quantisation consists of deforming the universal enveloping algebra of a semi-simple \Lieat, by using its root system and the quantum deformation of $U(\Lieft{sl}_2)$. The authors of \cite{Lukierski_1991} opted for a \namefont{Wigner-Inonu} contraction procedure, which consists in first building the \namefont{Drinfel'd-Jimbo} quantum version of the \namefont{de Sitter} algebra $\Oa{2,3}{\Real}$ and then taking the limit $q \to 0$ and $r \to + \infty$ with $\iCpx r \log(q) = \dpkM^{-1}$ constant. Here $q$ denotes the usual complex parameter in group quantisation, and $r$ is the \namefont{de Sitter} radius, that allows to go from $\Oa{2,3}{\Real}$ to $\Palg[1,3]$ when $r \to + \infty$. The computation for dimensions others than $4$ was done shortly after \cite{Lukierski_1994}.

From this computation, one obtains the \kPt algebra. The latter is given, in the so-called \namefont{Majid-Ruegg} basis \cite{Majid_1994} by

\begin{subequations}
	\label{eq:kM_kP_kP}
\begin{align}
	[J_j, J_k] &= i \tensor{\epsilon}{_{jk}^l} J_l, & 
	[J_j, K_k] &= i\tensor{\epsilon}{_{jk}^l} K_l, & 
	[K_j, K_k] &= -i\tensor{\epsilon}{_{jk}^l} J_l, \\
	[P_j, J_k] &= -i\tensor{\epsilon}{_{jk}^l} P_l, &
	[P_j, \kPE] &= [J_j, \kPE] = 0, &
	[P_j, P_k] &= 0,
	\label{eq:kM_kP_kP_alg_tran}
\end{align}%
\vspace{\dimexpr-\abovedisplayskip-\belowdisplayskip-\baselineskip+\jot + 14pt}%
\begin{align}
   	[K_j, \kPE] &= -\frac{i}{\dpkM} P_j \kPE, &
   	[P_j, K_k] = \frac{i}{2} \eta_{jk} 
   	\left( \dpkM(1-\kPE^2) + \frac{1}{\dpkM} P_l P^l \right) 
   	+ \frac{i}{\dpkM} P_j P_k,
   	\label{eq:kM_kP_kP_alg_rot}
\end{align}%
%\vspace{\dimexpr-\abovedisplayskip-\belowdisplayskip-\baselineskip+\jot}%
\begin{align}
   	\Delta P_0 &= P_0 \otimes 1 + 1 \otimes P_0, &
	\Delta P_j &= P_j \otimes 1 + \kPE \otimes P_j, 
	\label{eq:kM_kP_kP_coalg_tran} \\
	\Delta \kPE &= \kPE \otimes \kPE, &
	\Delta J_j &= J_j \otimes 1 + 1 \otimes J_j,
\end{align}%
\vspace{\dimexpr-\abovedisplayskip-\belowdisplayskip-\baselineskip+\jot + 14pt}%
\begin{align}
   	\Delta K_j = K_j \otimes 1 + \kPE \otimes K_j - \frac{1}{\dpkM} \tensor{\epsilon}{_j^{kl}} P_k \otimes J_l,
\end{align}%
%\vspace{\dimexpr-\abovedisplayskip-\belowdisplayskip-\baselineskip+\jot}%
\begin{align}
   	\varepsilon(P_0) = \varepsilon (P_j) = \varepsilon(J_j) = \varepsilon(K_j) = 0, &&
   	\varepsilon(\kPE) = 1,
\end{align}%
%\vspace{\dimexpr-\abovedisplayskip-\belowdisplayskip-\baselineskip+\jot}%
\begin{align}
	S(P_0) &= - P_0, &
	S(\kPE) &= \kPE^{-1}, &
	S(P_j) &= -\kPE^{-1} P_j,
\end{align}%
\vspace{\dimexpr-\abovedisplayskip-\belowdisplayskip-\baselineskip+\jot + 14pt}%
\begin{align}
	S(J_j) &= -J_j, &
	S(K_j) &= -\kPE^{-1}(K_j - \frac{1}{\dpkM} \tensor{\epsilon}{_j^{kl}} P_k J_l).
\end{align}%
%\vspace{\dimexpr-\abovedisplayskip-\baselineskip+\jot}%
\end{subequations}
where $\{P_\mu\}_{\mu = 0, \ldots, d}$ are the generators of the deformed translations, $\{J_j\}_{j = 1, \ldots, d}$ the generators of the deformed rotations, $\{K_j\}_{j = 1, \ldots, d}$ the generators of the deformed boosts, and $\eta$ is the \Minkt metric. Note that we introduced $\kPE = e^{-P_0/\dpkM}$ for convenience. $\dpkM$ is here the \dpt and already has a mass dimension.

\paragraph{}
The question of how to find a $\kP[1,d]$-module algebra to form the \kMt space-time is not trivial regarding the complex structure of \eqref{eq:kM_kP_kP}. However, the authors of \cite{Majid_1994} observed that \kPt has a bicrossproduct structure steaming from the cross-product structure of the \Palgt $\Palg[1,d] = \Tralg[1,d] \rtimes \SOa{1,d}$, where $\Tralg[1,d]$ denotes the algebra of translations and $\SOa{1,d}$ the algebra of rotations and boosts. Explicitly, if one defines the (\namefont{Hopf}) algebra of deformed translations $\kTran[1,d]$ as the set of $\{P_\mu\}_\mu$ and $U(\SOa{1,d})$ the universal enveloping algebra of the (deformed) rotations $\{J_j\}_j$ and boosts $\{K_j\}_j$, then one has
\begin{align}
	\kP[1, d]
	&= \kTran[1,d] \bicrosr U(\SOa{1,d}).
	\label{eq:kM_kP_kPa}
\end{align}
Definitions and examples of (co)actions on (co)algebra can be found in \secref{ha_rt}.

The \kMt space-time is defined as the (\Hfat) dual of the deformed translations, $\kM = \dual{(\kTran[1,d])}$, and is fully determined as such. One can show, as done in the proof of \exref{ha_as_kM}, that \kMt is generated by $d + 1$ elements $\{x^\mu\}_{\mu=0, \ldots, d}$ which satisfy
\begin{subequations}
	\label{eq:kM_kP_kM}
\begin{align}
	[x^0, x^j]
	&= \frac{\iCpx}{\dpkM} x^j, &
	[x^j, x^k]
	&= 0,
	\label{eq:kM_kP_kM_alg}
\end{align}
\vspace{\dimexpr-\abovedisplayskip-\belowdisplayskip-\baselineskip+\jot + 14pt}%
\begin{align}
	\Delta(x^\mu)
	&= x^\mu \otimes 1 + 1 \otimes x^\mu, &
	\varepsilon(x^\mu)
	&= 0, &
	S(x^\mu)
	&= - x^\mu.
	\label{eq:kM_kP_kM_coalg}
\end{align}
\end{subequations}
Thanks to this duality, one can define a (left) action of $\kTran[1,d]$ on $\kM$, via the action\footnote{
	Note that $\kM$ is a (right) $\kM$-comodule with the coaction given by the coproduct \eqref{eq:kM_kP_kM_coalg}.
}
\eqref{eq:ha_rt_dcomod}, and dualise the coaction of $U(\SOa{1,d})$ on $\kTran[1,d]$ to an action of $U(\SOa{1,d})$ on $\kM$. One computes \cite{Majid_1994} that
\begin{subequations}
	\label{eq:kM_kP_act}
\begin{align}
    (P_\mu \actl f)(x) &= -\iCpx \partial_\mu f(x), &
    (\kPE \actl f)(x) &= f(x^0 + \frac{\iCpx}{\dpkM}, x^j), \\
    (J_j \actl f)(x) &= \big( \tensor{\epsilon}{_{jk}^l} x^k P_l \actl f \big) (x), &&
\end{align}%
    \vspace{\dimexpr-\abovedisplayskip-\baselineskip+\jot + 4pt}%
\begin{align}
    (K_j \actl f)(x)
    &= \left( \Big( \frac{1}{2} x^j \big( \dpkM (1 - \kPE^2) + \frac{1}{\dpkM} P_l P^l \big) + x^0 P_j - \frac{\iCpx}{\dpkM} x^k P_k P_j  \Big) \actl f \right)(x).
\end{align}
\end{subequations}
for any $f \in \kM$.

\paragraph{}
With the action \eqref{eq:kM_kP_act}, \kMt is a $\kP[1,d]$-module algebra , thus allowing us to interpret \kPt as the (quantum) symmetries of the (quantum) deformed \Minkt space-time. The connection between a \qST and its symmetries has been discussed in \subsecref{gnc_qg_qst}. On the other hand, if one tries to determine the \Liegt of momenta by exponentiating the \Lieat of coordinates \eqref{eq:kM_kP_kM_alg}, as done in \subsecref{ncft_p4_dms} through the wave packet formalism, one finds that this group corresponds to the deformed translations $\kTran[1,d]$. The latter observation can be explained by the bicrossproduct structure \eqref{eq:kM_kP_kPa}, which give rise to a \Lieat double dual to a \Liegt double. Finally, \kPt gathers both the symmetries and the momentum space of our \qST.

\paragraph{}
Generalised $\dpkM$-deformations of \Minkt were studied in \cite{Lukierski_2002}, stemming from deformations of the \namefont{Poincar\'{e}-Weyl} algebra $\Lieft{W}_{1,d}$, that is the \Palgt with an extra generator of dilatation. The \Lieat part \eqref{eq:kM_kP_kM_alg} is changed to
\begin{align}
	[x^\mu, x^\nu]_{\star_\dpkM}
	&= \frac{\iCpx}{\dpkM} ( a^\mu x^\nu - a^\nu x^\mu ),
	\label{eq:kM_kP_gkM}
\end{align}
where $a^\mu \in \Mink$ is a constant vector. The \kMt \eqref{eq:kM_kP_kM_alg} is recovered when $a^\mu = \delta^\mu_0 = (1, 0, \ldots, 0)$. The latter case is called \enquote{time-like} \kMt since $a^\mu a_\mu = -1$ in this case, but one could also consider light-like ($a^\mu a_\mu = 0$) and space-like ($a^\mu a_\mu = +1$) deformations.

\paragraph{}
We now turn to the determination of a \spdtt on \kMt, which satisfies \eqref{eq:kM_kP_kM_alg}. We present briefly here the two methods of \secref{gnc_dq}.

The \calg method was first carried out in \cite{Durhuus_2013, Poulain_2018}. Starting from the \Lieat of coordinates \eqref{eq:kM_kP_kM}, called the affine algebra, one computes the associate \Liegt to be the affine group, corresponding to \eqref{eq:ncft_p4_kMm}. By computing the \Ftt of the convolution product and the involution on this group, one obtains
\begin{subequations}
	\label{eq:kM_kP_sp}
\begin{align}
	(f \star_\dpkM g)(x)
	&= \int \frac{\td p_0}{2 \pi} \tdl{}{y^0} e^{-\iCpx y^0 p_0} f(x^0 + y^0, x^j) g(x^0, e^{-p_0/\dpkM} x^j), 
    \label{eq:kM_kP_sps} \\
    f^\dagger(x)
    &= \int \frac{\td p_0}{2\pi} \tdl{}{y^0} e^{-\iCpx y^0 p_0} \overline{f} (x^0 + y^0, e^{-p_0/\kappa} x^j)
    \label{eq:kM_kP_spi}
\end{align}
\end{subequations}
for any $f, g \in \kM$, where $\overline{f}$ denotes the complex conjugation of $f$. The multiplier space of \eqref{eq:kM_kP_sps}, \ie the set of (smooth) functions stable under the latter \spdtt, was studied in \cite{Durhuus_2013}. It turns out that it contains at least functions whose derivatives to any order grow at most polynomially at infinity.

\begin{proof}
	Here, we perform the computation of the \spdtt \eqref{eq:kM_kP_sps}. The involution \eqref{eq:kM_kP_spi} follows from a similar computation with 
	\begin{align*}
		f^\dagger
		&= \Ft^{-1} \left( \overline{\Ft(f)}(\dminus \cdot) \ \Delta(\cdot) \right)
	\end{align*}
	where $\Ft$ is the \Ftt and $\Delta$ the modular function of the affine group \eqref{eq:ncft_p4_kMm}.
	
	The \spdtt is computed through the formula \eqref{eq:gnc_dq_ca_spdt}. The right-invariant \Haarm is considered on the group since it enables the construction of \kPt-invariant actions. Thus, one has
	\begin{align*}
		(f \star_{\dpkM} g)(x)
		&= \Ft^{-1} \big( \Ft(f) \cpdt \Ft(g) \big)(x) \\
		&= \int \rHm(p) e^{\iCpx p_\mu x^\mu} \big(\Ft(f) \cpdt \Ft(g) \big)(p) \\
		&= \int \tdl{d+1}{p} e^{\iCpx p_\mu x^\mu} \int \tdl{d+1}{q} \Ft(f)(p \dminus q) \, \Ft(g)(q) \\
		&= \frac{1}{(2\pi)^{2d+2}} \int \tdl{d+1}{p} \tdl{d+1}{q} e^{\iCpx p_\mu x^\mu} \int \tdl{d+1}{y} \tdl{d+1}{z} e^{-\iCpx (p \dminus q)_\mu y^\mu} e^{-\iCpx q_\mu z^\mu} f(y) g(z) \\
		&\begin{aligned}
			= \frac{1}{(2\pi)^{2d+2}} \int 
			& \tdl{d+1}p \tdl{d+1}q \tdl{d+1}y \tdl{d+1}{z} e^{\iCpx p_0 (x^0 - y^0)} e^{\iCpx q_0(y^0 - z^0)} \\
			& e^{\iCpx p_j (x^j - y^j)}  e^{\iCpx q_j (e^{(q_0-p_0)/\dpkM} y^j - z^j)} f(y) g(z)
		\end{aligned} \\
		&= \frac{1}{(2\pi)^2} \int \tdl{}{p_0} \tdl{}{q_0} \tdl{}{y^0} \tdl{}{z^0}  e^{\iCpx p_0 (x^0 - y^0)} e^{\iCpx q_0(y^0 - z^0)} \, f(y^0, x^j) \, g(z^0, e^{(q_0-p_0)/\dpkM} x^j) \\
		&\overset{(p_0 \to p_0 + q_0)}{=} \frac{1}{(2\pi)^2} \int \tdl{}{p_0} \tdl{}{q_0} \tdl{}{y^0} \tdl{}{z^0}  e^{\iCpx p_0 (x^0 - y^0)} e^{\iCpx q_0(x^0 - z^0)} \, f(y^0, x^j) \, g\left(z^0, e^{-p_0/\dpkM} x^j \right) \\
		&\overset{(y^0 \to x^0 + y^0)}{=} \frac{1}{2\pi} \int \tdl{}{p_0} \tdl{}{y^0} e^{-\iCpx p_0 y^0} f(x^0 + y^0, x^j) \, g\left(x^0, e^{-p_0/\dpkM} x^j \right).
		\qedhere
	\end{align*}
\end{proof}

The \Dt method for generating the \spdtt is less straightforward here than for other \qST{s}. The first expressions were obtained in \cite{Bu_2008, Meljanac_2007}, but shortly after a no-go theorem came out \cite{Borowiec_2014} stating that, in dimension $2$ and $4$, the \kPt algebra cannot be recovered by a twist deformation. In other words, there is no twist of $\Palg[1,3]$ that makes it possible to reconstruct the \kMt space-time. The former authors actually deformed the \namefont{Poincar\'{e}-Weyl} algebra $\Lieft{W}_{1,d}$, that is the \Palgt with an extra generator corresponding to dilatations. The considered \Dt is Abelian and writes
\begin{align}
	\Hoft{F}
	&= \exp\left(- \frac{\iCpx}{2 \dpkM} ( P_0 \otimes D - D \otimes P_0 ) \right)
	= \exp\left( \frac{\iCpx}{2 \dpkM} ( \partial_0 \otimes x^j \partial_j - x^j \partial_j \otimes \partial_0 ) \right),
	\label{eq:kM_kP_twiA}
\end{align}
where $D$ is the dilatation generator, which can be shown to act on $\kM$ as $- \iCpx x^j \partial_j$. Some authors \cite{Dimitrijevic_2014} alternatively considered the \Dt of \namefont{Jordan} type
\begin{align}
	\Hoft{F}
	&= \exp \left( - \iCpx D \otimes \ln \left(1 + \frac{1}{\dpkM} P_0 \right) \right).
	\label{eq:kM_kP_twiJ}
\end{align}

Contrary to the \Moy space, the three \spdtt{s} \eqref{eq:kM_kP_sps}, \eqref{eq:kM_kP_twiA} and \eqref{eq:kM_kP_twiJ} have different expressions. It is not known if the physics of a \qST described by different \spdtt{s} is the same or not. Studies on \dq methods have led to the notion of equivalent \spdtt{s}, \ie two \spdtt{s} $\star$ and $\tilde{\star}$ are said equivalent if there exists an invertible (formal power series of) differential operator $T$ such that
\begin{align}
	f \, \tilde{\star} \, g
	&= T^{-1} \big( T(f) \star T(g) \big).
	\label{eq:kM_kP_equisp}
\end{align} 
This notion of equivalence has further been refined to the notion of \namefont{Morita} equivalence, which basically compares the ($*$-)representations of two ($*$-)algebras. Therefore, two \opalg{s} which are \namefont{Morita} equivalent will give rise to the same quantum states and observables. One can find a more extensive discussion and the relevant references in \cite{Waldmann_2016}. It is not clear that this equivalence stands for more advanced notions, such as a \npf{n} of a field theory. Moreover, the different \namefont{Morita} equivalence classes of \kMt are not known up to date.

Note that the \spdtt \eqref{eq:kM_kP_sps} can be put under an exponential form with
\begin{align}
	\Hoft{F}
	&= \exp \left( - \frac{\iCpx}{\dpkM} \dpkM(1 - \kPE) \otimes D \right)
	\label{eq:kM_kP_Dtus}
\end{align}
but the latter expression does not satisfy the $2$-cocycle condition \eqref{eq:gnc_dq_dt_2co} and so is not a \Dt.
\begin{proof}
	We factorise here the \spdtt \eqref{eq:kM_kP_sps} as the action of \eqref{eq:kM_kP_Dtus}. In order to do so, we use the infinite regularity of functions on \kMt to write infinite \namefont{Taylor} expansions of the form
	\begin{align*}
		f(x^0 + y^0, x^j)
		&= \sum_{n=0}^{+\infty} \frac{(y^0)^n}{n!} \partial_0^n f(x), &
		g\big(x^0, e^{-p_0/\dpkM} x^j \big)
		&= \sum_{k = 0}^{+ \infty} \frac{(e^{-p_0/\dpkM} - 1)^k}{k!} (x^j \partial_j)^k g(x),
	\end{align*}
	where the decomposition $e^{-p_0/\dpkM} x^j = x^j + (e^{-p_0/\dpkM} - 1) x^j$ has been used for the second equality. Note that in the following computation integrals and infinite sums are swapped thanks to the convenient space in which $f$ and $g$ lives.
	\begin{align*}
		(f \star_\dpkM g)(x)
		&= \int \frac{\td p_0}{2\pi} \tdl{}{y^0} e^{-\iCpx p_0 y^0} f(x^0 + y^0, x^j) \, g\big(x^0, e^{-p_0/\dpkM} x^j \big) \\
		&= \int \frac{\td p_0}{2\pi} \tdl{}{y^0} e^{-\iCpx p_0 y^0} \left( \sum_{n=0}^{+\infty} \frac{(y^0)^n}{n!} \partial_0^n f(x) \right) \left( \sum_{k = 0}^{+ \infty} \frac{(e^{-p_0/\dpkM} - 1)^k}{k!} (x^j \partial_j)^k g(x) \right) \\
		&= \sum_{n=0}^{+\infty} \sum_{k=0}^{+\infty} \frac{\partial_0^n f(x)}{n!} \frac{(x^j \partial_j)^k g(x)}{k!} \int \frac{\td p_0}{2\pi} \tdl{}{y^0} e^{-\iCpx p_0 y^0} (y^0)^n (e^{-p_0/\dpkM} - 1)^k \\
		&= \sum_{n=0}^{+\infty} \sum_{k=0}^{+\infty} \frac{\partial_0^n f(x)}{n!} \frac{(x^j \partial_j)^k g(x)}{k!} \int \frac{\td p_0}{2\pi} \tdl{}{y^0} e^{-\iCpx p_0 y^0} (y^0)^n \sum_{s = 0}^k \binom{k}{s} e^{-s p_0 / \dpkM} (-1)^{k-s} \\
		&= \sum_{n=0}^{+\infty} \sum_{k=0}^{+\infty} \frac{\partial_0^n f(x)}{n!} \frac{(x^j \partial_j)^k g(x)}{k!} \sum_{s = 0}^k \binom{k}{s} (-1)^{k-s} \int \frac{\td p_0}{2\pi} \tdl{}{y^0} (y^0)^n e^{- \iCpx p_0 (y^0 - \iCpx  s / \dpkM)} \\
		&= \sum_{n=0}^{+\infty} \sum_{k=0}^{+\infty} \frac{\partial_0^n f(x)}{n!} \frac{(x^j \partial_j)^k g(x)}{k!} \sum_{s = 0}^k \binom{k}{s} (-1)^{k-s} \left( \frac{\iCpx s}{\dpkM} \right)^n \\
		&= \sum_{k=0}^{+\infty} \frac{(x^j \partial_j)^k g(x)}{k!} \sum_{s = 0}^k \binom{k}{s} (-1)^{k-s} \sum_{n=0}^{+\infty} \frac{\partial_0^n f(x)}{n!} \left( \frac{\iCpx s}{\dpkM} \right)^n \\
		&= \sum_{k=0}^{+\infty} \frac{(x^j \partial_j)^k g(x)}{k!} \sum_{s = 0}^k \binom{k}{s} (-1)^{k-s} f\left( x^0 + \frac{\iCpx s}{\dpkM}, x^j \right) \\
		&= \sum_{k=0}^{+\infty} \frac{(x^j \partial_j)^k g(x)}{k!} \sum_{s = 0}^k \binom{k}{s} (-1)^{k-s} \kPE^s(f)(x) \\
		&= \sum_{k=0}^{+\infty} \frac{(x^j \partial_j)^k g(x)}{k!} (\kPE - 1)^k(f)(x) \\
		&= \cdot \circ \exp \left( - \frac{\iCpx}{\dpkM} \dpkM(1 - \kPE) \otimes D \right) \actl (f \otimes g)
		\qedhere
	\end{align*}
\end{proof}

\paragraph{}
If one considers the integral on \Minkt \sT as an integral over \kMt, then one can show that it is not cyclic. More precisely, one has
\begin{align}
	\int \tdl{d+1}{x} (f \star_{\dpkM} g)(x)
	&= \int \tdl{d+1}{x} \left( \big( \kPE^d \actl g \big) \star_{\dpkM} f \right) (x),
	\label{eq:kM_kP_ncyc}
\end{align}
where $\kPE = e^{-P_0/\dpkM}$ is a generator of \kPt, see \eqref{eq:kM_kP_kP}. We say that the integral is a \enquote{twisted} trace, meaning that one of the factor gets transformed by the automorphism $\kPE^d$. Note that this formula could be obtained by \eqref{eq:ncft_p4_cyc}, as one can compare $\kPE^d = e^{- d P_0 / \dpkM}$ and $\Delta(p)^{-1} = e^{- d p_0 / \dpkM}$. On the other hand, this integral enables the building of \kPt-invariant field theories since one can compute that
\begin{align}
	X \actl \int \tdl{d+1}{x} \mathscr{L}(\phi)
	&= \varepsilon(X) \int \tdl{d+1}{x} \mathscr{L}(\phi),
\end{align}
for any $X \in \kP[1,d]$ and any Lagrangian density $\mathscr{L}$ of any field $\phi$.

The lost of cyclicity \eqref{eq:kM_kP_ncyc} was considered to be the main obstacle to construct a \gT on \kMt, as already mentioned in \secref{ncft_db}. We now go to the construction of a gauge theory on \kMt.

\section{Gauge theory on \tops{$\dpkM$}{kappa}-\namefont{Minkowski}}
\label{sec:kM_gt}
\paragraph{}
As already reviewed in \cite{Hersent_2023a}, several formulation of \gT on \kMt have been considered. They are mainly based on the \namefont{Seiberg-Witten} map and bypass the lost of cyclicity \eqref{eq:kM_kP_ncyc} by either taking a non-trivial integration measure (and so changing the notion of \enquote{integral} on \kMt) or by considering a deformed version of the \namefont{Hodge} duality.

\paragraph{}
In this section, we mainly focus on the model developed by \namefont{Mathieu} and \namefont{Wallet} \cite{Mathieu_2020, Mathieu_2021}. Its starting point is the observation that the cyclicity of the integral is not broken, but only twisted as shown in \eqref{eq:kM_kP_ncyc}. Should the field strength $F$ gauge transform in a \enquote{twisted} way accordingly, the integral of $F \star_{\dpkM} F^\dagger$ would be gauge invariant. It appears that if one considers a set of natural (twisted) derivations and adapts the procedure of the \dbdc (detailed in \secref{ncft_db}), one ends up with a gauge invariant action. However, the gauge invariance imposes a specific space-time dimension. The latter constraint stems from the fact that the twist in \eqref{eq:kM_kP_ncyc} depends on the spacial dimension $d$.

\paragraph{}
Let us first discuss how the straightforward \nCYM action \eqref{eq:ncft_db_YMa} behaves with respect to the integral with twisted cyclicity \eqref{eq:kM_kP_ncyc}. Performing the gauge transformation of the action similarly to \eqref{eq:ncft_db_YMgi}, one ends up at the last step with a term $\kPE^d(\ggft{u}) \star_{\dpkM} \ggft{u}^\ddagger$ in the action. Recall that $\kPE = e^{-P_0/\dpkM}$ is a generator of \kPt \eqref{eq:kM_kP_kP} appearing in \eqref{eq:kM_kP_ncyc}. We have further shortened $\kPE^d \actl \ggft{u}$ to $\kPE^d(\ggft{u})$. One could think of imposing that $\kPE^d(\ggft{u}) \star_{\dpkM} \ggft{u}^\ddagger = 1$ to restore gauge invariance. Therefore, the set of deformed gauge transformations $\ugt{n}$ should now satisfy
\begin{align}
	\ggft{u}^\ddagger \star_{\dpkM} \ggft{u}
	&= \ggft{u} \star_{\dpkM} \ggft{u}^\ddagger
	= 1, &
	\kPE^{-d}(\ggft{u})^\ddagger \star_{\dpkM} \ggft{u}
	&= \kPE^d(\ggft{u}) \star_{\dpkM} \ggft{u}^\ddagger
	= 1.
	\label{eq:kM_gt_Ed}
\end{align}
for any $\ggft{u} \in \kM$.

The authors of \cite{Mathieu_2020} advocated that such a requirement \eqref{eq:kM_gt_Ed} is too restrictive since it imposes that $\ggft{u}$ is independent of time. From \eqref{eq:kM_gt_Ed}, one has $\kPE^d(\ggft{u}) = \ggft{u}$, which writes $\ggft{u}(x^0 + d \frac{\iCpx}{\dpkM}, x^j) = \ggft{u}(x^0, x^j)$ so that $\ggft{u}$ needs to be periodic in the imaginary direction along $x^0$. If one considers that $\ggft{u}$ needs to be entire, then the \namefont{Liouville} theorem states that an entire bounded complex function needs to be constant, thus $\ggft{u}$ needs to be constant along $x^0$. The latter requirement would also impact the commutative limit and be of less physical relevance.

\subsection{The gauge theory on twisted derivations based differential calculus of deformed translation}
\label{subsec:kM_gt_td}
\paragraph{}
One starts by considering the generators of translations of the \kPt algebra $P_\mu$. In commutative gauge theory, it consists of the usual set of derivations (\ie the $\partial_\mu$). One could think of using the $P_\mu$'s as derivations to build the \gT along the lines of \secref{ncft_db}. In view of their coproduct \eqref{eq:kM_kP_kP_coalg_tran}, the $P_j$'s are not derivations, but rather \enquote{twisted}\footnote{
	The twist ($\kPE$) has here nothing to do with a \Dt. Despite the confusion that similar denomination can bring, the name \enquote{twist} for twisted derivations is coherent with respect to the name \enquote{twisted} spectral triple, the two sharing common grounds and properties. 
}
derivations, \ie
\begin{align}
	P_j \actl (f \star_{\dpkM} g)
	&= \big( P_j \actl f \big) \star_\dpkM g + \big( \kPE \actl f \big) \star_{\dpkM} \big(P_j \actl g \big).
	\label{eq:kM_gt_tderPj}
\end{align}
If one considers the set of functions on $\kM$ given by $X_0 = \dpkM (1 - \kPE) \actl $ and $X_j = P_j \actl $, one obtains $d+1$ twisted derivations 
\begin{align}
	X_\mu(f \star_{\dpkM} g)
	&= X_\mu(f) \star_{\dpkM} g + \kPE(f) \star_{\dpkM} X_\mu(g).
	\label{eq:kM_gt_tder}
\end{align}
In \eqref{eq:kM_gt_tder} and in the following, we abbreviate $\kPE \actl f$ as $\kPE(f)$. We denote the set of twisted derivations on \kMt as $\Der_\kPE(\kM)$. The latter derivations consist of an Abelian \Lieat and each $P_\mu$ boils down to $\partial_\mu$ at the commutative limit $\dpkM \to + \infty$, up to an irrelevant $- \iCpx$ factor. One can indeed check that $\kPE \to 1$ in this limit and that a development of the exponential gives $X_0 \to P_0$.

\paragraph{}
One should note that twisted structures already appear in twisted \st{s}.
\begin{Emph}{Twisted spectral triples}
	In the context of the noncommutative \SM, it was shown that some inconsistencies arises because the full \Do is not bounded as required by the \st axioms (see \defref{gnc_st}). That is, for $(\algft{A}, \Hilbft{H}, \Dirop)$ as \st, the element $[\Dirop, f]$ is not a bounded operator in $\Hilbft{H}$, for any $f \in \algft{A}$. \namefont{Connes} and \namefont{Moscovici} \cite{Connes_2008} cured the latter problem by considering instead a twisted \st. The latter is called twisted since it introduces an automorphism of $\algft{A}$, $\rho \in \Aut(\algft{A})$, such that
	\begin{align}
		[\Dirop, f]_\rho
		&= \Dirop f - \rho(f) \Dirop
		\label{eq:kM_gt_tDo}
	\end{align}
	is bounded. Correspondingly, the zeroth \eqref{eq:gnc_st_0th} and first \eqref{eq:gnc_st_1st} order conditions write now also with the twisted bracket $[\cdot, \cdot]_\rho$, so that the set of forms \eqref{eq:gnc_st_qc_nform} (and so the differential calculus) is also twisted.
\end{Emph}

In the case of \kMt, the twist $\rho$ is considered to be $\kPE \in \Aut(\kM)$. To make contact even further with the twisted \gT under study, one can show that $[\Dirop, \cdot]_\rho$ is a twisted derivation, that is
\begin{align}
	[\Dirop, f \star g]_\rho
	&= [\Dirop, f ]_\rho \star g + \rho(f) \star [\Dirop, g]_\rho.
	\label{eq:kM_gt_tDo_der}
\end{align}
The relation \eqref{eq:kM_gt_tDo_der} has to be confronted with \eqref{eq:kM_gt_tder}. Furthermore, $\rho$ is required to be regular, that is $\rho^\dagger = \rho^{-1}$, a property satisfied by $\kPE$.

\paragraph{}
Due to the fact that we are considering twisted derivations instead of the usual derivations, the procedure of \secref{ncft_db} must be adapted to the twisted setting, as described below. Indeed, if one wants to define a connection on a (right) $\kM$-module $\modft{X}$ similar to \eqref{eq:ncft_db_con} with twisted derivations, then one is confronted with inconsistencies, which result from the equality of  $\nabla_X(s \actr (f \star_{\dpkM} g))$ and $\nabla_X((s \actr f) \actr g)$. The problem lies in the \Lru \eqref{eq:ncft_db_con_Lr} and it can be shown \cite{Mathieu_2021} that the only possible twisted connection corresponds to
\begin{align}
	\nabla_\mu(s \actr f)
	&= \nabla_\mu(s) \actr f + \kPE(s) \actr X_\mu(f)
	\label{eq:kM_gt_tcon}
\end{align}
for $f, g \in \kM$ and $s \in \modft{X}$, where we noted $\nabla_{X_\mu} = \nabla_\mu$. In the previous expression \eqref{eq:kM_gt_tcon} $\kPE$ has been lifted to the module structure and one has $\kPE(s \actr f) = \kPE(s) \actr \kPE(f)$. In a similar fashion, the only twisted curvature, giving rise to a module homomorphism, writes
\begin{align}
	R_{\mu\nu}
	&= \kPE^{-1} \big( \nabla_\mu \kPE \nabla_\nu  - \nabla_\nu \kPE \nabla_\mu \big)
	\label{eq:kM_gt_tcur}
\end{align}
Recall that the module homomorphism property states that $R_{\mu\nu}(s \actr f) = R_{\mu\nu}(s) \actr f$, and is necessary in order to extract a field strength $F$ out of the full curvature $R$ through \eqref{eq:ncft_db_U1_cur}. It is important to note that there is no freedom in the choice of a (twisted) connection and its associated curvature, since expressions \eqref{eq:kM_gt_tcon} and \eqref{eq:kM_gt_tcur} are imposed by the choice of the twist.

\begin{Emph}{Twisted gauge theory}
	The latter result can be generalised to any twisted derivations. Consider two algebra automorphisms\footnote{
		The fact that the $\rho_j$ needs to be algebra homomorphisms is imposed by the \Lru \eqref{eq:kM_gt_gtder} applied to three or more elements.
}
	$\rho_j : \algft{A} \to \algft{A}$, such that
	\begin{align}
		X( f \star g)
		&= X(f) \star \rho_1(g) + \rho_2(f) \star X(g),
		\label{eq:kM_gt_gtder}
	\end{align}
	for $X$ a twisted derivation. Then, one can show that the connection and its associated curvature are given by
	\begin{subequations}
		\label{eq:kM_gt_gtcc}
	\begin{align}
		\nabla_X(s \actr f)
		&= \nabla_X(s) \actr \rho_1(f) + \tau_{2}(s) \actr X(f), \\
		R_{X,Y}
		&= \tau_2^{-1} \nabla_X \tau_2 \nabla_Y - \tau_2^{-1} \nabla_Y \tau_2 \nabla_X - \nabla_{[X,Y]}
	\end{align}
	\end{subequations}
	where $\tau_{2}$ is a twisted module automorphism $\tau_{2}(s \actr f) = \tau_{2}(s) \actr \rho_2(f)$ and $\rho_1$ and $\rho_2$ are assumed to commute with any (twisted) derivation $X$. In this case, the curvature $R$ is a twisted module homomorphism
	\begin{align}
		R_{X,Y}(s \actr f)
		&= R_{X,Y}(s) \actr \rho_1^2(f).
		\label{eq:kM_gt_gtcurth}
	\end{align}
	One should note that the \namefont{Lie} bracket changes the twist, in the sense that two twisted derivations $X$ and $Y$ with twists $\rho_1$ and $\rho_2$ as in \eqref{eq:kM_gt_gtder}, one can show that $[X,Y]$ is a twisted derivation with twists $\rho_1^2$ and $\rho_2^2$.
\end{Emph}

\paragraph{}
Now consider the specific case of noncommutative electrodynamics $\modft{X} = \kM$. One follows the steps of \secref{ncft_db} in the twisted case. By defining $A_X = \iCpx \nabla_X(1)$, one computes
\begin{subequations}
	\label{eq:kM_gt_cc}
\begin{align}
	\nabla_\mu(f)
	&= X_\mu(f) - \iCpx A_\mu \star_{\dpkM} f,
	\label{eq:kM_gt_cc_con}\\
	F_{\mu\nu}
	&= X_\mu(A_\nu) - X_\nu(A_\mu) - \iCpx \big( \kPE(A_\mu) \star_{\dpkM} A_\nu - \kPE(A_\nu) \star_{\dpkM} A_\mu \big).
	\label{eq:kM_gt_cc_curv}
\end{align}
\end{subequations}
Note that the $X_\mu$'s are not real but twisted real\footnote{
	The minus sign in \eqref{eq:kM_gt_treal} is matter of convention. Indeed, if one considers $\iCpx X_\mu$ instead of $X_\mu$, that stems for a representation $\partial_\mu$ instead of $- \iCpx \partial_\mu$, then the minus sign disappears. 
},
\ie
\begin{align}
	(X_\mu(f))^\dagger 
	&= - \kPE^{-1} X_\mu(f^\dagger).
	\label{eq:kM_gt_treal}
\end{align}
The Hermiticity condition \eqref{eq:ncft_db_Hcon} has also to be twisted to \cite{Hersent_2022b}
\begin{align}
	X_\mu \big( (s_1, s_2) \big)
	&= \big( \kPE^{-1} \nabla_\mu(s_1), s_2 \big) 
	+ \big( \kPE^{-1}(s_1), \nabla_\mu(s_2) \big)
\end{align}
for any $s_1, s_2 \in \modft{X}$, which imposes
\begin{align}
	A_\mu^\dagger 
	&= \kPE^{-1} ( A_\mu).
	\label{eq:kM_gt_hgf}
\end{align}
The latter condition \eqref{eq:kM_gt_hgf} has the same commutative limit then the usual (untwisted) Hermiticity condition $A_\mu^\dagger = A_\mu$, corresponding to a real valued gauge field $A$.

\paragraph{}
The gauge transformation of the connection, and so of the curvature, needs also to be twisted. As above, it can be shown \cite{Mathieu_2021} that it is imposed by the nature of the twist to be
\begin{align}
	A_\mu^{\ggft{u}}
	&= \kPE(\ggft{u}^\dagger) \star_{\dpkM} A_\mu \star_{\dpkM} \ggft{u} + \kPE(\ggft{u}^\dagger) \star_{\dpkM} X_\mu(\ggft{u}), &
	F_{\mu\nu}^\ggft{u}
	&= \kPE^2(\ggft{u}^\dagger) \star_{\dpkM} F_{\mu\nu} \star_{\dpkM} \ggft{u}.
	\label{eq:kM_gt_ccgt}
\end{align}
Therefore, if one considers an action of the form \eqref{eq:ncft_db_YMa}
\begin{align}
	S = \int \tdl{d+1}{x} F^{\mu\nu} \star_{\dpkM} F_{\mu\nu}^\dagger,
	\label{eq:kM_gt_YMa}
\end{align}	
its gauge transform is computed quite similarly as in \eqref{eq:ncft_db_YMgi}, except that the last prefactor is now twisted as
\begin{align}
	S^\ggft{u}
	&= \int \tdl{d+1}{x} \kPE^{d-2}(\ggft{u}) \star_{\dpkM} \kPE^2(\ggft{u}^\dagger) \star_{\dpkM} F^{\mu\nu} \star_{\dpkM} F_{\mu\nu}^\dagger.
	\label{eq:kM_gt_agt}
\end{align}
In view of the $\ugt{1}$ gauge group, which imposes $\ggft{u} \star_{\dpkM} \ggft{u}^\dagger = 1$, one obtains that the action $S$ is gauge invariant, that is $S^\ggft{u} = S$, when the powers of the $\kPE$'s in the prefactor are equal. The latter equality imposes 
\begin{align}
	d + 1
	&= 5.
	\label{eq:kM_gt_dim5}
\end{align}

\paragraph{}
Several comments of the latter result are in order.

In the context of \kMt, the action \eqref{eq:kM_gt_YMa} has a straightforward commutative limit, which correspond to the usual $\Ug{1}$ \namefont{Yang-Mills} action on the \Minkt \sT. The previous statement holds for any dimension, but the gauge invariance analysis requires that the space-time dimension is fixed to $5$ by \eqref{eq:kM_gt_dim5}. Therefore, the commutative limit of \eqref{eq:kM_gt_YMa} corresponds to the $4 + 1$-dimensional $\Ug{1}$ \namefont{Yang-Mills} action.

The action \eqref{eq:kM_gt_YMa} can be shown to be \kPt invariant, such that it has even been named \enquote{\kPt invariant \gT}. This implies that the \namefont{Poincar\'{e}} invariance is restored automatically at the commutative limit, but also that this $\ugt{1}$ \gT triggers a \LIV via a \Pdef. The previous observation sparks phenomenological considerations concerning this theory as discussed in \secref{qg_ph}.

The dimension constraint \eqref{eq:kM_gt_dim5} can be traced back to the fact that the twist in the integral cyclicity \eqref{eq:kM_kP_ncyc} depends on the spacial dimensions $d$. Furthermore, one should note that the latter constraint is very strong. Indeed, a twisted module structure of the form $f \actr g = f \star_{\dpkM} \rho(g)$, with $\rho$ an automorphism of $\kM$, does not affect the constraint \eqref{eq:kM_gt_dim5} \cite{Hersent_2022b}. Moreover, there is actually a freedom in the gauge transformation \eqref{eq:kM_gt_ccgt}, since one could also considered
\begin{align}
	F^\ggft{u}_{\mu\nu}
	&= \kPE^2 \tilde{\rho}(\ggft{u}^\dagger) \star_\dpkM F_{\mu\nu} \star_{\dpkM} \tilde{\rho}(\ggft{u})
\end{align}
where $\tilde{\rho} \in \Aut(\kM)$. However, the latter freedom leaves the requirement \eqref{eq:kM_gt_dim5} unchanged, so that we have put $\tilde{\rho} = \id$ previously.

\subsection{Quantisation and tadpole computation}
\label{subsec:kM_gt_tad}
\paragraph{}
We now turn to the study of the quantisation and the perturbation theory of \eqref{eq:kM_gt_YMa}. Note that the commutative limit of \eqref{eq:kM_gt_YMa} corresponds to the $5$-dimensional (quantum) electrodynamics action, which is known to be non-renormalisable. Therefore, one should be careful when considering the commutative limit in the context of the perturbative study of \eqref{eq:kM_gt_YMa}.

\paragraph{}
We perform a \BRST of the action \eqref{eq:kM_gt_YMa}, as in \cite{Hersent_2022a}, for which the corresponding twisted symmetry has been studied in \cite{Mathieu_2021b}. To do so, let us introduce the \namefont{Fadeev-Popov} ghost field $c$, anti-ghost field $\overline{c}$ and the  \namefont{Nakanishi–Lautrup} field $b$. The \namefont{Slavnov} operator $s$ is then defined as
\begin{align}
	s A_\mu 
	&= X_\mu(c) - \kPE(c) \star_\dpkM A_\mu + A_\mu \star_\dpkM c, &
	s c
	&= - c \star_\dpkM c, &
	s \overline{c}
	&= b, &
	s b
	&= 0,
\end{align}
from which one computes
\begin{align}
	s F_{\mu\nu}
	&= F_{\mu\nu} \star_\dpkM c - \kPE^2(c) \star_\dpkM F_{\mu\nu}, &
	s^2
	&= 0.
\end{align}
Then, one adds a gauge-fixing term to the action \eqref{eq:kM_gt_YMa} of the form
\begin{align}
	s \int \tdl{5}{x} \overline{c} \star_\dpkM \kPE^{-4} X_\mu(A^\mu), &&
	s \int \tdl{5}{x} \overline{c} \star_{\dpkM} \kPE^{-4} (A_0 - \lambda,)
	\label{eq:kM_gt_gft}
\end{align}
where the first one corresponds to the deformed  \namefont{Lorenz} gauge $X_\mu A^\mu = 0$ and the second one is the (parametrized) temporal gauge $A_0 = \lambda$, for $\lambda \in \Real$. The full action consisting of \eqref{eq:kM_gt_YMa} and \eqref{eq:kM_gt_gft} can be put under the form
\begin{align}
	S = S_{AA} + S_{c\overline{c}} + S_{AAA} + S_{AAAA} + S_{Ac\overline{c}}
\end{align}
where $S_{AA}$ correspond to the gauge field kinetic term, $S_{c\overline{c}}$ the ghost kinetic term, $S_{AAA}$ the gauge field $3$-vertex, $S_{AAAA}$ the $4$-vertex and $S_{Ac\overline{c}}$ the gauge field-ghost interaction.

We then use usual functional methods to compute the tadpole diagram, that is the one-loop \npf{1}.

\begin{Emph}{Functional methods in quantum field theory}
	We describe here a textbook functional method to obtain the \npf{n} in \qFT. If one considers an action function $S(A, c, \overline{c})$ depending on three fields $A$ (the gauge field) and $c$, $\overline{c}$ (the ghost fields). The latter splits into a kinetic part $S_{\text{kin}}(A, c, \overline{c})$ and an interaction part $S_{\text{int}}(A, c, \overline{c})$. The kinetic term is put under the form
	\begin{align}
		S_{\text{kin}}(A, c, \overline{c})
		&= \int \frac{1}{2} A_\mu K_A A^\mu + \overline{c} K_c c
		\label{kM_gt_QFT_kin}
	\end{align}
	where $K_A$ and $K_c$ are kinetic operators. We further introduce the source action $S_{\text{sou}} = \int A_\mu J^\mu + \overline{\eta} c + \overline{c} \eta$, where $J$, $\eta$, $\overline{\eta}$ are the source fields. 
	
	One defines the (resp.~free) generating functional of the connected correlation functions\footnote{
		Note that one has $W = e^Z$, where $Z$ is the generating functional of the connected \namefont{Green} functions introduced in \subsecref{ncft_p4_UVIR}.	
	} (resp.~$W_0(J, \overline{\eta}, \eta)$) $W(J, \overline{\eta}, \eta)$ as
	\begin{subequations}
	\begin{align}
		e^{W(J, \overline{\eta}, \eta)}
		&= \int \tdl{}{A} \tdl{}{c} \tdl{}{\overline{c}} e^{-S_{\text{int}}(A, c, \overline{c}) - S_{\text{kin}}(A, c, \overline{c}) + S_{\text{sou}}}, \\
		W_0(J, \overline{\eta}, \eta)
		&= \int \frac{1}{2} J_\mu K_A^{-1} J^\mu + \overline{\eta} K_c^{-1} \eta. 
	\end{align}
	\end{subequations}
	In the integral of $e^W$, one can perform the change of variables $A_\mu \to A_\mu + K^{-1}_A J_\mu$, $c \to c + K_c^{-1}\eta$ and $\overline{c} \to \overline{c} + K_c^{-1} \overline{\eta}$. Then, upon infinite expansion of $e^{S_{\text{int}}}$, one can compute that
	\begin{align}
		W(J, \overline{\eta}, \eta)
		&= \ln \left( e^{W_0} \left( 1 + e^{-W_0} \left( e^{-S_{\text{int}}\left( \frac{\partial}{\partial J}, \frac{\partial}{\partial \overline{\eta}}, \frac{\partial}{\partial \eta} \right) } - 1 \right) e^{W_0} \right) \right).
	\end{align}
	The first non-trivial term of the expansion of $e^{-S_{\text{int}}}$ yields
	\begin{align}
		W^1(J, \overline{\eta}, \eta)
		&= W_0(J, \overline{\eta}, \eta) - e^{-W_0(J, \overline{\eta}, \eta)} S_{\text{int}} \left( \frac{\partial}{\partial J}, \frac{\partial}{\partial \overline{\eta}}, \frac{\partial}{\partial \eta} \right) e^{W_0(J, \overline{\eta}, \eta)}
	\end{align}
	
	Finally, one computes the correlation functions thanks to the generating functional of proper vertices $\Gamma(A, c, \overline{c})$, that consists of the \namefont{Legendre} transform of $W(J, \overline{\eta}, \eta)$. Explicitly,
	\begin{align}
		\Gamma(A, c, \overline{c}) + W(J, \overline{\eta}, \eta) - \int A_\mu J^\mu + \overline{\eta} c + \overline{c} \eta
		&= 0, &
		A_\mu 
		&= \frac{\partial W}{\partial J^\mu}, &
		J_\mu 
		&= \frac{\partial \Gamma}{\partial A^\mu},
		\label{eq:kM_gt_QFT_Leg}
	\end{align}
	and similarly for $c$ with $\eta$ and $\overline{c}$ with $\overline{\eta}$. At first order, the middle expressions of \eqref{eq:kM_gt_QFT_Leg} boils down to $A_\mu = \frac{\partial W_0}{\partial J^\mu} = \int K_A^{-1} J_\mu$, with similar expressions for $c$ and $\overline{c}$. The latter expression implies that 	
	\begin{align}
		W_0(J, \overline{\eta}, \eta)
		&= \int A_\mu J^\mu + \overline{\eta} c + \overline{c} \eta
		\label{eq:kM_gt_QFT_W01}
	\end{align}
	Using \eqref{eq:kM_gt_QFT_W01}, a simplification occurs in \eqref{eq:kM_gt_QFT_Leg} so that the one-loop \npf{1} writes
	\begin{align}
		\langle A \rangle_{\text{1-loop}}
		= \Gamma^1(A, c, \overline{c})
		&= e^{-W_0(J, \overline{\eta}, \eta)} S_{\text{int}} \left( \frac{\partial}{\partial J}, \frac{\partial}{\partial \overline{\eta}}, \frac{\partial}{\partial \eta} \right) e^{W_0(J, \overline{\eta}, \eta)}
		\label{eq:kM_gt_1l1p}
	\end{align}
	where $J_\mu = \int K_A A_\mu$, and similarly for $\eta$ and $\overline{\eta}$.
\end{Emph}

One has all the ingredient to perform the computation of the tadpole \eqref{eq:kM_gt_1l1p}, which writes
\begin{align}
	\langle A \rangle_{\text{1-loop}}
	&= \int \tdl{5}{x} \mathcal{J}(\dpkM) \, A_0(x),
	\label{eq:kM_gt_tad}
\end{align}
where $\mathcal{J}$ is a gauge dependent divergent integral, to be regularised. The fact that \eqref{eq:kM_gt_tad} is proportional to $A_0$ can be linked to the fact that the time component has a peculiar role in \kMt. Moreover, it justifies the study of the tadpole in the temporal gauge $A_0 = \lambda$. The major result of this computation, beyond its non-vanishing, is the gauge dependence of the tadpole. This implies that the $\ugt{1}$ gauge symmetry has been broken in the quantisation process, as we started with a gauge invariant action. One should note however, that the commutative limit is correct since in any gauge, $\mathcal{J}$ vanishes in the limit $\kappa \to +\infty$.

\paragraph{}
Several comments are in order. It gathers the main elements of discussions of \cite{Hersent_2023b}.

First, one should know that non-zero tadpole has been experienced in other \qST, like the $2$-dimensional \Moy space \cite{Martinetti_2013} and $\Rcl$ \cite{Gere_2014}. The two previous computations followed the same quantisation procedure as above. This may imply that the quantisation method used here cannot be applied in the context of \nCG, as suggested in \cite{Camacho_2005}.

Then, the notion of vacuum on \qST{s} has not reach consensus, so that the expression \eqref{eq:kM_gt_tad} may not be the physical vacuum expectation value of the quantum electrodynamical theory on \kMt. On the one hand, the \namefont{Poincar\'{e}} symmetry is broken so that one cannot define particles as irreducible representation of the little group. This idea is known in \qFT on curved space-time, in which the vacuum state is only defined thanks to the asymptotic flatness hypothesis. Concerning \kMt, there has been several attempts in defining a physical vacuum, either from a deformation of the little group study, or by defining its energy. The latter notion is based on the \namefont{Casimir} operator of the \kPt algebra, but is not well-defined as it is coordinate dependent.

Finally, as discussed in \secref{ncft_db}, one could consider other physical variables then $A$ to encode the \enquote{noncommutative photon}, and so have a zero tadpole with this quantity.

\section{Causality on \tops{$\dpkM$}{kappa}-\namefont{Minkowski}}
\label{sec:kM_c}
\paragraph{}
Consider a single massive object made out of a spatial superposition of two states of different masses. The gravitational field will also be in a superposition of state thus leading to superposed space-time geometries. From there, one could argue that it is possible to superpose two geometries for which two events are causal (\ie time-like) in one geometry, but non-causal (\ie space-like) in the other. Therefore, the notion of classical causality beaks down at the quantum level. One could argue either that causality is not an intrinsic property of nature, but rather an emergent feature of some phenomenon, or that causality is deformed in the quantum setting. The latter consideration has pushed towards the search of the properties of the would be quantum causality. We refer to \cite{Brukner_2014} for an early review on the topic.

\paragraph{}
In the context of \qST{s}, the question of what becomes of causality is interesting in several aspects. A notion of causality on a \qST needs to give back the usual notion of causality at the commutative limit. In this sense, a causality on a \qST should be a deformation of the usual causality that could lead to an effective behaviour when considering the first order correction. The phenomenology of deformed causality could be of primordial importance to have experimental tests or constraints on such models. 

There is two algebraic formulation of causality which correspond to the causality on Lorentzian \st, developed by \namefont{Franco} and \namefont{Eckstein} \cite{Franco_2013} and the isocone-based approach of \namefont{Besnard} \cite{Besnard_2015}. We focus here on the first one that was applied to the \kMt space in \cite{Franco_2023}. Note that the latter was also constructed on the \enquote{quantum \namefont{Minkowski}} space (\ie the \Moy space with a \namefont{Minkowski} metric), as well as the \kMt with another \Do (see \cite{Franco_2023} for a short review).

\paragraph{}
One should note that models of causality have already been considered on the \kMt space-time by \namefont{Mercati} and \namefont{Sergola} \cite{Mercati_2018, Mercati_2018b}. The latter study is based on the observation that commutative causality may be defined thanks to \namefont{Pauli-Jordan} functions. Considering a scalar field theory with field $\phi$, the \namefont{Pauli-Jordan} function corresponds to $[\phi(x), \phi(y)]$ and encodes the light cone frontier. By implementing a noncommutative scalar field on \kMt, the latter authors could derive the \namefont{Pauli-Jordan} function of the scalar field and observe that the light cone was blurred: one does not go from time-like to space-like by an abrupt change, there is a smooth transition within which one is neither space nor time-like. By observing that the width of the blurred region increases with (space-time) distance, the authors advocate that this deformation of the light cone is close to the present measurement accuracy, if the effect is amplified by cosmological distance. Deformed light-cones could trigger time delays in photon travel from ultra-high energy cosmic rays toward Earth (see \secref{qg_ph} for more details on \qG phenomenology).

\paragraph{}
We introduce the formalism of Lorentzian \st as well as the formulation of causality on it.

\begin{Emph}{Lorentzian spectral triple}
	The Lorentzian \st is defined by the same set of data a the \st of \defref{gnc_st}, with $\algft{A}$ a \Csalg which represents on a \Hsp $\Hilbft{H}$ with inner product $\langle \cdot, \cdot \rangle$, and $\Dirop$ an operator on $\Hilbft{H}$. The \enquote{Lorentzian} property is introduced through the so-called fundamental symmetry $I \in \Hilbft{B(H)}$ which should satisfy
	\begin{subequations}
		\label{eq:kM_c_Lst}
	\begin{align}
		I^2
		&= 1, &
		I^\dagger
		&= I, &
		[I, f]
		&= 0,
		\label{eq:kM_c_Lst_fs}
	\end{align}
	for any $f \in \algft{A}_1$, where ${}^\dagger$ is the adjoint for $\langle \cdot, \cdot \rangle$. For purely technical reasons, one needs to consider $\algft{A}_1$, a unitalisation of $\algft{A}$. Note also that the representation $\pi : \algft{A} \to \Hilbft{H}$ has been dropped in \eqref{eq:kM_c_Lst_fs}.  $I$ transforms the positive form $\langle \cdot, \cdot \rangle$ of $\Hilbft{H}$ on an indefinite form $\langle \cdot, \cdot \rangle_I = \langle \cdot, I \cdot \rangle$, which is not necessarily positive. The latter $I$ may be seen as a generalisation of the complex $\iCpx$ to \Hsp{s}, and the form $\langle \cdot, \cdot \rangle_I$ as a \namefont{Wick} rotation of $\langle \cdot, \cdot \rangle$. The latter indefinite product is called a \namefont{Krein} product, and $\Hilbft{H}$ equipped with it, a \namefont{Krein} space.
	
	From there, the Lorentzian \st follows the axioms of the \st adapted to the \namefont{Krein} space. Explicitly, the self-adjointness of $\Dirop$ for $\langle \cdot, \cdot \rangle$ is replaced by a self-adjointness for $\langle \cdot, \cdot \rangle_I$, which writes
	\begin{align}
		\Dirop^\dagger I
		&= - I \Dirop.
		\label{eq:kM_c_Lst_saDo}
	\end{align}
The compact resolvent condition writes
	\begin{align}
		f \, (1 + \tilde{\Dirop}^2)^{-\frac{1}{2}}
		\ \text{ is compact},
	\end{align}
	for any $f \in \algft{A}$, where $\tilde{\Dirop}^2 = \frac{1}{2}(\Dirop^\dagger \Dirop + \Dirop \Dirop^\dagger)$. Finally, the \Do should satisfy
	\begin{align}
		[\Dirop, f] \in \Hilbft{B(H)},
	\end{align}	 
	for any $f \in \algft{A}$, as in the usual \st case.
	
	Finally, there is an additional condition requiring that it exists a self-adjoint operator $T$ and a positive element $N \in \algft{A}_1$ such that $\Dom(T) \cap \Dom(N)$ is dense in $\Hilbft{H}$ and satisfies
	\begin{align}
		(1 + T)^{-\frac{1}{2}}
		&\in \algft{A}_1, &
		I
		&= - N [\Dirop, T].
		\label{eq:kM_c_Lst_Ls}
	\end{align}
	\end{subequations}
	The operator $T$ corresponds to the (noncommutative) generalisation of a global time function, while the right hand side of \eqref{eq:kM_c_Lst_Ls} ensures a Lorentzian type signature. In this sense, note that $-I$ could satisfy all the axioms \eqref{eq:kM_c_Lst} if one changes $T$ to $-T$. The latter symmetry can be directly linked to the choice of a signature.
	
	\paragraph{}
	The equation \eqref{eq:kM_c_Lst_Ls} is not the only possible way to construct a fundamental symmetry $I$. Furthermore, one can define a Lorentzian \st corresponding to a globally hyperbolic classical \sT. We refer to \cite{Franco_2013} for more details on these points. 
\end{Emph}

\begin{Emph}{Causality on Lorentzian spectral triples}
	\noindent%
	\begin{minipage}{.49\textwidth}
		\paragraph{}
		If one considers a classical \sT $\manft{M}$, with metric $\metft{g}$, and one wants to know if two points $x, y \in \manft{M}$ are causally connected, then one can consider a causal curve relating the two. A causal curve is a smooth curve $\gamma: \Real \to \manft{M}$ satisfying
		\begin{align}
			\metft{g}(\gamma'(t), \gamma'(t)) \leqslant 0.
			\label{eq:kM_c_ccu}
		\end{align}
		The previous relation states that, at any $t \in \Real$, the tangent vector to the curve $\gamma$ is future directed, \ie time-like and pointing to increasing time (see \figref{causal_c}). Therefore, $y$ is in the causal future of $x$ if there exists a causal curve $\gamma$ such that $\gamma(t_1) = x$ and $\gamma(t_2) = y$ for some $t_1 \leqslant t_2 \in \Real$.
	\end{minipage}%
	\hfill%
	\begin{minipage}{.49\textwidth}
		\begin{Figure}
			[label={fig:causal_c}]%
			{
				Space-time diagram of a causal curve $\gamma$ (in red) linking two points $x$ and $y$ of flat \sT. The derivative $\gamma'$ along the curve is pictured to be always time-like.
			}%
			\begin{tikzpicture}[scale = 1.1]
		\draw[black, dashed, thin] (0,0) to (4,4);
		\draw[black, dashed, thin] (-.35,.35) to (-2,2);
		\draw[black, -To, thin] (0,0) to (0,5) 
			node[anchor= north east]{$x^0$};
		\draw[black, -To, thin] (-2,0) to (4,0) 
			node[anchor= south east]{$x^j$};

		\filldraw[black] (0,0) node[anchor= south east]{$x$} circle (.05);
		\filldraw[black] (1.5,5) node[anchor= north east]{$y$} circle (.05);	
	
		\draw[darkred, thick] (0,0) to[out=90, in=-90]
			(1,2) to[out=90, in=-90]
			(2,4) to[out=90, in=-90]
			(1.5,5);
		\draw[darkred, dashed] (.5,1) to (0,1.5);
		\draw[darkred, dashed] (.5,1) to (1, 1.5);
		\draw[darkred, -To] (.5,1) to (1.1, 1.7);
		\draw[darkred, dashed] (1,2) to (.5,2.5);
		\draw[darkred, dashed] (1,2) to (1.5,2.5);
		\draw[darkred, -To] (1,2) to (1,3);
		\draw[darkred, dashed] (2,4) to (1.5,4.5);
		\draw[darkred, dashed] (2,4) to (2.5,4.5);
		\draw[darkred, -To] (2,4) node[anchor= north east]{$\gamma$} to (1.9, 5);
\end{tikzpicture}%
		\end{Figure}
	\end{minipage}

	\vspace{4pt}
	Within this setting the causal structure of $\manft{M}$ is fully determined by its set of so-called causal functions. Explicitly, a causal function is a function $f : \manft{M} \to \Real$ which is non-decreasing along every future directed causal curve $\gamma$. Therefore, 
	\begin{align}
		y \text{ is in the causal future of } x
		\text{ if and only if } f(x) \leqslant f(y)
		\label{eq:kM_c_cf}
	\end{align}
	for any causal function $f$ \cite{Franco_2013}. The graph of a causal function $f$ corresponds to the region of causal simultaneity.

	To go to the (noncommutative) algebraic setting, one can replace \sT points by pure states of $\algft{A}$, $\Sppst{\algft{A}}$ (see \defref{oa_rt_st}). As $\algft{A}$ contains the noncommutative analogue of the smooth functions, the causal curves will be generalised as a subset of $\algft{A}$, called the causal cone.

	\paragraph{}
	A causal cone $\mathcal{C} \subset \algft{A}_1$ is defined as
	\begin{subequations}
		\label{eq:kM_c_cco}
	\begin{align}
		f^\dagger = f, && 
		f + g \in \mathcal{C}, &&
		\lambda f \in \mathcal{C}, &&
		x 1 \in \mathcal{C},
		\label{eq:kM_c_cco_con}
	\end{align}
	for any $f, g \in \mathcal{C}$, $\lambda \in \pReal$ and $x \in \Real$. These stability conditions correspond to the mathematical definition of an (iso)cone. Furthermore, one should have that the linear span of $\mathcal{C}$ form all $\algft{A}_1$, so that there are no disconnected region. Finally, the \enquote{causal} feature of the causal cone is due to the requirement that\footnote{
		We abbreviated the mathematical notation $\langle \ket{\psi}, \ket{\psi} \rangle$ to the physical one $\braket{\psi}{\psi}$, in accordance with \exref{oa_rt_stHsp}.
	}
	\begin{align}
		\braket{\psi}{[\Dirop, f] \psi}_I
		&= \braket{\psi}{I [\Dirop, f] \psi}
		\leqslant 0
		\label{eq:kM_c_cco_cau}
	\end{align}
	\end{subequations}
	for any $f \in \mathcal{C}$ and $\ket{\psi} \in \Hilbft{H}$. For commutative Lorentzian \st, it is possible to link \eqref{eq:kM_c_cco_cau} directly to \eqref{eq:kM_c_ccu}. Moreover, the sign of \eqref{eq:kM_c_cco_cau} is a signature choice, similarly to \eqref{eq:kM_c_ccu}. One could require it to be positive, if one changes $I$ to $-I$.

	\paragraph{}
	To summarise, when considering a Lorentzian \st, an element $f \in \algft{A}$ is in the causal cone $\mathcal{C}$ if and only if it satisfies \eqref{eq:kM_c_cco_cau}, for any $\ket{\psi} \in \Hilbft{H}$. From there, we dispose of \enquote{causal time charts} $f$. Then, for any two (pure) states $\psi_1, \psi_2 \in \Sppst{\algft{A}}$
	\begin{align}
		\psi_2 \text{ is in the causal future of } \psi_1
		\text{ if and only if } \psi_1(f) \leqslant \psi_2(f),
		\label{eq:kM_c_cst}
	\end{align}
	for any $f \in \mathcal{C}$. The latter relation \eqref{eq:kM_c_cst} is the noncommutative analogue of \eqref{eq:kM_c_cf}.
\end{Emph}

\paragraph{}
The previous mathematical framework was applied to $1+1$-dimensional \kMt $\algft{A} = \kM[1]$ in \cite{Franco_2023}. The considered \Hsp is given by
\begin{align}
	\Hilbft{H} 
	&= \Hilbft{H}_+ \oplus \Hilbft{H}_0 \oplus \Hilbft{H}_-, &
	\Hilbft{H}_a
	&= \Cpx^2 \otimes L^2(\Real)
	\label{eq:kM_c_Hsp}
\end{align}
where $a = 0, \pm$ and $L^2(\Real)$ consist of the square integrable functions on $\Real$ (with the \namefont{Lebesgue} measure). On each summand $\Hilbft{H}_a$, the representation $\pi_a$ is inherited from the unitary irreducible representation of the affine group (the \Liegt of momenta) on $L^2(\Real)$ given by
\begin{align}
	\big(\pi_a(f) \psi^{(a)} \big)(p_0)
	&= \int \tdl{}{q_0} f(q_0 - p_0, a e^{-p_0/\dpkM}) \, \psi^{(a)}(q_0)
	\label{eq:kM_c_rep}
\end{align}
for any $f \in \kM[1]$ and $\psi^{(a)} \in L^2(\Real)$. The fact that $\Cpx^2$ appear in \eqref{eq:kM_c_Hsp} comes from the fact that we use a $2$-dimensional \Do below. The full representation on $\Hilbft{H}$ is given by $\pi = (\pi_+ \oplus \pi_0 \oplus \pi_-) \otimes \Matid{2}$. In the following, the representations are clearly stated to avoid confusion. Considering two elements $\psi_1 = \bigoplus_a \psi_1^{(a)}, \psi_2 = \bigoplus_a \psi_2^{(a)} \in \Hilbft{H}$, one defines the inner product as
\begin{align}
	\langle \psi_1, \psi_2 \rangle
	&= \sum_{a = +, 0, -} \int \tdl{}{p_0} \big(\psi_1^{(a)}\big)^\dagger(p_0) \ \psi_2^{(a)}(p_0).
\end{align} 
One can note that the representation \eqref{eq:kM_c_rep} implies that $\pi_\pm(x^0) = - \iCpx \frac{\td}{\td p_0}$ and $\pi_\pm(x^1) = \pm e^{-p_0 / \dpkM}$. Therefore, one has that $x^0 = \hat{x}$ and $x^1 = \pm e^{-\hat{p} / \dpkM}$, where $\hat{x}$, $\hat{p}$ are the \namefont{Schr\"{o}dinger} representation of the position and momentum operators respectively of the $1$-dimensional \qM[al] system.

The considered Lorentzian \st writes 
\begin{align}
	\Dirop
	&= - \iCpx \gamma^\mu X_\mu \otimes \Matid{3}
	= \begin{pmatrix} 0 & X_- \\ X_+ & 0 \end{pmatrix} \otimes \Matid{3}, &
	I
	&= \iCpx \gamma^0 \otimes \Matid{3}
	\label{eq:kM_c_Do}
\end{align}
where $\gamma^0 = \begin{pmatrix} 0 & \iCpx \\ \iCpx & 0 \end{pmatrix}$ and $\gamma^1 = \begin{pmatrix} 0 & -\iCpx \\ \iCpx & 0 \end{pmatrix}$ are the $2$-dimensional \namefont{Dirac} gamma matrices, $X_\mu$ corresponds to the twisted derivations \eqref{eq:kM_gt_tder}, and $X_\pm = X_0 \pm X_1$. Note that $\Matid{3}$ is here to express that the same \Do and fundamental symmetry are considered on each $+$, $0$ and $-$ representations. Furthermore, the \Do expression \eqref{eq:kM_c_Do} is the straightforward generalisation of the commutative \Do on the \Minkt \sT $\Dirop = - \iCpx \gamma^\mu \partial_\mu$.

\paragraph{}
It is important to point out that the considered derivations in the \Do are twisted, so that we have to consider a twisted (Lorentzian) \st, as introduced in \subsecref{kM_gt_td}. Therefore, one has to consider the twisted bracket $[\cdot, \cdot]_\kPE$ in the axioms \eqref{eq:kM_c_Lst}, instead of the usual bracket. The twisted bracket expresses as
\begin{align}
	[\Dirop, f]_\kPE
	&= \Dirop f - \kPE(f) \Dirop
\end{align}
in accordance with \eqref{eq:kM_gt_tDo}.

By considering 
\begin{align}
	T
	&= \bigoplus_a \big( \pi_a(x^0) \otimes \Matid{2} \big), &
	N
	&= 1,
	\label{eq:kM_c_gtime}
\end{align}
one can compute, thanks to \eqref{eq:kM_c_rep}, that \eqref{eq:kM_c_Lst_Ls} is satisfied. The equation \eqref{eq:kM_c_gtime} states that the global time $T$ corresponds to the representation of $x^0$. It seems quite straightforward to consider that indeed $x^0$ is a global time, even in the deformed theory.

\paragraph{}
It remains to consider how the conditions \eqref{eq:kM_c_cf} and \eqref{eq:kM_c_cco_cau} writes in this case.

One can show that a relevant set of pure states is given by
\begin{align}
	\psi^\pm(f)
	&= \langle \psi, \pi_{\pm}(f) \psi \rangle
\end{align}
where $\ket{\psi} \in \Hilbft{H}_\pm$. Therefore, the condition \eqref{eq:kM_c_cf} imposes that $\psi_2 \in \Sppst{\kM[1]}$ is in the causal future of $\psi_1 \in \Sppst{\kM[1]}$ if and only if
\begin{align}
	\int \tdl{}{p_0} \tdl{}{q_0} f(q_0 - p_0, \pm e^{-p_0/\dpkM}) \, \frac{\td}{\td t} \big( \overline{\psi}_t(p_0) \psi_t(q_0) \big)
	\geqslant 0 
	\label{eq:kM_c_kM_cs}
\end{align}
for all $f \in \mathcal{C}$, where $t \in [1,2]$ is a continuous parameter that interpolates between $\psi_1$ and $\psi_2$.

The condition for a function $f$ to be in the causal cone $\mathcal{C}$ \eqref{eq:kM_c_cco_cau} can be computed to be
\begin{align}
	\int \tdl{}{p_0} \tdl{}{q_0} \big( \iCpx (1 - e^{-(q_0-p_0)/\dpkM}) \, f(q_0 - p_0, a e^{- p_0 / \dpkM}) \pm \partial_1 f(q_0 - p_0, a e^{-p_0/\dpkM}) \big) \overline{\psi}(p_0) \psi(q_0)
	\geqslant 0
	\label{eq:kM_c_kM_cf}
\end{align}
for all $\psi \in L^2(\Real)$ and $a = \pm, 0$, where $\partial_1$ denotes the derivative with respect to the second (spacial) variable. Note that any function $f = x^0 + v x^1$, with $v \in [-1,1]$ satisfies \eqref{eq:kM_c_kM_cf} and therefore is in the causal cone, as one could expect. Merging the two condition \eqref{eq:kM_c_kM_cs} and \eqref{eq:kM_c_kM_cf} gives a non-trivial transport equation that was considered too tedious to be solved.

\paragraph{}
Even if the causality on $\kM[1]$ has not been fully characterised, two comments are in order. First of all, the commutative limit of the former model has few physical interest since the considered representation \eqref{eq:kM_c_rep} is $1$-dimensional when $\dpkM \to + \infty$. Thus, it raises the question of what happens when one considers a $2$-dimensional representation with a good commutative limit, like the \namefont{Gel'fand-Na\u{i}mark-Segal} representation (see \subsecref{oa_rt_sr}). One could also wonder how much this causality model depends on the considered \Hsp.

Moreover, if one writes the condition that two states $\psi_1, \psi_2$ to be causally related \eqref{eq:kM_c_cco_cau} with $f(p_0, \pm e^{-q_0/\dpkM}) = p_0 \pm e^{-q_0 / \dpkM}$, one obtains
\begin{align}
	\braket{\psi_2}{x^0 \psi_2} - \braket{\psi_1}{x^0 \psi_1}
	\geqslant \big\vert \braket{\psi_2}{x^1 \psi_2} - \braket{\psi_1}{x^1 \psi_1} \big\vert.
	\label{eq:kM_c_sll}
\end{align}
Therefore, the equations that determine the causal evolution of states can be loosely written as $\langle \delta x^0 \rangle \geqslant \vert \langle \delta x^1 \rangle \vert$, which is the expectation value of the speed-of-light limit on \Minkt: $\delta x^0 \geqslant \vert \delta x^1 \vert$. The expression \eqref{eq:kM_c_sll} was thus considered to be the analogue of the speed-of-light limit. The presence of expectation values suggests that the speed-of-light limit has to be satisfied on average but that it could be broken locally. This observation could lead to important phenomenological considerations. However, the expression \eqref{eq:kM_c_sll} is not telling how much or how often such a violation could occur. This would require a deepened analysis. Furthermore, one could consider more complex causal functions and see what becomes of \eqref{eq:kM_c_sll}.

\section{Other deformations of \tops{$\kappa$}{kappa}-\namefont{Minkowski} space-time}
\label{sec:kM_rM}
\paragraph{}
Despite that \kMt has been the first and well-most studied deformation of the \Minkt \sT, there exists other deformations of the \Pogt \cite{Zakrzewski_1994, Zakrzewski_1997}. Mainly three have been studied as reviewed in \cite{Mercati_2023}. First, the $\dpMoy$-deformation of \Minkt corresponds to a \Moy-like deformation (see \secref{ncft_Moy}) with a \Minkt metric. Second, the main topic of this \chapref{kM} correspond to the $\dpkM$-deformation. Finally, there was recently a proposal for a new deformation of \enquote{angular} type \cite{Dimitrijevic_2018} called $\dprM$-deformation. In this \secref{kM_rM}, we introduce the \rMt space-time and discuss the first result of field and gauge theory on it.

Note that other deformations of \Minkt have been considered in the literature with fewer manpower of research, like the generalisation of \eqref{eq:kM_kP_gkM} \cite{Ballesteros_2003} or the very recent T-\namefont{Minkowski} \cite{Mercati_2023}.

\paragraph{}
The \rMt space-time was first derived by a \Dt deformation (see \subsecref{gnc_dq_dt}) with the twist \cite{Dimitrijevic_2018}
\begin{align}
\begin{aligned}
	\Hoft{F}
	&= \exp \left( \frac{\iCpx \dprM}{2} \big( P_0 \otimes J_3 - J_3 \otimes P_0 \big) \right) \\
	&= \exp \left( - \frac{\iCpx \dprM}{2} \big( \partial_0 \otimes (x^1 \partial_2 - x^2 \partial_1) - (x^1 \partial_2 - x^2 \partial_1) \otimes \partial_0 \big) \right).
\end{aligned}
	\label{eq:kM_rM_Dt}
\end{align}
The latter is called \enquote{angular twist} and corresponds to a \Dt of the \Palgt $\Palg[1,3]$: there is no need to extend the algebra as discussed in \secref{kM_kP} for $\dpkM$-deformation. One calculates the \Lieat of coordinates to be
\begin{align}
	[x^0, x^1]_{\star_\dprM}
	&= \iCpx \dprM x^2, &
	[x^0, x^2]_{\star_\dprM}
	&= - \iCpx \dprM x^1, &
	[x^1, x^2]
	&= 0
	\label{eq:kM_rM_co}
\end{align}
with $x^3$ a central element, \ie it commutes with the other coordinates. The \dpt $\dprM$ has the dimension of a length. One can express the latter relations \eqref{eq:kM_rM_co} in cylindrical coordinates $(x^0, x^r, x^\varphi, x^3)$ to obtain
\begin{align}
	[x^0, x^\varphi]_{\star_\dprM}
	&= \dprM x^\varphi
	\label{eq:kM_rM_cco}
\end{align}
and the other pairs of coordinates commute. One should note the similarity between \eqref{eq:kM_rM_cco} and a $1+1$-dimensional \kMt \eqref{eq:kM_kP_kM_alg} with two central coordinates when considering the change $\dprM \to \frac{\iCpx}{\dpkM}$.

The \rPt, derived thanks to the \Dt \eqref{eq:kM_rM_Dt}, was shown to have a bicrossproduct structure \cite{Fabiano_2023b} quite similar to the one of \kPt.

\paragraph{}
A derivation of the \spdtt through \calg technique (see \subsecref{gnc_dq_ca}) has been done in \cite{Hersent_2023c}. The sketchy reasoning of the \spdtt construction is made here with coordinates \eqref{eq:kM_rM_co}, but one could equivalently perform the construction with \eqref{eq:kM_rM_cco} and obtain a similar result.

The non-trivial part of \eqref{eq:kM_rM_co} corresponds to the Euclidean \Lieat, for which the associated \Liegt is the Euclidean group. The latter can be derived to satisfy
\begin{subequations}
	\label{eq:kM_rM_eg}
\begin{align}
	p_0 \dplus q_0
	&= p_0 + q_0, &
	\vec{p} \dplus \vec{q}
	&= \vec{p} + R(\dprM p_0) \vec{q}, 
	\label{eq:kM_rM_eg_dp} \\
	 \dminus p_0
	 &= - p_0, &
	 \dminus \vec{p}
	 &= - R(- \dprM p_0) \vec{p},
	 \label{eq:kM_rM_eg_dm}
\end{align}
\end{subequations}
where we noted $\vec{p} = \begin{pmatrix} p_1 \\ p_2 \end{pmatrix}$ and $R(\dprM p_0) = \begin{pmatrix} \cos(\dprM p_0) & - \sin(\dprM p_0) \\ \sin(\dprM p_0) & \cos(\dprM p_0) \end{pmatrix}$ corresponds to the rotation matrix of angle $\dprM p_0$ around the $x^3$ axis. This group is unimodular $\Delta = 1$ and has a \Haarm corresponding to the \namefont{Lebesgue} measure, \ie $\lHm(p) = \td^{4}p$. From there, one derives the \spdtt and involution to be
\begin{subequations}
	\label{eq:kM_rM_sp}
\begin{align}
	(f \star_\dprM g)(x)
	&= \int \frac{\td p_0}{2 \pi} \tdl{}{y^0} e^{- \iCpx y^0 p_0} \, f(x^0 + y^0, \vec{x}, x^3) \, g(x^0, R(\dprM p_0) \vec{x}, x^3), 
	\label{eq:kM_rM_sps} \\
	f^\dagger(x)
	&= \int \frac{\td p_0}{2 \pi} \tdl{}{y^0} e^{- \iCpx y^0 p_0} \, \overline{f}(x^0 + y^0, R(\dprM p_0) \vec{x}, x^3),
	\label{eq:kM_rM_spi}
\end{align}
\end{subequations}
where we again used the shorthand notation $\vec{x} = (x^1, x^2)$.

From the unimodularity of the group \eqref{eq:kM_rM_eg}, one can deduce (see \eqref{eq:ncft_p4_cyc}) that the integral is cyclic
\begin{align}
	\int \tdl{4}{x} (f \star_\dprM g)(x)
	&= \int \tdl{4}{x} (g \star_\dprM f)(x).
	\label{eq:kM_rM_cyci}
\end{align}
The latter property can also be derived by direct computation with \eqref{eq:kM_rM_sps}, or by the \spdtt derived from the \Dt \eqref{eq:kM_rM_Dt}.

\paragraph{}
The analysis of a charged \phif on \rMt, based on the \spdtt \eqref{eq:kM_rM_sp}, has been studied in \cite{Hersent_2023c}. The \enquote{charged} property means that the considered action has $\phi^\dagger$ term, so that $\phi$ and $\overline{\phi}$ represents the field and its charge conjugated field respectively. The considered action is of the form
\begin{align}
	S(\phi, \overline{\phi})
	&= \int \tdl{4}{x} (\partial_\mu \phi)^\dagger \star_\dprM \partial^\mu \phi
	+ m^2 \phi^\dagger \star_\dprM \phi
	+ \frac{\cC^2}{4!} V_{\text{int}}(\phi, \overline{\phi})
	\label{eq:kM_rM_phi4a}
\end{align}
where $V_{\text{int}}$ is the interaction term considered to be 
\begin{align}
	V_{\text{int}}(\phi, \overline{\phi})
	&= \phi^\dagger \star_\dprM \phi \star_\dprM \phi^\dagger \star_\dprM \phi, &
	V_{\text{int}}(\phi, \overline{\phi})
	&= \phi^\dagger \star_\dprM \phi^\dagger \star_\dprM \phi \star_\dprM \phi.
	\label{eq:kM_rM_int}
\end{align}
The first interaction term is referred to as \enquote{orientable} and the right one as \enquote{non-orientable}. Thanks to the cyclicity of the integral \eqref{eq:kM_rM_cyci}, the two interactions \eqref{eq:kM_rM_int} are the only $4$-interaction one can write with two $\phi$ and two $\phi^\dagger$ fields.

In order to quantise the action \eqref{eq:kM_rM_phi4a}, one can perform an analysis very similar to the one of \subsecref{ncft_p4_UVIR}. One should be careful, however, to the fact that $\phi$ and $\overline{\phi}$ have to be considered as distinct fields. Therefore, one need to introduce a source $\overline{J}$ coupled to $\overline{\phi}$.

The \npf{2} at one-loop is computed very similarly to \eqref{eq:ncft_p4_nc2p}, except that not all diagrams of \figref{ncft_p} and \figref{ncft_np} are accessible. This stems from the fact that a $\phi$ has to be linked to a $\overline{\phi}$ from the kinetic term of \eqref{eq:kM_rM_phi4a} together with the fact that $\phi$ and $\overline{\phi}$ cannot be interchanged. The latter observation makes the number of diagrams boils down from twelve to six, in each case. The former observation makes it go from six to four. Explicitly, the orientable interaction give rise to diagrams of \figref{ncft_p1}, \ref{fig:ncft_p2}, \ref{fig:ncft_p3} and \ref{fig:ncft_p4}, while the non-orientable interaction has diagrams of \figref{ncft_p1}, \ref{fig:ncft_p2}, \ref{fig:ncft_np1}, \ref{fig:ncft_np3}. From there, one computes that the orientable interaction has no \UVIR in the \npf{2} at one-loop, but the non-orientable has. This can be traced back to the fact that the orientable theory has no non-planar \namefont{Feynman} diagrams, but the non-orientable has. This observation has already been made on \kMt \cite{Poulain_2018}. Moreover, the analysis of \cite{Hersent_2023c} shows that the \npf{4}, at one-loop, of the orientable theory has a \UVIR, stemming from the fact that now some non-planar diagram arises.

\paragraph{}
Finally, a deformed $\Ug{1}$ gauge theory was considered on \rMt \cite{Maris_2024}. As the integral is cyclic \eqref{eq:kM_rM_cyci}, one has \aprio less troubles than for \kMt (see \secref{kM_gt}). However, the coproduct of the deformed translations writes
\begin{align}
	\Delta(P_0)
	&= P_0 \otimes 1 + 1 \otimes P_0, &
	\Delta(P_3)
	&= P_3 \otimes 1 + 1 \otimes P_3, &
	\Delta(P_\pm)
	&= P_\pm \otimes 1 + \kPE^{\mp 1} \otimes P_\pm,
	\label{eq:kM_rM_tder}
\end{align}
where $P_\pm = P_1 \pm \iCpx P_2$ and $\kPE = \exp(\iCpx \dprM P_0)$. Because of \eqref{eq:kM_rM_tder}, twisted derivations are still considered, even if the integral is cyclic. We refer to \subsecref{kM_gt_td} for a discussion on why the twisted cyclicity of the integral hinted for the use of twisted derivations. Note that this approach considered derivations with different twists so that one has to be careful when deriving the \gT.

The considered action is of the \nCYM type \eqref{eq:ncft_db_YMa}, with 
\begin{align}
	F_{\mu\nu}
	&= \kPE_\nu^{-1} \nabla_\mu \kPE_\nu \nabla_\nu
	- \kPE_\mu^{-1} \nabla_\nu \kPE_\mu \nabla_\mu
\end{align}
where $\mu, \nu = 0, \pm, 3$, $\kPE_\mu = (1, \kPE^{-1}, \kPE, 1)$ and $\nabla_\mu$ is a twisted connection. The latter action has been shown to be \rPt invariant and $\ugt{1}$ gauge invariant.

\chapter{Quantum gravity and quantum space-times}
\label{ch:qg}

\paragraph{}
As discussed in \secref{gnc_found} and \secref{gnc_phys}, \nCG has lots of motivations arising from \qG. The aim of this \chapref{qg} is threefold. First, we introduce the main issues and challenges of \qG as a whole. Then, we depicts the main topics of \qG phenomenology and the impact of the physics of \qST{s} in this field. Moreover, we also present the various constraints on the \sT deformations obtained by \qG phenomenology and its experimental results. Finally, we depict the obstacles to build models of gravity on \qST{s}, models which are generically named \enquote{noncommutative gravity}. More specifically, we present a recent attempt in constructing an analogue of \sT for which the tangent space is \kMt, rather then the \Minkt \sT.

\paragraph{}
There has been many proposal for \qG theories. For a historical note on \qG research see for example \cite{Rovelli_2000}. We wish to argue, in this \chapref{qg}, that \nCG, at least as a mathematical tool, is a serious candidate for expressing a \qG theory, as it has inspired many promising phenomenological framework.

\section{The motivations to study quantum gravity}
\label{sec:qg_mot}

\subsection{Theoretical mismatches of quantum and gravity}
\label{subsec:qg_mot_mis}
\paragraph{}
The theory of \gR has not been challenged by experiments yet, but there are theoretical reasons to think that \gR is to be interpreted as an \eFT of some more fundamental theory, valid below some energy scale. Indeed, when merged with \qM or \qFT, some inconsistencies arises.

\paragraph{}
The first one is the black hole information paradox. When considering a \qFT in \namefont{Schwarzschild} \sT, \namefont{Hawking} \cite{Hawking_1976} found that a black hole emits radiations, called \namefont{Hawking} radiations, which causes the black hole to evaporate over time. The final state of the black hole is only determined by the total mass, charge and angular momentum of the initial state, so that different initial states may lead to the same final state. Therefore, by only knowing the information of the final state, one can reconstruct at most a class of initial states, but not the full initial state: some information is lost. Therefore, there is a mismatch with quantum theories as information loss is prevented in \qFT via unitarity.

\paragraph{}
The second puzzle concerns the interpretation of time. In \qM, the time is considered as universal and absolute. It can be compared to a thermodynamical time or a parameter that controls the evolution of the quantum system. Whereas in \gR, there is not a single notion of time since it is observer dependant. General covariance even requires that it is relative, in the sense that the evolution in space affects the evolution in time: one can think for example to time dilatation in special relativity. The interpretation of time in the two theories is so different that it is not even clear how a theory of \qG should solve this so-called \enquote{problem of time}.

\paragraph{}
The third inconsistency lies in the computation of the vacuum energy (cosmological constant) of the Universe, if interpreted as the mean energy of all its elementary constituents described by \qFT. The discrepancy between the theoretical and experimental value is more then $56$ orders of magnitude \cite{Koksma_2011}, highlighting the fact that the vacuum energy cannot be explained by \qFT alone.

\paragraph{}
The fourth discrepancy arises when one tries to apply \qFT methods on gravity. The value of the \Plent, corresponding to a \namefont{Schwarzschild} black hole having a radius of order of its \namefont{Compton} wavelength, suggests that quantum fluctuations of gravity could become relevant at small scale. Therefore, one would need a quantum (or at least semi-classical) formulation of \gR to capture these effects. As detailed in \subsecref{qg_mot_pi}, the \pIQ of \gR is non-renormalisable so that we cannot make sense of the straightforward version of \enquote{quantum \gR}. Therefore, gravity cannot be considered similarly as the other fundamental forces and needs to be quantised by other ways, if it needs to be quantised at all.

\paragraph{}
Many authors consider other puzzles that \qG should solve.
\begin{nclist}
	\item The nature of \sT at the \Plent is thought to be very different from a usual smooth manifold. The latter assertion takes its root in different aspects of \qG be it loop quantum gravity, group field theory or the \namefont{Doplicher}, \namefont{Fredenhagen} and \namefont{Roberts} argument \cite{Doplicher_1995} (detailed in \subsecref{gnc_phys_qg}).
	\item The geodesic completeness, broken in singular \sT{s}, is supposed to be restored in a \qG theory thanks to the smearing out of the curvature singularity. Note that this singularity can correspond to a black hole singularity. In this sense, a \qG theory is expected to grasp more insight on the black hole interior. This singularity can also be a primordial one, so that \qG should resolve the constitution of the early Universe.
	\item The actual theoretical physics supposed to explain all what we know is made of two distinct theories: one for particle physics and the other for gravity. As the first is intrinsically quantum, a theory of \qG is expected to shed light on some common ground for a unified theory. 
	\item Some authors are hoping for a theory of \qG to account for dark matter phenomenons. 
	\item As already discussed in \secref{kM_c}, the notion of causality is questioned when both quantum theory and (curved) \sT happen to coexist. A theory of \qG should be able to settle if causality is an intrinsic property of nature, and if so, what quantum causality is. 
\end{nclist}

\subsection{The perturbative quantisation of gravity}
\label{subsec:qg_mot_pi}
\paragraph{}
We gather here the sketch of computation for the non-renormalisability of \gR. We refer to \cite{Hamber_2009} for more complete computations and an enhanced set of references.

\paragraph{}
The \namefont{Einstein-Hilbert} action on a $d+1$-dimensional \sT $\manft{M}$ is given by
\begin{align}
	S_{\text{EH}}
	&= \frac{1}{16 \pi} \int \tdl{d+1}{x} \sqrt{\metft{g}(x)} \, R(x),
	\label{eq:qg_mot_EHa}
\end{align}
where $\metft{g}$ is the metric of $\manft{M}$ and $R = \metft{g}^{\mu\nu} R_{\mu\nu}$ is the \namefont{Ricci} scalar. In perturbation theory, one consider small fluctuations $\metft{h}$ of a fixed background metric $\eta$ (usually considered to be the \Minkt metric) as
\begin{align}
	\metft{g}_{\mu\nu}(x)
	&= \eta_{\mu\nu} + \sqrt{16 \pi} \, \metft{h}_{\mu\nu}(x).
\end{align}
It is important to note that this weak field expansion is not unique and that the perturbative expansion highly depends on it. Regardless of the gauge fixing, one can have different sets of \namefont{Feynman} rules stemming from different weak field expansions. The integrand of the action can be shown to reduce, at lowest order in $\metft{h}$, to
\begin{align}
	- \frac{1}{4} \partial_\mu \metft{h}_{\nu\rho} \, \partial^\mu \metft{h}^{\nu\rho}
	+ \frac{1}{8} (\partial_\mu \tensor{\metft{h}}{^\nu_\nu})^2
	+ \frac{1}{2} \big( \partial_\nu \tensor{\metft{h}}{^\nu_\mu} - \frac{1}{2} \partial_\mu \tensor{\metft{h}}{^\nu_\nu} \big)^2,
	\label{eq:qg_mot_Lag0}
\end{align}
where traces and inverse are taken with respect to the background metric. In order to perform the \BRST, one has to introduce a gauge fixing. A convenient gauge fixing correspond to the \namefont{de Donder} gauge
\begin{align}
	\partial_\nu \metft{h}^{\nu\mu} - \frac{1}{2} \partial^\mu \tensor{\metft{h}}{^\nu_\nu} = 0.
	\label{eq:qg_mot_DDg}
\end{align}
The gauge \eqref{eq:qg_mot_DDg} removes the zero modes $\metft{h}_{\mu\nu} \sim \partial_\mu \zeta_\nu + \partial_\nu \zeta_\mu$ which makes the propagator ill defined, in a similar fashion that the \namefont{Lorenz} gauge $\partial^\mu A_\mu = 0$ removes the zero modes and so the divergence in the \namefont{Yang-Mills} propagator. The gauge fixing \eqref{eq:qg_mot_DDg} cancels with the last term of \eqref{eq:qg_mot_Lag0} so that the full Lagrangian, at second order in $\metft{h}$, is given by
\begin{align}
	- \frac{1}{4} \partial_\mu \metft{h}_{\nu\rho} \, \partial^\mu \metft{h}^{\nu\rho}
	+ \frac{1}{8} (\partial_\mu \tensor{\metft{h}}{^\nu_\nu})^2
	- \partial_\mu \overline{\metft{c}}_\nu \, \partial^\mu \metft{c}^\nu
	\label{eq:qg_mot_Lag}
\end{align}
where $\metft{c}$ is the ghost field. One can read from \eqref{eq:qg_mot_Lag} the propagator for the graviton and the ghost. The graviton $3$-vertex and $4$-vertex are given respectively by the expansion of \eqref{eq:qg_mot_Lag} to third and fourth order in $\metft{h}$ respectively. The \namefont{Feynman} rules are quite tedious to obtain and we refer to \cite{Hamber_2009} for their expressions.

The one-loop analysis in dimensional regularization shows that the needed counterterm for pure gravity writes
\begin{align}
	\frac{\sqrt{\metft{g}}}{8 \pi^2 (d+1-4)} \left( \frac{1}{120} R^2 + \frac{7}{20} R_{\mu\nu} R^{\mu\nu} \right),
	\label{eq:qg_mot_cc}
\end{align}
where $R$ is the \namefont{Ricci} scalar and $R_{\mu\nu}$ the \namefont{Ricci} tensor. Therefore, in $d+1 = 4$ dimensions, \gR is not perturbatively renormalisable, at least in the usual sense. One should note that adding minimally coupled matter to the action does not radically change the counterterm \eqref{eq:qg_mot_cc}. The one-loop divergence can be removed either on-shell (\ie when $R_{\mu\nu} = 0$), or by a field redefinition. However, the two-loop analysis shows that this magical behaviour does not occur for higher loops.

\paragraph{}
The later analysis could have been hinted by a dimensional analysis. Indeed, if one performs a power counting, the degree of divergence of a \namefont{Feynman} diagram is given by
\begin{align}
	(d+1) L + 2 V - 2 I
	\label{eq:qg_mot_ddp}
\end{align}
where $L$ is the number of loop in the diagram, $V$ the number of vertices and $I$ the number of internal lines. The equation \eqref{eq:qg_mot_dd} stems from the fact that each loop involves an momentum integration $\td^{d+1}p$, each vertex goes like $p^2$ and each internal line involves the propagator that goes as $p^{-2}$. For any diagram, one has the relation that $L = 1 + I - V$ which comes from a topological constraint on the graphs. Indeed, if one wants to have a single loop which goes through exactly $V$ vertices, one need $V$ internal lines. Each extra internal lines generates another loop, or needs another vertex not to form a loop. Therefore, there is a balance between $L - 1$ and $I - V$.

Merging this topological constraint with \eqref{eq:qg_mot_ddp} imposes that the divergence degree of a diagram is
\begin{align}
	(d - 1) L + 2.
	\label{eq:qg_mot_dd}
\end{align} 
One should note that the latter number only depends on the number of loops $L$ and that for dimensions $d+1 > 2$, the divergence degree is growing with the number of loops. The higher the loop order, the worse the divergence. Therefore, \gR is not perturbatively renormalisable.

One should note that the loop prefactor ($d-1$) in \eqref{eq:qg_mot_dd} is linked to the \namefont{Newton}'s constant dimension ($1-d$ length dimension).

\paragraph{}
The main pool of \qG theories have tried to tackle the previous renormalisability problem by different ways. Concerning \nCG, some authors are hoping that the \dpt would act as a ultra-violet cut-off in the spirit of \namefont{Snyder}. However, the renormalisability (or sometimes finiteness) of a field theory is far more tricky on \qST{s}, even at the level of a scalar field theory, as exemplified by the \UVIR. The study of \nCFT[ies] have some way to go before tackling the quantisation of a full noncommutative gravity theory.

Still, the main aspect we want to put forward in this manuscript is that, even if \nCG cannot give rise to a full theory of \qG straightforwardly, it has been and still is an essential tool for \qG phenomenology. In other words, the physics of \qST{s} could pave the way for early \qG tests. In the following \secref{qg_ph}, we discuss recent results in \qG phenomenology and the constraint that one already has on \qST{s}.

\section{Phenomenological considerations}
\label{sec:qg_ph}

\subsection{Phenomenology of quantum gravity}
\label{subsec:qg_ph_qg}
At the time this manuscript is written, there is no probe of any \qG effects, and even beyond \gR ones\footnote{
	The flourishing phenomenology of modified gravity goes far beyond the scope of this manuscript.
}. The usual cause pointed out for this experimental loophole is that energy scales involved are simply too high, the \Pmast being of order $\Pmas \sim 10^{19} \text{ GeV}$. However, the phenomenology of \qG has already reached such energies by considering highly energetic astrophysical sources and possible effects that got amplified within cosmological distances.

We give here the main trends and results of \qG phenomenology. Note that one can find many more details and an extensive set of references in the recent review of the COST Action CA18108 \enquote{Quantum Gravity Phenomenology in the multi-messenger approach} \cite{Addazi_2022}. The collaboration gathered the most recent results of \qG tests in their online catalogue \cite{COST_2021}.

\paragraph{}
The supposed \qG effect that gathering the most hope for detection is the modification of particle dynamics in vacuum, called \mDR{s}. 
\begin{Emph}{Modified dispersion relations}
The modified kinematics is mostly thought to be energy dependant so that it writes
\begin{align}
	E
	&= \vert \vec{p} \vert \, \left( 1 + \epsilon \frac{n+1}{2} \frac{E^n}{E_{\text{QG}}^n} \right)
	\label{eq:qgph_mdr}
\end{align}
where $E$ is the energy of the particle and $\vec{p}$ its momentum, $n$ is the order of deformation, $E_{\text{QG}}$ is the energy scale at which the modification is relevant and $\epsilon = \pm 1$ parametrises if the modification is superluminal ($+1$) or subluminal ($-1$). Note that a \mDR can also trigger a modified interaction dynamics for particles. We refer to \cite{Addazi_2022} for more details on this part.
\end{Emph}

As propagation is modified, two photons with two different energies would travel at different velocities\footnote{
	The term \enquote{velocity} here is pictorial but may be misleading. Some theoretical frameworks breaks the speed-of-light limit for particle so that the velocities of two photons are actually different. However, some other framework may require that the two photons travel at the same speed but in different (energy dependant here) backgrounds/\sT{s} such that the arrival time of two simultaneously emitted photon is different. 
}.
One can then think of a thought experiment of an astrophysical source emitting simultaneously two photons with different energies toward Earth. Since they travel differently, there will be a time delay in the detection of the two photons. The higher the energy discrepancy and the higher the distance to the source, the bigger the time delay. Note that one can perform the same thought experiment and change the cosmic messenger. In other words, one can consider gravitational waves or neutrinos instead of photons.

This has led to the search of time delays in highly energetic astrophysical events, mainly gamma ray bursts, active galactic nuclei and pulsars. For some events, the lower bound for $E_{\text{QG}}$ for a first order \mDR (\ie $n=1$) has even reached the \namefont{Planck} scale. However, one has to handle a high systematic error coming from different sources: the expansion of the Universe, the \enquote{simultaneity} in time and energy measurements, the mechanism of emission at the source, scattering with the interstellar medium, \etc The latter errors are thought to be reduced by a statistical treatment, which would need more events.

\paragraph{}
There are other, possibly testable, effects that a \mDR triggers. First, if the messenger carries a polarisation, then a birefringence effect could occur. Specifically, the polarisation of light could be rotated of some tiny amount accumulating with distance, and gravitational wave could have different speed for $+$ and $\times$ polarisations. Second, there could be modifications in the redshift due to the Universe expansion. Third, the gravitational lensing would become frequency dependent so that a black hole shadow would be affected by some rainbow effect. Fourth, the \namefont{Greisen-Zatsepin-Kuz'min} cut-off of ultra-high-energy cosmic rays, due to their interaction with the cosmic microwave background, would be smaller as the cosmic microwave background could interact with less energetic photons. Fifth, the oscillation of neutrinos would be affected by a modification of propagation. One can find an extended list of effects and discussions in \cite{Addazi_2022}.

\paragraph{}
The \mDR has been experienced in several \qG phenomenological frameworks, the most studied ones being \dSR and \LIV. 

\begin{Emph}{Doubly special relativity}
	There is an apparent contradiction between some \qG framework claiming an observer-independent smallest length (of order of the \Plent) and the length dilatation and contraction of special relativity. The \dSR framework aims at solving this issue by considering both an observer-independent upper limit for the speed (the speed of light) and an observer-independent lower limit for lengths, called $\ell_{\text{DSR}}$. This implies that \enquote{inertial} observers are no more related by a \namefont{Poincar\'{e}} transformation but rather a deformation of it. \DSR models are thought to describe effectively some flat \sT limit of \qG.
\end{Emph}
A \dSR theory imposes a deformed momentum space via a deformation of the composition law of the momenta. Therefore, it may imply a \mDR for the photons, where the energy scale $E_{\text{QG}}$ is to be linked with $\ell^{-1}_{\text{DSR}}$.

\begin{Emph}{\namefont{Lorentz} invariance violation}
	\LIV is a generic term for all effective theories that breaks \namefont{Lorentz} invariance. It may be done by deforming the \namefont{Lorentz} symmetry, as for example in \dSR, or by imposing a preferred direction for field propagation. Indeed, one can add in the fermionic action a term of the form
	\begin{align}
		\iCpx u_\mu u_\nu \overline{\psi} \gamma^\mu \nabla^\nu \psi
	\end{align}	
	where $\psi$ is the fermionic field, $\nabla$ the covariant derivative and $u$ the preferred direction of \sT encoded as a vector field. While some authors have considered $u$ to be non-dynamical, it appears that nice results may be derived by a dynamical $u$ coupled to gravity and having a non-zero expectation value.
\end{Emph}
\LIV could generate time delays via non-covariant dispersion relations. The mixing of dynamical tensors in the action leads to new gravitational wave polarisations. Finally, birefringence can appear for CPT-odd operators (in the context of \CPTV).

The \LIV framework is vast and has already been tightly constrained by (minimal) \SM extensions, up to dimension five operators. Still, the constrains remains far below the \namefont{Planck} scale.

\paragraph{}
Other phenomenological settings arises in the context of \qG phenomenology as detailed below.

\paragraph{}
Similarly to the Galilean time becoming relative in special relativity, it appears that locality becomes relative in \dSR theories. More explicitly, if a process is local for a (close) observer, then for another observer (obtained after a \dSR transformation) the process may be non-local. The notion of locality of, say, an interaction of particle becomes observer dependent. This is called the relative locality principle. 

The notion of \namefont{Born} geometry tries to settle a \enquote{covariant relative locality}. In the same spirit that space or time is observer dependant in special relativity but space-time is not, \namefont{Born} geometries tries to identify the global observer independent concept of causality that would lead to relative locality.

\paragraph{}
In \dSR, the momentum space is curved due to a modification of the momentum addition law. It can be shown to be of constant curvature corresponding to $\ell_{\text{DSR}}^{-1}$. The generalisation of \dSR to generally covariant framework has been to consider curved momentum spaces. One can ask the question of what becomes of the usual dispersion relation $p^2 + m^2 =0$, since the latter is coordinate dependent and therefore not (momentum) covariant. In order to solve this issue, one introduces a covariant Hamiltonian for which the level sets $H(x,p) = \text{constant}$ correspond to dispersion relations. The Hamiltonian is defined on a curved phase space $(x,p)$, which is defined by a metric $\metft{g}^{\mu\nu}(x,p)$.

\paragraph{}
As already formalised in \subsecref{gnc_phys_qg}, the \gUP studies the effects of a momentum dependent \HUP. The \HUP does not impose one to have a minimal length, since one can have $\Delta x$ as small as wanted if $\Delta p$ is big enough. Therefore, even in its simplest form
\begin{align}
	\Delta x \, \Delta p
	\geqslant \frac{\hbar}{2} \, \big( 1 + \ell_{\text{GUP}}^2 \, p^2 \big),
\end{align}
the \gUP imposes a smallest length $\ell_{\text{GUP}}$. Constraints coming from gravitational wave detection, neutrino physics and cosmological studies affects the \gUP parameters but has not reached \namefont{Planck} scale yet. Besides, the \gUP imposes the black hole evaporation to stop at $\ell_{\text{GUP}}$ and so to leave a remnant behind. The latter consideration lead to a possible solution in the black hole information paradox discussed in \subsecref{qg_mot_mis}, as the \enquote{lost} information could be carried by this remnant.

\subsection{The phenomenology of quantum space-times}
\label{subsec:qg_ph_qst}
\paragraph{}
Most of the phenomenological frameworks depicted above can emerge naturally in the context of \nCG. We detail below how these frameworks appear on some example, as well as some constraints it imposes on the \dpt.

The \kPt algebra \eqref{eq:kM_kP_kP} realises a \dSR with $\ell_{\text{DSR}} = \dpkM^{-1}$. In other words, if one interpret the \kPt generators as frame transformations, then those transformations conserves both an observer independent speed and an observer independent length $\dpkM^{-1}$.

\LIV occurs via deformations of the \Pogt. Therefore, a \qST having a deformed \namefont{Poincar\'{e}} symmetry may break the usual \namefont{Lorentz} invariance. Note that all forms of \LIV are not necessarily linked to a symmetry deformation, since one could simply consider that the breaking of \namefont{Lorentz} invariance does not give rise to another symmetry.

The \rMt space (discussed in \secref{kM_rM}) was first considered for its close relation to relative locality \cite{Amelino-Camelia_2011}. The latter also has a deformed version of \Palgt as its space of symmetries and, therefore, can be also linked with \dSR and \LIV.

Curved momentum space arises at least in \Lieat-type noncommutative space-times, as developed in \secref{ncft_p4}. Considering the momentum space to be the exponentiation of the coordinate space, a \Lieat-type noncommutativity imposes a deformed composition law of momenta through the \BCH. From the knowledge of this deformed law, one can compute a connection, and its associated curvature, of the momentum space.

Finally, \gUP are present in some \qST{s}, like the \namefont{Snyder} space-time.

Note that \CPTV is thought to appear in \qST{s}. However, there are no unambiguous definitions of the discrete symmetries in a noncommutative context.

\paragraph{}
In view of all the physical considerations associated with \qST{s}, one could ask what kind of \mDR are been generated. However, when considering deformed symmetries, as in the case of \kMt, the derived dispersion relation is highly dependent of the coordinate choice in the \Hfat of symmetries. For example, if one considers that the generalised derivative $\partial_\mu$ of the kinetic operator are $P_\mu$, or $X_\mu$ (defined in \subsecref{kM_gt_td}), one ends up with two different dispersion relations which writes
\begin{align}
	P_\mu: \quad
	E^2 &= \vert \vec{p} \vert^2, &
	X_\mu: \quad
	E^2 = \vert \vec{p} \vert^2 + \frac{1}{\dpkM} E^3 + \BigO{\frac{1}{\dpkM^2}}.
	\label{eq:qg_ph_drkM}
\end{align}
Other expressions have been derived throughout the literature. If an action with a kinetic term involving the $X_\mu$'s is considered, like in \eqref{eq:kM_gt_YMa}, the phenomenological constraints on time delays for high-energetic astrophysical photons imposes that $\kappa \gtrsim 10^{17}-10^{19} \text{ GeV}$.

Choosing the appropriate kinetic term has been discussed extensively. It can be directly linked to the question of momentum coordinate invariance of \secref{ncft_p4}.

\paragraph{}
Other physical constraints can be put on the \dpt{s}, see  \cite{Hersent_2023a} for a review. When considering a \nCYM \eqref{eq:ncft_db_YMa}, the expansion of the action with respect to the \dpt generates contributions from higher dimensional operator, which have already been studied in the context of \SM extensions. The strongest constraints comes from \namefont{Lorentz} invariance violating operator which imposes $\kbar \gtrsim 10^{17} \text{ GeV}$. The \Moy space has been intensively studied in this sense, so that $\kbar = \sqrt{\dpMoy}$ in this context.

Another consideration one can have on the $5$-dimensional model of $\dpkM$-deformed \namefont{Yang-Mills} theory developed in \secref{kM_gt}, is to relate the \dpt $\dpkM$ with the extra dimension size. If one assumes a simple compactification scheme on the simple orbifold $\mathbb{S}^1/\Int_2$, then constraints from the Large Hadron Collider gives $\dpkM \gtrsim 10^{13} \text{ GeV}$.

\paragraph{}
Other recent proposal for \nCG tests have been proposed, see for example \cite{Kalita_2022, Napolitano_2024, Liu_2023, Iorio_2023, Kozak_2023}.

\section{Toy model of noncommutative gravity}
\label{sec:qg_ncg}
\paragraph{}
There is a flourishing literature of formulations of gravity on \qST{s}. For a review see \cite{Hersent_2023a} and for an extensive set of references see \cite{Hersent_2022c}. Most of the studies focus on the noncommutative black hole, since this object is so compact that \qG fluctuations are thought to critically change our understanding of black holes. For an early review of $\dpMoy$-deformed black holes see \cite{Nicolini_2009}.

\paragraph{}
The question of what becomes of the metric is a major issue in noncommutative gravity. There have been many proposals for metric formulations on \qST{s} (bilinear or sesquilinear, symmetric or hermitian, invertible or nondegenerate, real or not, on derivations or on forms, \etc) and none has reached consensus. 

A noncommutative metric faces other difficulties when one considers its commutative limit. Most of the noncommutative settings work with complex entries so that a noncommutative metric is necessarily complex. If no reality constraints is imposed, the corresponding commutative metric would also be complex. Furthermore, the noncommutativity of the \spdtt $\star$ generally imposes that the metric is not symmetric, which has also been shown to be pathological in a commutative setting.

Finally, it is not known if the \enquote{noncommutative metric} is the relevant object to study gravity on \qST{s}. For a classical \sT, the data of the metric fully determines the latter \sT and so the gravity behaviour. It is further obtained by solving the \namefont{Einstein} equations. It is not know if such a property extends to the \qST{s} on which many metrics could be defined \aprio. Besides, if noncommutative metrics would be in one to one correspondence with \qST{s}, then one would further need an analogue of the \namefont{Einstein} equation, which up to date has not been derived.

\paragraph{}
We detail here the construction of the toy model of noncommutative gravity introduced in \cite{Hersent_2023d}. It consists of the first of a series and tries to put some mathematical grounds on the noncommutative analogue of a partition of unity. The main idea is to consider a \sT on which the \Minkt tangent space is changed to be the \kMt \enquote{tangent space}. If one imposes that \kMt is a local tangent space, then one needs a notion of partition of unity to glue all the local pieces together and form global objects. In the following, the \namefont{Einstein} summation convention applies for \sT indices (corresponding to the end of the Greek alphabet $\mu$, $\nu$, $\rho$, $\sigma$, $\tau$) and does not for the covering indices (corresponding to the beginning of the Greek alphabet $\alpha$, $\beta$). In the latter case, the sum is always made explicit.

The previous idea takes its root in the commutative setting for which the \Minkt \sT arise as a local tangent space. Indeed, if one is given a patch of local open sets $\{U_\alpha\}_\alpha$ covering the \sT $\manft{M}$, then 
\begin{align}
	TU_\alpha \simeq U_\alpha \times \Mink,
	\label{eq:qg_ncg_ctsp}
\end{align}
where $TU_\alpha$ is the tangent space of $U_\alpha$. The relation \eqref{eq:qg_ncg_ctsp} gives rise to the vielbein formalism, where one can compare different frames of the tangent bundle at any point of $U_\alpha$ by a \namefont{Poincar\'{e}} transformation, meaning a transformation of $\Mink$. The property \eqref{eq:qg_ncg_ctsp} is referred to as the local triviality of $\manft{M}$ in the mathematical literature, stemming from the fact that locally the tangent bundle is only layers of copies of \Minkt. Note that in the physical literature, the need of an open set $U_\alpha$ where local coordinates can be defined is frequently not mentioned. One usually exports objects defined on each $U_\alpha$'s to the full \sT $\manft{M}$ by the use of a partition of unity $\chi$, also generally omitted in the physics literature.

Several questions now arise when trying to implement this framework. What is the analogue $\algft{A}_\alpha$ of the local set $U_\alpha$ in the \nCG? How can one implement that $\kM$ is the \enquote{tangent space} of $\algft{A}_\alpha$, similarly to \eqref{eq:qg_ncg_ctsp}? How does one glue elements of $\algft{A}_\alpha$ to the full $\algft{A}$, that is how can one define a noncommutative partition of unity? Possible answers to those questions are the specific topic of this \secref{qg_ncg}.

\paragraph{}
The question of generalising open covers $\{U_\alpha\}_\alpha$ of a \sT $\manft{M}$ to the noncommutative setting was addressed in \cite{Calow_2000}, through the notion of ideals and covering of algebras. If one considers $\algft{A}$ to be the set of smooth functions $\funcs(\manft{M})$, then $\algft{A}_\alpha$ would correspond to $\funcs(U_\alpha)$. The two are related via the restriction on $U_\alpha$, noted $\vert_\alpha: \funcs(\manft{M}) \to \funcs(U_\alpha)$.

The kernel of the latter restriction $\Ker(\vert_\alpha)$ corresponds to the set of (smooth) functions that vanish on $U_\alpha$. It is an ideal of $\funcs(\manft{M})$ since a product of two functions, one from $\Ker(\vert_\alpha)$ the other from $\funcs(\manft{M})$, vanishes necessarily on $U_\alpha$. Furthermore, one has
\begin{align}
	\funcs(U_\alpha) 
	&= \funcs(\manft{M}) / \Ker(\vert_\alpha),
	\label{eq:qg_ncg_cqa}
\end{align}
in terms of quotient of ideals (see \defref{ha_as_qa}). The latter result is detailed in \exref{ha_as_lfqa}.

Furthermore, the property that $\{U_\alpha\}_\alpha$ forms a covering of $\manft{M}$ translates, for the restrictions, to
\begin{align}
	\bigcap_\alpha \Ker(\vert_\alpha) 
	= \{0\}.
	\label{eq:qg_ncg_cci}
\end{align}
Indeed, if a (smooth) function is zero on every $U_\alpha$, it is zero on the all $\manft{M}$ as the $U_\alpha$'s cover all $\manft{M}$.

Finally, concerning two open covers $\{U_\alpha\}_\alpha$ and $\{\tilde{U}_{\tilde{\beta}}\}_{\tilde{\beta}}$, the link between $\funcs(U_\alpha)$ and $\funcs(U_\alpha \cap \tilde{U}_{\tilde{\beta}})$ can also follow from \eqref{eq:qg_ncg_cqa} considering the restriction $\vert_{\tilde{\beta}}$ (and similarly for $\funcs(\tilde{U}_{\tilde{\beta}})$ with $\vert_\alpha$). On the other hand, one can go directly from $\funcs(\manft{M})$ to $\funcs(U_\alpha \cap \tilde{U}_{\tilde{\beta}})$ through
\begin{align}
	\funcs(U_\alpha \cap \tilde{U}_{\tilde{\beta}})
	&= \funcs(\manft{M}) / \big(\Ker(\vert_\alpha) + \Ker(\vert_{\tilde{\beta}}) \big),
	\label{eq:qg_ncg_cii}
\end{align}
where $+$ denotes here the smallest closed $*$-ideal containing both elements. The relation \eqref{eq:qg_ncg_cii} holds for an intersection of two open sets, but it can be straightforwardly generalised to an intersection involving an arbitrary number of open sets.

The way one generalises the previous aspects to the noncommutative setting goes as follows.
\begin{Emph}{Covering of algebras}
	Let $\algft{A}$ be a \Salg, possibly obtained by \dq of a classical \sT. We define a family of $*$-ideals $\{J_\alpha\}_\alpha$ to be a covering of $\algft{A}$ if it satisfies
	\begin{align}
		\bigcap_\alpha J_\alpha
		&= \{0\}.
		\label{eq:qg_ncg_ci}
	\end{align}
	One can directly read that this is the noncommutative analogue of \eqref{eq:qg_ncg_cci}. Having defined such a covering, one considers the \enquote{local} algebra $\algft{A}_\alpha$ to be the quotient of the \enquote{global} algebra $\algft{A}$ by the ideal $J_\alpha$, similarly to \eqref{eq:qg_ncg_cqa}, \ie
	\begin{align}
		\algft{A}_\alpha
		&= \algft{A} / J_\alpha.
		\label{eq:qg_ncg_qa}
	\end{align}
	Finally, the quotient of $\algft{A}$ by several (here two) coverings $\{J_\alpha\}_\alpha$ and $\{\tilde{J}_{\tilde{\beta}}\}_{\tilde{\beta}}$ is given by the generalisation of \eqref{eq:qg_ncg_cii}
	\begin{align}
		\algft{A}_{\alpha\tilde{\beta}}
		&= \algft{A} / \big( J_\alpha + \tilde{J}_{\tilde{\beta}} \big).
		\label{eq:qg_ncg_ii}
	\end{align}
	We say that a set $\{\algft{A}_\alpha\}_\alpha$ satisfying \eqref{eq:qg_ncg_ci} and \eqref{eq:qg_ncg_qa} is a covering of algebras, the ideals $J_\alpha$ being implied.
\end{Emph}

It is important to note that any $\algft{A}_\alpha$ can be made a \Salg via the involution $[f]_\alpha^{\dagger_\alpha} = [f^\dagger]_\alpha$, where $[f]_\alpha$ denotes the representative of the equivalent class of $f \in \algft{A}$. In the previous equation ${}^{\dagger_\alpha}$ denotes the involution of $\algft{A}_\alpha$ and ${}^\dagger$ the involution on $\algft{A}$. Besides, we note the canonical projection $\pi_\alpha : \algft{A} \to \algft{A}_\alpha$, such that $\pi_\alpha(f) = [f]_\alpha$. The latter projection is surjective by definition. By considering the projection of $\algft{A}$ on $\algft{A}_\alpha$ and then on $\algft{A}_{\alpha\tilde{\beta}}$, or directly from $\algft{A}$ to $\algft{A}_{\alpha\tilde{\beta}}$, one can show that the diagram
\begin{center}
	\noindent%
	\begin{tikzpicture}
	%Nodes
	\node (Aa)  at ( 0, 1) {$\algft{A}_\alpha$};
	\node (Ab)  at ( 0,-1) {$\algft{A}_{\tilde{\beta}}$};
	\node (A)   at (-2, 0) {${}_{\phantom{0}} \algft{A}$};
	\node (Aab) at ( 2, 0) {$\algft{A}_{\alpha\tilde{\beta}}$};
	%Lines
	\draw[thick, ->] (A.north east) -- 
	    node[anchor=south east]{$\pi_\alpha$} (Aa.west);
	\draw[thick, ->] (A.south east) -- 
	    node[anchor= north east]{$\pi_{\tilde{\beta}}$} (Ab.west);
	\draw[thick, ->] (A.east) --
	    node[anchor=south]{$\pi_{\alpha\tilde{\beta}}$} (Aab.west);
	\draw[thick, ->] (Aa.east) -- 
	    node[anchor=south west]{$\pi^\alpha_{\tilde{\beta}}$} (Aab.north west);
       \draw[thick, ->] (Ab.east) -- 
	    node[anchor=north west]{$\pi_\alpha^{\tilde{\beta}}$} (Aab.south west);
\end{tikzpicture}%
\end{center}
commutes, so that $\pi^\alpha_{\tilde{\beta}} \circ \pi_\alpha = \pi_\alpha^{\tilde{\beta}} \circ \pi_{\tilde{\beta}} = \pi_{\alpha\tilde{\beta}}$.

\paragraph{}
We now turn to the implementation of $\kM$ as the local tangent space. The idea we develop here shares some similarities with the quantum fiber bundle model developed by \namefont{Brzezi\'{n}ski} and \namefont{Majid} \cite{Brzezinski_1993}. If one realises that the coordinates on \kMt has trivial coproduct $\Delta(x_\mu) = x_\mu \otimes 1 + 1 \otimes x_\mu$ (see \eqref{eq:kM_kP_kM_coalg}), then the $x_\mu$'s behaves as derivations on any $\kM$-module algebra. Explicitly, if one assumes that $(\algft{A}_\alpha, \star_\alpha)$ is a $\kM$-module algebra, the coproduct yields
\begin{align}
	x_\mu \actl (f \star_\alpha g)
	&= (x_\mu \actl f) \star_\alpha g + f \star_\alpha (x_\mu \actl g),
	\label{eq:qg_ncg_kMder}
\end{align} 
for any $f, g \in \algft{A}_\alpha$. Note that in the case of \Salg{s}, the relation 
\begin{align}
	(x \actl f)^{\dagger_\alpha}
	&= S(x)^\dagger \actl f^{\dagger_\alpha}
	\label{eq:qg_ncg_sact}
\end{align}
holds, for any $x \in \kM$ and $f \in \algft{A}_\alpha$, where $S$ is the antipode of $\kM$, and ${}^\dagger$ its involution. As in \cite{Brzezinski_1993}, we require that
\begin{align}
	\algft{A}_\alpha \text{ is a }
	\kTran[1,d] \text{-comodule algebra}
	\label{eq:qg_ncg_kTcom}
\end{align}
(see \defref{ha_rt_comodalg}), with coaction $\coactr_\alpha : \algft{A}_\alpha \to \algft{A}_\alpha \otimes \kTran[1,d]$. From \eqref{eq:qg_ncg_kTcom}, one can derive that $\algft{A}_\alpha$ is a $\kM$-module algebra with the action
\begin{align}
	x \actl f
	&= \sum \langle f_{(1)}, x \rangle f_{(0)}
	\label{eq:qg_ncg_dact}
\end{align}
for any $x \in \kM$ and $f \in \algft{A}_\alpha$, where we used the \namefont{Sweedler} notations $\coactr_\alpha = \sum f_{(0)} \otimes f_{(1)}$ and where $\langle \cdot, \cdot \rangle : \kTran[1,d] \times \kM \to \Cpx$ denotes the dual pairing between $\kTran[1,d]$ and $\kM$.

Considering the set
\begin{align}
	\mathfrak{D}_\dpkM
	&= \Span \big( \{x_\mu\}_{\mu = 0, \ldots, d} \big),
	\label{eq:qg_ncg_Dk}
\end{align}
consisting of the linear span of the $x_\mu$'s, one has from \eqref{eq:qg_ncg_kMder}, stemming from \eqref{eq:qg_ncg_kTcom} and \eqref{eq:qg_ncg_dact}, that $\mathfrak{D}_\dpkM$ is a sub-\Lieat of $\Der(\algft{A}_\alpha)$. In the spirit of the restricted \dbdc, we consider the set of restricted derivations (see discussion of \secref{ncft_Moy})
\begin{align}
	\Der_R(\algft{A}_\alpha)
	&= \Cen{\algft{A}_\alpha} \otimes \mathfrak{D}_\dpkM
	\label{eq:qg_ncg_rder}
\end{align}
on $\algft{A}_\alpha$. It can be shown that $\Der_R(\algft{A}_\alpha)$ is a sub-\Lieat of $\Der(\algft{A}_\alpha)$ and a $\Cen{\algft{A}_\alpha}$-module. One can further build a (restricted) \dc $(\Omega^\bullet_R(\algft{A}_\alpha), \td_\alpha, \wedge_\alpha)$ by following the steps of \secref{gnc_dbdc}. It can also be shown that $\Omega^1_R(\algft{A}_\alpha) \simeq \algft{A}_\alpha \otimes \dual{\mathfrak{D}}_\dpkM$. Here $\dual{\mathfrak{D}}_\dpkM$ stands as the dual of $\mathfrak{D}_\dpkM$ and corresponds to the linear span of the deformed translations $P^\mu \in \kTran[1,d]$.

The situation above corresponds quite closely to the commutative setting \eqref{eq:qg_ncg_ctsp}. Indeed, the derivations correspond to the sections of the tangent bundle, that is $\Der(\funcs(U_\alpha)) = \Gamma(U_\alpha)$, but from \eqref{eq:qg_ncg_ctsp}, one has
\begin{align}
	\Gamma(U_\alpha) \simeq \funcs(U_\alpha) \otimes \Mink
	\label{eq:qg_ncg_cder}
\end{align}
which exactly match \eqref{eq:qg_ncg_rder} at the commutative limit. The previous statement is usually written, in the physics literature, as $\zeta = \zeta^\mu \partial_\mu$ for any \vf $\zeta$ (the tensor product being implied). Following these notation, we denote below $\zeta = \zeta^\mu x_\mu$, for any $\zeta \in \Der_R(\algft{A}_\alpha)$, where $\zeta^\mu \in \Cen{\algft{A}_\alpha}$. Correspondingly, the set of one-forms writes $\Omega^1(U_\alpha) \simeq \funcs(U_\alpha) \otimes \kM$, which is the commutative limit of $\Omega^1_R(\algft{A}_\alpha) \simeq \funcs(U_\alpha) \otimes \dual{\mathfrak{D}}_\dpkM$.

\paragraph{}
The latter construction of local derivations, because of its close relation to the commutative setting, has some similarities with the \gR setting for gravity. Indeed, if one defines a connection $\nabla$, as in \secref{ncft_db}, on the module $\modft{X} = \Der_R(\algft{A}_\alpha)$, then it is fully determined by its components on $\mathfrak{D}_\dpkM$, that is
\begin{align}
	\nabla_{x_\mu}(x_\nu)
	&= \Gamma_{\mu\nu}^\rho x_\rho
	\label{eq:qg_ncg_con}
\end{align}
with $\Gamma_{\mu\nu}^\rho \in \Cen{\algft{A}_\alpha}$. If one requires that $\nabla$ is hermitian, that is $\nabla_{\zeta^{\dagger_\alpha}}(\xi^{\dagger_\alpha}) = (\nabla_\zeta(\xi))^{\dagger_\alpha}$ for any $\zeta, \xi \in \Der_R(\algft{A}_\alpha)$, then one has
\begin{align}
	( \Gamma_{\mu\nu}^\rho )^{\dagger_\alpha}
	&= - \Gamma_{\mu\nu}^\rho
	\label{eq:qg_ncg_hcon}
\end{align}
by making use of \eqref{eq:kM_kP_kM_coalg} together with \eqref{eq:qg_ncg_sact}. Moreover, the associated curvature, defined in \eqref{eq:ncft_db_cur}, writes
\begin{align}
	\tensor{R}{_{\mu\nu\rho}^\sigma}
	&= ( x_\mu \actl \Gamma_{\nu\rho}^\sigma )
	- ( x_\nu \actl \Gamma_{\mu\rho}^\sigma )
	+ \Gamma_{\nu\rho}^{\tau} \Gamma_{\mu\tau}^\sigma
	- \Gamma_{\mu \rho}^\tau \Gamma_{\nu\tau}^\sigma
	- \tensor{\sC}{_{\mu\nu}^\tau} \Gamma_{\tau \rho}^\sigma 
	\label{eq:qg_ncg_cur}
\end{align}
where $\tensor{\sC}{_{\mu\nu}^\tau} = \frac{\iCpx}{\dpkM} (\delta_\mu^0 \delta_\nu^\tau - \delta_\nu^0 \delta_\mu^\tau)$ is the structure constant of the $\mathfrak{D}_\dpkM$ algebra, that one can read from \eqref{eq:kM_kP_kM_alg}. The expression \eqref{eq:qg_ncg_cur} is very akin to the usual expression for the \sT curvature.

\paragraph{}
Finally, we tackle the last question on how to define a coherent noncommutative partition of unity. It is quite instructive to start with the commutative definition of a partition of unity
\begin{Def}{Partition of unity}
	{qg_ncg_cpu}
	Let $\manft{M}$ be a \sT. A partition of unity on $\manft{M}$ is a set of functions $\{\chi_\alpha\}_\alpha$ satisfying
	\begin{clist}
		\item \label{it:cpu_f}
		$\chi_\alpha \in \funcs(\manft{M})$,
		\item \label{it:cpu_s}
		$\{\Supp(\chi_\alpha)\}_\alpha$ is locally finite,
		\item \label{it:cpu_p}
		$ \chi_\alpha \geqslant 0$,
		\item \label{it:cpu_u}
		$\sum_\alpha \chi_\alpha = 1$,
	\end{clist}
	where $\Supp$ denotes the support. The partition is further said to be subordinate to an open cover $\{U_\alpha\}_\alpha$ of $\manft{M}$, if for every $\beta$ there exists $\alpha$ such that $\Supp(\chi_\beta) \subset U_\alpha$. We say that the partition is adapted to the cover whenever $\beta = \alpha$. In this case, the cover and the partition of unity share the same set of indices.
\end{Def}

A partition of unity is manufactured to glue pieces of objects, defined only locally, to have a globally defined object. Requirement \ref{it:cpu_f} ensures that the partition of unity does not deteriorate the smoothness of the objects we work with. \ref{it:cpu_s} allows for $\chi_\alpha$ to be only non-zero inside $U_\alpha$ (in the case of an adapted partition) and thus makes $\chi_\alpha$ \enquote{select} the desired $U_\alpha$. Finally, \ref{it:cpu_p} together with \ref{it:cpu_u} ensure that no information is lost: gluing all the pieces together gives a coherent object. As an example, if one considers $f \in \funcs(\manft{M})$, then the identity
\begin{align}
	f = \sum_\alpha \chi_\alpha f\vert_\alpha
	\label{eq:qg_ncg_cpuf}
\end{align}
holds. The different local pieces $f\vert_\alpha$ are reconstructed to be $f$ via the partition of unity $\chi_\alpha$. The previous statement \eqref{eq:qg_ncg_cpuf} is at the very basis of what we want to achieve with a noncommutative partition of unity, that is to define objects on the local algebras $\algft{A}_\alpha$ and to glue local pieces together to have a global object on $\algft{A}$. Note that the latter gluing should not depend on the choice of the cover or the partition of unity. The independence relies on the property that the product of a partition of unity is still a partition of unity. Therefore, we need to export this property to the noncommutative setting also. More explicitly, if $\{U_\alpha\}_\alpha$ and $\{\tilde{U}_{\tilde{\beta}}\}_{\tilde{\beta}}$ are open covers of $\manft{M}$ with adapted partition of unity $\{\chi_\alpha\}_\alpha$ and $\{\tilde{\chi}_{\tilde{\beta}}\}_{\tilde{\beta}}$ respectively, then $\{ \chi_\alpha \tilde{\chi}_{\tilde{\beta}} \}_{\alpha, \tilde{\beta}}$ is an adapted partition of unity for $\{ U_\alpha \cap \tilde{U}_{\tilde{\beta}} \}_{\alpha, \tilde{\beta}}$. 

We define the noncommutative partition of unity as
\begin{Def}{Noncommutative partition of unity}
	{qg_ncg_ncpu}
	Let $\algft{A}$ be a \Salg. A partition of unity on $\algft{A}$ is a set of elements $\{\chi_\alpha\}_\alpha$ satisfying
	\begin{clist}
		\item \label{it:ncpu_f}
		$\chi_\alpha \in \algft{A}$,
		\item \label{it:ncpu_s}
		$\{\Supp(\chi_\alpha)\}_\alpha$ is locally finite,
		\item \label{it:ncpu_p}
		$\chi_\alpha \geqslant 0$, \ie it exists $\varsigma_\alpha \in \algft{A}$ such that $\chi_\alpha = \varsigma_\alpha \star \varsigma_\alpha^{\dagger}$,
		\item \label{it:ncpu_u}
		For any $f \in \algft{A}$, $\sum_\alpha \chi_\alpha \star f = f$,
	\end{clist}
	where $\Supp(f) = \{ \varphi \in \Spch{\algft{A}}, \varphi(f) \neq 0\}$ and $\Spch{\algft{A}}$ is the space of characters of $\algft{A}$ (see \subsecref{oa_rt_cCsalg}). The partition of unity is further said to be subordinate to an algebra cover $\{\algft{A}_\alpha\}_\alpha$ of $\algft{A}$, if for every $\beta$ there exists $\alpha$ such that
	\begin{align}
		\Supp(\chi_\alpha) \subset \Ker(J_\beta) = \{ \varphi \in \Spch{\algft{A}}, \varphi(f) = 0 \ \forall f \in J_\beta \}
		\label{eq:qg_ncg_subcov}
	\end{align}
	We say that the partition is adapted to the cover whenever $\beta = \alpha$.
\end{Def}

Several comments are in order.

First, the \defref{qg_ncg_ncpu} perfectly match the \defref{qg_ncg_cpu} at the commutative limit. The latter statement mainly holds thanks to the \nameref{thm:oa_rt_cGN}, which gives a one to one correspondence between the space of characters and the set of points of the space. Note that the requirement \ref{it:ncpu_u} has been changed to match the case of a non-unital algebra. However, one could work with an approached unit or a unitalisation of the algebra and define \ref{it:ncpu_u} as in \defref{qg_ncg_cpu}.

The condition of subordinate partition \eqref{eq:qg_ncg_subcov} is akin to the one of \defref{qg_ncg_cpu} in the commutative case. Indeed, in this case, one has that $J_\beta = \Ker(\vert_\beta)$ so that $\Ker(J_\beta) = \{ x \in \manft{M}, f(x) = 0 \ \forall f \in \Ker(\vert_\beta) \} = U_\beta$.

Furthermore, one has to specify a topology in order to define the notion of local finiteness in \ref{it:ncpu_s}. Indeed, the local finiteness imposes that there are a finite number of indices $\alpha$ for which $\Supp(\chi_\alpha) \cap V \neq \emptyset$, given any neighbourhood $V$. Therefore, one needs to specify what are the neighbourhoods in $\Spch{\algft{A}}$. We take the latter to be given by the weak* topology.

One can show that given two partitions of unity $\{\chi_\alpha\}_\alpha$ and $\{\tilde{\chi}_{\tilde{\beta}} \}_{\tilde{\beta}}$, adapted to some covering of algebras $\{\algft{A}_\alpha\}_\alpha$ and $\{\tilde{\algft{A}}_{\tilde{\beta}} \}_{\tilde{\beta}}$ of $\algft{A}$, then $\{ \chi_\alpha \star \tilde{\chi}_{\tilde{\beta}} \}_{\alpha, \tilde{\beta}} $ is a partition of unity adapted to the covering $\{\algft{A}_{\alpha \tilde{\beta}}\}_{\alpha, \tilde{\beta}}$. This can be shown using that 
\begin{align}
	\chi_\alpha \star \tilde{\chi}_{\tilde{\beta}}
	&= \tilde{\chi}_{\tilde{\beta}} \star \chi_\alpha
	= \chi_\alpha \bullet \tilde{\chi}_{\tilde{\beta}}
	= \tilde{\chi}_{\tilde{\beta}} \bullet \chi_\alpha,
	\label{eq:qg_ncg_puprop}
\end{align}
where $\chi_\alpha \bullet f = \varsigma_\alpha \star f \star \varsigma_\alpha^{\dagger}$, for any $f \in \algft{A}$ and where $\varsigma_\alpha$ is defined by \ref{it:ncpu_p}.

Finally, one can show that the diagonal elements of the matrix basis on the \Moy space $\{f_{mm}\}_{m \in \NInt}$ (see \secref{ncft_Moy}), forms a partition of unity.

\paragraph{}
Having built the noncommutative partition of unity, we know how to export the local derivations and the local forms to define global objects. Thus, if one considers
\begin{align}
	\Der_R(\algft{A}) 
	&= \left\{ \sum_\alpha \chi_\alpha \star (\zeta_\alpha \circ \pi_\alpha), \ \zeta_\alpha \in \Der_R(\algft{A}_\alpha) \right\},
	\label{eq:qg_ncg_l2gder}
\end{align}
then it consists of a $\Cen{\algft{A}}$-submodule of $\Der(\algft{A})$. Using the explicit expression of $\Der_R(\algft{A}_\alpha)$ \eqref{eq:qg_ncg_rder}, one has that any $\zeta \in \Der_R(\algft{A})$ is of the form
\begin{align}
	\zeta
	&= \sum_\alpha (\chi_\alpha \star \zeta^\mu_\alpha) \, x_\mu^\alpha
	\label{eq:qg_ncg_l2gvf}
\end{align}
where $\zeta^\mu_\alpha \in \Cen{\algft{A}_\alpha}$ and $x_\mu^\alpha = x_\mu \circ \pi_\alpha$. The latter form \eqref{eq:qg_ncg_l2gvf} corresponds to the usual formula for \vf{s} on a differential manifold.

Similarly as above, one can define
\begin{align}
	\Omega^n_R(\algft{A})
	&= \left\{ \sum_\alpha \chi_\alpha \star \omega^\alpha \big( \pi_\alpha( \cdot \star \chi_\alpha), \ldots, \pi_\alpha( \cdot \star \chi_\alpha) \big), \ \omega^\alpha \in \Omega^n_R(\algft{A}_\alpha) \right\}
	\label{eq:qg_ncg_l2gform}
\end{align}
and show that it is a $\algft{A}$-submodule of $\Omega^n(\algft{A})$. This noncommutative analogue seems also to conserve the \enquote{local} property of forms. This follows from
\begin{align}
	\omega^\alpha \wedge_\alpha \eta^\alpha
	&= \pi_\alpha ( \omega \wedge \eta ), &
	(\td \omega)\vert_\alpha
	&= \td_\alpha \omega\vert_\alpha,
\end{align} 
where in the first equality $\omega^\alpha \in \Omega^n(\algft{A}_\alpha)$ (resp.~$\eta^\alpha \in \Omega^m(\algft{A}_\alpha)$) denotes a convenient projection of $\omega \in \Omega^n(\algft{A})$ (resp.~$\eta \in \Omega^m(\algft{A})$), and in the second equality $\omega\vert_\alpha \in \Omega^n(\algft{A}_\alpha)$ denotes an appropriate notion of restriction of $\omega \in \Omega^n(\algft{A})$. Using the explicit structure of $\Omega^1_R(\algft{A}_\alpha) \simeq \algft{A}_\alpha \otimes \dual{\mathfrak{D}}_\dpkM$, any $\omega \in \Omega^1_R(\algft{A})$ is of the form
\begin{align}
	\omega
	&= \sum_\alpha (\chi_\alpha \star \omega^\alpha_\mu) \, P^\mu_\alpha,
	\label{eq:qg_ncg_l2g1f}
\end{align}
where $\omega^\alpha_\mu \in \algft{A}_\alpha$ and $P^\mu_\alpha = P^\mu( \pi_\alpha (\cdot \star \chi_\alpha ))$. Again, the component form \eqref{eq:qg_ncg_l2g1f} corresponds to the usual formula for $1$-forms on a differential manifold.

\paragraph{}
The previous formalism offers a convenient generalisation of locally trivial tangent bundle with \kMt arising as the space of derivations. Furthermore, it brings a formalism close to the tensorial formalism on commutative manifolds. However, a gravity theory would need further an action functional to be defined. In order to define an action, one needs a notion of integral (or trace) that has not been defined in this context. The partition of unity could be of use to export an integral on a local algebra ($\algft{A}_\alpha$) to an integral on the global algebra ($\algft{A}$), as in the commutative case. One should then explore the impact of this local triviality on the theory defined on $\algft{A}_\alpha$.
\chapter*{Conclusion}
\markboth{Conclusion}{}
\addcontentsline{toc}{chapter}{Conclusion}

\paragraph{}
The physics of \qST{s} is very promising for the study of \qG phenomenology and has not yet delivered all its secrets. Most of the puzzles of \qST{s} arise when considering field theories and particle states. The possible alternatives or redefinitions of such objects could pave the way for new physics and maybe experimental tests. Even if \nCG is not the fitted formalism for a full theory of \qG, it remains one of the best framework that may describe \qG effects, at least in some regime.

\paragraph{}
As the \chapref{gnc} illustrates, \enquote{\nCG} is a generic word which designates intertwined mathematical frameworks that tries to quantise the geometry. The field and gauge formalisms described in \chapref{ncft} also convey the actual lack of a global framework or at least some criterion to distinguish between all the possible choices. Constructing noncommutative analogues can be performed by several ways and introduces a huge, and for now unconstrained, freedom to write \nCFT[ies].

\paragraph{}
Beyond the plural nature of \qST{s}, some common features arise. Indeed, in \secref{ncft_p4} the \UVIR phenomenon is tackled. Vaguely described in the literature as the appearance of mixed ultra-violet and infra-red divergences in field and gauge theories, an unambiguous definition \cite{Hersent_2024a} of the \UVIR is proposed in this manuscript. It consists of three propositions: the UV divergence of the planar diagrams, the IR singularity of the non-planar diagrams and the UV finiteness of the non-planar diagrams. It was shown that the two first points is equivalent to the divergence of the propagator, thus designated as a criterion for \UVIR. However, the third point has not been addressed since the non-planar diagrams involve a non-trivial conservation of momenta that mixes the external momenta with the internal momenta. The study of the UV finiteness of the non-planar diagrams requires both the behaviour of the propagator and the solution of this non-trivial momentum conservation. The previous computation works for any \Lieat-type noncommutative space-times and so consists of the first attempt to generalise \UVIR studies.

The path to the understanding of the mixing has still some way to go. First, it consists of a one-loop analysis of the \npf{2}. One could ask what become of the mixing to higher loop orders and for generic \npf{n}{s}. The presence of other new phenomenon for higher loops and higher correlation functions needs to be better characterised. The \UVIR has been experienced mostly for \Lieat-type noncommutative space-times, but may be generalised to other \qST{s}. The \Lieat-type allows one to define properly a momentum space and therefore to work with it. In the case of \Hfat formalism, one could also consider a momentum space through a \namefont{Hopf} duality with the \qST. This idea needs to be further explored. Moreover, the analysis of \secref{ncft_p4} is performed for scalar field theory, but the mixing has also been experienced in \nCYM[ies] on \Moy. Therefore, one could think of broadening this study to \gT[ies]. Eventually, the ordering ambiguity defining the momentum space may be of primordial importance in the study of \nCFT[ies]. Indeed, it is not clear whether physical predictions depend or not on the chosen ordering prescription, which is only present in the noncommutative setting. The question of choosing a \enquote{physical} ordering or working with an ordering invariant theory\footnote{
	The link between ordering invariance and momentum diffeomorphism invariance has only been made for the \kMt space-time. Whether this correspondence can be exported to other \qST{s} is still to be determined.
}
has not been settled and could have a dramatic impact on the physical interpretation of these theories.

\paragraph{}
The study of \gT[ies] on \qST{s} has been explored through several examples. The case of \kMt is discussed in \secref{kM_gt}. In order to solve the loss of cyclicity of the integral, this \gT relies on the notion of twisted derivations. Still, the gauge invariance of the \kPt-invariant \namefont{Yang-Mills}-like action imposes that the \sT dimension is $5$. This can be directly traced back to the fact that the deformed cyclicity of the integral involves the \sT dimension. The \npf{1} (tadpole) of the latter action is computed at one-loop order. It is shown to be both non-vanishing and gauge dependant, as experienced on other \qST{s}, like \Moy and $\Rcl$. The gauge dependence of the tadpole is very puzzling since the considered action is gauge invariant and since the gauge dependence is not present in the commutative theory. Note that the tadpole still vanishes at the commutative limit in any gauge.

The origin of this noncommutative gauge anomaly has not been traced back and could stem from different sources. First, it is not clear that the $A$ variable corresponds to the physical (or at least mathematically convenient) variable to encode the \enquote{noncommutative photon}. One of the promising alternative variable on \Moy was the so-called covariant coordinate $\mathcal{A}$. Alternatives to $A$ have not been explored outside \Moy. Second, the gauge anomaly could be due to an unsuitable quantisation procedure. The path followed in this manuscript consists of the usual \BRST method. The latter may need to be adapted to the noncommutative case for a consistent treatment.

\paragraph{}
In \secref{kM_c}, a toy model of causality on \kMt is constructed. It relies on the formulation of Lorentzian \st and establishes the causality relation of two (pure) states, as inspired by the commutative setting. Even if this model relies on a too small \namefont{Hilbert} space, an analogue of the speed-of-light limit has been derived. It corresponds to an inequality constraint between the expectation value of the time and space operators for causally related states.

This model consists of one of the first established causality on \qST{s} and has therefore room for improvements. One of the most phenomenologically relevant investigation consists of deriving this speed-of-light limit for more complex operators. This could lead to a full set of constraints.

\paragraph{}
Finally, a toy model for noncommutative gravity is derived in \secref{qg_ncg}. It aims at characterising the \qST{s} through \kMt which arises as the local tangent space. The notion of locality is here defined in the mathematical sense of open sets. Once objects are defined on the local algebras via the use of \kMt, they are glued together via a noncommutative partition of unity to define global objects. One can thus build the tensorial formalism on the global algebra by only making use of properties of \kMt.

This rather mathematical framework needs a physical follow up in which an action is implemented. One has first to define an integral (or trace) on the global algebra, which could express through the integral over \kMt. 

\appendix
\chapter{\namefont{Hopf} algebras and quantum groups}
\label{ch:ha}

\paragraph{}
There is no well-posed definition of what \qg{s} are. But it is commonly acknowledged that \Hfat{s} give a common structure to most of the quantum groups encountered. In this sense, \Hfat{s} seems to be an interesting mathematical structure that we focus on in this Appendix. More insights on \qg{s} are given in \secref{gnc_qg}, and we refer to \cite{Montgomery_1993, Majid_1995, Klimyk_1997} for more details on these algebraic structures.

\section{Algebraic structures}
\label{sec:ha_as}
\paragraph{}
We here introduce the algebraic structure needed in the context of \qg{s}. They are defined mainly as vector space over $\Cpx$, but one could also consider $\Real$ or other generic fields.

\subsection{Algebras}
\label{subsec:ha_as_alg}
\paragraph{}
In \nCG, the classical notion of \sT is replaced by the algebraic notion of algebra, as motivated by \exref{ha_as_algf} and the commutative \nameref{thm:oa_rt_cGN}.

\begin{Def}{Algebra}
	{ha_as_alg}
	An associative algebra is a vector space $\algft{A}$ over $\Cpx$ with a multiplicative intern law $\star: \algft{A} \otimes \algft{A} \to \algft{A}$, which satisfies
	\begin{subequations}
		\label{eq:ha_as_alg}
	\begin{align}
		f \star (g + h)
		&= f \star g + f \star h, &
		\text{(distributive over the sum)}&
		\label{eq:ha_as_alg_d}\\
		(z f) \star g 
		&= f \star (zg)
		=z (f \star g), &
		\text{(compatible with the scalar product)}&
		\label{eq:ha_as_alg_csp}\\
		f \star (g \star h)
		&= (f \star g) \star h, &
		\text{(associativity)}&
		\label{eq:ha_as_alg_ass}
	\end{align}
	for any $f,g,h \in \algft{A}$ and $z \in \Cpx$.
	
	The algebra $\algft{A}$ is said unital if it exists $1_\algft{A} \in \algft{A}$ such that
	\begin{align}
		f \star 1_\algft{A}
		&= 1_\algft{A} \star f
		= f.
		\label{eq:ha_as_alg_uni}
	\end{align}
	
	The algebra $\algft{A}$ is further said commutative if $\star$ satisfies
	\begin{align}
		f \star g
		&= g \star f.
		\label{eq:ha_as_alg_com}
	\end{align}
	\end{subequations}
\end{Def}

As straightforward examples, $(\Real, \cdot, 1)$ and $(\Cpx, \cdot, 1)$ are unital commutative algebras. Despite, this genuine remark, one can build an algebra out of any set $\manft{X}$ through the space of functions over $\manft{X}$ going into $\Cpx$, as detailed \exref{ha_as_algf}.

The most simple and known example of a noncommutative algebra is the space of $n \times n$ matrices $\Mat{n}{\Cpx}$ (with $n > 1$), equipped with the matrix product and a unit given by the identity matrix $\Matid{n}$. In the context of \nCG, the space of matrices was quite studied as it allows one to mimic a $\SUg{n}$ gauge theory. See the \nCYM of \secref{ncft_db}.

\begin{Ex}{Algebra of functions}
	{ha_as_algf}
	Consider $\manft{X}$ to be any set of any kind of objects. Now consider the space of functions $f: \manft{X} \to \Cpx$, noted $\dual{\manft{X}}$, called the dual of $\manft{X}$. From the algebra structure of $\Cpx$, one can endow $\dual{\manft{X}}$ with an algebra structure with product and sum given by
	\begin{align*}
		(zf + g)(x)
		&= z f(x) + g(x), &
		(f \cdot g)(x)
		&= f(x) g(x)
	\end{align*}
	for any $f, g \in \dual{\manft{X}}$, $x \in \manft{X}$ and $z \in \Cpx$. The product of $\dual{\manft{X}}$ is called the \enquote{commutative product} throughout the manuscript and corresponds to the product \eqref{eq:gnc_found_cpdt}.
\end{Ex} 
Note that the \exref{ha_as_algf} could be done also considering functions on $\Real$ instead of $\Cpx$.

\paragraph{}
One could express the \defref{ha_as_alg} by considering the maps $\star: \algft{A} \otimes \algft{A} \to \algft{A}$ and $1_\algft{A} : \Cpx \to \algft{A}$ and require that the following diagrams commute

\begin{center}
\noindent%
	\begin{tikzpicture}
		%Nodes
		\node (HHH) {$\algft{A} \otimes \algft{A} \otimes \algft{A}$};
		\node (HHu) [node distance=1.4cm, right=of HHH] {$\algft{A} \otimes \algft{A}$};
		\node (HHd) [below=of HHH] {$\algft{A} \otimes \algft{A}$};
		\node (H)   [below=of HHu] {$\algft{A} \!\!\!\!\!\phantom{\otimes}$};
		%Lines
		\draw[thick, <-] (HHu.west) -- 
		    node[anchor=south]{$\star \otimes \id$} (HHH.east);
		\draw[thick, <-] (HHd.north) -- 
		    node[anchor=west]{$\id \otimes \star$} (HHH.south);
		\draw[thick, <-] (H.north) --
		    node[anchor=west]{$\star$} (HHu.south);
		\draw[thick, <-]  (H.west) -- 
		    node[anchor=south]{$\star$} (HHd.east);
	\end{tikzpicture}%
	\hspace{2cm}
	\begin{tikzpicture}
	    %Nodes
		\node (HH) {$\algft{A} \otimes \algft{A}$};
		\node (HC) [node distance=1.2cm, right=of HH] {$\algft{A} \otimes \Cpx$};
		\node (CH) [node distance=1.2cm, left=of HH] {$\Cpx \otimes \algft{A}$};
		\node (H) [below=of HH] {$\algft{A}$};
		%Lines
		\draw[thick, <-] (HH.west) -- 
		    node[anchor=south]{$1_\algft{A} \otimes \id$} (CH.east);
		\draw[thick, <-] (HH.east) -- 
		    node[anchor=south]{$\id \otimes 1_\algft{A}$} (HC.west);
		\draw[thick, <-]  (H.north west) --
		    node[anchor=north east]{$\id$} (CH.south);
		\draw[thick, <-] (H.north east) -- 
		    node[anchor=north west]{$\id$} (HC.south);
		\draw[thick, <-] (H.north) --
		    node[anchor=west]{$\star$} (HH.south);
	\end{tikzpicture}
\end{center}

In this sense, one can have the correspondence
\begin{align*}
	(\star \otimes \id) \circ \star
	&= (\id \otimes \star) \circ \star
	&&\Longleftrightarrow &
	& \eqref{eq:ha_as_alg_ass} \\
	(1_\algft{A} \otimes \id) \circ \star
	&= (\id \otimes 1_\algft{A}) \circ \star
	= \id
	&&\Longleftrightarrow &
	& \eqref{eq:ha_as_alg_uni} \\
	\star \circ \tau
	&= \star
	&&\Longleftrightarrow &
	& \eqref{eq:ha_as_alg_com}
\end{align*}
where $\tau : \algft{A} \otimes \algft{A} \to \algft{A} \otimes \algft{A}$ is the flip map, \ie $\tau(f \otimes g) = g \otimes f$. Recall that \eqref{eq:ha_as_alg_ass} corresponds to the associativity of $\star$, \eqref{eq:ha_as_alg_uni} expresses that $1_\algft{A}$ is a unit for $\star$ and \eqref{eq:ha_as_alg_com} that $\star$ is commutative.

\paragraph{}
The elements of $\algft{A}$ commuting with any other elements of $\algft{A}$ corresponds to the center of $\algft{A}$, denoted $\Cen{\algft{A}}$, \ie $z \in \Cen{\algft{A}}$ if and only if
\begin{align}
	z \star f = f \star z,
	\label{eq:ha_as_alg_cen}
\end{align}
for any $f \in \algft{A}$.

\paragraph{}
The notions of the charge conjugation of a particle or of adjoint operators are defined by involutions. This is implemented via the notion of $*$-algebra.
\begin{Def}{$*$-algebra}
	{ha_as_salg}
	An algebra $\algft{A}$ is said to be a \Salg if it has an antihomomorphism involution ${}^\dagger : \algft{A} \to \algft{A}$, \ie
	\begin{subequations}
		\label{eq:ha_as_salg}
	\begin{align}
		(z_1 f + z_2 g)^\dagger
		&= \overline{z_1} f^\dagger + \overline{z_2} g^\dagger, &
		&\text{(antilinear)}
		\label{eq:ha_as_salg_al}\\
		(f \star g)^\dagger
		&= g^\dagger \star f^\dagger, &
		&\text{(antimorphism)}
		\label{eq:ha_as_salg_ah}\\
		(f^\dagger)^\dagger
		&= f, &
		&\text{(idempotent)}
		\label{eq:ha_as_salg_id}
	\end{align}
	\end{subequations}
	for any $f, g \in \algft{A}$ and $z_1, z_2 \in \Cpx$, where $\overline{z}$ denotes the complex conjugate of $z$. 
\end{Def}
If $\algft{A}$ is unital, then one has $1^\dagger = 1$.

\paragraph{}
The algebra of \exref{ha_as_algf} can be turned into a \Salg thanks to the complex conjugation $f^\dagger(x) = \overline{f(x)}$. In this example, it is important to note that the algebra structure of $\dual{\manft{X}}$ does not bring any information on the elements of $\manft{X}$, since it is fully exported from $\Cpx$. Similarly, the $*$-structure does not bring additional information on the space $\manft{X}$, since it also comes from $\Cpx$. In the context of \dq, one can deform the product of $\dual{\manft{X}}$ into a noncommutative one to obtain a (noncommutative) algebra of functions noted $\dual{\manft{X}}\llbracket\kbar\rrbracket$. Here, $\kbar$ denotes the \dpt. One can also deform the involution ${}^\dagger$ such that $\dual{\manft{X}}\llbracket\kbar\rrbracket$ becomes a \Salg.

\paragraph{}
The structure of algebra carries also a notion of ideals, that is of great use in \secref{qg_ncg}. An ideal is a subset of elements that is stable under multiplication. It can be seen as a set of elements sharing a property and propagating this property to other elements. Its importance relies in the notion of quotient algebra. The quotient algebra correspond to the packing of the algebra elements, which are grouped with respect to the latter property. A simple example corresponds to $\Int_2 = \Int / 2 \Int$ which correspond to packing integers to two \enquote{groups}: odd and even.

\begin{Def}{Ideal of algebra}
	{ha_as_id}
	Let $\algft{A}$ be an algebra. A sub-vector space $\idft{J}$ is a left (resp.~right) ideal if it is stable by left (resp.~right) multiplication, \ie for any $f \in \algft{A}$ and $g \in \idft{J}$, one has $f \star g \in \idft{J}$ (resp.~$g \star f \in \idft{J}$). It is often written as $\algft{A} \idft{J} \subset \idft{J}$ (resp.~$\idft{J} \algft{A} \subset \idft{J}$).
	
	A two-sided ideal is both a left and right ideal.
\end{Def}

\begin{Def}{Quotient algebra}
	{ha_as_qa}
	Let $\idft{J}$ be a two-sided ideal of an algebra $\algft{A}$. We define the relation $f \sim g$ if and only if $f - g \in \idft{J}$, so that its equivalent class are written $[f] = f + \idft{J}$. The quotient space, that is the space of equivalent classes, $\algft{A} / \idft{J}$ forms an algebra with product $[f \star g] = [f] \star [g]$.
	
	We further define the so-called canonical projection $\pi : \algft{A} \to \algft{A} / \idft{J}$ by $\pi(f) = [f]$. It is surjective by definition.
\end{Def}

\begin{proof}
	We need to verify that the product in $\algft{A} / \idft{J}$ is well-defined:
	\begin{align*}
		[f] \star [g]
		&= (f + \idft{J}) \star (g + \idft{J})
		= f \star g + f \star  \idft{J} + \idft{J} \star g + \idft{J}^2
		= f \star g + \idft{J}
		= [f \star g]
		\qedhere
	\end{align*}
\end{proof}

An example of algebra quotient is given by the universal enveloping algebra, in \defref{ha_as_uealg}. We give another example with some importance for this manuscript.

\begin{Ex}{Local functions as a quotient algebra}
	{ha_as_lfqa}
	Let $\manft{M}$ be a space-time and $\func^\infty(\manft{M})$ the algebra of smooth functions over $\manft{M}$. Consider a covering of open set $\{U_\alpha\}_\alpha$ of $\manft{M}$, that is $U_\alpha \subset \manft{M}$ for any $\alpha$ and $\bigcup \limits_{\alpha} U_\alpha = \manft{M}$. In the context of differential manifolds, this patch of open sets $\{U_\alpha\}_\alpha$ always exist and $U_\alpha$ can further be equipped with coordinates in $\Mink[d]$, where $d$ is the spacial dimension of $\manft{M}$. In other words, $U_\alpha$ corresponds to a region of space that is \enquote{local} and that we can describe with coordinates.
	
	We want to characterise algebraically the space of local functions, that is $\func^\infty(U_\alpha)$, by only knowing $\func^\infty(\manft{M})$. This is done by considering some quotient algebra. Indeed, let $\vert_\alpha : \func^\infty(\manft{M}) \to \func^\infty(U_\alpha)$ denote the restriction from $\manft{M}$ to $U_\alpha$, \ie $f\vert_\alpha (x) = f(x)$ for any $x \in U_\alpha$. Then, $\Ker(\vert_\alpha)$ is a (two-sided) ideal of $\func^\infty(\manft{M})$ and we have
	\begin{align}
		\func^\infty(U_\alpha)
		&= \func^\infty(\manft{M}) / \Ker(\vert_\alpha).
		\label{eq:ha_as_lfga}
	\end{align}
	Finally, the equivalent classes corresponds to the further restriction, that is $[f]_\alpha = f\vert_\alpha$, for any $f \in \func^\infty(\manft{M})$.
\end{Ex}

\begin{proof}
	We prove here that 
	\begin{align*}
		\Ker(\vert_\alpha) 
		&= \big\{ f \in \func^\infty(\manft{M}), f\vert_\alpha = 0 \big\}
		= \big\{ f \in \func^\infty(\manft{M}), f(x) = 0 \text{ for any } x \in U_\alpha \big\}
	\end{align*}
	is a two-sided ideal. Indeed, let $g \in \func^\infty(\manft{M})$ and $f \in \Ker(\vert_\alpha)$, then for any $x \in U_\alpha$, 
	\begin{align*}
		(f \cdot g)(x)
		&= f(x) g(x)
		= 0, &
		(g \cdot f)(x)
		&= g(x) f(x)
		= 0,
	\end{align*}
	so that $f \cdot g \in \Ker(\vert_\alpha)$ and $g \cdot f \in \Ker(\vert_\alpha)$. 
\end{proof}

\subsection{\namefont{Lie} algebras}
\label{subsec:ha_as_Lalg}
\paragraph{}
The notion of group of symmetries is rather mathematically expressed as elements of a \Liegt. Any \Liegt is in one-to-one correspondence with a \Lieat. The latter corresponds physically to the infinitesimal transformations associated with the former symmetries.
\begin{Def}{\namefont{Lie} algebra}
	{ha_as_Lalg}
	A \Lieat is a vector space $\Lieft{g}$ over $\Cpx$ equipped with a bracket $[\cdot, \cdot]_\Lieft{g}: \Lieft{g} \times \Lieft{g} \to \Lieft{g}$ such that
	\begin{subequations}
		\label{eq:ha_as_Lalg}
	\begin{align}
		[z_1 X + z_2 Y, Z]_\Lieft{g}
		&= z_1 [X, Z]_\Lieft{g} + z_2 [Y, Z]_\Lieft{g},
		& \text{(bilinearity)}&
		\label{eq:ha_as_Lalg_bil}\\
		[X, Y]_\Lieft{g}
		&= - [Y, X]_\Lieft{g},
		& \text{(antisymmetric)}&
		\label{eq:ha_as_Lalg_asym}\\
		[X, [Y, Z]_\Lieft{g} ]_\Lieft{g}
		+ [Y, [Z, X]_\Lieft{g} ]_\Lieft{g}
		&+ [Z, [X, Y]_\Lieft{g} ]_\Lieft{g}
		= 0,
		& \text{(\namefont{Jacobi} identity)}&
		\label{eq:ha_as_Lalg_Jac}
	\end{align}
	\end{subequations}
	for any $X, Y, Z \in \Lieft{g}$ and $z_1, z_2 \in \Cpx$.
\end{Def}
Note that the subscript $\Lieft{g}$ is often omitted when writing the bracket.

A simple physically motivated example of a \Lieat is the \namefont{Heisenberg} algebra. It corresponds to the algebra generated by $3$ elements $\hat{x}$, $\hat{p}$ and $1$ such that
\begin{align*}
	[\hat{x}, \hat{p}] = i \hbar\, 1, && 
	[\hat{x}, 1] = 0, &&
	[\hat{p}, 1] = 0.
\end{align*}

We give another physically motivated example.
\begin{Ex}{\namefont{Poincar\'{e}} algebra}
	{ha_as_Palg}
	The \Palgt $\Palg[1,d]$ in $d$ spacial dimensions corresponds to a \Lieat generated by the translations $\{P_\mu\}_{\mu =0, \ldots, d}$, the rotations $\{J_j\}_{j=1, \ldots, d}$ and the boosts $\{K_j\}_{j=1, \ldots, d}$ via
	\begin{align*}
		[P_0, J_j]
		&= 0, &
		[P_j, J_k]
		&= - i \tensor{\epsilon}{_{jk}^l} P_l, &
		[P_0, K_j]
		&= i P_j, &
		[P_j, K_k]
		&= i \eta_{jk} P_0, \\
		[P_\mu, P_\nu]
		&= 0, &
		[J_j, J_k]
		&= i \tensor{\epsilon}{_{jk}^l} J_l, &
		[J_j, K_k]
		&= i \tensor{\epsilon}{_{jk}^l} K_l, &
		[K_j, K_k]
		&= - i \tensor{\epsilon}{_{jk}^l} K_l,
	\end{align*}
	where $\eta$ is the \namefont{Minkowski} metric with signature $(- + \cdots +)$, and $\epsilon$ is the \namefont{Levi-Civita} fully antisymmetric tensor.
\end{Ex}

\begin{Rmk}{Algebras and \namefont{Lie} algebras}
	{ha_as_alg_Lalg}
	An algebra is always a \Lieat. This is done by considering the bracket $[f, g]_\star = f \star g - g \star f$, where $\star$ is the product of the algebra. One can show that such a bracket (often without subscripts) satisfies \eqref{eq:ha_as_Lalg}.
	
	The converse is not necessarily true. Therefore, the \Lieat structure is broader then the algebra one in a sense. However, from any \Lieat $\Lieft{g}$, one can construct an algebra called the universal enveloping algebra $U(\Lieft{g})$ as detailed in \defref{ha_as_uealg}.
\end{Rmk}

From a \Lieat\footnote{
	Actually, this construction can be made for any vector space.
}
$\Lieft{g}$, one can consider tensor powers of $\Lieft{g}$, that is $\Lieft{g}^{\otimes n} = \Lieft{g} \otimes \cdots \otimes \Lieft{g}$ ($n$ times), for any $n \in \NInt$, with the convention $\Lieft{g}^{\otimes 0} = \Cpx$. Then, the sum $T(\Lieft{g}) = \bigoplus \limits_{n = 0}^{+\infty} \Lieft{g}^{\otimes n}$ is an algebra, called the tensor algebra, with the product $XY = X \otimes Y \in \Lieft{g}^{\otimes (j + k)}$, for any $X \in \Lieft{g}^{\otimes j}$ and $Y \in \Lieft{g}^{\otimes k}$.

\begin{Def}{Universal enveloping algebra}
	{ha_as_uealg}
	Given a \Lieat $(\Lieft{g}, [\cdot, \cdot])$, one defines
	\begin{align*}
		U(\Lieft{g})
		= T(\Lieft{g}) / \{X \otimes Y - Y \otimes X - [X, Y], X,Y \in \Lieft{g}\}
	\end{align*}
\end{Def}

The definition above corresponds to the quotient of $T(\Lieft{g})$ by the two-sided ideal $\idft{J} = \{X \otimes Y - Y \otimes X - [X, Y], X,Y \in \Lieft{g}\}$. For simplicity, the definition of $U(\Lieft{g})$ can be understood as the tensor algebra, \ie formal polynomial of elements of $\Lieft{g}$, for which the bracket of $\Lieft{g}$ is realised by the product, \ie $[X, Y] = XY - YX$.

\subsection{Coalgebras}
\label{subsec:ha_as_coalg}
\paragraph{}
Motivated by the structure of the dual of a group (see \exref{gnc_gpf} for the motivations and detailed explanations), we define here the notion of coalgebra.

\begin{Def}{Coalgebra}
	{ha_as_coalg}
	A coalgebra is a vector space $\algft{A}$ over $\Cpx$ with two linear mappings called the coproduct $\Delta : \algft{A} \to \algft{A} \otimes \algft{A}$ and the counit $\varepsilon : \algft{A} \to \Cpx$, satisfying
	\begin{subequations}
		\label{eq:ha_as_coalg}
	\begin{align}
		(\Delta \otimes \id) \circ \Delta
		&= (\id \otimes \Delta) \circ \Delta, &
		\text{(coassocitivity)}&
		\label{eq:ha_as_coalg_coass} \\
		(\varepsilon \otimes \id) \circ \Delta
		&= (\id \otimes \varepsilon) \circ \Delta
		= \id. &
		\text{(counit property)}&
		\label{eq:ha_as_coalg_couni}
	\end{align}
	It is further said cocommutative if $\Delta$ satisfies
	\begin{align}
		\tau \circ \Delta
		&= \Delta.
		\label{eq:ha_as_coalg_cocom}
	\end{align}
	\end{subequations}
\end{Def}
The coproduct $\Delta$ is often written in terms of \namefont{Sweedler} notations
\begin{align}
	\Delta(f)
	&= \sum f_{(1)} \otimes f_{(2)}
	\label{eq:ha_as_Sn}
\end{align}
for any $f \in \algft{A}$. In these notations $f_{(1)}, f_{(2)} \in \algft{A}$ are the \enquote{components} of $f$ with respect to $\Delta$ and the sum sign $\sum$ is here to remind that $\Delta$ corresponds to the sum of all the components.

Using these notations, the axioms \eqref{eq:ha_as_coalg_coass} and \eqref{eq:ha_as_coalg_couni} becomes respectively
\begin{subequations}
	\label{eq:ha_as_coalg_S}
\begin{align}
	\sum f_{(1)(1)} \otimes f_{(1)(2)} \otimes f_{(2)}
	&= \sum f_{(1)} \otimes f_{(2)(1)} \otimes f_{(2)(2)},
	\label{eq:ha_as_coalg_coass_S} \\
	\sum \varepsilon(f_{(1)}) f_{(2)}
	&= \sum \varepsilon(f_{(2)}) f_{(1)}
	= f,
	\label{eq:ha_as_coalg_couni_S}
\end{align}
\end{subequations}
for any $f \in \algft{A}$. The equation \eqref{eq:ha_as_coalg_coass_S}  states that the element on which we take subscripts does not matter, only the order does. Therefore, one usually also writes \eqref{eq:ha_as_coalg_coass_S} as
\begin{align}
	\sum f_{(1)} \otimes f_{(2)} \otimes f_{(3)}.
	\tag{\ref{eq:ha_as_coalg_coass_S}}
\end{align}

\paragraph{}
The requirements \eqref{eq:ha_as_coalg_coass} and \eqref{eq:ha_as_coalg_couni} can be put under the form of commutative diagrams (respectively)
\begin{center}
	\noindent%
	\begin{tikzpicture}
		%Nodes
		\node (HHH) {$\algft{A} \otimes \algft{A} \otimes \algft{A}$};
		\node (HHu) [node distance=1.4cm, right=of HHH] {$\algft{A} \otimes \algft{A}$};
		\node (HHd) [below=of HHH] {$\algft{A} \otimes \algft{A}$};
		\node (H)   [below=of HHu] {$\algft{A} \!\!\!\!\!\phantom{\otimes}$};
		%Lines
		\draw[thick, ->] (HHu.west) -- 
		    node[anchor=south]{$\Delta \otimes \id$} (HHH.east);
		\draw[thick, ->] (HHd.north) -- 
		    node[anchor=west]{$\id \otimes \Delta$} (HHH.south);
		\draw[thick, ->] (H.north) --
		    node[anchor=west]{$\Delta$} (HHu.south);
		\draw[thick, ->] (H.west) -- 
		    node[anchor=south]{$\Delta$} (HHd.east);
	\end{tikzpicture}%
	\hspace{2cm}
	\begin{tikzpicture}
	    %Nodes
		\node (HH) {$\algft{A} \otimes \algft{A}$};
		\node (HC) [node distance=1.2cm, right=of HH] {$\algft{A} \otimes \Cpx$};
		\node (CH) [node distance=1.2cm, left=of HH] {$\Cpx \otimes \algft{A}$};
		\node (H) [below=of HH] {$\algft{A}$};
		%Lines
		\draw[thick, ->] (HH.west) -- 
		    node[anchor=south]{$\varepsilon \otimes \id$} (CH.east);
		\draw[thick, ->] (HH.east) -- 
		    node[anchor=south]{$\id \otimes \varepsilon$} (HC.west);
		\draw[thick, ->] (H.north west) --
		    node[anchor=north east]{$\id$} (CH.south);
		\draw[thick, ->] (H.north east) -- 
		    node[anchor=north west]{$\id$} (HC.south);
		\draw[thick, ->] (H.north) --
		    node[anchor=west]{$\Delta$} (HH.south);
	\end{tikzpicture}
\end{center}
Note that these commutative diagrams are very similar to the ones of the algebra definition, to the exception that the coproduct and counit are replaced by the product and the unit respectively and that the arrow are reversed. This is due to the fact that algebras and coalgebras are dual structures, as depicted in \figref{ha_as_dha}.

\paragraph{}
The previous statement can be made more general. Let $(\algft{A}, \Delta_{\algft{A}})$ be a coalgebra and $(\algft{B}, \star_{\algft{B}})$ be an algebra. Let $\Hilbft{L}(\algft{A}, \algft{B})$ denote the set of linear functions from $\algft{A}$ to $\algft{B}$. Then, $\Hilbft{L}(\algft{A}, \algft{B})$ is an algebra with the product
\begin{align}
	(f \star g)(a)
	&= \star_\algft{B} \circ (f \otimes g) (\Delta_\algft{A} (a))
	= \sum f(a_{(1)}) \star_\algft{B} g(a_{(2)})
\end{align}
for any $f, g \in \Hilbft{L}(\algft{A}, \algft{B})$ and $a \in \algft{A}$. The case of the dual of $\algft{A}$ is obtained when considering $\algft{B} = \Cpx$.

\paragraph{}
Finally, a $*$-coalgebra is a colagebra $\algft{A}$ equipped with an involution ${}^\dagger: \algft{A} \to \algft{A}$ satisfying \eqref{eq:ha_as_salg_al}, \eqref{eq:ha_as_salg_id} and
\begin{align}
	\Delta(f^\dagger) = \Delta(f)^\dagger
	\label{eq:ha_as_scolag}
\end{align}
where $(f \otimes g)^\dagger = f^\dagger \otimes g^\dagger$. In this case, one has $\varepsilon(f^\dagger) = \overline{\varepsilon(f)}$.

\subsection{\namefont{Hopf} algebras}
\label{subsec:ha_as_ha}
\paragraph{}
Algebras and coalgebras can be merged into one single structure called the bialgebras. The latter requires that the algebra sector (product, unit) and the coalgebra sector (coproduct, counit) are merged coherently.
\begin{Def}{Bialgebra}
	{ha_as_bialg}
	A bialgebra $\algft{A}$ is both an algebra and a coalgebra such that the product and unit are coalgebra homomorphisms, or equivalently, such that the coproduct and the counit are algebra homomorphisms, \ie
	\begin{align}
		\Delta(f \star g)
		&= \Delta(f) \star \Delta(g), &
		\varepsilon(f \star g)
		&= \varepsilon(f) \varepsilon(g), &
		\Delta(1)
		&= 1 \otimes 1, &
		\varepsilon(1)
		&= 1.
		\label{eq:ha_as_bialg}
	\end{align}
	It is further said to be a $*$-bialgebra if it is both a \Salg and a $*$-colagebra.
\end{Def}

\paragraph{}
In the context of a group dual of \exref{gnc_gpf}, one reads that the bialgebra structure is not sufficient to render all the group structure since it does not contain the information about the inverse of the group. In order to do so, one can make the structure of bialgebra grow to the one of \Hfat by defining a generalised notion of inverse $S$, called antipode or coinverse.
\begin{Def}{\namefont{Hopf} algebra}
	{ha_as_ha}
	A bialgebra\footnote{
		Note that we changed notations between \defref{ha_as_bialg} and \defref{ha_as_ha}. The one from \defref{ha_as_bialg} was chosen so that it matches with the \spdtt formalism of the deformed algebra of functions. However, the notations of \defref{ha_as_ha} were chosen to be coherent with \secref{gnc_qg}.
	}
	$(\Hoft{H}, \cdot, 1, \Delta, \varepsilon)$ is called a \Hfat, if there exists a linear map $S: \Hoft{H} \to \Hoft{H}$ satisfying
	\begin{align}
		\cdot \circ (S \otimes \id) \circ \Delta
		&= \cdot \circ (\id \otimes S) \circ \Delta
		= 1 \circ \varepsilon
		\label{eq:ha_as_ha_coinv}
	\end{align}
	
	If $\Hoft{H}$ is further a $*$-bialgebra, then it is said to be a \namefont{Hopf} $*$-algebra.
\end{Def}
In \namefont{Sweedler} notations \eqref{eq:ha_as_Sn}, the condition \eqref{eq:ha_as_ha_coinv} writes
\begin{align}
	\sum S(X_{(1)}) X_{(2)}
	&= \sum X_{(1)} S(X_{(2)})
	= \varepsilon(X) 1
	\label{eq:ha_as_coinv_Sn}
\end{align}
for any $X \in \Hoft{H}$.

\paragraph{}
The structure of \Hfat has many physical applications, part of which are discussed in \secref{gnc_qg}. We refer to \cite{Majid_1995} for more details. Note that more contemporary use of \qg{s} are made in machine learning or quantum computing \cite{Ercolessi_2023}.

\paragraph{}
One can derive several properties from this definition.

First, the antipode $S$ is an algebra anti-homomorphism and a coalgebra anti-homomorphism, that is
\begin{subequations}
	\label{eq:ha_as_ant_prop}
\begin{align}
	S(XY)
	&= S(Y) S(X), &
	S(1)
	&= 1, 
	\label{eq:ha_as_ant_prop_alg}\\
	\Delta \circ S
	&= \tau \circ (S \otimes S) \circ \Delta, &
	\varepsilon \circ S
	&= \varepsilon,
	\label{eq:ha_as_ant_prop_coalg}
\end{align}
for any $X, Y \in \Hoft{H}$. If $\Hoft{H}$ is further a \namefont{Hopf} $*$-algebra then
\begin{align}
	S \circ {}^\dagger \circ S \circ {}^\dagger = \id,
	\label{eq:ha_as_ant_prop_s}
\end{align} 
which implies that $S$ is invertible and $S^{-1} = {}^\dagger \circ S \circ {}^\dagger$.
\end{subequations}
Note that, in \namefont{Sweedler} notations, left hand side of \eqref{eq:ha_as_ant_prop_coalg} becomes
\begin{align}
	\sum S(X)_{(1)} \otimes S(X)_{(2)}
	= \sum S(X_{(2)}) \otimes S(X_{(1)}),
	\tag{\ref{eq:ha_as_ant_prop_coalg}}
\end{align}
which states that applying $S$ to a decomposition reverts the elements of the latter decomposition. This property is the reason why $S$ is sometimes called a braiding.

\begin{proof}
	First, given $X \in \Hoft{H}$, applying \eqref{eq:ha_as_coinv_Sn} to $X^\dagger$ gives
	\begin{align*}
		\sum X_{(1)}^\dagger S(X_{(2)}^\dagger)
		&= \sum S(X_{(1)}^\dagger) X_{(2)}^\dagger
		= \overline{\varepsilon(X)} 1
	\end{align*}
	and applying ${}^\dagger$ again gives
	\begin{align}
		\sum S(X_{(2)}^\dagger)^\dagger X_{(1)}
		&= \sum X_{(2)} S(X_{(1)}^\dagger)^\dagger
		= \varepsilon(X) 1.
		\label{eq:ha_as_ant_proof_*} 
	\end{align}
	Thus, we have
	\begin{align*}
		S(S(X^\dagger)^\dagger)
		&= S\left(S\left(\left(\sum \varepsilon(X_{(1)})X_{(2)}\right)^\dagger\right)^\dagger\right)
		&& \eqref{eq:ha_as_coalg_couni_S} \\
		&= S \left( S\left( \sum \overline{\varepsilon(X_{(1)})} X_{(2)}^\dagger \right)^\dagger \right)
		&& \\
 		&= S\left( \left( \sum \overline{\varepsilon(X_{(1)})} S(X_{(2)}^\dagger) \right)^\dagger \right)
 		&& \text{(lineraity of $S$)} \\
 		&= S \left( \sum \varepsilon(X_{(1)}) S(X_{(2)}^\dagger) ^\dagger \right)
 		&& \\
 		&= \sum \varepsilon(X_{(1)}) S(S(X_{(2)}^\dagger)^\dagger) 
 		&& \text{(lineraity of $S$)} \\
 		&= \sum X_{(1)} S(X_{(2)}) S(S(X_{(3)}^\dagger) ^\dagger) 
 		&& \text{(\eqref{eq:ha_as_coinv_Sn} on $X_{(1)}$)} \\
 		&= \sum X_{(1)} S(S(X_{(3)}^\dagger)^\dagger X_{(2)})
 		&& \text{(left equation of \eqref{eq:ha_as_ant_prop_alg})} \\
 		&= \sum X_{(1)} S(\varepsilon(X_{(2)}) 1)
 		&& \text{\eqref{eq:ha_as_ant_proof_*}} \\
 		&= \sum X_{(1)} \varepsilon(X_{(2)})
 		&& \text{(linearity of $S$ and \eqref{eq:ha_as_ant_prop_alg})} \\
 		&= X
 		&& \eqref{eq:ha_as_coinv_Sn}
	\end{align*}
	which proves \eqref{eq:ha_as_ant_prop_s}.
	
	Then, given $X, Y \in \Hoft{H}$,
	\begin{align*}
		S(Y) & S(X) \\ 
		&= \sum S(Y_{(1)} \varepsilon(Y_{(2)})) S(X_{(1)}\varepsilon(X_{(2)})) 
		&& \eqref{eq:ha_as_coalg_couni_S} \\
		&= \sum S(Y_{(1)}) S(X_{(1)}) \varepsilon(X_{(2)} Y_{(2)})
		&& \text{(\eqref{eq:ha_as_bialg} and  lineraity of $S$)} \\
		&= \sum  S(Y_{(1)}) S(X_{(1)}) (X_{(2)} Y_{(2)})_{(1)} S((X_{(2)} Y_{(2)})_{(2)})
		&& \eqref{eq:ha_as_coinv_Sn} \\
		&= \sum S(Y_{(1)}) S(X_{(1)}) X_{(2)} Y_{(2)} S(X_{(3)}Y_{(3)}) 
		&& \eqref{eq:ha_as_coalg_coass_S} \\
		&= \sum S(Y_{(1)}) (\varepsilon(X_{(1)})1) Y_{(2)} S(X_{(2)}Y_{(3)})
		&& \eqref{eq:ha_as_coinv_Sn} \\
		&= \sum \varepsilon(X_{(1)})\varepsilon(Y_{(1)})S(X_{(2)}Y_{(2)})
		&& \eqref{eq:ha_as_coinv_Sn} \\
		&= S(XY)
		&& \text{(following the reverse two first steps)}
	\end{align*}
	This shows the left identity of \eqref{eq:ha_as_ant_prop_alg}. The right one is obtained by applying \eqref{eq:ha_as_ha_coinv} to $1$, combined with $\Delta(1) = 1 \otimes 1$ and $\varepsilon(1) = 1$ \eqref{eq:ha_as_bialg}.

	Similarly, for $X \in \Hoft{H}$,
	\begin{align*}
		\sum & S(X_{(2)}) \otimes S(X_{(1)}) \\
		&= \sum S(X_{(2)} \varepsilon(X_{(3)})) \otimes S(X_{(1)})
		&& \text{(\eqref{eq:ha_as_coalg_couni_S} on $X_{(2)})$} \\	
		&= \sum (S(X_{(2)}) \otimes S(X_{(1)})) (\varepsilon(X_{(3)}) 1 \otimes 1) 
		&& \text{(tensor product definition)} \\
		&= \sum (S(X_{(2)}) \otimes S(X_{(1)})) (\Delta(X_{(3)} S(X_{(4)})))
		&& \eqref{eq:ha_as_coinv_Sn} \\
		&= \sum (S(X_{(2)}) \otimes S(X_{(1)})) (X_{(3)} \otimes X_{(4)}) \Delta(S(X_{(5)}))
		&& \eqref{eq:ha_as_bialg} \\
		&= \sum (S(X_{(2)}) X_{(3)} \otimes S(X_{(1)}) X_{(4)}) (\Delta(S(X_{(5)}))) 
		&& \text{(tensor product definition)} \\
		&= \sum (\varepsilon(X_{(2)}) 1 \otimes S(X_{(1)}) X_{(3)}) (\Delta(S(X_{(4)}))) 
		&& \eqref{eq:ha_as_coinv_Sn} \\
		&= \sum (1 \otimes S(X_{(1)}) X_{(2)}) (\Delta(S(X_{(3)})))
		&& \text{(\eqref{eq:ha_as_coalg_couni_S} applied to $\varepsilon(X_{(2)})X_{(3)}$)} \\
		&= \sum (1 \otimes \varepsilon(X_{(1)}) 1) (\Delta(S(X_{(2)}))) 
		&& \eqref{eq:ha_as_coinv_Sn} \\
		&= \Delta(S(X))
		&& \text{(using \eqref{eq:ha_as_coalg_couni_S} applied to $\varepsilon(X_{(1)})X_{(2)}$)}
	\end{align*}
which proves the left relation of \eqref{eq:ha_as_ant_prop_coalg}. The right relation comes from
\begin{align*}
		\varepsilon(S(X))
		&= \varepsilon \left(S \left( \sum X_{(1)} \varepsilon(X_{(2)}) \right) \right)
		= \varepsilon \left(\sum S(X_{(1)} X_{(2)}) \right)
		= \varepsilon( \varepsilon(X) 1 )
		= \varepsilon(X).
		\qedhere
	\end{align*}
\end{proof}

\paragraph{}
Given a \Lieat $\Lieft{g}$, one can endow $U(\Lieft{g})$ with a \Hfat structure as done in \exref{gnc_uealg}. It is the most simple example of \Hfat. It is used to generate non-trivial \Hfat via a \Dt deformation (see \subsecref{ha_as_Dt}).

Another example, that is relevant for this manuscript, is the \Hfat of \kPt. More details on this algebra are given in \secref{kM_kP}.

\begin{Ex}{The \namefont{Hopf} algebra of $\dpkM$-\namefont{Poincar\'{e}}}
	{ha_as_kP}
	In \cite{Lukierski_1991, Majid_1994}, the quantum deformation of the \Palgt is constructed and is given by the following structure:
	\begin{subequations}
		\label{eq:ha_as_kP}
	\begin{align}
		[J_j, J_k] &= i \tensor{\epsilon}{_{jk}^l} J_l, & 
		[J_j, K_k] &= i\tensor{\epsilon}{_{jk}^l} K_l, & 
		[K_j, K_k] &= -i\tensor{\epsilon}{_{jk}^l} J_l, \\
		[P_j, J_k] &= -i\tensor{\epsilon}{_{jk}^l} P_l, &
		[P_j, \kPE] &= [J_j, \kPE] = 0, &
		[P_j, P_k] &= 0,
		\label{eq:ha_as_kP_alg_tran}
	\end{align}%
    \vspace{\dimexpr-\abovedisplayskip-\belowdisplayskip-\baselineskip+\jot + 4pt}%
	\begin{align}
    	[K_j, \kPE] &= -\frac{i}{\dpkM} P_j \kPE, &
    	[P_j, K_k] = \frac{i}{2} \eta_{jk} 
    	\left( \dpkM(1-\kPE^2) + \frac{1}{\dpkM} P_l P^l \right) 
    	+ \frac{i}{\dpkM} P_j P_k,
    	\label{eq:ha_as_kP_alg_rot}
	\end{align}%
    %\vspace{\dimexpr-\abovedisplayskip-\belowdisplayskip-\baselineskip+\jot}%
	\begin{align}
    	\Delta P_0 &= P_0 \otimes 1 + 1 \otimes P_0, &
		\Delta P_j &= P_j \otimes 1 + \kPE \otimes P_j, 
		\label{eq:ha_as_kP_coalg_tran} \\
		\Delta \kPE &= \kPE \otimes \kPE, &
		\Delta J_j &= J_j \otimes 1 + 1 \otimes J_j,
	\end{align}%
    \vspace{\dimexpr-\abovedisplayskip-\belowdisplayskip-\baselineskip+\jot + 4pt}%
	\begin{align}
    	\Delta K_j = K_j \otimes 1 + \kPE \otimes K_j - \frac{1}{\dpkM} \tensor{\epsilon}{_j^{kl}} P_k \otimes J_l,
	\end{align}%
    %\vspace{\dimexpr-\abovedisplayskip-\belowdisplayskip-\baselineskip+\jot}%
	\begin{align}
    	\varepsilon(P_0) = \varepsilon (P_j) = \varepsilon(J_j) = \varepsilon(K_j) = 0, &&
    	\varepsilon(\kPE) = 1,
	\end{align}%
    %\vspace{\dimexpr-\abovedisplayskip-\belowdisplayskip-\baselineskip+\jot}%
	\begin{align}
		S(P_0) &= - P_0, &
		S(\kPE) &= \kPE^{-1}, &
		S(P_j) &= -\kPE^{-1} P_j,
	\end{align}%
    \vspace{\dimexpr-\abovedisplayskip-\belowdisplayskip-\baselineskip+\jot + 4pt}%
	\begin{align}
		S(J_j) &= -J_j, &
		S(K_j) &= -\kPE^{-1}(K_j - \frac{1}{\dpkM} \tensor{\epsilon}{_j^{kl}} P_k J_l).
	\end{align}%
    %\vspace{\dimexpr-\abovedisplayskip-\baselineskip+\jot}%
	\end{subequations}
	where $\kPE = e^{-P_0/\dpkM}$ and $\dpkM$ is the \dpt.
\end{Ex}

\begin{Ex}{The \namefont{Hopf} algebra of deformed translation}
	{ha_as_kT}
	We here consider the algebra of deformed translations, noted $\kTran[1,d]$, generated by $\{P_\mu\}_{\mu=0, \ldots, d}$. One can read from \eqref{eq:ha_as_kP} that it is a \namefont{Hopf} subalgebra of \kPt.
\end{Ex}

\paragraph{}
We now go back to the statement of \secref{gnc_found} that \Hfat{s} are extensions of algebras in a \enquote{self-dual} form. As coalgebras and algebras are dual to one another, the \Hfat is \enquote{self-dual} in the sense that it contains the two in a single structure. But this duality goes a bit beyond this vague argument, since one can define the dual of a \Hfat $\Hoft{H}$, noted $\dual{\Hoft{H}}$.

\begin{Def}{Dual \namefont{Hopf} algebra}
	{ha_as_dha}
	Let $\Hoft{H}$ be a \Hfat. One defined the dual \Hfat $(\dual{\Hoft{H}}, \cdot, 1_{\dual{\Hoft{H}}}, \Delta_{\dual{\Hoft{H}}}, \varepsilon_{\dual{\Hoft{H}}}, S_{\dual{\Hoft{H}}})$ of $\Hoft{H}$ by
	\begin{subequations}
		\label{eq:ha_as_dha}
	\begin{align}
		(fg) (X)
		= (f \otimes g) \big(\Delta(X) \big)
		&= \sum f(X_{(1)}) g(X_{(2)}), &
		1_{\dual{\Hoft{H}}} (X)
		&= \varepsilon(X),
		\label{eq:ha_as_dha_alg} \\
		\Delta_{\dual{\Hoft{H}}}(f) (X \otimes Y)
		&= f(X Y), &
		\varepsilon_{\dual{\Hoft{H}}}(f)
		&= f(1),
		\label{eq:ha_as_dha_coalg}
	\end{align}%
		 \vspace{\dimexpr-\abovedisplayskip-\belowdisplayskip-\baselineskip+\jot}%
	\begin{align}
		\big( S_{\dual{\Hoft{H}}}(f) \big) (X)
		= f \big( S(X) \big),
		\label{eq:ha_as_dha_coi}
	\end{align}
	\end{subequations}
	for any $f, g \in \dual{\Hoft{H}}$ and $X, Y \in \Hoft{H}$. One can read the similarities with \exref{gnc_gpf}.
\end{Def}

In \defref{ha_as_dha}, the algebra sector of $\dual{\Hoft{H}}$ corresponds to the dual of the coalgebra sector of $\Hoft{H}$. Conversely, the coalgebra sector of $\dual{\Hoft{H}}$ corresponds to the dual of the algebra sector of $\Hoft{H}$, as pictured in \figref{ha_as_dha}.

\begin{Figure}%
	[label={fig:ha_as_dha}]%
	{
		Schematic representation of the correspondence between a \Hfat $\Hoft{H}$ and its dual $\dual{\Hoft{H}}$.
	}%
	\begin{tikzpicture}
	\node (alg)
		{algebra sector of $\Hoft{H}$};
	\node (coalg) [node distance=-1pt, below=of alg]
		{coalgebra sector of $\Hoft{H}$};
	\node (coi)   [node distance=-3pt, below=of coalg]
		{coinverse of $\Hoft{H}$};
	\node (dalg)   [node distance=100pt, right=of alg]
		{algebra sector of $\dual{\Hoft{H}}$};
	\node (dcoalg) [node distance=-1pt, below=of dalg]
		{coalgebra sector of $\dual{\Hoft{H}}$};
	\node (dcoi)   [node distance=-3pt, below=of dcoalg]
		{coinverse of $\dual{\Hoft{H}}$};
		
	\draw[{To}-{To}] (alg.east) -- (dcoalg.west);
	\draw[{To}-{To}] (coalg.east) -- (dalg.west);
	\draw[{To}-{To}] (coi.east) -- (dcoi.west);
\end{tikzpicture}
\end{Figure}

\paragraph{}
Note that the relations \eqref{eq:ha_as_dha} and the correspondence depicted in \figref{ha_as_dha} are even clearer when using a dual pairing notation. Let us define $\langle \cdot, \cdot \rangle: \dual{\Hoft{H}} \otimes \Hoft{H} \to \Cpx$, defined by
\begin{align*}
	\langle f, X \rangle = f(X)
\end{align*}
Then, one has
\begin{align}
	\langle f g, X \rangle
	&= \langle f \otimes g, \Delta(X) \rangle, &
	\langle 1_{\dual{\Hoft{H}}}, X \rangle
	&= \varepsilon(X),
	\tag{\ref{eq:ha_as_dha_alg}} \\
	\langle \Delta(f), X \otimes Y \rangle
	&= \langle f, X Y \rangle, &
	\varepsilon_{\dual{\Hoft{H}}}(f)
	&= \langle f, 1 \rangle,
	\tag{\ref{eq:ha_as_dha_coalg}}
\end{align}%
	\vspace{\dimexpr-\abovedisplayskip-\belowdisplayskip-\baselineskip+\jot}%
\begin{align}
	\langle S_{\dual{\Hoft{H}}}(f), X \rangle
	&= \langle f, S(X) \rangle.
	\tag{\ref{eq:ha_as_dha_coi}}
\end{align}

\begin{Ex}{The \kMt space}
	{ha_as_kM}
	We here define the \kMt space as the dual \Hfat of $\kTran[1,d]$, similarly to \cite{Majid_1994}. One can show that \kMt, noted $\kM[d]$ has $d+1$ generators noted $\{x^\mu\}_{\mu = 0, \ldots, d}$ satisfying
	\begin{subequations}
		\label{eq:ha_as_kM}
	\begin{align}
		[x^0, x^j]
		&= \frac{\iCpx}{\dpkM} x^j, &
		[x^j, x^k]
		&= 0,
		\label{eq:ha_as_kM_alg}
	\end{align}
	\vspace{\dimexpr-\abovedisplayskip-\belowdisplayskip-\baselineskip+\jot + 4pt}%
	\begin{align}
		\Delta(x^\mu)
		&= x^\mu \otimes 1 + 1 \otimes x^\mu, &
		\varepsilon(x^\mu)
		&= 0, &
		S(x^\mu)
		&= - x^\mu.
	\end{align}
	\end{subequations}
\end{Ex}

\begin{proof}
	Let $\{x^\mu\}_{\mu = 0, \ldots, d}$ be the set of generators dual to $\{P_\mu\}_{\mu = 0, \ldots, d}$, that is 
	\begin{align*}
		\langle P_\mu, x^\nu \rangle
		= \iCpx \delta_\mu^\nu
	\end{align*}
	where $\langle \cdot, \cdot \rangle : \kTran[1,d] \times \kM[d] \to \Cpx$ is the dual pairing. Recall from \eqref{eq:ha_as_kP_alg_tran} that the $P_\mu$'s are commutative, \ie $[P_\mu, P_\nu] = 0$. This implies that the $x^\mu$ are cocommutative. Explicitly,
	\begin{align*}
		0 
		&= \langle [P_\mu, P_\nu], x^\rho \rangle 
		&& \\
		&= \langle P_\mu \otimes P_\nu, \Delta(x^\rho) \rangle - [ \mu \leftrightarrow \nu]
		&& \eqref{eq:ha_as_dha_alg} \\
		&= \langle P_\mu \otimes P_\nu, \Delta(x^\rho) - \tau \circ \Delta(x^\rho) \rangle,
		&&
	\end{align*}
	so that $\Delta = \tau \circ \Delta$. One considers for simplicity that $x^\mu$ has a trivial coproduct, \ie $\Delta x^\mu = x^\mu \otimes 1 + 1 \otimes x^\mu$. This takes root first because this space is seen as some deformed universal enveloping algebra of the \Minkt space-time (with commuting coordinates) and also because the trivial coproduct is cocommutative.
	
	From there, one computes the counit via \eqref{eq:ha_as_coalg_couni} as
	\begin{align*}
		(\varepsilon \otimes \id) \circ \Delta(x^\mu)
		&= \varepsilon(x^\mu) 1 + \varepsilon(1) x^\mu 
		= \varepsilon(x^\mu) 1 + x^\mu \\
		&= \id(x^\mu)
		= x^\mu
	\end{align*}
	so $\varepsilon(x^\mu) 1 = 0$ and $\varepsilon(x^\mu) = 0$.
	
	Then, one computes
	\begin{align*}
		\langle P_k, [x^0, x^j] \rangle
		&=  \langle P_k, x^0 x^j - x^j x^0 \rangle 
		&& \\
		&= \langle \Delta(P_k), x^0 \otimes x^j \rangle - [j \leftrightarrow 0] 
		&& \eqref{eq:ha_as_dha_coalg} \\
		&= \langle P_k \otimes 1 + \kPE \otimes P_k, x^0 \otimes x^j \rangle - [j \leftrightarrow 0]
		&& \eqref{eq:ha_as_kP_coalg_tran} \\
		&= - \delta^0_k \varepsilon(x^j) + \frac{1}{\kappa} \delta^j_k - [j \leftrightarrow 0]
		&& \\
		&= \frac{1}{\dpkM} \delta_k^j 
		&& \text{($k,j \neq 0$)}
	\end{align*}
	where we have used $\langle \kPE, x^\nu \rangle = - \frac{i}{\dpkM} \delta^\nu_0$. As $\Delta(P_0)$ does not involve $\kPE$, one computes similarly that $\langle P_0, [x^0, x^j] \rangle = 0$. With a similar computation, one shows that $\langle P_\mu, [x^j, x^k] \rangle = 0$ due to the fact that $\langle \kPE, x^j \rangle = 0$. Finally,
	\begin{align*}
		[x^0, x^j]
		&= \langle P_\mu, [x^0, x^j] \rangle \, x^\mu 
		= \frac{\iCpx}{\dpkM} x^j
	\end{align*}
	which is left hand side of \eqref{eq:ha_as_kM_alg}. The right hand side follows similarly.
\end{proof}

\subsection{\namefont{Drinfel'd} twist and \tops{\Rmat}{R-matrix}}
\label{subsec:ha_as_Dt}
\paragraph{}
In the \exref{gnc_gpf}, the dual of the group is given a \Hfat structure which is cocommutative if and only if the group is commutative. There exists a class of \Hfat{s}, called quasitriangular \Hfat{s}, that are almost cocommutative. The term \enquote{almost} means here that their non-cocommutativity is measured by some element $\Rma$, called the \Rmat. This induces a parametrized noncommutativity on the would-be group, dual to the \Hfat, and is the precise reason why some authors consider (strict) \qg{s} to be quasitriangular \Hfat{s} \cite{Majid_1995}.

\begin{Def}{(Quasi)triangular \namefont{Hopf} algebra}
	{ha_as_qtHalg}
	Let $\Hoft{H}$ be a \Hfat. It is said quasitriangular if it exists some invertible element $\Rma \in \Hoft{H} \otimes \Hoft{H}$, called the \Rmat, such that
	\begin{subequations}
		\label{eq:ha_as_qtHalg}
	\begin{align}
		(\Delta \otimes \id)(\Rma)
		&= \Rma_{13} \Rma_{23}, &
		(\id \otimes \Delta)(\Rma)
		&= \Rma_{13} \Rma_{12},
		\label{eq:ha_as_qtHalg_cocy}
	\end{align}%
	\vspace{\dimexpr-\abovedisplayskip-\belowdisplayskip-\baselineskip+\jot}%
	\begin{align}
		\tau \circ \Delta(X)
		&= \Rma \Delta(X) \Rma^{-1},
		\label{eq:ha_as_qtHalg_dcoco}
	\end{align}
	\end{subequations}%
	for any $X \in \Hoft{H}$, where $\tau$ is the flip map and
	\begin{align*}
		\Rma_{12}&
		= \sum \Rma_1 \otimes \Rma_2 \otimes 1, &
		\Rma_{13}&
		= \sum \Rma_1 \otimes 1 \otimes \Rma_2, &
		\Rma_{23}&
		= \sum 1 \otimes \Rma_1 \otimes \Rma_2.
	\end{align*}
	Note that we used the notation $\Rma = \sum \Rma_1 \otimes \Rma_2$.
	
	$\Hoft{H}$ is further said triangular if
	\begin{align}
		\Rma_{21} \Rma
		&= 1 \otimes 1
	\end{align}
	where $\Rma_{21} = \sum \Rma_2 \otimes \Rma_1 = \tau(\Rma)$.
\end{Def}

The equation \eqref{eq:ha_as_qtHalg_dcoco} precisely states that $\Rma$ deforms the cocommutativity condition $\tau \circ \Delta = \Delta$ \eqref{eq:ha_as_coalg_cocom}.

One can show that the \Rmat satisfy the following identities
\begin{subequations}
	\label{eq:ha_as_Rma}
\begin{align}
	(\varepsilon \otimes \id)(\Rma)
	&= (\id \otimes \varepsilon)(\Rma)
	= 1
	\label{eq:ha_as_Rma_cou}
\end{align}%
	\vspace{\dimexpr-\abovedisplayskip-\belowdisplayskip-\baselineskip+\jot}%
\begin{align}
	(S \otimes \id)(\Rma)
	&= \Rma^{-1}, &
	(\id \otimes S)(\Rma^{-1})
	&= \Rma,
	\label{eq:ha_as_Rma_ant}
\end{align}%
\vspace{\dimexpr-\abovedisplayskip-\belowdisplayskip-\baselineskip+\jot}%
\begin{align}
	\Rma_{12} \Rma_{13} \Rma_{23}
	&= \Rma_{23} \Rma_{13} \Rma_{12}.
	\label{eq:ha_as_Rma_qYB}
\end{align}
\end{subequations}
The equation \eqref{eq:ha_as_Rma_qYB} is called the quantum \namefont{Yang-Baxter} equation.

\paragraph{}
Another way to deform cocommutativity is to use a \Dt. The main interest of \Dt relies in the fact that one can build a new \Hfat by deforming a \Hfat. This new \Hfat has a deformed coproduct and antipode with respect to the former one. This is of great physical relevance since one can start with a commutative and cocommutative \Hfat corresponding to some symmetries over a classical \sT and deform the latter via a \Dt. The new (quantum) symmetries obtained will now act on a deformed (quantum) version of the latter \sT, for which all the deformation is controlled by the \Dt. This procedure for generating a \qST is detailed in \subsecref{gnc_qg_qst} and \subsecref{gnc_dq_dt}.

\begin{Def}{\namefont{Drinfel'd} twist}
	{ha_as_Dt}
	Let $\Hoft{H}$ be a \Hfat. A \Dt is an invertible element $\Hoft{F} \in \Hoft{H} \otimes \Hoft{H}$ that satisfies
	\begin{subequations}
		\label{eq:ha_as_Dt}
	\begin{align}
		(\Hoft{F} \otimes 1) (\Delta \otimes \id) (\Hoft{F})
		&= (1 \otimes \Hoft{F}) (\id \otimes \Delta) (\Hoft{F}), &
		\text{($2$-cocycle condition)}&
		\label{eq:ha_as_Dt_2co} \\
		(\id \otimes \varepsilon) (\Hoft{F})
		&= (\varepsilon \otimes \id) (\Hoft{F})
		= 1, &
		\text{(normalisation)}&
		\label{eq:ha_as_Dt_norm}
	\end{align}
	\end{subequations}
\end{Def}

In the previous definition of the \Dt, we omitted the semi-classical condition \eqref{eq:gnc_dq_dt_scl} since it more physically motivated, and therefore mainly do not appear in mathematical textbooks.

\begin{Thm}{Twisted \namefont{Hopf} algebra}
	{ha_as_Dt_tHalg}
	Let $\Hoft{H}$ be a \Hfat with a \Dt $\Hoft{F}$. If one defines 
	\begin{align}
		\Delta^\Hoft{F}
		&= \Hoft{F} \Delta \Hoft{F}^{-1}, &
		S^{\Hoft{F}}
		&= \chi S \chi^{-1},
		\label{eq:ha_as_Dt_tHalg}
	\end{align}
	where $\chi = \Hoft{F}_1 S(\Hoft{F}_2)$, then the set $\Hoft{H}^\Hoft{F} = \big( \Hoft{H}, \cdot, 1, \Delta^\Hoft{F}, \varepsilon, S^\Hoft{F} \big)$ is a \Hfat.
\end{Thm}

We detail the proof here to show the importance of the $2$-cocycle condition \eqref{eq:ha_as_Dt_2co} in this theorem.

\begin{proof}
	One can check point by point that $\Hoft{H}^\Hoft{F}$ satifies almost all the \Hfat axioms because $\Hoft{H}$ is a \Hfat. The main requirement to check is that $\Delta^\Hoft{F}$ satisfies the coassociativity property \eqref{eq:ha_as_coalg_coass}. By using the expression \eqref{eq:ha_as_Dt_tHalg}, one can write
	\begin{align*}
		(\Delta^\Hoft{F} \otimes \id) \circ \Delta^\Hoft{F}(X)
		&= \big( (\Hoft{F} \otimes 1) (\Delta \otimes \id) (\Hoft{F}) \big)
		\big( (\Delta \otimes \id) \circ \Delta(X) \big) 
		\big( (\Delta \otimes \id) (\Hoft{F}^{-1}) (\Hoft{F} \otimes 1)^{-1} \big) \\
		(\id \otimes \Delta^\Hoft{F}) \circ \Delta^\Hoft{F}(X)
		&= \big( (1 \otimes \Hoft{F}) (\id \otimes \Delta) (\Hoft{F}) \big)
		\big( (\id \otimes \Delta) \circ \Delta(X) \big) 
		\big( (\id \otimes \Delta) (\Hoft{F}^{-1}) (1 \otimes \Hoft{F})^{-1} \big)
	\end{align*}
	for any $X \in \Hoft{H}$. Both lines of the previous equation are equal thanks to the coassociativity property of $\Delta$, the fact that $\Delta$ is a homomorphism and the $2$-cocycle condition of $\Hoft{F}$. 
\end{proof}

\paragraph{}
Finally, these two deformations of the coalgebra sector are linked to one another by the following theorem.

\begin{Thm}{Twisted (quasi)triangular \namefont{Hopf} algebra}
	{ha_as_Dt_tqtHalg}
	Let $\Hoft{H}$ a quasitriangular \Hfat with \Rmat $\Rma$, and a \Dt $\Hoft{F}$. Then, the twisted \Hfat $\Hoft{H}^\Hoft{F}$ of \thmref{ha_as_Dt_tHalg} is also quasitriangular with \Rmat $\Rma^\Hoft{F} = \Hoft{F}_{21} \Rma \Hoft{F}^{-1}$.
	
	$\Hoft{H}^\Hoft{F}$ is further triangular if and only if $\Hoft{H}$ is.
\end{Thm}

\begin{proof}
	We refer to proof of Theorem 2.3.4 of \cite{Majid_1995}.
\end{proof}

\section{Representation theory}
\label{sec:ha_rt}
\paragraph{}
When working with a complicated structure, one would rather want to transform it into a simpler structure easier work with. This is precisely the goal of a representation. Instead of working with a group, one represents it on some set of matrices, for which there are known results. The main tool to have a representation is called a module. The module is a vector space on which our complicated structure will act. Taking back the example of the group, one can interpret it as a physical group of symmetries which acts on some space. The latter space is precisely the module.

In the present case, the study of modules for algebras and \Hfat{s} is two-fold. First, it is the way to implement symmetries as discussed above. Therefore, one implements some (quantum) symmetries, gathered in a \Hfat, on some space precisely via the module structure. Second, the \namefont{Serre-Swan} theorem states that sections over a fiber bundle are in one-to-one correspondence with a module of the algebra of smooth functions. A gauge theory, relying on the notion of fiber bundle, is thus implemented in \nCG via the module structure. This is rather detailed in \chapref{ncft}.

\begin{Def}{Module over an algebra}
	{ha_rt_mod}
	A left (resp.~right) module $\modft{X}$ over an algebra $(\algft{A}, \star, 1)$ is a vector space together with a linear action $\actl : \algft{A} \otimes \modft{X} \to \modft{X}$ (resp.~$\actr : \modft{X} \otimes \algft{A} \to \algft{A}$) such that 
	\begin{subequations}
		\label{eq:ha_rt_mod}
	\begin{align}
		f \actl ( g \actl s)
		&= (f \star g) \actl s,
		& \big(\text{resp.~}
		(s \actr f) \actr g
		&= s \actr (f \star g)
		\big)
		\label{eq:ha_rt_mod_ass} \\
		1 \actl s
		&= s,
		& \big(\text{resp.~}
		s \actr 1
		&= s
		\big)
		\label{eq:ha_rt_mod_uni}
	\end{align}
	\end{subequations}
	for any $f, g \in \algft{A}$ and $s \in \modft{X}$. We say that $\modft{X}$ is a left (resp.~right) $\algft{A}$-module.
	
	If $\modft{X}$ is both a right and left $\algft{A}$-module such that 
	\begin{align}
		(f \actl s) \actr g
		= f \actl (s \actr g),
		\label{eq:ha_rt_bimod}
	\end{align}
	then $\modft{X}$ is said to be a $\algft{A}$-bimodule.
\end{Def}

A physically motivated example of such a module structure can be found in \exref{gnc_qg_qst_modex}.

Another example is the module given by the algebra itself, that is $\modft{X} = \algft{A}$, with the action given by the product $\actl = \star$. In this case, \eqref{eq:ha_rt_mod_ass} is fulfilled thanks to the associativity property of $\star$ \eqref{eq:ha_as_alg_ass}, and \eqref{eq:ha_rt_mod_uni} is satisfied by the definition of the unit $1$ \eqref{eq:ha_as_alg_uni}.

One can generalise the previous example to
\begin{Ex}{The module of $n$ copies of the algebra}
	{ha_rt_ncop}
	Let $n \in \NInt$ and $\algft{A}$ be an algebra. Consider the tensor product of $n$ copies of the algebra: $\algft{A}^{\otimes n} = \algft{A} \otimes \cdots \otimes \algft{A}$. Then, $\modft{X} = \algft{A}^{\otimes n}$ is a (left) $\algft{A}$-module, with the action $\actl = \star \otimes \cdots \otimes \star$.
	
	Indeed, let $f = f_1 \otimes \cdots \otimes f_n \in \algft{A}^{\otimes n}$. Then,
	\begin{align*}
		g_1 \actl (g_2 \actl f)
		&= g_1 \actl \big( g_2 \star f_1 \otimes \cdots \otimes g_2 \star f_n \big) \\
		&= g_1 \star g_2 \star f_1 \otimes \cdots \otimes g_1 \star g_2 \star f_n
		= (g_1 \star g_2) \actl f \\
		1 \actl f
		&= 1 \star f_1 \otimes \cdots \otimes 1 \star f_n
		= f
	\end{align*}
	for any $g_1, g_2 \in \algft{A}$, where we used the associativity of $\star$ \eqref{eq:ha_as_alg_ass} and the unit property \eqref{eq:ha_as_alg_uni}. One proceeds similarly in the case of the right module.
\end{Ex}

\paragraph{}
One can also define modules over a coalgebra, that are called comodules.
\begin{Def}{Comodule over a coalgebra}
	{ha_rt_comod}
	A left (resp.~right) comodule $\modft{X}$ over a coalgebra $(\algft{A}, \Delta, \varepsilon)$ is a vector space together with a linear coaction $\coactl: \modft{X} \to \algft{A} \otimes \modft{X}$ (resp.~$\coactr : \modft{X} \to \modft{X} \otimes \algft{A}$) such that
	\begin{subequations}
		\label{eq:ha_rt_comod}
	\begin{align}
		( \id \otimes \coactl ) \circ \coactl
		&= (\Delta \otimes \id) \circ \coactl,
		& \big(\text{resp.~}
		( \coactr \otimes \id) \circ \coactr
		&= (\id \otimes \Delta) \circ \coactr
		\big)
		\label{eq:ha_rt_comod_ass} \\
		( \id \otimes \varepsilon) \circ \coactl
		&= \id,
		& \big( (\varepsilon \otimes \id) \circ \coactr
		&= \id
		\big)
		\label{eq:ha_rt_comod_uni}
	\end{align}
	\end{subequations}
	We say that $\modft{X}$ is a left (resp.~right) $\algft{A}$-comodule.
\end{Def}

A simple example of comodule is the coalgebra itself $\modft{X} = \algft{A}$, with the coaction given by the coproduct $\coactl = \Delta$.

Note that the coaction can be written in \namefont{Sweedler} notations through
\begin{align}
	\coactl s
	&= \sum s_{(1)} \otimes s_{(0)},
	\label{eq:ha_rt_comod_Sn}
\end{align}
where $s_{(0)} \in \modft{X}$ and $s_{(1)} \in \algft{A}$. Thanks to \eqref{eq:ha_rt_comod_ass}, which states
\begin{align*}
	\sum (s_{(1)})_{(1)} \otimes (s_{(1)})_{(2)} \otimes s_{(0)}
	&= \sum s_{(1)} \otimes (s_{(0)})_{(1)} \otimes (s_{(0)})_{(0)}
\end{align*}
denoted by $\sum s_{(2)} \otimes s_{(1)} \otimes s_{(0)}$ for simplicity, the \namefont{Sweedler} notation of the coaction is coherent with the \namefont{Sweedler} notation for the coproduct \eqref{eq:ha_as_Sn}. It is also quite straightforward since the zeroth component is the one of the module and the others are the one of the algebra.

One can write commutative diagrams that stands for \eqref{eq:ha_rt_comod} and show that they are dual to the ones representing \eqref{eq:ha_rt_mod}, in a similar fashion that the axioms of the coalgebra are dual to the ones of the algebra. One has to \enquote{reverse the arrows}.

This duality goes a bit beyond this observation. If one consider $\modft{X}$ to be a left (resp.~right) $\Hoft{H}$-comodule, for $\Hoft{H}$ a \Hfat, then $\modft{X}$ is a right (resp.~left) $\dual{\Hoft{H}}$-module with the action
\begin{align}
	s \actr f
	&= \sum \langle f, s_{(1)} \rangle \, s_{(0)}, &
	\big( \text{resp.~} f \actl s
	&= \sum \langle f, s_{(1)} \rangle \, s_{(0)} \big)
	\label{eq:ha_rt_dcomod}
\end{align}
using \namefont{Sweedler} notations \eqref{eq:ha_rt_comod_Sn}, for any $s \in \modft{X}$ and $f \in \dual{\Hoft{H}}$.

\begin{proof}
	We do the proof for the case of a right comodule, the left case being similar. We have to verify the module axioms \eqref{eq:ha_rt_mod}. Let $f, g \in \dual{\Hoft{H}}$ and $s \in \modft{X}$,
	\begin{align*}
		(s \actr g) \actr f
		&= \sum \langle g, s_{(1)} \rangle \, (s_{(0)} \actr f) \\
		&= \sum \langle g, s_{(1)} \rangle \, \langle f, s_{(2)} \rangle \, s_{(0)} \\
		&= \left\langle g \otimes f, \sum s_{(1)} \otimes s_{(2)} \right\rangle \, s_{(0)} \\
		&= \langle gf, s_{(1)} \rangle \, s_{(0)} \\
		&= s \actr \, (gf) 
	\end{align*}
	where we used \eqref{eq:ha_rt_comod_ass} implied by the \namefont{Sweedler} notations and the property \eqref{eq:ha_as_dha_alg} of the dual pairing.
	
	The counit property goes as
	\begin{align*}
		s \actr 1
		&= \sum \langle 1, s_{(1)} \rangle s_{(0)}
		= \sum \varepsilon(s_{(1)}) s_{(0)}
		= s
	\end{align*}	
	where the last equality corresponds to \eqref{eq:ha_rt_comod_uni} in \namefont{Sweedler} notations.
\end{proof}
However, generically, a module structure does not give a comodule structure on the dual. This is only true in the finite dimensional case (see Lemma 1.6.3 of \cite{Montgomery_1993}).

\paragraph{}
In the context of \nCG, one studies an algebra $\algft{A}$ that corresponds to the deformed functions over some space-time. We further want to implement that a \Hfat $\Hoft{H}$ corresponds to the (quantum) symmetries of this space-time. This can be done in two ways, depending whether $\Hoft{H}$ stands as the deformation of the group of symmetries or as the deformations of the algebra of infinitesimal transformations corresponding to these symmetries.

If we consider a group of symmetries $G$ on a \sT $\manft{M}$, then one can build a \Hfat on $\dual{G}$ as detailed in \exref{gnc_gpf}. From there, $\func^\infty(\manft{M})$ is a $G$-module (see \exref{gnc_qg_qst_modalgex} for an explicit case). By duality, this may be implemented by giving to $\func^\infty(\manft{M})$ a $\dual{G}$-comodule structure. As $\func^\infty(\manft{M})$ is also an algebra we say that it is a $\dual{G}$-comodule algebra. Therefore, when considering the noncommutative counterpart of this picture, we say that the \qST $\algft{A}$ has quantum symmetries $\Hoft{H}$ if $\algft{A}$ is a $\Hoft{H}$-comodule algebra.

Now if we consider a \Lieat $\Lieft{g}$ of infinitesimal symmetries on a \sT $\manft{M}$, then one can endow $U(\Lieft{g})$ with a \Hfat structure as in \exref{gnc_uealg}. The infinitesimal symmetries acts on functions, \ie $\func^\infty(\manft{M})$ is a $U(\Lieft{g})$-module.  As $\func^\infty(\manft{M})$ is also an algebra, we say that it is a $U(\Lieft{g})$-module algebra. Thus, going noncommutative, the \qST $\algft{A}$ has quantum (infinitesimal) symmetries $\Hoft{H}$ if $\algft{A}$ is a $\Hoft{H}$-module algebra.

The latter meaning of quantum symmetry is the one we consider mainly in this manuscript, as in \subsecref{gnc_qg_qst}. However, the two are coherent and actually can be seen as dual to one another.

\paragraph{}
We introduce below the two notions of module algebra and comodule algebra.

\begin{Def}{Module algebra}
	{ha_rt_modalg}
	Let $(\algft{A}, \star, 1)$ be an algebra and $\Hoft{H}$ be a \Hfat. $\algft{A}$ is said to be a left (resp.~right) $\Hoft{H}$-module algebra if $\algft{A}$ is a left (resp.~right) $\Hoft{H}$-module and the action $\actl$ (resp.~$\actr$) links the coproduct of $\Hoft{H}$ to the product of $\modft{A}$, \ie
	\begin{subequations}
		\label{eq:ha_rt_modalg}
	\begin{align}
		X \actl (f \star g)
		&= \Delta(X) \actl (f \otimes g),
		& \big(\text{resp.~}
		(f \star g) \actr X
		&= (f \otimes g) \actr \Delta(X)
		\big)
		\label{eq:ha_rt_modalg_ass} \\
		X \actl 1
		&= \varepsilon(X) 1,
		& \big(\text{resp.~}
		1 \actr X
		&= \varepsilon(X) 1
		\big)
		\label{eq:ha_rt_modalg_uni}
	\end{align}
	\end{subequations}
	for any $X \in \Hoft{H}$ and $f, g \in \algft{A}$.
\end{Def}

This definition can be extended to the \Salg case by enforcing the condition
\begin{align}
	(X \actl f)^\dagger &= S(X)^\ddagger \actl f^\dagger
	\label{eq:ha_rt_modalg_salg}
\end{align}
where ${}^\dagger$ is the involution on $\algft{A}$, ${}^\ddagger$ the involution on $\Hoft{H}$ and $S$ the antipode of $\Hoft{H}$.

\begin{Def}{Comodule algebra}
	{ha_rt_comodalg}
	Let $(\algft{A}, \star, 1)$ be an algebra and $\Hoft{H}$ be a \Hfat. $\algft{A}$ is said to be a left (resp.~right) $\Hoft{H}$-comodule algebra if $\algft{A}$ is a left (resp.~right) $\Hoft{H}$-comodule and the coaction $\coactl$ (resp.~$\coactr$) links the product of $\Hoft{H}$ and the product of $\algft{A}$, \ie
	\begin{subequations}
		\label{eq:ha_rt_comodalg}
	\begin{align}
		\coactl (f \star g)
		&= (\coactl f) (\star \otimes \cdot) (\coactl g),
		& \big( \text{resp.~}
		\coactr (f \star g)
		&= (\coactr f) (\cdot \otimes \star) (\coactr g) \big)
		\label{eq:ha_rt_comodalg_ass} \\
		\coactl 1
		&= 1 \otimes 1,
		& \big( \text{resp.~} 
		\coactr 1
		&= 1 \otimes 1 \big)
		\label{eq:ha_rt_comodalg_uni}
	\end{align}
	\end{subequations}
	for any $f, g \in \algft{A}$.
\end{Def}

Again, comodule algebras can be extended to the \Salg case, by requiring that the coaction is compatible with the involution ${}^\dagger$ of $\algft{A}$ as
\begin{align}
	\coactl (f^\dagger)
	= \sum f_{(1)}^\dagger \otimes f_{(0)}^\ddagger
	= (\coactl f)^\dagger
\end{align}
where ${}^\ddagger$ is the involution of $\Hoft{H}$.

\paragraph{}
Note that one can define plenty of other algebraic structures in the same spirit, like module coalgebras, comodule coalgebras, module bialgebras, comodule bialgebras, \etc
\chapter{Operator algebras}
\label{ch:oa}

\paragraph{}
\Opalg{s} have been studied as a mathematical framework for \qM \cite{Landsmann_1998}. They have been inspired from early works of \namefont{Weyl} \cite{Weyl_1927} and \namefont{von Neumann} \cite{von_Neumann_1931}. These works first evolved around \opalg{s} called \namefont{von Neumann} algebras, formerly called $W^*$-algebras. A \namefont{von Neumann} algebra corresponds to an algebra of bounded operator on a \Hsp, as in \exref{oa_tas_boHsp}, \ie the operators are viewed as functions on some \Hsp of states.

In this section, we focus on a more general class of \opalg called the \Csalg{s}. The \Csalg{s} are broader in the sense that they do not specify a \Hsp to be defined but can still be represented on \Hsp{s}. Therefore, it is further used for applications in physics as in algebraic \qFT \cite{Fewster_2019} or in \qFT on curved \sT \cite{Wald_1994}. We refer to \secref{gnc_phys} for an outline of the role of \Csalg in quantum physics.

One can go through textbooks like \cite{Gracia-Bondia_2001, Landsman_1998, Takesaki_2002, Takesaki_2003a, Takesaki_2003b, Blackadar_2006} for more details on \opalg{s}.

\paragraph{}
In \nCG, the philosophy and use of \Csalg{s} may be thought differently since the \Csalg is supposed to be the (functions on the) space-time itself, while the above mentioned approaches consider \Csalg on a specific (curved) classical \sT. Given a space $\manft{M}$, which at this point could be finite, discrete, fractal or infinite, the space of smooth functions $\func^\infty(\manft{M})$ can be endowed with a \Csalg structure. Moreover, the topology of $\manft{M}$ is preserved in $\func^\infty(\manft{M})$ through its space of states. The latter statement corresponds to the (commutative) \nameref{thm:oa_rt_cGN} and is the principal motivation of studying \Csalg in \nCG.

Indeed, if one considers now a noncommutative \Csalg, from an extrapolation of the \nameref{thm:oa_rt_cGN}, this algebra can be considered as the space of functions over some geometry, given by its space of states. This geometry is the \nCG. We refer to \secref{gnc_found} for a more explicit guideline of what a \nCG is.

\section{Topological algebraic structures}
\label{sec:oa_tas}
\paragraph{}
Some algebraic structure required in the study of \opalg{s} are gathered in this section.

\subsection{\namefont{Hilbert} space}
\label{subsec:oa_tas_Hsp}
\paragraph{}
Considering some space $V$, the notion of \enquote{distance between objects} of $V$ is introduced through the norm.

\begin{Def}{Normed vector space}
	{oa_tas_nvs}
	Let $V$ be a vector space. It is called a normed vector space if it is equipped with a function $\Vert\cdot\Vert: V \to \pReal$, called the norm, which satisfies
	\begin{subequations}
		\label{eq:oa_tas_nvs}
	\begin{align}
		\Vert u + v \Vert 
		&\leqslant \Vert u \Vert + \Vert v \Vert,
		&& \text{(triangle inequality)}
		\label{eq:oa_tas_nvs_ti} \\
		\Vert z u \Vert
		&= \vert z \vert \, \Vert u \Vert,
		&& \text{(homogeneity)}
		\label{eq:oa_tas_nvs_h} \\
		\Vert u \Vert = 0
		&\Leftrightarrow u = 0,
		&& \text{(non-degenerate)}
		\label{eq:oa_tas_nvs_d}
	\end{align}
	\end{subequations}
	for any $u, v \in V$ and $z \in \Cpx$.
\end{Def}

Note that the previous notion of \enquote{distance between objects} is to be understood in a broad sense. For example, consider a spinless particle on the real line $\Real$. Then, the particle has a probability amplitude given by a wave function $\psi(t, x)$, for $x \in \Real$ a point on the line and $t \in \Real$ a time. For a given region of the line, say $0 \leqslant x \leqslant L$, the integral 
\begin{align*}
	\Vert \psi(t) \Vert_{L}^2
	= \int_0^L \vert\psi(t, x)\vert^2 \tdr{}{x}
\end{align*}
is interpreted as the probability that the particle is in $[0,L]$ at time $t$. Note that the previous statement requires the normalisation condition
\begin{align*}
	\Vert \psi(t) \Vert^2
	= \int_{-\infty}^{+\infty} \vert\psi(t, x)\vert^2 \tdr{}{x}
	= 1
\end{align*}
which can also express that the probability that the particle is on the line $\Real$ is $1$, for any time $t$.

This follows from the fact that the considered (\namefont{Hilbert}) space is the one of square integrable functions on $\Real$, noted $L^2(\Real^2)$, and that $\Vert\cdot\Vert$ and $\Vert \cdot \Vert_L$ are norms for this space.

\paragraph{}
This notion of norm can also help us to make sense of other quantities, like sequential measurements.

Consider a particle in a space that moves along an infinite number of displacement $x_n$ for $n \in \NInt$. The total distance travelled by the particle would correspond to $\sum \limits_{n=0}^\infty x_n$. However, one needs a structure so that this sum has actually a meaning. This can be done in a normed vector spaces through absolute convergence, \ie the convergence of $\sum \Vert x_n\Vert$. A space where the convergence of $\sum \Vert x_n\Vert$ is equivalent to the convergence of $\sum x_n$ is called a \namefont{Banach} space and is said to be complete\footnote{
	Completeness is more often defined by the convergence of \namefont{Cauchy} sequences, but in the case of a \namefont{Banach} space, this condition is equivalent to the absolute convergence implying the convergence.
}.

\begin{Def}{\namefont{Banach} space}
	{oa_tas_Bs}
	A complete normed vector space is called a \namefont{Banach} space.
\end{Def}

\paragraph{}
A convenient way to construct a norm is through a Hermitian sesquilinear form.

\begin{Def}{Sesquilinear form}
	{oa_tas_Hsf}
	Let $V$ be a vector space. A sesquilinear form on $V$ is a map $\langle \cdot, \cdot \rangle : V \times V \to \mathbb{C}$ such that
	\begin{subequations}
		\label{eq:oa_tas_Hsf}
	\begin{align}
		\langle z_1 u_1 + z_2 u_2, v \rangle
		&= z_1 \langle u_1, v \rangle + z_2 \langle u_2, v \rangle,
		&& \text{(linearity over the first variable)}
		\label{eq:oa_tas_Hsf_lin} \\
		\langle u, z_1 v_1 + z_2 v_2 \rangle
		&= \overline{z_1} \langle u, v_1 \rangle + \overline{z_2} \langle u, v_2 \rangle,
		&& \text{(semilinearity over the second variable)}
		\label{eq:oa_tas_Hsf_slin}
	\end{align}
	for any $u, v, u_1, u_2, v_1, v_2 \in V$ and $z_1, z_2 \in \Cpx$.
	
	One can further add some properties to this form:
	\begin{align}
		\langle u, v \rangle
		&= \overline{\langle v, u \rangle},
		&& \text{(Hermitianity)}
		\label{eq:oa_tas_Hsf_her} \\
		\langle u, u \rangle
		&\geqslant 0,
		&& \text{(positivity or positive definite)}
		\label{eq:oa_tas_Hsf_pos} \\
		\langle u, u \rangle = 0 
		&\Leftrightarrow u = 0.
		&& \text{(non-degenerate)}
		\label{eq:oa_tas_Hsf_nd}		
	\end{align}
	\end{subequations}
\end{Def}
If one considers a vector space equipped with a positive non-degenerate Hermitian sesquilinear form $(V, \langle \cdot, \cdot \rangle)$, then $\Vert u \Vert = \sqrt{\langle u, u \rangle}$, for any $u \in V$ defines a norm on $V$. This means $(V, \Vert \cdot \Vert)$ is a normed vector space as defined in \defref{oa_tas_nvs}.

Note that the sesquilinear form satisfies the \namefont{Cauchy-Schwarz} inequality
\begin{align}
	\big\vert \langle u, v \rangle \big\vert
	\leqslant \Vert u \Vert \, \Vert v \Vert.
	\label{eq:oa_tas_CSi}
\end{align}

\paragraph{}
We are now ready to define a \Hsp.
\begin{Def}{\namefont{Hilbert} space}
	{oa_tas_Hsp}
	A vector space $\Hilbft{H}$ equipped with a positive non-degenerate Hermitian sesquilinear form $\langle \cdot, \cdot \rangle$, such that $\Hilbft{H}$ is separated (or \namefont{Hausdorff}) and complete for the norm $\Vert \cdot \Vert = \sqrt{\langle \cdot, \cdot \rangle}$, is called a \Hsp.
\end{Def}
Note that, in this context, the form $\langle \cdot, \cdot \rangle$ is more commonly called scalar product or inner product. We now give the case in point example used in \qM.

\begin{Ex}{Quantum particle on the line}
	{oa_tas_qpl}
	We consider a spinless quantum particle on a $1$-dimensional space, that is on the line $\Real$. It is described by square integrable functions $\psi \in L^2(\Real^2)$, which are interpreted as probability amplitude wave functions. One then defines
	\begin{align}
		\langle \psi_1(t), \psi_2(t) \rangle
		&= \int_{-\infty}^{+\infty} \psi_1(t,x) \, \overline{\psi_2}(t,x) \tdr{}{x}
		\label{eq:oa_tas_qpl_ip}
	\end{align}
	which can be shown to be an inner product on $L^2(\Real^2)$. It has an associated norm which writes
	\begin{align}
		\Vert \psi(t) \Vert^2
		&= \int^{+\infty}_{-\infty} \vert \psi(t, x)\vert^2 \tdr{}{x}
		\label{eq:oa_tas_qpl_n}
	\end{align}
	interpreted to be the probability of finding the particle on the line $\Real$ at time $t$. One can show that $L^2(\Real^2)$ with the norm \eqref{eq:oa_tas_qpl_n} is complete. Therefore, $L^2(\Real^2)$ equipped with the inner product \eqref{eq:oa_tas_qpl_ip} is a \Hsp.
\end{Ex}

\subsection{\tops{$C^*$}{C*}-algebras}
\label{subsec:oa_tas_Csa}
\paragraph{}
We now go to \namefont{Banach} algebras which are the first step toward operator algebras. Indeed, one wants the structure of a \namefont{Banach} space for the reasons stated in \subsecref{oa_tas_Hsp}. Moreover, one wants the structure of algebra in order to consider functions (operators) over some space. Therefore, one ends up with a
\begin{Def}{\namefont{Banach} algebra}
	{oa_tas_Ba}
	A \namefont{Banach} space $\opaft{A}$ is a \namefont{Banach} algebra, if it is an algebra and its norm satisfies
	\begin{align}
		\Vert f g \Vert 
		\leqslant \Vert f \Vert \, \Vert g \Vert
		\label{eq:oa_tas_Ba}
	\end{align}
	for any $f, g \in \opaft{A}$. In this case, one has $\Vert 1 \Vert = 1$.
	
	It is further said to be a \namefont{Banach} $*$-algebra, or $B^*$-algebra, if $\opaft{A}$ is a \Salg and that the involution ${}^\dagger$ is an isometry for the norm, \ie
	\begin{align}
		\Vert f^\dagger \Vert
		= \Vert f \Vert
		\label{eq:oa_tas_sBa}
	\end{align}
\end{Def}

The notion of \namefont{Banach} $*$-algebra is all the more important for physics then physical observables are considered to be self-adjoint operators, that is operators $f$ such that
\begin{align}
	f^\dagger = f.
	\label{eq:oa_tas_Csa_sa}
\end{align}

We give here an example of such a \namefont{Banach} algebra.
\begin{Ex}{Bounded operators on a \namefont{Banach} space}
	{oa_tas_boBa}
	Let $V$ be a \namefont{Banach} space. Consider $\Hilbft{B}(V)$ to be the bounded linear operators from $V$ to $V$, that we equip with the norm
	\begin{align}
		\Vert f \Vert_{\Hilbft{B}(V)}
		&= \underset{u \neq 0}{\sup} \frac{\Vert f(u) \Vert}{\Vert u \Vert}
		= \underset{\Vert u \Vert = 1}{\sup} \Vert f(u) \Vert,
		\label{eq:oa_tas_boBa_n}
	\end{align}
	for any $f \in \Hilbft{B}(V)$, where $\Vert \cdot \Vert$ is the norm of $V$. This is often called the operator norm. It exists because we work with bounded operators. Then, $\Hilbft{B}(V)$ equipped with the composition law $\circ$ is a \namefont{Banach} algebra.
	
	Indeed, by definition \eqref{eq:oa_tas_boBa_n}, one has that for any $u \neq 0$, $\Vert f(u) \Vert \leqslant \Vert f \Vert_{\Hilbft{B}(\Hilbft{H})} \, \Vert u \Vert$. Therefore, if $f, g \in \Hilbft{B}(V)$, for any $u \in V$, such that $u \neq 0$, one has
	\begin{align*}
		\frac{\Vert (f \circ g)(u) \Vert}{\Vert u \Vert}
		&= \frac{\Vert f(g(u)) \Vert}{\Vert g(u) \Vert} \frac{\Vert g(u) \Vert}{\Vert u \Vert}
		\leqslant \Vert f \Vert_{\Hilbft{B}(V)} \, \Vert g \Vert_{\Hilbft{B}(V)}.
	\end{align*}
	Note that the previous equality can be written because $g(u) \neq 0$ thanks to $u \neq 0$ and the continuity property of $g$. Now, as the upper bounds found does not depend on $u$ we can take the supremum and obtain that $\Vert f \circ g \Vert_{\Hilbft{B}(V)} \leqslant \Vert f \Vert_{\Hilbft{B}(V)} \, \Vert g \Vert_{\Hilbft{B}(V)}$, which is \eqref{eq:oa_tas_Ba}.
	
	In most cases, the subscript ${}_{\Hilbft{B}(V)}$ for the norm is omitted.
\end{Ex}

\paragraph{}
We now gather all these structure into a single one, the \Csalg.
\begin{Def}{$C^*$-algebra}
	{oa_tas_Csalg}
	A \namefont{Banach} $*$-algebra $\opaft{A}$ is called a \Csalg if
	\begin{align}
		\Vert f f^\dagger \Vert
		&= \Vert f \Vert^2,
		& \big(\text{or equivalently }
		\Vert f^\dagger f \Vert
		&= \Vert f \Vert^2
		\big)
		\label{eq:oa_tas_Csalg}
	\end{align}
	for any $f \in \opaft{A}$.
\end{Def}

The previous definition is both motivated and explained by the important example of bounded operators on a \Hsp. It is at the very basis of why \opalg{s} are considered in the context of \qM.
\begin{Ex}{Bounded operators on a \namefont{Hilbert} space}
	{oa_tas_boHsp}
	Let $\Hilbft{H}$ be a \Hsp and $\Hilbft{B}(\Hilbft{H})$ denote the bounded linear operators on $\Hilbft{H}$. $\Hilbft{H}$ has a norm associated to its inner product and, therefore, one can follow \exref{oa_tas_boBa} to show that $\Hilbft{B}(\Hilbft{H})$ is a \namefont{Banach} algebra.
	
	We introduce the involution given by the adjoint operator. Given $f \in \Hilbft{B}(\Hilbft{H})$, one defines $f^\dagger \in \Hilbft{B}(\Hilbft{H})$ as the unique operator satisfying
	\begin{align}
		\langle f(\psi_1), \psi_2 \rangle
		&= \langle \psi_1, f^\dagger(\psi_2) \rangle,
	\end{align}
	for any $\psi_1, \psi_2 \in \Hilbft{H}$. It is an involution thanks to the uniqueness property and it preserves the norm. Indeed, for any $\psi \neq 0$, the \namefont{Cauchy-Schwarz} inequality \eqref{eq:oa_tas_CSi} writes
	\begin{align*}
		\Vert f(\psi) \Vert^2
		&= \langle f(\psi), f(\psi) \rangle
		= \langle \psi, (f^\dagger \circ f)(\psi) \rangle
		&& \\
		&\leqslant \Vert \psi \Vert \, \Vert (f^\dagger \circ f)(\psi) \Vert
		&& \eqref{eq:oa_tas_CSi} \\
		&\leqslant \Vert f^\dagger \circ f \Vert \, \Vert \psi \Vert^2
		&& \eqref{eq:oa_tas_boBa_n} \\
		&\leqslant \Vert f^\dagger \Vert \, \Vert f \Vert \, \Vert \psi \Vert^2.
		&& \eqref{eq:oa_tas_Ba}
	\end{align*}
	Dividing by $\Vert \psi \Vert^2$ and taking the supremum over $\psi$ imposes that $\Vert f \Vert \leqslant \Vert f^\dagger \Vert$. One can then invert the role of $f$ and $f^\dagger$ to have the other inequality leading to $\Vert f \Vert = \Vert f^\dagger \Vert$ which corresponds to \eqref{eq:oa_tas_sBa}.
	
	Finally, in the previous computation we have shown that $\Vert f \Vert^2 \leqslant \Vert f \circ f^\dagger \Vert$, using mainly \eqref{eq:oa_tas_CSi}. But using \eqref{eq:oa_tas_Ba} and \eqref{eq:oa_tas_sBa}, one has that $\Vert f \circ f^\dagger \Vert \leqslant \Vert f \Vert\, \Vert f^\dagger \Vert = \Vert f \Vert^2$ so that \eqref{eq:oa_tas_Csalg} is satisfied. Therefore, $\Hilbft{B}(\Hilbft{H})$ is a \Csalg.
\end{Ex}

To make contact with usual notations of physics, $\Hilbft{B}(\Hilbft{H})$ denotes the quantum operators, but physical observables correspond to self-adjoint operators, \ie satisfying \eqref{eq:oa_tas_Csa_sa}. The elements of the \Hsp, the states, are more often denoted with the \enquote{bra-ket} notation, \ie $\ket{\psi} \in \Hilbft{H}$. The bra-ket notation is actually here to render the inner product structure of $\Hilbft{H}$ since it is more often denoted $\langle \psi_1, \psi_2 \rangle = \braket{\psi_1}{\psi_2}$. Moreover, operators are often written in terms of hatted capital letters so that $f$ should be $\hat{A}$. 

The inner product and the norm are expressed as integrals like in \eqref{eq:oa_tas_qpl_ip} and \eqref{eq:oa_tas_qpl_n} in the context of wave functions, and they are interpreted as probability amplitudes. In this sense, the operator norm $\Vert \hat{A} \Vert$, defined in \eqref{eq:oa_tas_boBa_n}, can be interpreted as the maximum probability of $\hat{A}\ket{\psi}$ for a normalised state $\ket{\psi}$, or equivalently the amount of definitely lost information when applying $\hat{A}$ to the system.

\paragraph{}
Another important example, more motivated in the context of \nCG, is of order. It corresponds to the algebra of complex-valued continuous functions, which vanish at infinity.

\begin{Ex}{Functions over a locally compact space}
	{oa_tas_sfcs}
	Let $\manft{M}$ be a locally compact topological \namefont{Hausdorff} space. Let $\func_0(\manft{M})$ be the complex-valued continuous functions over $\manft{M}$ that vanish at infinity. One can equip the functions with the sup-norm $\Vert f \Vert = \underset{x \in \manft{M}}{\sup} \vert f(x)\vert$, which is well defined thanks to the continuity of $f$ and the vanishing of $f$ at infinity together with the local compactness of $\manft{M}$. One can also define an involution using the complex conjugation through $f^\dagger(x) = \overline{f(x)}$.
	
	The relation \eqref{eq:oa_tas_Ba} can be proved using that $\vert f(x) g(x) \vert = \vert f(x) \vert \, \vert g(x) \vert$, for any $f, g \in \func_0(\manft{M})$. And the presence of the complex modulus $\vert f(x) \vert$ implies that the norm does not distinguish $f^\dagger$ from $f$, \ie \eqref{eq:oa_tas_sBa} and \eqref{eq:oa_tas_Csalg} are satisfied. This means that $\func_0(\manft{M})$ is a \Csalg. 
\end{Ex} 

Note that the \nameref{thm:oa_rt_cGN} can be thought as the converse statement of \exref{oa_tas_sfcs}. Explicitly, if $\opaft{A}$ is a commutative \Csalg, then the \nameref{thm:oa_rt_cGN} states that there exists a topological \namefont{Hausdorff} space such that $\opaft{A} \simeq \func_0(\manft{M})$.

\section{Representation theory}
\label{sec:oa_rt}
\paragraph{}
We here introduce tools to understand and manipulate \opalg{s}. First, the case of commutative \Csalg is quite well established, especially thanks to the \nameref{thm:oa_rt_cGN}. In the noncommutative case, one can either go to \Hsp{s}, by representing the \Csalg, or try to tackle the \Csalg by its own. The \Hsp approach is always possible thanks to the \nameref{thm:oa_rt_sGN} and somewhat well paved via the \namefont{Gel'fand-Na\u{i}mark-Segal} construction. Note that part of the considered objects throughout the manuscript may have close links with the modular theory or \namefont{Tomita-Takesaki} theory, but the latter is not presented here as it goes far beyond the scope of this thesis.

\subsection{Commutative \tops{$C^*$}{C*}-algebras and characters}
\label{subsec:oa_rt_cCsalg}
\paragraph{}
The goal of this subsection is mainly to introduce the commutative \nameref{thm:oa_rt_cGN} which is at the very basis of \nCG.

\paragraph{}
The first notion we need is the space of characters of a \Csalg $\opaft{A}$, denoted $\Spch{\opaft{A}}$. It consists of continuous functions $\varphi : \opaft{A} \to \Cpx$, such that  $\varphi \neq 0$, and
\begin{align}
	\varphi(fg) = \varphi(f) \, \varphi(g)
	\label{eq:oa_rt_char}
\end{align}
for any $f, g \in \opaft{A}$. $\varphi$ is called a character and $\Spch{\opaft{A}}$ is sometimes named the structure space of $\opaft{A}$. If $\opaft{A}$ is unital, then \eqref{eq:oa_rt_char} and $\varphi \neq 0$ imposes that $\varphi(1) = 1$.

One can relate the algebra to the functions on the characters via the \namefont{Gel'fand} transform.
\begin{Def}{\namefont{Gel'fand} transform}
	{oa_rt_Gt}
	For $\opaft{A}$ a \Csalg, we define the \namefont{Gel'fand} transform $\Gt{f}$ of an element $f \in \opaft{A}$ as $\Gt{f}(\varphi) = \varphi(f)$.
\end{Def}
The latter transform is an algebra homomorphism quite straightforwardly since $\Gt{fg}(\varphi) = \varphi(fg) = \varphi(f) \varphi(g) = \Gt{f}(\varphi) \Gt{g}(\varphi)$.

\begin{Thm}{}
	{oa_rt_char}
	Let $\opaft{A}$ be a commutative \Csalg, then the following hold true.
	\begin{nclist}
		\item $\Spch{\opaft{A}}$ is a locally compact \namefont{Hausdorff} space.
		\item If $\opaft{A}$ is unital, then $\Spch{\opaft{A}}$ is compact.
		\item For any $f \in \opaft{A}$, the \namefont{Gel'fand} transform $\Gt{f}$ is continuous and vanish at infinity, \ie $\Gt{f} \in \func_0(\Spch{\opaft{A}})$.
		\item For any $f \in \opaft{A}$, $\Vert \Gt{f} \Vert_{\Spch{\opaft{A}}} \leqslant \Vert f \Vert$, so the \namefont{Gel'fand} transform is continuous.
	\end{nclist}
\end{Thm}

This theorem is quite important to understand the link between all the notions used in the \nameref{thm:oa_rt_cGN}. First, the space of characters is locally compact and \namefont{Hausdorff} so that it makes a suitable choice as a space for which $\opaft{A}$ is the space of functions. This choice is made clear thanks to the third point. Then, unitality is linked to compactness. Finally, the continuity of the transform paves the way for it to be an isometry.

\begin{Thm}{\namefont{Gel'fand-Na\u{i}mark} theorem}
	{oa_rt_cGN}
	Let $\opaft{A}$ be a commutative \Csalg. The \namefont{Gel'fand} transform $\opaft{A} \to \func_0(\Spch{\opaft{A}})$ is an isometric $*$-isomorphism. In particular, $\Vert \Gt{f} \Vert_{\Spch{\opaft{A}}} = \Vert f \Vert$ and $\varphi(f^\dagger) = \overline{\varphi(f)}$, for any $f \in \opaft{A}$ and $\varphi \in \Spch{\opaft{A}}$.
	
	One further has that $\Spch{\opaft{A}}$ is compact if and only if $\opaft{A}$ is unital. 
\end{Thm}

We refer to \cite{Landsman_1998} for a complete proofs of these two theorems.

\paragraph{}
The \nameref{thm:oa_rt_cGN} states two important things.

First and foremost, any commutative \Csalg actually corresponds to functions over some topological space and allows alone to characterise this topology. Therefore, the knowledge of the set of points is not relevant since all the information is stored in the space of functions. This precise reason is at the foundation of \nCG, which characterise a geometry via the space of functions.

Consider $\opaft{A}_{\kbar}$ to be a noncommutative \Csalg, where $\kbar$ is a real parameter, such that $\opaft{A}_0$ (\ie $\kbar \to 0$) is commutative. This is the usual framework of \dq as detailed in \secref{gnc_dq}. Then, the \nCG defined by $\opaft{A}_{\kbar}$ always have a \enquote{commutative limit} which links it to some classical space (corresponding here to $\Spch{\opaft{A}_0}$).

Note that the points of the space underlying a commutative \Csalg are fully determined in the theorem and correspond to the characters. Therefore, the theorem implies that if $\manft{M}$ is a (locally compact \namefont{Hausdorff}) topological space, then for any point $x \in \manft{M}$, there exist one and only character, that we note $\varphi_x$, such that $\varphi_x(f) = f(x)$ for all $f \in \func_0(\manft{M})$. Furthermore, no other character exists. In other words, there is no $\varphi$ that is not a $\varphi_x$. This enlightens the correspondence \eqref{eq:gnc_GN_corr}.

\subsection{States and representations}
\label{subsec:oa_rt_sr}
\paragraph{}
In a general context, a \Csalg can be associated with a \Hsp in several ways. The first one is that we can represent a \Csalg on a \Hsp, just as it is done for \qM with \eqref{eq:gnc_qm_rep}. The \Hsp is seen as a convenient substrate on which the \opalg is modelled and renders its properties. Given a specific state, that can be interpreted as a vacuum state, there is even an explicit construction of a representation, called the \namefont{Gel'fand-Na\u{i}mark-Segal} construction, or GNS construction for short. This highlight the importance of the notion of states in the study of \Csalg{s}.

Furthermore, we may be interested in how \enquote{accurate} this representation is. The faithfulness of a representation imposes that an element of the algebra $f \in \opaft{A}$ is associated to only one operator on the \Hsp $\hat{f} \in \Hilbft{B}(\Hilbft{H})$. In the context of \Csalg the \nameref{thm:oa_rt_sGN} precisely states that such a faithful representation always exists. One can also read in the proof the kind of states required for the representation to be faithful. Finally, the representation from the \namefont{Gel'fand-Na\u{i}mark-Segal} construction can also have the property to be irreducible if the considered state is pure. The pure states correspond to points, in the commutative theory.

We here construct objects in the case of a unital \Csalg for simplicity, but all this construction adapts to the non-unital case. For the non-unital case, one can always consider an approximated unit.

\paragraph{}
Let us begin by defining a representation.
\begin{Def}{Representation of $C^*$-algebras}
	{oa_rt_repCa}
	Given a \Csalg $\opaft{A}$ and a \Hsp $\Hilbft{H}$, a representation $\pi : \opaft{A} \to \Hilbft{B}(\Hilbft{H})$ is a linear map satisfying
	\begin{align}
		\pi(f g)
		&= \pi(f) \circ \pi(g), &
		\pi(f^\dagger)
		&= \pi(f)^\dagger
		\label{eq:oa_rt_repCa}
	\end{align}
	for any $f, g \in \opaft{A}$. In \eqref{eq:oa_rt_repCa}, $\circ$ denotes the composition of operators and $\pi(f)^\dagger$ corresponds to the adjoint operator of $\pi(f)$ as defined in \exref{oa_tas_boHsp}.
	
	The representation is said faithful is $\pi$ is injective. It is said irreducible if a (closed) subspace of $\Hilbft{H}$ which is stable under $\pi(\opaft{A})$ is either $\Hilbft{H}$ or $\{0\}$.
\end{Def}	

\paragraph{}
As expressed above, the notion of state is central in the representation theory of \opalg{s}. We define here the notion of state in the context of \Csalg.

An element $f$ of a \Csalg $\opaft{A}$ is said to be positive if it exists $g \in \opaft{A}$ such that $f = g^\dagger g$. We denote it by $f \geqslant 0$. Given a linear form $\psi : \opaft{A} \to \Cpx$, it is said positive if $\psi(f) \geqslant 0$, for any $f \geqslant 0$, or equivalently, $\psi(g^\dagger g) \geqslant 0$, for any $g \in \opaft{A}$. Again we note $\psi \geqslant 0$.

Positive linear forms satisfy the following properties
\begin{nclist}
	\item $\psi$ is continuous with $\Vert \psi \Vert = \psi(1)$, where the norm is defined similarly as in \eqref{eq:oa_tas_boBa_n}.
	\item Conversely, any continuous linear form $\psi : \opaft{A} \to \Cpx$ satisfying $\Vert \psi \Vert = \psi(1) = 1$ is positive.
	\item Given two positive linear forms $\psi_1$ and $\psi_2$ such that $\Vert \psi_1 \Vert = \Vert \psi_2 \Vert$ and $\psi_1 - \psi_2 \geqslant 0$, then $\psi_1 = \psi_2$. 
\end{nclist}

\begin{proof}
	\begin{nclist}
		\item For any $f \geqslant 0$, one has $0 \leqslant f \leqslant \Vert f \Vert 1$, in the sense $f - \Vert f \Vert 1 \geqslant 0$. Thus, applying $\psi$ gives $0 \leqslant \psi(f) \leqslant \Vert f \Vert \psi(1)$. Therefore, $\Vert \psi \Vert \leqslant \psi(1)$ and the supremum is reach for $f = 1$ so that equality actually stands.
		\item Let $f \geqslant 0$. Let us first prove that $\psi(f) \in \Real$. We write $\psi(f) = \alpha + \iCpx \beta$ with $\alpha, \beta \in \Real$. Let $g = f - \alpha 1$, then $\psi(g) = \iCpx \beta$, using that $\psi(1) = 1$. We compute for any $t \in \Real$,
		\begin{align*}
			\vert \psi(g + it 1) \vert^2
			&= \beta^2 + 2t\beta + t^2 \\
			&\leqslant \Vert g + it 1 \Vert ^2
			= \Vert (g + it 1)^\dagger (g+ it 1) \Vert
			= \Vert g^2 + t^2 \Vert
			\leqslant \Vert g \Vert^2 + t^2
		\end{align*}
		where we used that $f = f^\dagger$ so that $g = g^\dagger$ and $\psi(h) \leqslant \Vert h \Vert \psi(1) = \Vert h \Vert$, for any $h \in \opaft{A}$, from the previous point. This implies that $\beta^2 + \beta t \leqslant \Vert g \Vert^2$ for any $t \in \Real$, which can only be fulfilled for $\beta \leqslant 0$. Repeating the argument for $-g$, one gets $\beta = 0$ and so $\psi(f) \in \Real$.
		
		Now, choose $\epsilon > 0$ small enough so that $\Vert 1 - \epsilon f \Vert \leqslant 1$. Then, one computes
		\begin{align*}
			1
			&\geqslant \Vert 1 - \epsilon f \Vert
			= \frac{\Vert \psi \Vert}{\psi(1)} \Vert 1 - \epsilon f \Vert
			\geqslant \frac{\vert \psi(1) - \epsilon \psi(f) \vert}{\psi(1)}.
		\end{align*}
		Thus, $\vert \psi(1) - \epsilon \psi(f) \vert \leqslant \psi(1)$ which can only be true if $\psi(f) \geqslant 0$.
		\item $\Vert \psi_1 - \psi_2 \Vert = (\psi_1 - \psi_2)(1) = \Vert \psi_1 \Vert - \Vert \psi_2 \Vert = 0$.
		\qedhere
	\end{nclist}
\end{proof}

\begin{Def}{States of $C^*$-algebras}
	{oa_rt_st}
	A positive linear form $\psi$ of norm one, that is $\psi(1) = 1$, on a \Csalg $\opaft{A}$ is called a state of $\opaft{A}$. We denote that space of states of $\opaft{A}$ as $\Spst{\opaft{A}}$.
	
	A state $\psi \in \Spst{\opaft{A}}$ is said to be faithful if, for any $f \geqslant 0$, $\psi(f) = 0$ implies $f = 0$.
	
	$\Spst{\opaft{A}}$ is a convex set, \ie for any $\psi_1, \psi_2 \in \Spst{\opaft{A}}$ and $t \in [0,1]$, $(1 - t) \psi_1 + t \psi_2 \in \Spst{\opaft{A}}$. The extreme point of this convex sets are called the pure states. The space of pure states is denoted $\Sppst{\opaft{A}}$.
\end{Def}

\begin{proof}
	Considering that $\psi_1$ and $\psi_2$ are positive linear functional, so is $(1-t) \psi_1 + t \psi_2$. One need to check that it has norm $1$, that is
	\begin{align*}
		\Vert (1-t) \psi_1 + t \psi_2 \Vert
		&= (1-t) \psi_1(1) + t \psi_2(1)
		= 1 - t + t
		= 1.
		\qedhere
	\end{align*}
\end{proof}

\paragraph{}
The previous definition of state may be far from the physical definition of a state. However, one can link the two through the following example.

\begin{Ex}{States of operators on a \namefont{Hilbert} space}
	{oa_rt_stHsp}
	Let $\opaft{A}$ be a \Csalg and $\Hilbft{H}$ a \Hsp on which $\opaft{A}$ is represented with inner product $\langle \cdot, \cdot \rangle$. Then, any normalised element $\ket{\psi} \in \Hilbft{H}$, $\Vert \ket{\psi} \Vert = 1$ give rise to a state $\psi \in \Spst{\opaft{A}}$ defined by\footnote{
		We take, in \eqref{eq:oa_rt_stHsp},	the physical notation for simplicity (as it was made for). Indeed, the associated mathematical equation would be $\psi(f) = \big\langle \ket{\psi}, \pi(f) \ket{\psi} \big\rangle$, for $\ket{\psi} \in \Hilbft{H}$, where $\pi : \opaft{A} \to \Hilbft{B}(\Hilbft{H})$ is the representation. $\pi$ was also implied in \eqref{eq:oa_rt_stHsp}.
	}
	\begin{align}
		\psi(f)
		&= \braket{\psi}{f \psi} 
		\label{eq:oa_rt_stHsp}
	\end{align}
	for any $f \in \opaft{A}$. We thereby justify our notation for $\psi$ in both cases.
	
	If $f = g^\dagger g \geqslant 0$, then $\psi(f) = \braket{g \psi}{g \psi} = \Vert g \ket{\psi} \Vert^2 \geqslant 0$, so that the linear form $\psi$ defined in \eqref{eq:oa_rt_stHsp} is positive. Moreover, it is of norm $1$ since $\ket{\psi}$ is of norm $1$, \ie $\Vert \psi \Vert = \psi(1) = \braket{\psi}{\psi} = \Vert \ket{\psi} \Vert ^2 = 1$.
\end{Ex}

However, the link goes even further.

\begin{Rmk}{States and probability}
	{oa_rt_sp}
	One can show that in the commutative case, \ie $\opaft{A} = \func_0(\manft{M})$ for $\manft{M}$ some topological space, the space of states $\Spst{\opaft{A}}$ consist of all probability measures on $\manft{M}$ (see Theorem 2.8.2 of \cite{Landsman_1998}, which mainly comes from \namefont{Riesz} theorem of measure theory). This means that there is a direct link between the states we defined in \defref{oa_rt_st} and the probability amplitude of a quantum particle on the space $\manft{M}$. This also means that the normalisation condition $\Vert \psi \Vert = 1$ can directly be interpreted as the conservation of probability, or equivalently stated, that the probability to find the particle associated to $\psi$ anywhere on $\manft{M}$ is $1$.
\end{Rmk}

Finally, one can show that, in the commutative theory, points are given by the pure states. In other words, if $\opaft{A} = \func_0(\manft{M})$ for $\manft{M}$ some topological space, then $\Sppst{\opaft{A}}$ is homeomorphic to $\manft{M}$. This is done by showing that the space of pure states $\Sppst{\opaft{A}}$ is in one-to-one correspondence with the space of characters $\Spch{\opaft{A}}$, when the algebra is commutative.

\paragraph{}
One of the main interest of states is that it allows to construct a representation through the \namefont{Gel'fand-Na\u{i}mark-Segal} construction.

\begin{Emph}{\namefont{Gel'fand-Na\u{i}mark-Segal} construction}
	Let $\opaft{A}$ be a \Csalg and $\psi \in \Spst{\opaft{A}}$. First, observe that $\langle f, g \rangle_{\psi} = \psi(f^\dagger g)$ is a positive sesquilinear form on $\opaft{A}$. Indeed, it is sesquilinear thanks to linearity of $\psi$ and positive thanks to positivity of $\psi$. It is further non-degenerate if $\psi$ is faithful (by definition). Therefore, in a general case, one needs to remove all functions for which this would-be inner product vanishes in order to have a non-degenerate form. Explicitly, one considers
	\begin{align*}
		\idft{J}_\psi
		&= \big\{ f \in \opaft{A}, \psi(f^\dagger f) = 0 \big\}
		= \big\{ f \in \opaft{A}, \psi(g^\dagger f) = 0 \text{ for any } g \in \opaft{A} \big\}
	\end{align*}
	which corresponds to the set of positive elements on which $\psi$ vanishes. The second equality is provided by the \namefont{Cauchy-Schwarz} inequality $\vert \psi(g^\dagger f) \vert^2 \leqslant \psi(g^\dagger g) \, \psi(f^\dagger f)$. If $\psi$ is faithful $\idft{J}_\psi = \{0\}$. 
	
	$\idft{J}_\psi$ is a closed (left) ideal in $\opaft{A}$ (see \defref{ha_as_id}), indeed for any $g, h \in \opaft{A}$ and $f \in \idft{J}_\psi$, $\psi(h^\dagger (g f)) = \psi ((g^\dagger h)^\dagger f) = 0$, \ie $gf \in \idft{J}_\psi$. Thus, one can consider the quotient $\opaft{A} / \idft{J}_\psi$ (as vector spaces), such that all positive elements for which $\psi$ vanishes are now grouped in a single equivalence class, which is $[0]$. Therefore, $\langle \cdot, \cdot \rangle_\psi$ is non-degenerate on $\opaft{A} / \idft{J}_\psi$. We finally consider the completion\footnote{
		See discussion of \subsecref{oa_tas_Hsp} about complete spaces. This completion may be physically irrelevant in this construction. 	
	} of $\opaft{A} / \idft{J}_\psi$, which is by definition a \Hsp, denoted $\Hilbft{H}_\psi$. If $\psi$ is faithful, we can simply take $\Hilbft{H}_\psi = \opaft{A}$.
	
	In order to have a representation, we need a map $\pi_\psi : \opaft{A} \to \Hilbft{B}(\Hilbft{H}_\psi)$. One consider here simply the left multiplication map, \ie $\pi_\psi(f) : [g] \mapsto [fg]$, for any $[g] \in \Hilbft{H}_\psi$. First, for any $f \in \opaft{A}$, one has $f^\dagger f \leqslant \Vert f^\dagger f \Vert 1$. As conjugation preserves positivity, one obtains that $g^\dagger f^\dagger fg \leqslant \Vert f^\dagger f \Vert g^\dagger g$. Applying $\psi$, one obtains $\psi(g^\dagger f^\dagger f g) \leqslant \Vert f^\dagger f \Vert \psi(g^\dagger g)$. Therefore, 
	\begin{align*}
		\Vert \pi_\psi(f) \Vert
		&= \underset{[g] \in \Hilbft{H}_\psi}{\sup} \frac{\Vert \pi_\psi(f)([g]) \Vert_\psi}{\Vert [g] \Vert_\psi}
		= \underset{[g] \in \Hilbft{H}_\psi}{\sup} \sqrt{\frac{\psi(g^\dagger f^\dagger f g)}{\psi(g^\dagger g)}}
		\leqslant \underset{[g] \in \Hilbft{H}_\psi}{\sup} \sqrt{\frac{\Vert f^\dagger f \Vert \psi(g^\dagger g)}{\psi(g^\dagger g)}}
		= \Vert f \Vert
	\end{align*}
	so that $\pi_\psi(f)$ is indeed a bounded operator with $\Vert \pi_\psi(f) \Vert \leqslant \Vert f \Vert$. Note that, in the previous computation, we used the associated norm $\Vert f \Vert_\psi = \sqrt{\langle f, f \rangle_\psi} = \sqrt{\psi(f^\dagger f)}$.
	
	Moreover, one has for any $f, g, h \in \opaft{A}$,
	\begin{align*}
		\pi_\psi(fg)([h])
		&= [fgh]
		= \pi_\psi(f) ([gh])
		= \big(\pi_\psi(f) \circ \pi_\psi(g)\big) ([h]), \\
		\big\langle [g], \pi_\psi(f)([h]) \big\rangle_\psi
		&= \psi(g^\dagger f h )
		= \big\langle \pi_\psi(f^\dagger)([g]), [h] \big\rangle_\psi
	\end{align*}
	so that \eqref{eq:oa_rt_repCa} is satisfied.
	
	Finally, $(\Hilbft{H}_\psi, \pi_\psi)$ is a representation of $\opaft{A}$.
\end{Emph}

An important property of this construction is that the representation $(\Hilbft{H}_\psi, \pi_\psi)$ is irreducible if and only if the state $\psi$ is pure. More generally, any irreducible representation of a \Csalg is the \namefont{Gel'fand-Na\u{i}mark-Segal} representation of a pure state.

\paragraph{}
Using this construction, one obtain the
\begin{Thm}{second \namefont{Gel'fand-Na\u{i}mark} theorem}
	{oa_rt_sGN}
	Any \Csalg has an isometric representation as a closed sub-algebra of the algebra $\Hilbft{B}(\Hilbft{H})$ of bounded operators on some \Hsp $\Hilbft{H}$.
\end{Thm}

\begin{proof}
	We follow here the sketch of proof given in \cite{Gracia-Bondia_2001}. First, one uses the \namefont{Hahn-Banach} theorem to show that for any non-zero positive elements $g^\dagger g \in \opaft{A}$, there exist a state $\psi \in \Spst{\opaft{A}}$ such that  $\psi(g^\dagger g) = \Vert g^\dagger g \Vert = \Vert g \Vert^2$. Then, we construct, through \namefont{Gel'fand-Na\u{i}mark-Segal}, the representation $\Hilbft{H}_\psi$  with inner product $\langle \cdot, \cdot \rangle_\psi$. We denote by $\Vert \cdot \Vert_\psi$ the associated norm, which relates to the norm of $\opaft{A}$ via $\Vert \pi_\psi(g) ([1]) \Vert_\psi = \Vert g \Vert$.
	
	The previous construction allows one to consider a family of state $\Spst{0}$, such that $\psi(g^\dagger g) = 0$ implies $g = 0$, for any $\psi \in \Spst{0}$ and $g \in \opaft{A}$ (if necessary we can take all states of $\opaft{A}$, \ie $\Spst{0} = \Spst{\opaft{A}}$).
	
	Finally, one consider the representation formed by the direct sum of the \namefont{Gel'fand-Na\u{i}mark-Segal} representations of all states in $\Spst{0}$, that is $\pi = \bigoplus \limits_{\psi \in \Spst{0}} \pi_\psi$ and $\Hilbft{H} =  \bigoplus \limits_{\psi \in \Spst{0}} \Hilbft{H}_\psi$. In doing so, we have for any $f \in \opaft{A}$, $\Vert \pi(f) \Vert = \Vert f \Vert$, and so the representation is isometric.
\end{proof}

The \Hsp constructed in the proof of the \nameref{thm:oa_rt_sGN} is mainly of mathematical interest, since it might be far too big for physical purpose. Still, this theorem helps us understanding the nature of \Csalg{s} and the interplay it has with \namefont{von Neumann} algebras.

The \nameref{thm:oa_rt_sGN} underlines the interest of the \Csalg{s} for physical purposes. It can render all the needed properties of quantum operators without necessarily specifying a \Hsp. Still, if one wants to work in the \Hsp formalism, there always exists nice representations of a \Csalg on some \Hsp and such representations can be constructed explicitly.

%%%%%%%%%%%%%%%%%%%%%%%%
%     BACKMATTER       %
%%%%%%%%%%%%%%%%%%%%%%%%
\backmatter

%%%%% BIBLIOGRAPHY %%%%%

%%%%%%%% INDICES %%%%%%%%
\printindex

%%%%%%%% FIGURES %%%%%%%%%
%\listoffigures

%%%%%%%%% TABLES %%%%%%%%%
%\listoftables

\end{document}